\documentclass[twoside,12pt]{article}
\usepackage{epsfig}
\usepackage{cite}
\usepackage[ulem=normalem]{changes}
\pdfoutput=1
\def\2ih{$2\Im/\hbar^2$}
\def\h2i{$\hbar^2/2\Im$}
\def\beq{\begin{equation}}
\def\eeq{\end{equation}}

\def\i13{i$_{13/2}$}

\def\Journal#1#2#3#4{{#1} {#2} (#4) #3 }
\def\PST{{\em Phys. Scr.} T}
\def\EPA{{\em Eur. Phys. J.} A}
\def\NIM{{\em Nucl. Instr. Meth.} A}

\def\NC{{\em Nuovo Cimento}}

\def\NPA{{\em Nucl. Phys.} A}

\def\PRO{{\em Prog. Theor. Phys.}}

\def\NP{{\em Nucl. Phys.}}

\def\PLB{{\em Phys. Lett.} B}

\def\PRL{\em Phys. Rev. Lett.}
\def\PREV{\em Phys. Rev.}
\def\PREP{\em Phys. Rep.}

\def\PPNP{{\em Prog.  Part. Nucl. Phys.}}

\def\PRC{{\em Phys. Rev.} C}
\def\PRB{{\em Phys. Rev.} B}
\def\PR{{\em Phys. Rev.}  }

\def\ANNP{\em Ann. Phys. (N.Y.)}
\def\RMP{{\em Rev. Mod. Phys.}}

\def\INT{{\em Int. J. Mod. Phys.} E}
\def\ANP{{\em Adv. Nucl. Phys.}} 
\def\NAT{{\em Nature}}
\def\ARN{{\em Annu. Rev. Nucl. Part. Sci.}}
\def\PRS{{\em Proc. Roy. Soc. }A}
\def\JPG{{\em J. Phys. }G}
\def\PAT{{\em  Phys. At. Nucl. }}
\def\ANUDT{{\em  Atomic Dat. and Nucl. Dat. Tables}}

\newcommand{\be}{\begin{equation}}
\newcommand{\ee}{\end{equation}}
\newcommand{\bea}{\begin{eqnarray}}
\newcommand{\eea}{\end{eqnarray}}

\begin{document}

\title{ \vspace{1cm} Overview of  Neutron-Proton Pairing}
\author{S. Frauendorf$^1$, and A. O. Macchiavelli$^2$ 
\\
$^1$Department of Physics, University Notre Dame\\
$^2$Nuclear Science Division \\ Lawrence Berkeley National Laboratory, Berkeley CA 94720}
\maketitle
\begin{abstract} 
The role of neutron-proton pairing correlations on the structure of nuclei along the $N=Z$ line is reviewed.
Particular emphasis is placed on the competition between isovector ($T=1$) and isoscalar $(T=0$) pair fields.   The expected 
properties of these systems, in terms of pairing collective motion, are assessed by different theoretical frameworks including 
schematic models,  realistic Shell Model and mean field approaches.  The results are contrasted with experimental data with the goal
of establishing  clear signals for the existence of neutron-proton ($np$)  condensates.
We will show that there is clear evidence for an  isovector $np$ condensate as expected from isospin invariance.  However, 
and contrary to early expectations,  a condensate of deuteron-like pairs appears quite elusive and pairing collectivity in the $T=0$ channel may
only show in the form of a phonon.  Arguments are presented for the use of direct reactions, adding or removing an $np$ pair, as the most
promising tool to provide a definite answer to this intriguing question. 
\end{abstract}
%\eject
%\tableofcontents
\section{Introduction}

%Start AOM 11/4/2013
More than 50 years ago
 Bohr, Mottelson and Pines \cite{bmp}, suggested a pairing mechanism in the atomic nucleus 
analogous
 to that observed in superconductors. Since then, a wealth of experimental data has
 been accumulated, supporting the important role played by $nn$ and $pp$
 ``Cooper pairs'' in modifying many nuclear properties such as deformation, moments of inertia, 
 alignments, etc.   \cite{BrogliaBrink,50yearsBCS,DeanMorten}.
Due to advances in experimental techniques and sensitive detector systems, 
together with the new possibilities  available with radioactive beams, we are seeing a renaissance of nuclear structure 
studies, in particular,  along the 
$N=Z$ line.   Of interest here is the role played by the
 isoscalar $(T=0)$ and isovector $(T=1)$ pairing correlations.

\begin{figure}[tb]
\begin{center}
%\begin{minipage}[t]{8 cm}
\epsfig{file=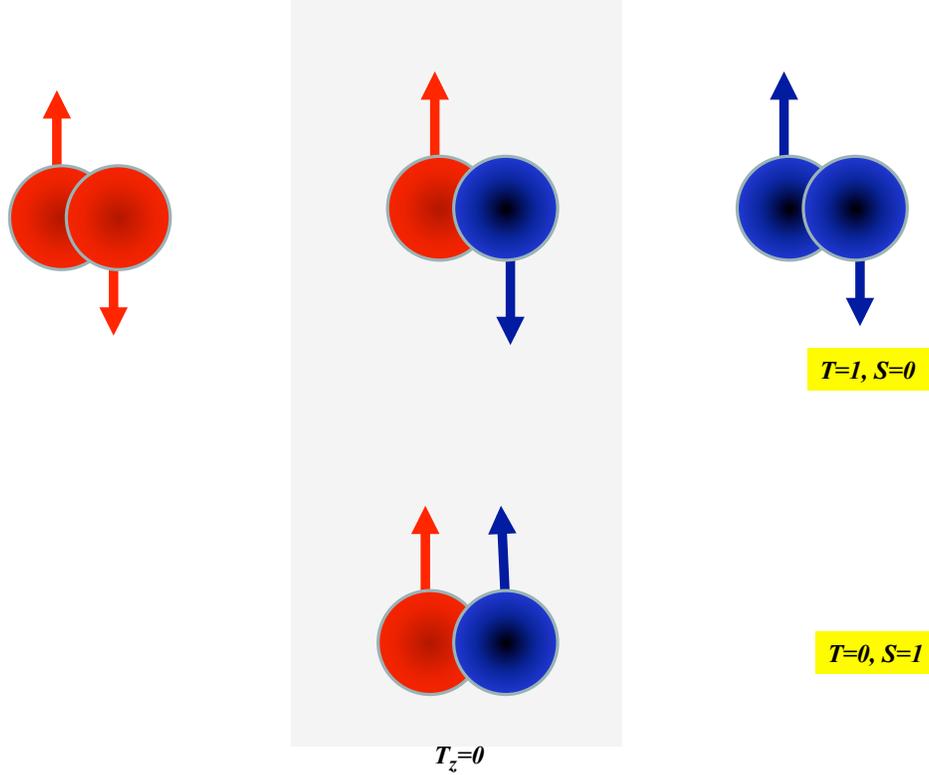,scale=0.6}
%\end{minipage}
%\begin{minipage}[t]{16.5 cm}
\caption{Schematic picture showing the possible pair arrangements in $N=Z$ nuclei.  \label{fig:IVIScartoon2}}
%\end{minipage}
\end{center}
\end{figure}

\begin{figure}[h]
\begin{center}
\begin{minipage}[t]{8 cm}
\epsfig{file=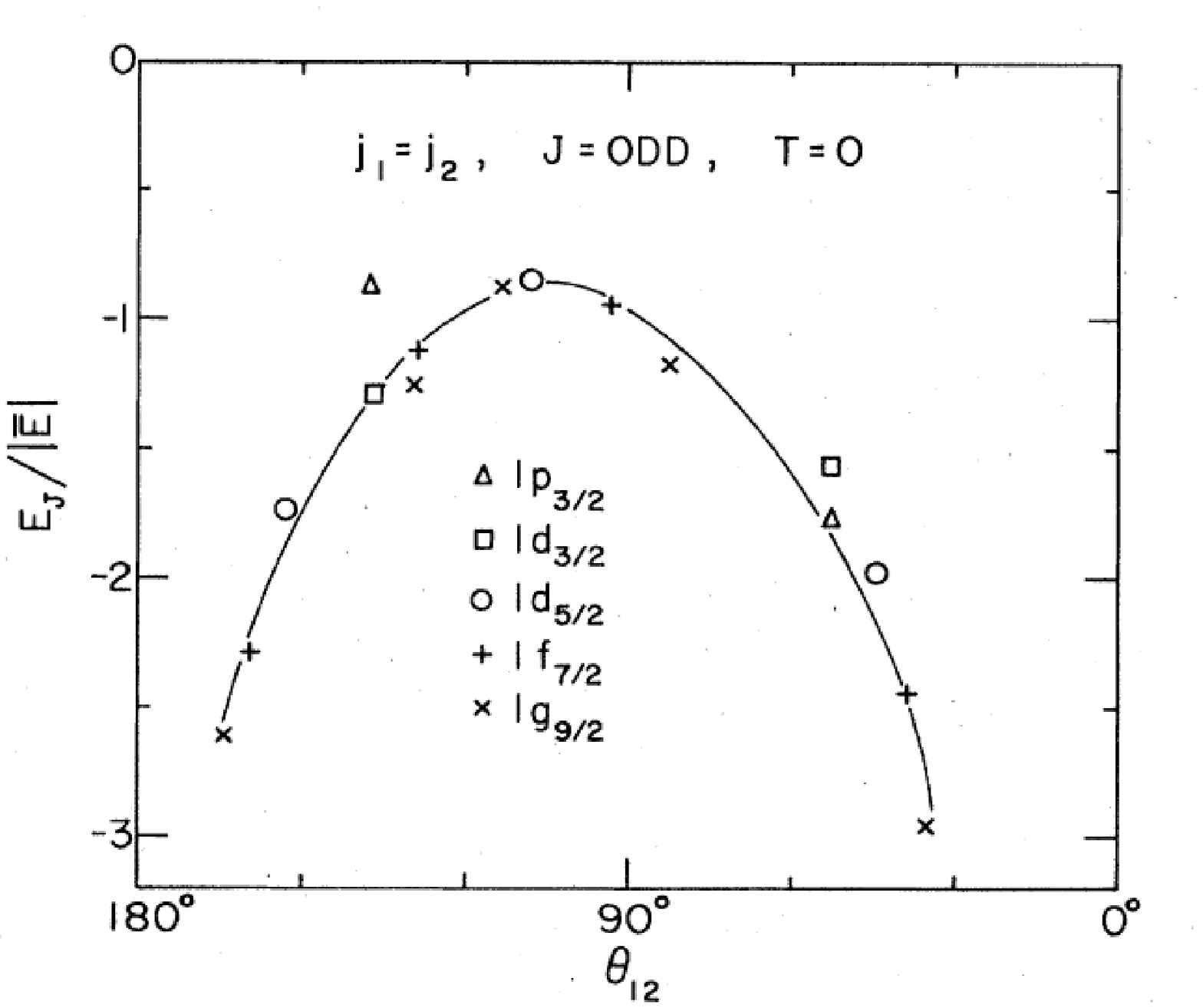,scale=0.4}
\end{minipage}
\begin{minipage}[t]{8 cm}
\epsfig{file=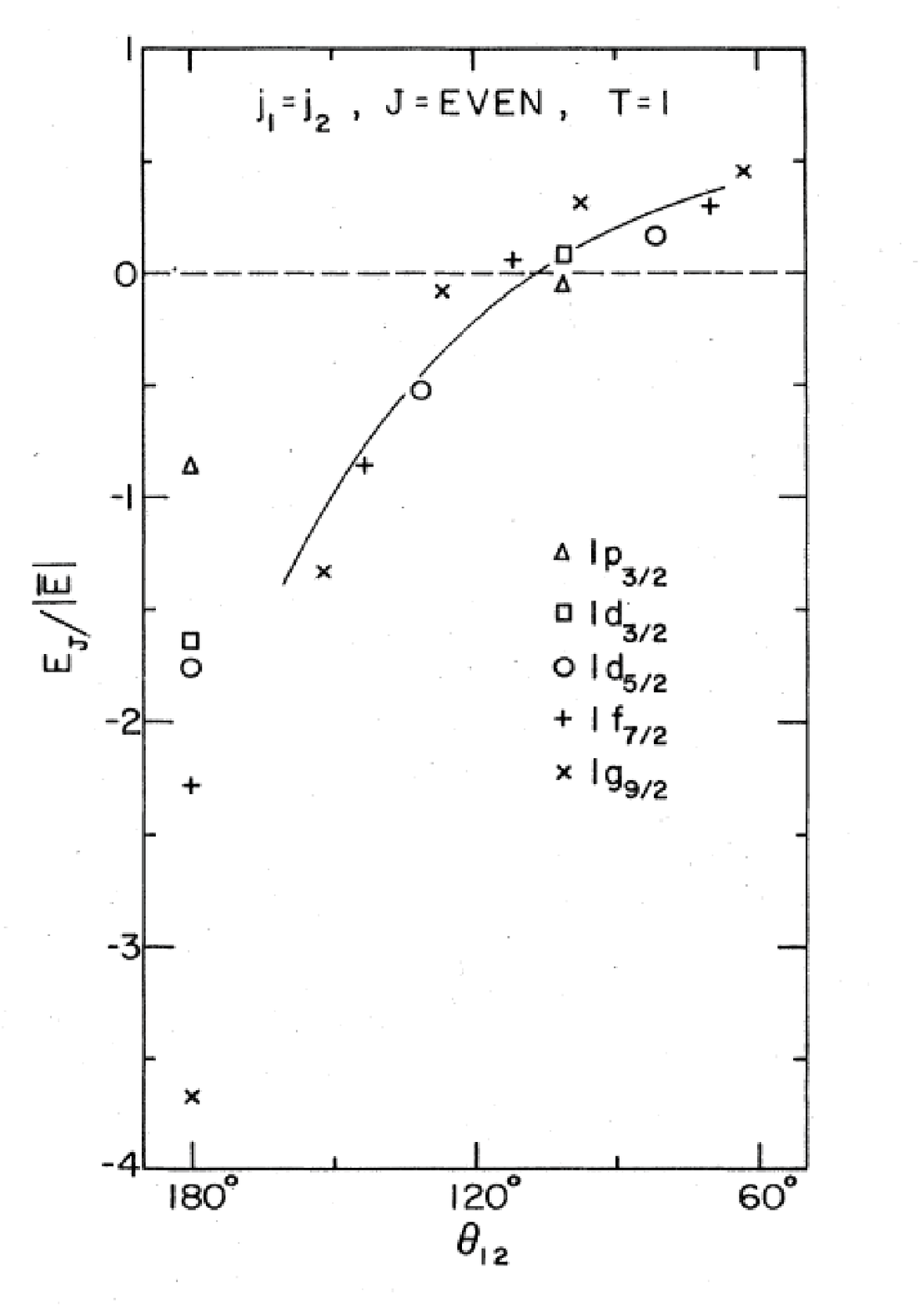,scale=0.5}
\end{minipage}
\caption{ The experimental interaction matrix  elements $E_J$ between two   nucleons in $j$-orbitals forming a $T=0$ pair (left panel)
and a $T=1$ pair (right panel). The angle between the angular momenta $\vec j$ of the two nucleons  is denoted by $\theta_{12}$. 
A scaling factor $E$ is applied such that different $j$-orbitals fall on the same curve. For each $j$-shell,  the first point to the left 
corresponds to $J=1$ and the last to the right to $J=2j$ in the case $T=0$, and the first point to the left corresponds to $J=0$ in the case $T=1$. 
   From \cite{Schiffer76 }.\label{fig:SchTru}}
\end{center}
\end{figure}

For almost all known nuclei, i.e.~those with $N>Z$, the pair
correlated state consists of neutron ($nn$) and/or proton ($pp$) pairs
coupled to angular momentum zero and isospin $T=1$.  For nuclei with
$N\approx Z$,  the protons and neutrons near the Fermi surface occupy identical orbitals, which
allows for a different type of pairs consisting of a neutron and a
proton ($np$).  The $np$ pairs can couple to angular momentum zero and
isospin $T=1$ (isovector), or, since they are no longer restricted by
the Pauli exclusion principle, they can couple to $T=0$ (isoscalar)
and $J=1$.   Fig. \ref{fig:IVIScartoon2} illustrates the two types of pairs for the LS coupling scheme,
which applies only for light nuclei. For medium-mass nuclei the spin-orbit potential induces 
the $jj$ coupling. Fig. \ref{fig:SchTru} shows the experimental effective  interaction between two nucleons
derived by {Molinari {\it et al.} \cite{Molinari75} and} Schiffer and True  \cite{Schiffer76}. As expected for a short range interaction, 
 the favored angular momenta are  $J=0$ for $T=1$ pairs and $J=1$ or $J=J_{max}$ for $T=0$ pairs.
   Charge independence of the nuclear force implies
that for $N=Z$ nuclei, $J=0,~T=1$ $np$ pairing should exist on an equal
footing with $J=0,~T=1$ $nn$ and $pp$ pairing.  There is convincing  evidence for this expectation that will be reviewed.
However, it is an open question whether strongly correlated $J=1,~T=0$ $np$ pairs 
also exist. Another interesting question relates to  the consequences of the strong attraction between the proton and neutron in a  $J=J_{max}, ~T=0$ pair.
Both questions are the focus of this review.

We believe it is important, right at the start, to clarify what we mean by "pairing" in the context of this work.
While it is obvious that a strong pairing force is present in the $T=0$ channel, the question is whether or not a
correlated state can be formed.   This correlated state is that in analogy to the pair phase of superconductors and superfluids.
Far from the thermodynamic limit, the definition of such state is to be discussed in detail.
 In fact, as simple models and general arguments indicate, the control
parameter that induces the transition between a normal system and a pair correlated  depends not only on the strength
of the force $G$ but also on the available degeneracy $\Omega$ and the single particle spacings $D$, usually in the form $G\Omega/D$.
In a mean field approach there is a pair field and a correlation energy that comes with it. 
In the context of exact solutions (either simple soluble models, or the Shell Model,) a careful definition is needed.  
Many authors have used the same terms as "correlation energy", "number of pairs", "order parameter", "paired phase" but meant something different. 
This has lead to  a certain degree of confusion, in particular with respect to the question: "Is there isoscalar $np$ pairing?" 
We adopt the following point of view:
The "paired" wave function must substantially deviate from the "unpaired" one by correlating nucleons into pairs.  The amount of
such correlations  can be quantified  by the expectation value the order parameter, which is the pair transfer matrix element,  or the correlation energy,
which is the difference between the expectation values of  the  pair interaction between the correlated state and some  uncorrelated reference state. 
 Such  a definition relates the nuclear pairing phenomena with superconductivity and superfluidity, which represent the limit of very large particle number.
However, one has always to keep in mind that the nucleus is a relatively small -mesoscopic- system. The neutron-proton pair phenomena which will be discussed
appear in nuclei with $A<100$, for which the number of valence particles/holes is less than 22. 
 For such small systems the concepts of "phase", "phase transition", "condensate", which have a clear meaning for the thermodynamic limit
of macroscopic systems, have to be used with some care.  In a small system as the nucleus,  the singularity of a phase transition is washed out into an
extended  cross-over region. Its onset is signaled by the occurrence of pair vibrations, which are the precursors of the strong pair correlations that characterize
the pair condensate of macroscopic superconductors / superfluids.   In Section \ref{sec:theory},
we will discuss these aspects in some detail, because we feel that the terminology of macroscopic systems is often used in a too loose  way.

The present article does not intend to provide a comprehensive review of the theoretical work. Rather, we shall only cover the theoretical 
background relevant for the discussion of predictions concerning the strength of the correlations and their experimental evidence. 
From the experimental point of view, the most important aspect is recognizing the possible signals for
$np$ pairing.  
We will review recent experimental information that sheds light on the question
of $np$ pairing and discuss further opportunities with the use of direct reactions.  
As we will show,  there is no clear evidence, so far, for $T=0$ collective pairing behavior in the form
of a condensate, as established for $T=1$ pairs.  If present,  
the $T=0$ "deuteron" condensate appears elusive, and theoretical studies point to collective $T=0$ 
correlation of a vibrational type. However,  more sensitive probes  may be required to
extract the pertinent  experimental signatures.  From our discussion above, transfer reactions of  $np$ pairs are perhaps the ultimate
"smoking-gun", as $(t,p)$ and $(p,t)$ reactions played a similar role to elucidate the role of $nn$ pairing
 correlations.
 
The article will proceed as follows:  in Section \ref{sec:theory} we review important theoretical aspects that are critical to the interpretation
of the experimental data. These include simple models, Shell Model calculations and mean field approaches.   Section \ref{sec:Wigner} is devoted to phenomenological 
analyses of binding energies in parallel with theoretical expectations.   The important effects of rotational motion  are discussed in detail in Section \ref{sec:rotation} 
with particular reference to the recently suggested "spin-aligned pairing phase" in $^{92}$Pd.
In Sections \ref{sec:lowoo}, \ref{sec:GT}, and  \ref{sec:PV} we show experimental data related to the structure of odd-odd $N=Z$ nuclei,  Gamow-Teller strengths and pairing vibrations respectively .
We end our review in Section  \ref{sec:transfer} by discussing the unique role that $np$ transfer reactions may play  in providing a definite answer to this question.

%End AOM 11/4/2013

\section{Expectations from theory}\label{sec:theory}

There has been extensive theoretical work on $np$ pairing in the framework of a variety of different approaches. 
In this section we review the major types of theories with the goal to provide the background for discussing 
the comparison with experiment in subsequent sections.   We discuss their 
predictions of the strength $np$ pair correlations and discuss conceptual issues  concerning the conservation of isospin.
We start with a discussion of qualitative features of the $np$ pairing in Section \ref{sec:schematic}. It is based on 
schematic models of one and two spherical $j$-shells, for which exact solutions are available.    Most of the  region $A<100$, where $N\approx Z$ nuclei can
be studied experimentally,  is accessible to modern large-scale Shell Model calculations in the spherical single particle basis, which will be discussed in Section \ref{sec:SM}.   
Shell Model studies in a deformed single particle basis will be  reviewed as well. Section \ref{sec:sigmf}  clarifies the meaning of the mean field solutions,
which spontaneously break the isospin  symmetry.  Their interpretation as intrinsic states of isorotational bands will be discussed. The mean field studies  of $np$ 
pairing will be reviewed in Section \ref{sec:MF}, which introduces the pertinent version of the  Hartree-Fock-Bogoliubov approximation in Section \ref{sec:HFBtheory}. 
  Section \ref{sec:SP} presents  the variational approaches based on symmetry projected  mean field type solutions. 

\subsection{Schematic models}\label{sec:schematic}
\subsubsection{Isovector pairing }\label{sec:2jIV}
To convey the essential aspects of the problem, we consider the schematic isovector pairing Hamiltonian,  
\be\label{eq:HIV}
 H=\sum_j\varepsilon_j\hat N(j)-G\sum_\mu P^\dagger_\mu P_\mu,
  \ee
which is defined by the $J=0$, $T=1$ pair operators 
\bea
&P^\dagger_1=\sum_{jm>0}(-)^{j-m}n^\dagger_{jm}n^\dagger_{j-m}&=\sum_{j}\sqrt{j+1/2}A^{11\dagger}_{00}(j,j),\nonumber\\
&P^\dagger_{-1}=\sum_{jm>0}(-)^{j-m}p^\dagger_{jm}p^\dagger_{j-m}&=\sum_{j}\sqrt{j+1/2}A^{1-1\dagger}_{00}(j,j),\nonumber\\
&P^\dagger_0=\frac{1}{\sqrt{2}}\sum_{jm>0}(-)^{j-m}\left(n^\dagger_{jm}p^\dagger_{j-m}+p^\dagger_{jm}n^\dagger_{j-m}\right)&
=\sum_{j}\sqrt{j+1/2}A^{10\dagger}_{00}(j,j) \label{eq:P},
\eea
and the number operators
\be\label{eq:NpNn}
\hat N_p(j)=\sum_mp^\dagger _{jm}p_{jm},~~\hat N_n(j)=\sum_mn^\dagger _{jm}n_{jm},~~\hat N(j)=\hat N_p(j)+\hat N_n(j),
\ee
where $p^\dagger_{jm}=c^\dagger_{jm,-1/2}$ creates a proton and $n^\dagger_{jm}=c^\dagger_{jm,1/2}$ a neutron. 
For later use we introduced the general pair operators
\be\label{eq:A}
A^{TM_T \dagger}_{JM_J}(j,j')=\frac{1}{\sqrt{1+\delta(j,j')}}\sum_{mm'\tau\tau'}\left<jmj'm'\vert JM_J\right>\left<\frac{1}{ 2}\tau \frac{1}{ 2} \tau'\vert TM_T\right>
c^\dagger_{jm,\tau}c^\dagger_{j'm',\tau'},
\ee
which generate a nucleon pair of isospin $T,T_z=M_T$ and angular momentum $J,J_z=M_J$.
The even particle number  $A=2M$, the isospin $T$, and its projection $T_z=(N-Z)/2$ are good quantum numbers.

We start with the discussion of pair correlations in two degenerated shells, which provides
qualitative insight in essential features of the $np$ pair correlations, because the solutions  can be
fond without approximation.

\subsubsection{Two-shell model}\label{sec:IV2shell}

Dussel {\it et al.} \cite{Dussel70} studied the case of two identical $j$-shells of degeneracy $\Omega=j+1/2$
separated by $\varepsilon_2-\varepsilon_1=D$, which are approximately  half filled ($M= 2\Omega,~2\Omega+2,2\Omega+4$,
where $M=A/2$ is the number of pairs).
The model  is generic for the realistic case of non-degenerate levels showing a phase transition between the 
paired and  unpaired regime. 
 The control parameter $y=G/G_c=2\Omega G/D<1$ or $>1$ decides, respectively, whether there is an isovector pair condensate or not. 
 As for any finite system, the transition is not sharp, there is a gradual cross-over between the two regimes.  
 We will use the term "condensate" for the finite pair field $\Delta$ of the mean field solution and
we will speak of a "paired phase" or "superfluid phase" if the mean field solution contains such a condensate.

We restate here that in nuclei there is no well defined phase transition that can be classified by its type of singularity (first order, second order)
as in  macroscopic systems. In a relatively small system as the nucleus there are two regimes  and an extended cross-over region between them.
When the particle number becomes large, the two regimes approach the two phases  of the  thermodynamic limit and the cross-over region
 shrinks to the singularity of the phase transition. With increasing particle number the mean field approximation becomes more and more accurate.
 Thus it seems  appropriate to use the terminology of macroscopic systems in a more general way to characterize the two regimes and draw an
 artificial border line in the transition region.  Moreover, the terminology is quite commonly used, often too careless in our our view.        
 
First consider the paired regime.  
The limit $D\rightarrow0$, i.e $y\rightarrow\infty$,  corresponds to one shell with
degeneracy $2\Omega$. The states are representations of the group SO(5) which are classified with respect to the particle number $A=2M$, isospin $T$,
its projection $T_z=M_T=(N-Z)/2$, the seniority $v$, and reduced isospin $t$ \cite{Flowers52}. The energy is 
\be
E_{v,t,M,T}=-\frac{1}{2}G\left[(M-v/2)(2\Omega+3-M-v/2)+t(t+1)\right]+\frac{1}{2}GT(T+1)
\ee 
which does not depend on $M_T$, i.e. there are isospin multiplets of isobaric nuclei with different $N$ and $Z$.
The states show a rotational pattern with the isospin $T$ playing the role of the angular momentum $I$  
of the analogous spatial rotation of deformed nuclei. Each isorotational band is determined by the quantum numbers $v,\  t,\  M$ of the 
intrinsic state and has  the energy \mbox{$E(v,t,M)+T(T+1)/2\theta$}, where the moment of inertia $\theta=1/G$.  
The isorotational bands are evidence for the presence of an isovector condensate, which is deformed in isospace. 
The presence of the condensate additionally induces pair rotational bands, which are manifest by the quadratic dependence
of the energy on the particle number $A=2M$.

The seniority $v$ counts the number of broken pairs. Only $T=0$, 2, 4, ... is allowed  in the case of  $v=0$ states in even-even nuclei (even $M$)
and $T=1$, 3, 5, ... in the case of $v=0$ states in odd-odd nuclei (odd $M$). The energy difference between states $v=2$ and $v=0$ and the same $T$ and $t$ is
$G(\Omega+1)$. The $v=0$ ground  state in the even-even system has $T=0$. The lowest $T=1$ state has $v=2$ and lies at $G(\Omega+1)+G$. The 
$v=0$ ground state of the odd-odd system has $T=1$. The lowest $T=0$ state has $v=2$ and lies at $G(\Omega+1)-G = G\Omega$.   
\begin{figure}[tb]
\begin{center}
%\begin{minipage}[t]{8 cm}
\epsfig{file=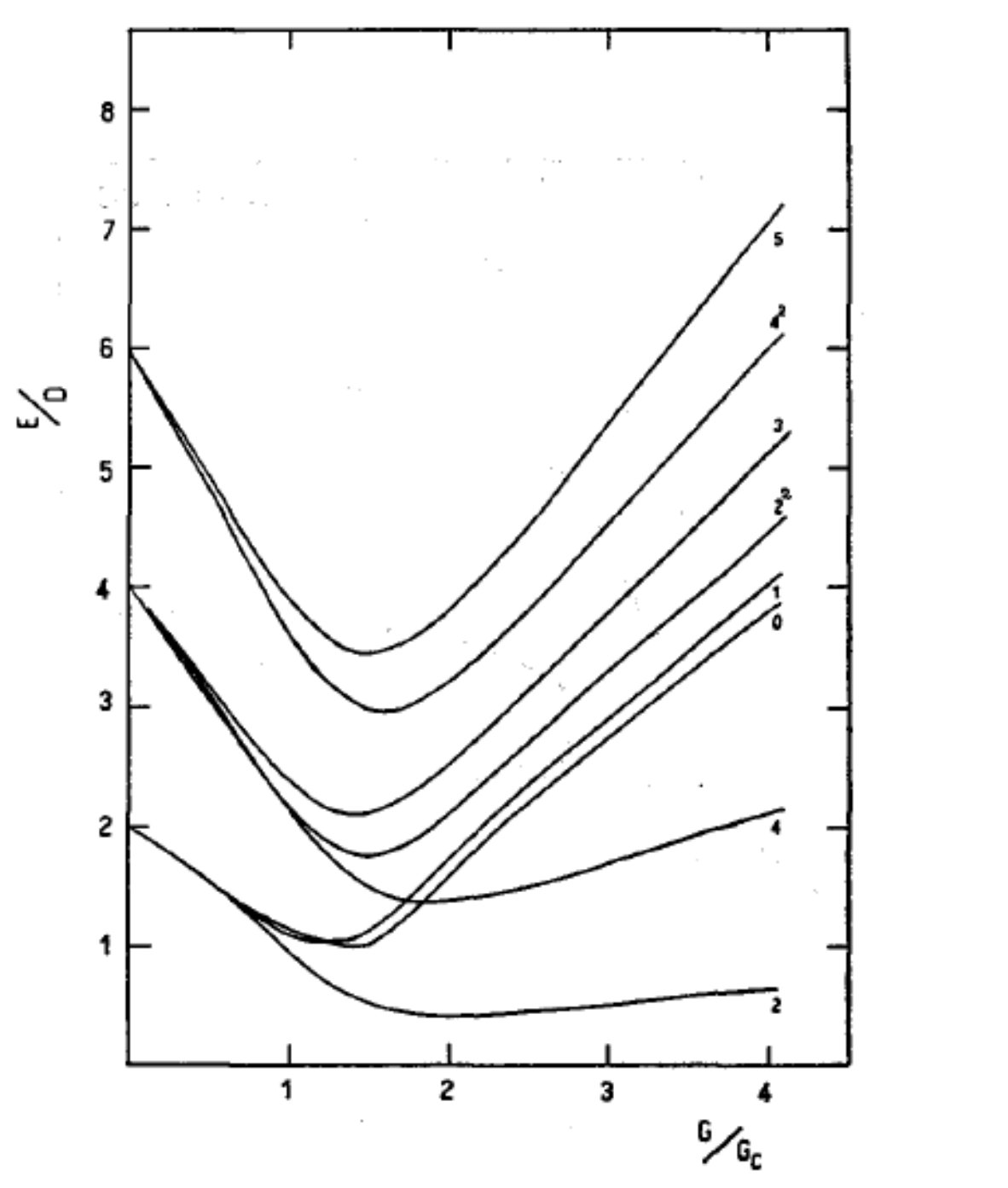,scale=1}
\caption{The energy of the states belonging to the first two isorotational bands (with $T \leq 5$ ) 
as  functions of $G/G_c$ for $\Omega = 10$ and $M = 20$. From \cite{Dussel70}.\label{fig:NPA153f1}}
%\end{minipage}
\end{center}

\end{figure}
%\begin{minipage}[t]{8 cm}
\begin{figure}[tb]
\begin{center}
\epsfig{file=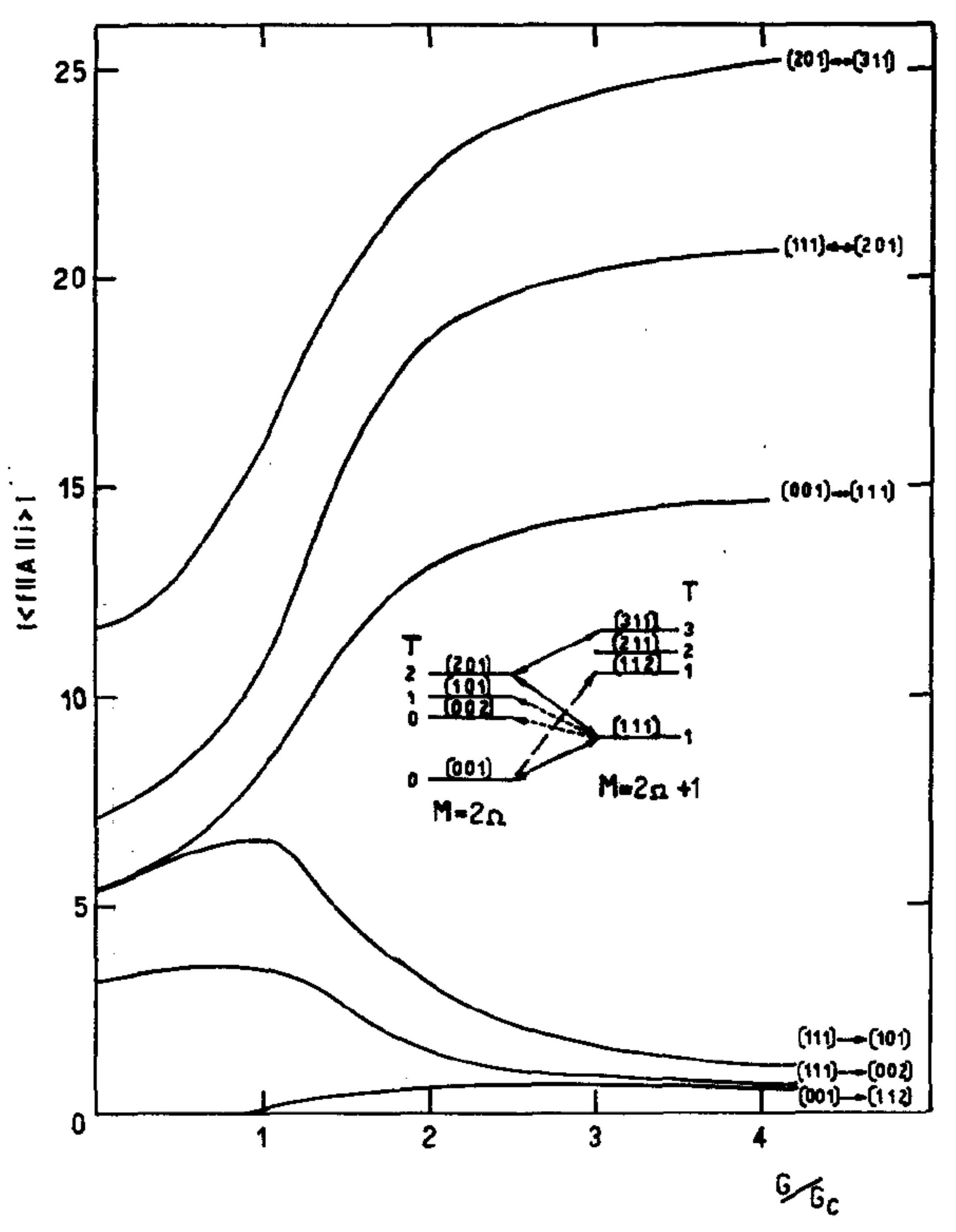,scale=0.4}
\caption{The absolute value of the reduced transition matrix elements between $M = 20$ and $M = 21$
as a function of $G/G_c$ and $\Omega = 10$. A level scheme clarifies the transitions that are considered. The
full double arrows are allowed-allowed, the dashed one is forbidden-forbidden connecting a ground
state of $M = 20$ with an excited state of $M = 21$; the dotted double arrows are the allowed-forbidden
transitions populating the pairing vibration of the $M = 20$ system. From \cite{Dussel70}.\label{fig:NPA153f2}}
%\end{minipage}
\end{center}
\end{figure}

Fig. \ref{fig:NPA153f1} demonstrates that the described isorotational pattern prevails down to $G/G_c \sim 2$.  The 
strength of the condensate is measured by the gap parameter $\Delta$ of the BCS solution for the model, 
which is $\Delta=\sqrt{(2G\Omega)^2-D^2}/2=D\sqrt{y^2-1}/2$ for $M=2\Omega$ (see Section \ref{sec:HFBsch}). 
The cranking value of the isorotational
moment of inertia  $\theta=(1-1/y^2)/G$ is smaller than 
 the asymptotic  value $1/G$, which accounts for the up bend of the energies when approaching the critical coupling $G_c$. 
Below the critical point the vibrational spectrum emerges.  The lowest group combines a $T=1$ pair addition with a $T=1$ pair removal
phonon, which couple to $T=$0, 1, 2. The excitation energy for both modes  is $\hbar \omega=2D\sqrt{1-y}$ in Random Phase Approximation (RPA) approximation,
which describes small-amplitude harmonic vibration in a microscopic way. 
Naturally, the spectrum strongly deviates  from the RPA and cranking estimates in the cross-over region $1<y<2$.   

The operators $P^\dagger _\mu$ represent the matrix elements for pair transfer. Due to isospin invariance the relative strength for two-proton ($\mu=-1$), 
two-neutron ($\mu=1$),
and $np$ transfer ($\mu=0$) are related by  Clebsch-Gordan coefficients,
\be
\left<T'M'_T\vert P^\dagger _\mu\vert TM_T\right>=\frac{1}{\sqrt{2T'+1}}\left<TM_T1\mu\vert T'M'_T\right>\left<T'\Vert P^\dagger \Vert T\right>.
\ee  
In the limit $D=0$ the reduced matrix elements $\left<T'v'\Vert P^\dagger \Vert Tv\right>=0$, as transitions between states of different seniority $v$ are forbidden.
Fig. \ref{fig:NPA153f2} displays how  some of the reduced matrix elements $  \left<T'\Vert P^\dagger \Vert T\right>$ change with the pair strength.
States are labeled by $(T, m, \alpha)$, where $\alpha$ denotes if it is the first, second etc.
time that a state with isospin $T$ and $ M = 2\Omega+ m $ appears.

The matrix element $(001)\leftrightarrow(111)$ between the $T=0$ ground state of an even-even nucleus and the $T=1$ ground state of 
its odd-odd neighbor strongly increases
 with the appearance of the pair condensate in the cross-over region. Its  mean field value is
 \be
   \left<1\Vert P^\dagger \Vert 0\right>=\frac{\sqrt{2}\Delta}{G}=\frac{\Omega}{\sqrt{2}}\left(\frac{y^2-1}{y^2}\right)^{1/2}.
  \ee  
  The expression  applies only sufficiently above the cross-over region, where the transition matrix elements rearrange into the rotational-vibrational pattern of a nucleus that is deformed in isospace (in analogy to the rotational and $\beta$- and $\gamma$- vibrational excitation of a deformed nucleus).  
  In the region $G<<G_c$ the  $(001)\leftrightarrow(111)$ transitions correspond to creating or annihilating  a pair addition phonon. The 
     $(111)\leftrightarrow(002)$ transitions correspond to creating or annihilating  a pair removal phonon. (The state 2 is composed of a pair removal and
     a pair addition phonon.) In the considered symmetric case both matrix elements are equal. The matrix element rapidly decreases with appearance of the
     pair condensate, which reflects the approach to the limit  $D=0$.
     %, where the reduced matrix elements $\left<T'v'\Vert P^\dagger \Vert Tv\right>=0$ between 
    % states of different seniority $v$ are forbidden.

 \begin{figure}[tb]
\begin{center}
\begin{minipage}[t]{8 cm}
\epsfig{file=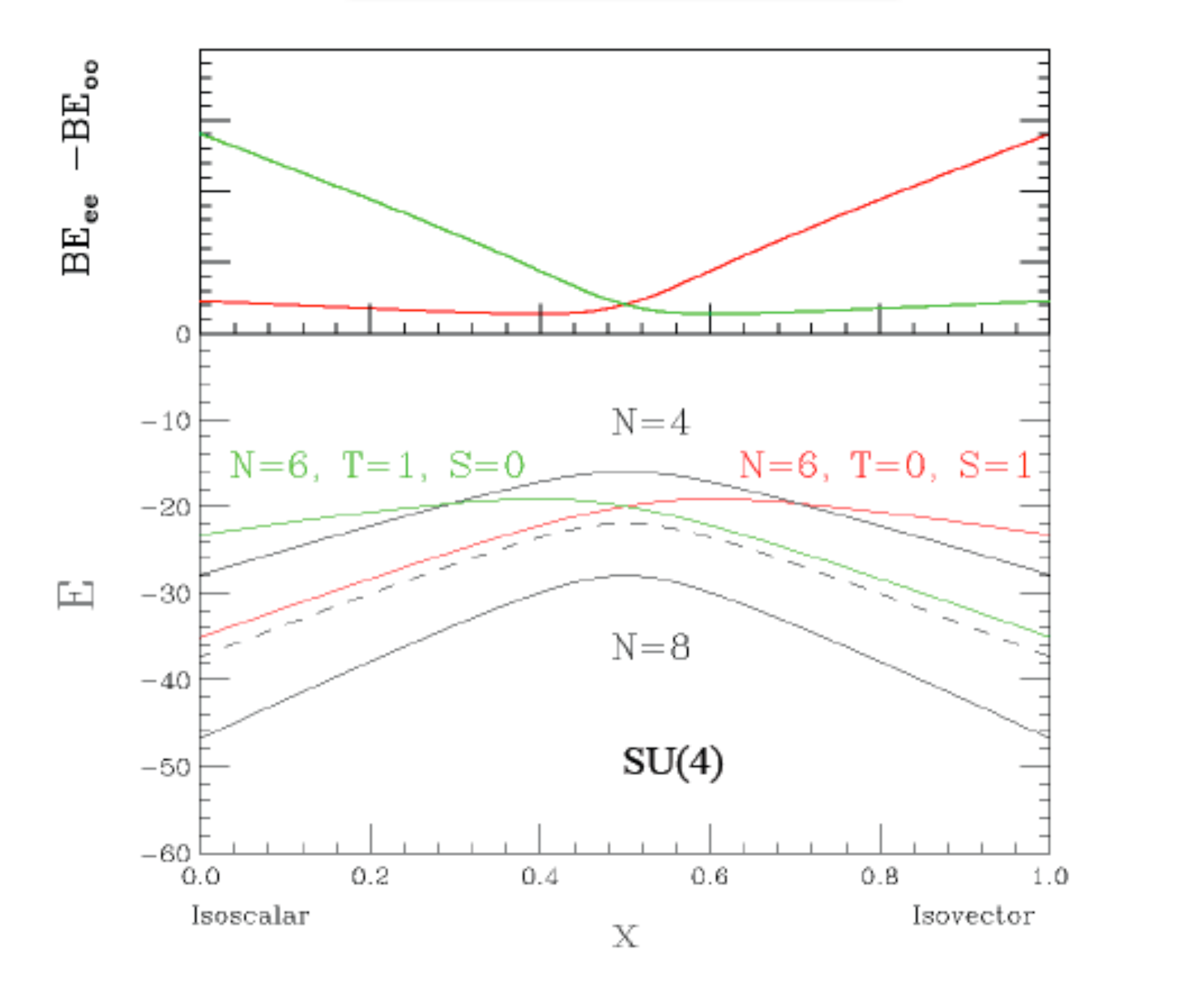,scale=0.45}
\caption{Energies of the lowest $T=1,~S=0$ and $T=0,~S=1$ for a half filled $l$-shell. 
Note that here the isoscalar and isovector coupling constants
are $G_{IV}=xG$ and $G_{IS}=(1-x)G$.  From \cite{AOM} .\label{fig:ee-ooIVIS}}
\end{minipage}
\hspace{0.5cm}
\begin{minipage}[t]{8 cm}
\hspace*{0.5cm}
\epsfig{file=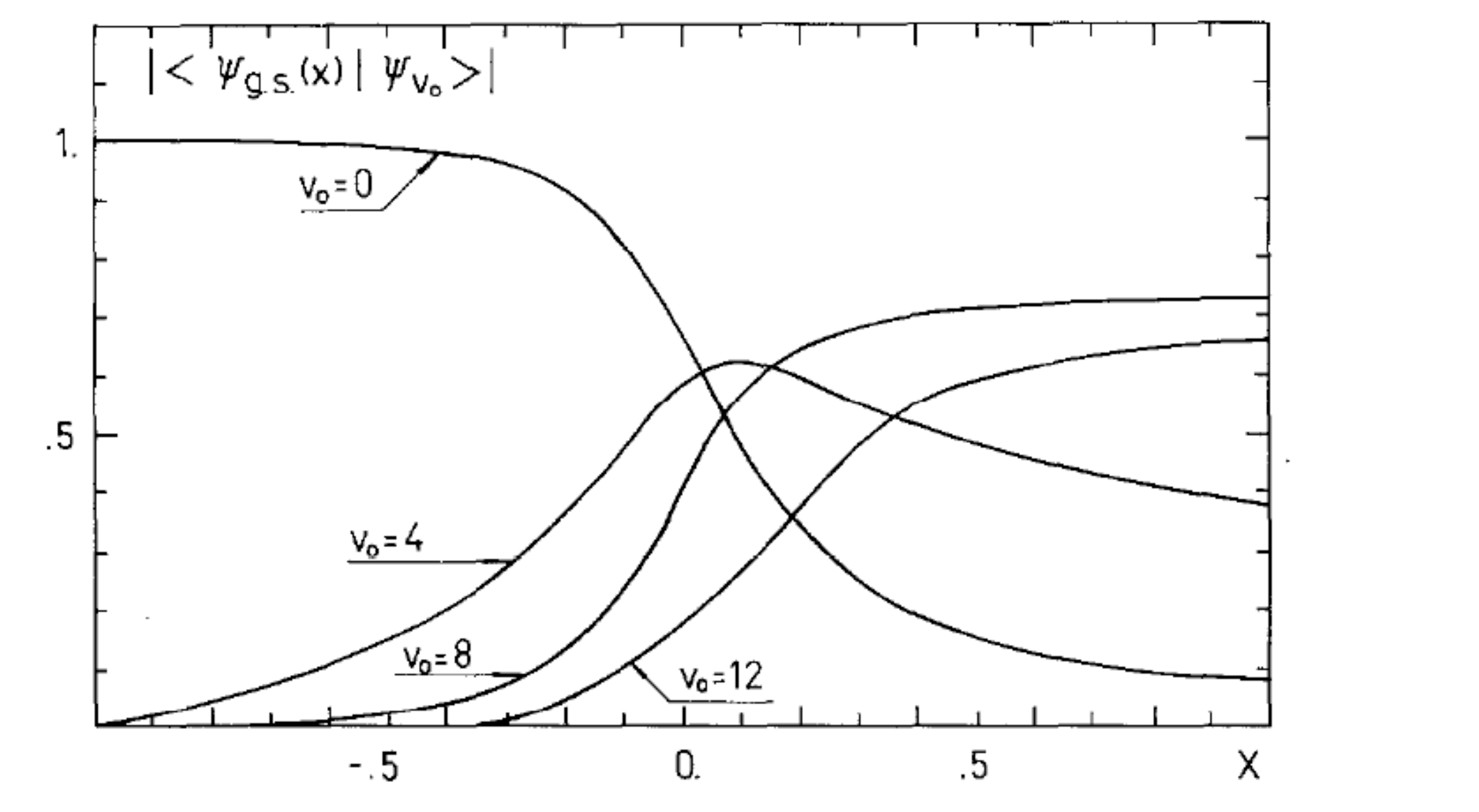,scale=0.3}
\caption{Absolute value of the components of the $(S, T) = (0, 0)$ ground-state wave function for $N = 12$
in the basis of good isoscalar seniority  $v_o$. The isovector and isoscalar coupling constants
are $G_{IV}=(x+1)G/2$ and $G_{IS}=(1-x)G/2$. From \cite{Evans81}.\label{fig:NPA367f4}}
\end{minipage}
\end{center}
\end{figure}

\subsubsection{Pure isoscalar pairing \label{sec:2jIS}}

Let us  consider now the schematic isoscalar pairing Hamiltonian,  
\be\label{eq:HISls}
 H=\sum_l\varepsilon_l\hat N(l)-G\sum_\mu D^\dagger _\mu D_\mu,
  \ee
which is defined by the $S=1$, $T=0$ pair operators 
\bea
D^\dagger _1=\frac{1}{\sqrt{2}}\sum_{lm>0}(-)^{l-m}\left(n^\dagger _{lm\uparrow}p^\dagger _{l-m\uparrow}-p^\dagger _{lm\uparrow}n^\dagger _{l-m\uparrow}\right),
&\label{eq:Dls1}\\ 
D^\dagger _0=\frac{1}{2}\sum_{lm>0}(-)^m\left(n^\dagger _{lm\uparrow}p^\dagger _{l-m\downarrow}+n^\dagger _{lm\downarrow}p^\dagger _{l-m\uparrow}
                                                                      -p^\dagger _{lm\uparrow}n^\dagger _{l-m\downarrow}-p^\dagger _{lm\downarrow}n^\dagger _{l-m\uparrow}\right),
   &\label{eq:Dls0}\\ 
D^\dagger _{-1}=\frac{1}{\sqrt{2}}\sum_{lm>0}(-)^{l-m}\left(n^\dagger _{lm\downarrow}p^\dagger _{l-m\downarrow}-p^\dagger _{lm\downarrow}n^\dagger _{l-m\downarrow}\right)
 &\label{eq:Dls-1}\eea
and the number operators
\be
\hat N(l)=\sum_m\left(p^\dagger _{lm\uparrow}p_{lm\uparrow}+p^\dagger _{lm\downarrow}p_{lm\downarrow}+n^\dagger _{lm\uparrow}n_{lm\uparrow}+n^\dagger _{lm\downarrow}n_{lm\downarrow}\right),
\ee
where $p^\dagger _{lm\uparrow}$ creates a proton and $n^\dagger _{lm\uparrow}$ a neutron with orbital momentum $l$ and spin up 
and  $p^\dagger _{lm\downarrow}$ and $n^\dagger _{lm\downarrow}$ with spin down, respectively.
The even particle number  $A=2M$, the total spin $S$, and its projection $S_z=M_S$ are good quantum numbers. 
The isoscalar pair Hamiltonian (\ref{eq:HISls})  is equivalent to the isovector pair Hamiltonian 
(\ref{eq:HIV}) when the isospin is replaced by the spin, and the results of the two-level model reinterpreted accordingly. 
In the case of a strong isoscalar condensate, the ground states of the even-even and odd-odd $N=Z$ nuclei are both $v=0$. They constitute   a pair rotational band,
which is the consequence of breaking gauge symmetry. This is analogous to pair rotational bands of a neutron pair condensate: The ground state energies of 
even-$N$ nuclei change in a gradual way when correlated neutron pairs are added. In the case of  isoscalar pairing,  $np$ pair carrying spin one are
added, and  one expects that the even-even and odd-odd $N=Z$ nuclei join into a smooth sequence. 

The odd-odd nuclei have  $I^\pi=1^+ $ ground states, which are slightly shifted up relative to the even-even neighbors
by the rotational energy of the condensate, which is $S(S+1)/2{\cal J}=1/{\cal J}=G/(1-1/y^2)$.   For the two level model and $G=4G_c$ the shift amounts to 
1/7 of the shell distance $D$ to be compared with about $D$ in the case of a weak isoscalar correlation.
%(See Fig. \ref{fig:ee-ooIVIS} for $x=0$.)

The spin rotational bands expected to appear with an isoscalar condensate will be difficult to be identified among the excited states, because there is no selective operator.
The spin-orbit interaction will couple them to the orbital rotation. One expects an increase of  the combined moment of inertia due to the spin alignment.

\begin{figure}[tb]
\begin{center}
\epsfig{file=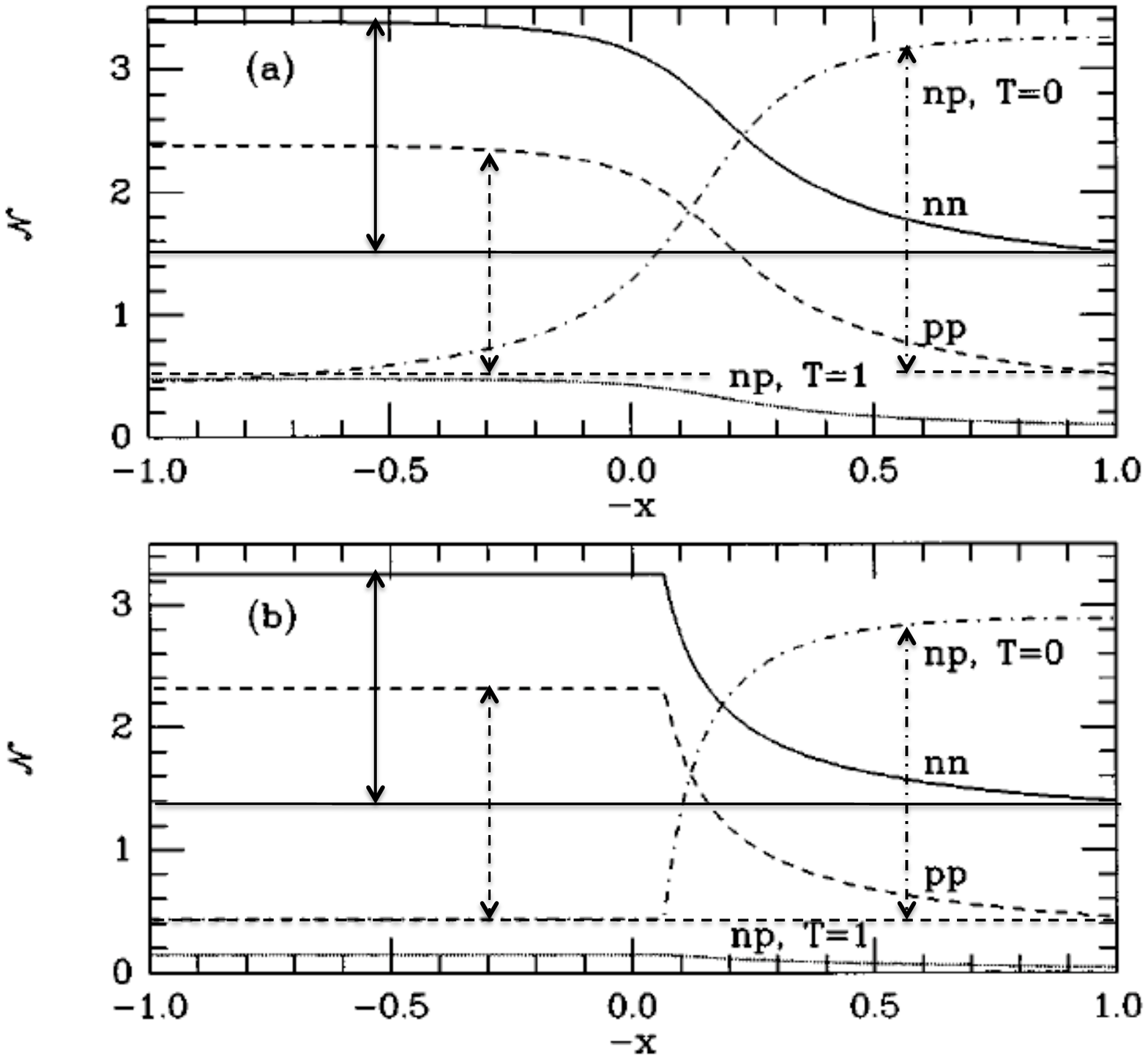,scale=0.6}
\caption{Expectation values of the number pairs of each type for 6 protons and 14 neutrons in a shell of $2\Omega=24$ degeneracy. 
The upper upper shows the exact values and the lower  the mean field results. 
 The isovector and isoscalar coupling constants
are $G_{IV}=(x+1)G/2$ and $G_{IS}=(1-x)G/2$. The values for pure isovector and isoscalar interaction are included as
horizontal lines. The differences indicated by double arrows measure the strength of the respective pair correlation.
Adapted from \cite{Engel97}.\label{fig:PRC55f11}}
\end{center}
\end{figure}

\subsubsection{Isoscalar and isovector pairing \label{sec:1jIVIS}}
Combining the isovector and isoscalar pair Hamiltonians (\ref{eq:HIV}) and (\ref{eq:HISls},\ref{eq:HISjj}) allows one to study  the competition between the two pair modes.
 Evans {\em et al.} \cite{Evans81}, Engel {\em et al.} \cite{Engel97}, and Macchiavelli {\em et al.} \cite{AOM} investigated
  the limiting case $D=0$, and Dussel {\em et al.} \cite{Dussel86} studied the general case $D\not=0$.

  \paragraph{One shell $D=0$}\label{sec:2jIVISD0}
  The transition between the two regimes is controlled by another  parameter $x$
   that measures the relative strength of the isovector and isoscalar paring interactions, $G_{IV}$ and $G_{IS}$,
    respectively, $\it i.e$  $x=(G_{IV}-G_{IS})/(G_{IV}+G_{IS})$ \footnote{The reader should be made aware of the different definitions 
    used in the literature. When appropriate we have added a note to that effect.}.  In addition to the 
 discussed  solutions for $x=\pm1$, represented by the group SU(5), the states are representations of the group SU(4) (Wigner super multiplets \cite{Wigner37}) 
 for  $x=0$, i.e. , $G_{IV}=G_{IS}$. For intermediate values of 
 $x$ solutions were found by numerical diagonalization using the representations of the group SO(8) as a basis. 
 
Fig. \ref{fig:ee-ooIVIS} shows the ground states of  adjacent even-even, odd-odd, even-even $N=Z$ systems. 
Around $x=0$ there is a cross-over from the isoscalar regime to the isovector regime. The discussed  characteristics of the pure phases
appear already for $|x|>0.2$.     On the isoscalar side, the energy of the $T=0$ ground state of the odd-odd nucleus  is close to the mean of the ground state energies of the
even-even neighbors, which is the hallmark of an isoscalar pair field that is stronger than the single particle level spacings. 
On the isovector side, the  energy of the $T=1$ ground state of the odd-odd nucleus  is close to the mean of the ground state energies of the
even-even neighbors, which is the hallmark of an isovector pair field that is stronger than the single particle spacings. The small displacement is of the
order of the pairing strength $G$ for the degenerate shell.  {It will be substantially larger, of the order of twice the average single particle level spacing,
$\approx 50$ MeV$/A$, in the absence of strong correlations, when the pairing gap is comparable with the  level distance.}
 Fig. \ref{fig:NPA367f4} shows that  the ground state wave function is dominated by the pure isoscalar ground state ($v_0=0$)
up to $x$=-0.2. Then components with higher seniority  ($v_0>0$) herald the cross-over to the isovector regime. The transition from the isovector regime is 
mirror symmetric. 
 
\begin{figure}[tb]
\begin{center}
\epsfig{file=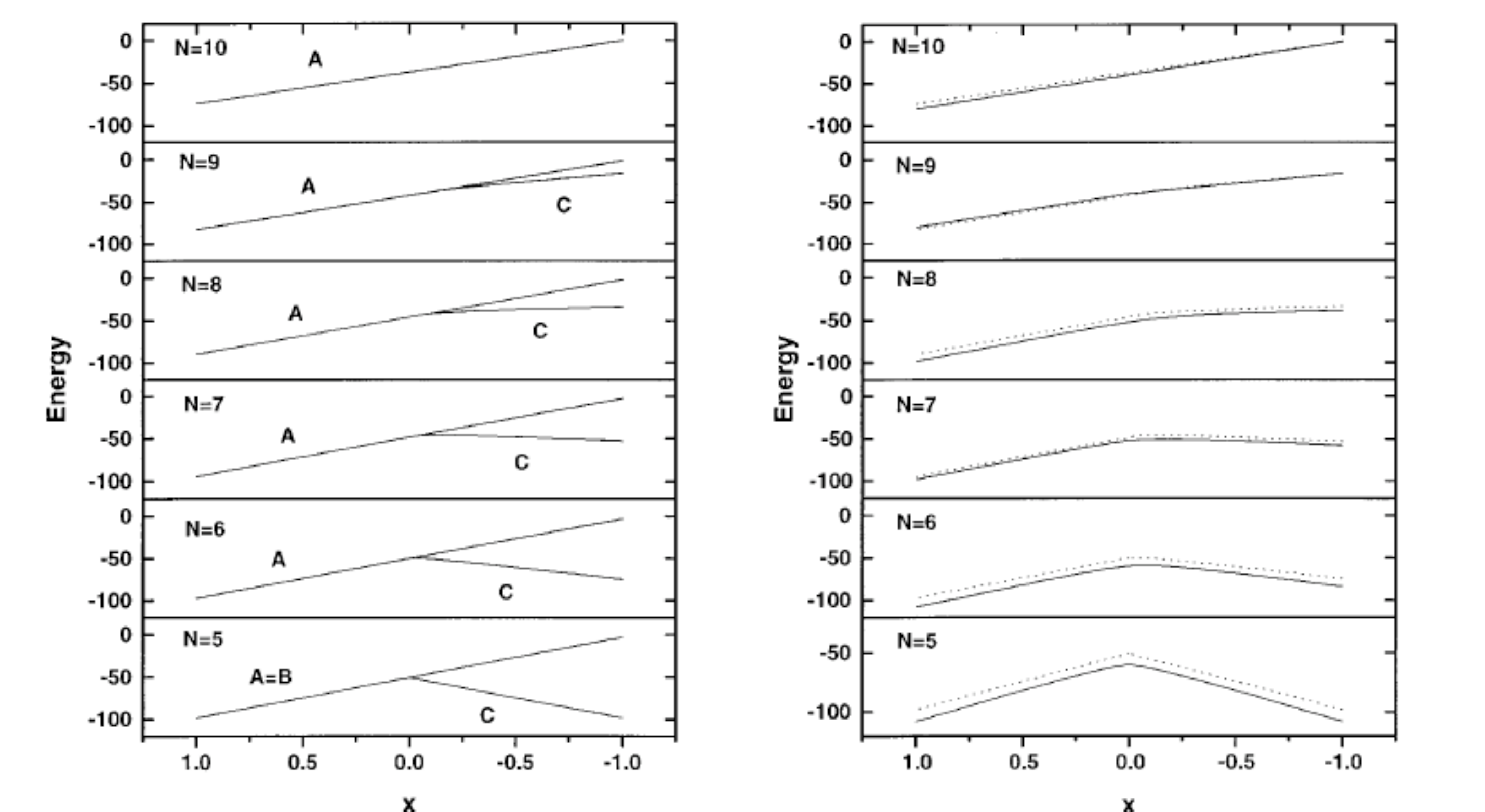,scale=0.55}
\caption{ Ground state energies (in arbitrary units)  of 5 nucleon pairs in a shell with degeneracy $2\Omega=24$. 
The isovector and isoscalar coupling constants
are $G_{IV}=(x+1)G/2$ and $G_{IS}=(1-x)G/2$. The left panel shows  the mean field solutions, the right panel compares the exact energies 
with the mean field approximation. The number of neutrons is indicated in each sub panel.
Solution A has zero $np$ pair field. Solution B corresponds to a pure $T=1$ $np$ pair field.
Solution C comprises a $T=0$ $np$ pair field coexisting with equal $T=1$ $nn$ and $pp$ pair fields. 
From \cite{Engel97}.\label{fig:PRC55f78}}
\end{center}
\end{figure}

Engel {\em et al.} \cite{Engel97} studied the relation between the mean field solutions and the exact ones for a shell with 
degeneracy $2\Omega=24$. They applied the Hartree-Fock-Bogoliubov (HFB) mean field theory (see Section \ref{sec:MF})
to the model. 
As a measure for the pair correlations  of each type they used the pair counting operators
\be\label{eq:pairnumber}
\Omega{\cal N}_{J=0,T=1\mu}= P^\dagger _\mu P_\mu, ~~~ \Omega{\cal N}_{J=1,T=0}=\sum_\mu D^\dagger _\mu D_\mu.
\ee
These are special cases of the operators
\be\label{eq:pairnumberJT}
{\cal N}_{J,T\mu}=\sum_{jj'M}A^{T\mu\dagger}_{JM}(j,j')A^{T\mu}_{JM}(j,j'),~~~{\cal N}_{J,T}=\sum_\mu{\cal N}_{J,T\mu},
\ee
which count the number of pairs of given angular momentum $J$ and isospin $T$.
The operators count the number uncorrelated pairs as well. We take as a measure for the isovector pair correlations the difference
of the expectation value of the pair counting operator $\langle   {\cal N}_{J=0,T\mu}\rangle$ and its expectation value
for $G_{IV}$=0, which we consider as the number of uncorrelated pairs. The  number of correlated isoscalar pairs is
estimated in the same way. 

Fig . \ref{fig:PRC55f11} compares the strength of the four different pair types for the exact and mean field solutions. 
The case $N-Z=8$ is displayed. (Note, the abscissa is $-x$.) The lower panel shows the mean field results. 
On the isovector side $x>0$, there is practically no isovector $np$
field. The increase from  the values for $x=\pm1$ measures the strength of the respective correlations.   
The mean field solution has no $np$ component on the isovector side, i. e.  
the isovector pairfield $\vec \Delta$ is oriented along the y-axis, as expected for isocranking about the z-axis (see discussion in Sec. \ref{sec:sigmf}). 
 On the isoscalar side,  isoscalar field decreases  below $-x=0.5$ and disappears at $x=0$. In this regime it coexists with 
 with $nn$ and $pp$ fields of equal strength, which disappear above $-x=0.5$.  
The upper panel shows the exact results. In essence, they reflect the mean field results.  As expected, the phase  transition of the mean field solutions
 gives way to a more gradual cross-over.       
The isoscalar $np$ pair field  persists  to substantial values  $|N-Z|$ (8 in this case).  This is in contrast to the isovector  pair field $\vec\Delta$,
which can always be oriented such that its $np$ component is zero. The isovector $np$ correlations appear as a consequence
of the restoration of good isospin, i.e. by rotation of $\vec\Delta$ in isospace. The two panels demonstrate that for the shown case of $T_z=4$ the
isovector $np$ correlations remain negligible after isospin restoration. However, for $T=T_z=0$ isovector $np$, $pp$, 
and $nn$  correlations have equal strength (see Section \ref{sec:sigmf}).

Fig. \ref{fig:PRC55f78} demonstrates that  the exact ground state energies of  5 nucleon pairs in the shell are well reproduced 
by the mean field values for the whole isobaric chain 
up to the pure neutron system. On the isovector side ($x>0$), the degenerate solutions A ( no $T$=1 pairing) and B (only $T$=1  pairing) of  the $N=Z=5$ system are 
two members of the infinite family of solutions corresponding to the orientation of  $\vec\Delta$ in isospace (see Section \ref{sec:sigmf}). 
On the isoscalar side ($x<0$), the $T=0$ pair field exists
for all $T_z$ as long as there are protons.

\begin{figure}[tb]
\begin{center}
\epsfig{file=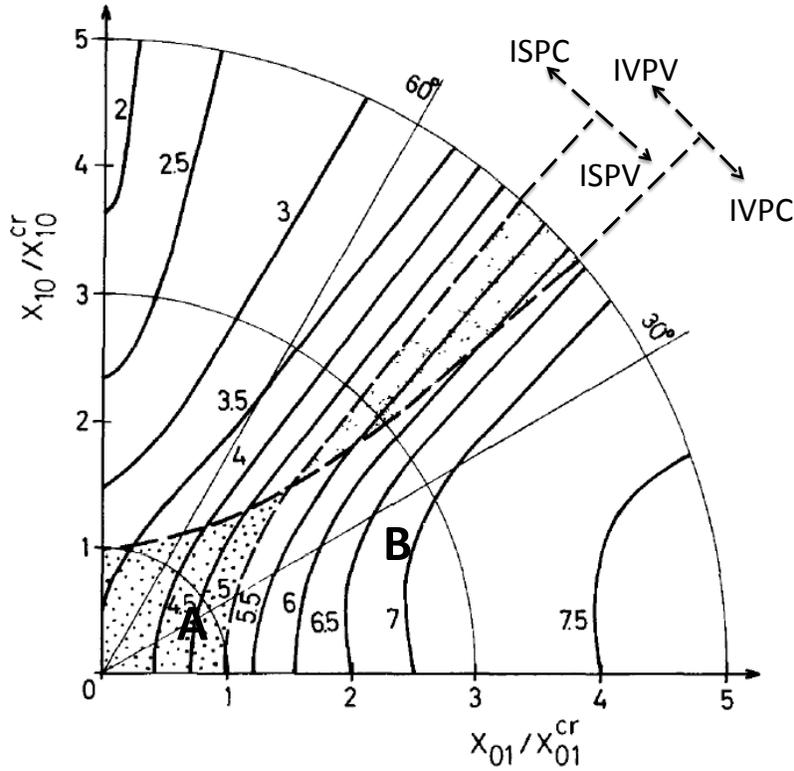,scale=0.55}
\caption{ Phase diagram for $T=1$ and $T=0$ pairing in two  $\ell$-shells with degeneracy $2\Omega_1=12$ and $2\Omega_2=8$ 
occupied by 12 protons and 12 neutrons.  Using polar coordinates, $\rho$ and $\gamma$, 
the angle $\gamma$ determines the relative strength of the isovector and isoscalar interactions. The radial distance $\rho$ measures 
the total strength in units of the critical strength for the appearance  of a condensate for pure isovector or isoscalar 
interaction ($\gamma=0,~90^\circ$). The contour lines show the values of isovector the transfer matrix element 
between the $A=24$ $T=0$ ground state and the lowest  $A=26$ $T=1$ state. The dashed lines delineate the 
various phases, where IVPV: isovector pair vibrations, ISPV: isoscalar pair vibration, IVPC: isovector pair condensate, 
ISPC: isoscalar pair condensate.  From \cite{Dussel86}.\label{fig:NPA450f4}}
\end{center}
\end{figure}

 \begin{figure}[tb]
\begin{center}
\begin{minipage}[t]{8 cm}
\hspace*{-0.5cm}
\epsfig{file=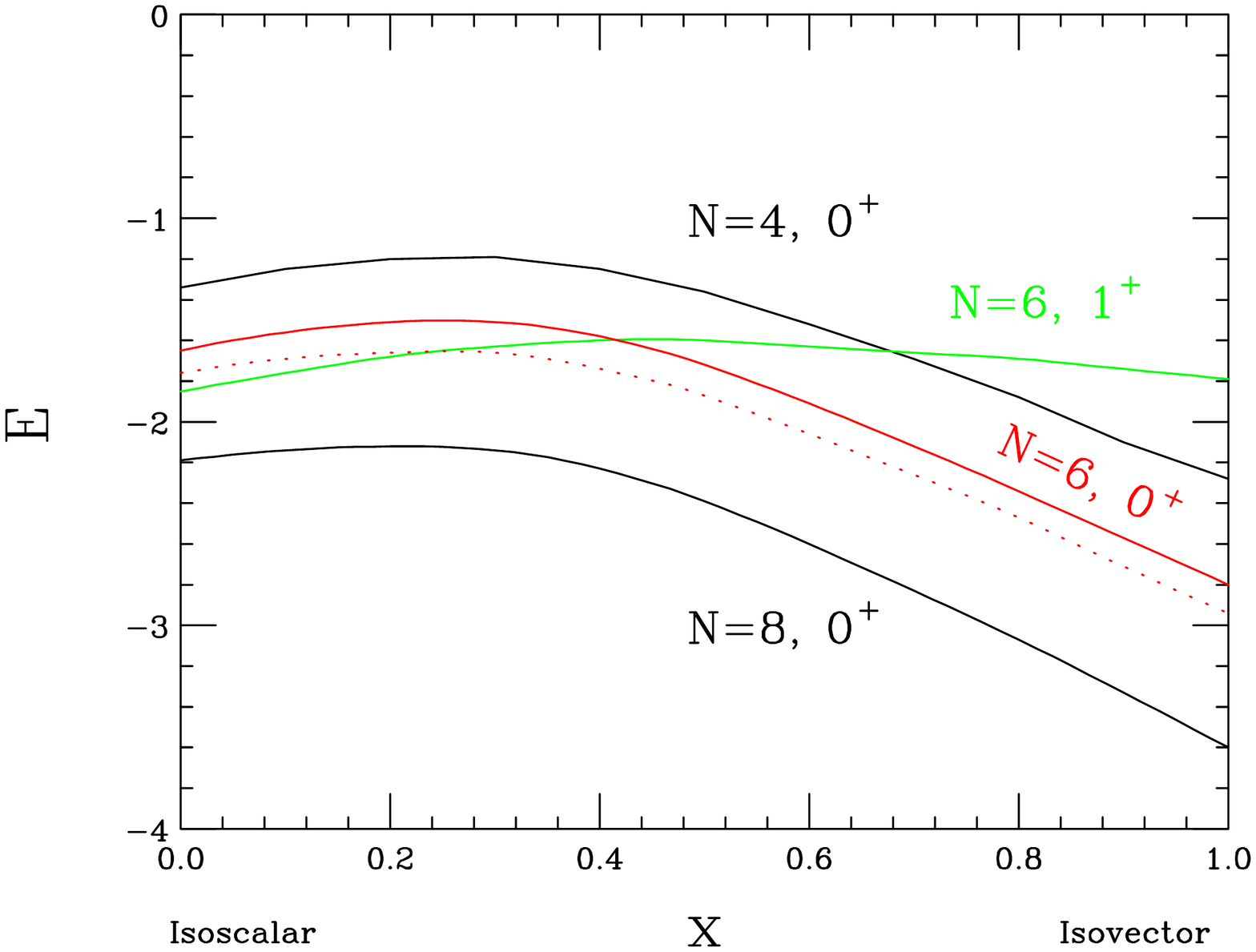,scale=0.35, angle=90}
\caption{Energies of the lowest states for $N=4,6,8$ in a single $j$ shell
as a function of the relative mixture parameter $x$ of the isovector and isoscalar parts in
the  interaction  (\ref{eq:oxbash}).  This is the equivalent to Fig. \ref{fig:ee-ooIVIS} for a $\ell$-shell.\label{fig:2jIVIS}}
\end{minipage}
\hspace{0.5cm}
\begin{minipage}[t]{8 cm}
\hspace*{-0.5cm}
\epsfig{file=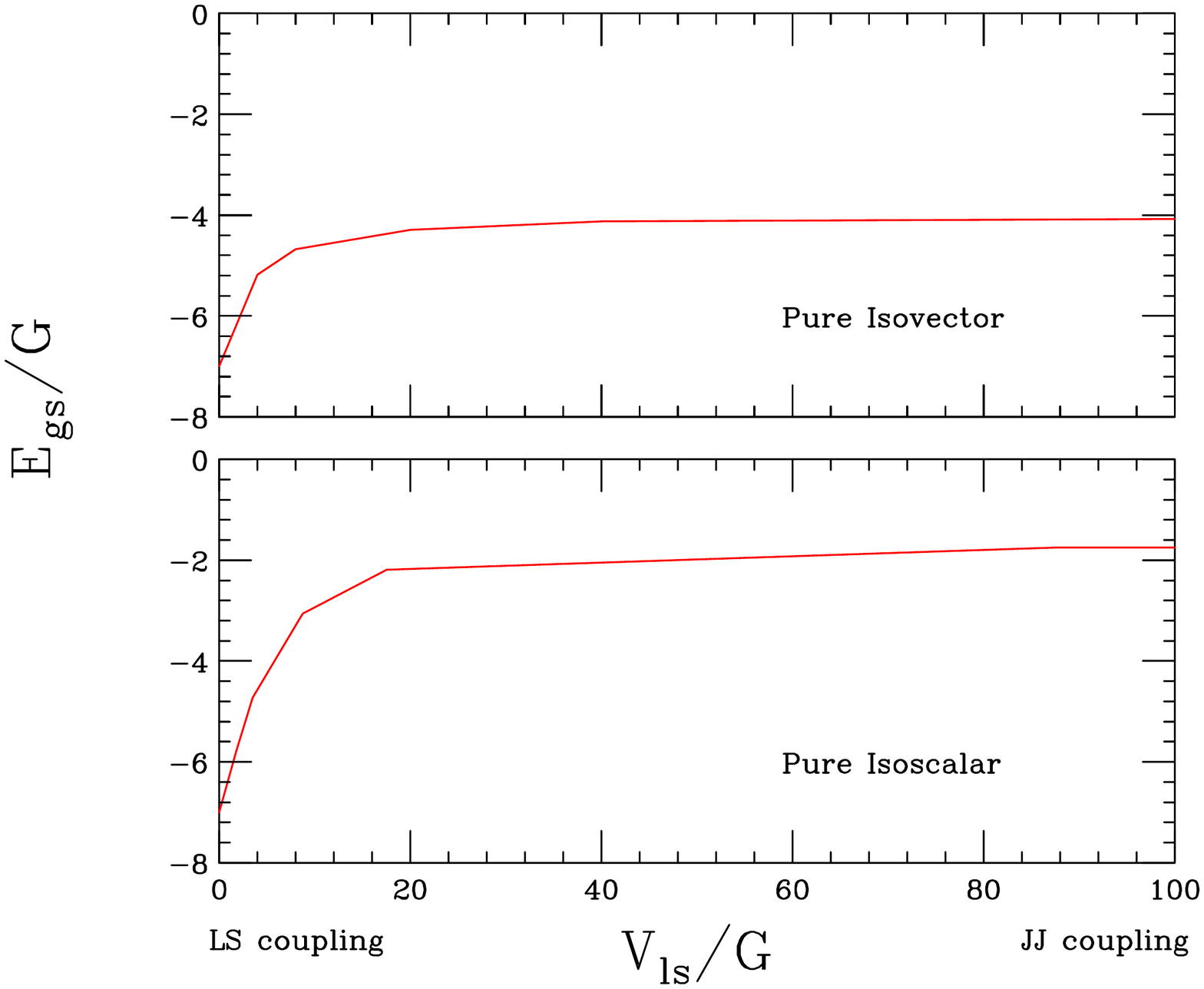,scale=0.33, angle=90}
\caption{Energies of the lowest states for $N=2$ in both isovector and isoscalar
limits as a function of the spin-orbit splitting. \label{fig:lsIVIS}}
\end{minipage}
\end{center}
\end{figure}

 \paragraph{Two shells $D\not=0$}\label{sec:2jIVISD1}
The  competition between isoscalar and isovector paring in two $\ell$-shells was studied by Dussel {et al.} \cite{Dussel86}
and in two $j$-shells by B\`es {\it et al.} \cite{Bes00} and ourselves (see below).
The phase diagram Fig. \ref{fig:NPA450f4} for two $\ell$-shells maps  out  the possible regimes to be expected in real nuclei.    
As discussed separately for the $T=1$ and $T=0$ pairing  (c.f. \ref{sec:IV2shell} and \ref{sec:2jIS}) the correlation strength 
is determined by the ratio $y=G\Omega/D $. The critical value $y=1$ marks the boundary between the phase 
where  a pair condensate is present, which can rotate, and the phase, where only  pair vibrations appear. 
Now the ratio between the coupling constants in  the $T=1$ and $T=0$ Hamiltonians is a second control parameter, which is 
fixed by the angle $\gamma$, such that $G_{IS}=G\sin\gamma$ and $G_{IV}=G \cos\gamma$. The  radial coordinate
$\rho=G\Omega/ D$ controls the overall strength of the pair correlations. The two axes $X_{ST}$ correspond to 
pure isovector ($X_{01}$) and isoscalar ($X_{10}$) Hamiltonians, respectively.   Fig. \ref{fig:NPA450f4} shows contours of
the transfer matrix element $\left<A=24,T=0\vert P^\dagger _1\vert A=26,T=1\right>$, between the respective  lowest states.
As seen in Fig. \ref{fig:NPA153f2}, the location of  largest slope of this matrix element can be taken to separate
the phase with an isovector  pair condensate from the one without it, where isovector pair vibrations are present.
This contour is dashed Fig. \ref{fig:NPA450f4}. The contours for the isocalar transfer matrix element $\left<A=24,T=0\vert D^\dagger _0\vert A=26,T=0\right>$
are obtained by reflection of the displayed contours on the $\gamma=45^\circ$ line. Only the contour is shown  that separates 
the isoscalar condensate from isoscalar pair vibration regime. The two  boundaries delineate four phases: weak pairing, only 
$T=0$ and $T=1$ pair vibrations exist (ISPV+IVPV),  $T=1$ condensate combined with $T=0$ pair vibrations (ISPV+IVPC),
$T=0$ condensate combined with $T=1$ pair vibrations (ISPC+IVPV), and coexistence of the two condensates (ISPC+IVPC).
Coexistence of isovector and isocalar condensates appears only in a narrow region above $\rho =2$, which widens as the model approaches
the limit of one degenerate shell, $\rho \rightarrow \infty$.  
For the parameter combination {\bf A}, the spectrum shown in Fig. 2 of \cite{Dussel86} is composed of near independent boson excitations
of the two types of pair vibrations. For the parameter combination   {\bf B}, the spectrum shown in Fig. 3 of \cite{Dussel86} combines isovector
rotational bands built on intrinsic excitation of isovector type ($\Delta$ and $\Gamma$) and isoscalar pair vibrations.
The distinct appearance of isoscalar vibrational states  at point {\bf B} is expected to be restricted to spherical nuclei around   $Z=N=20, ~28$,  
 because in open shell nuclei deformation tends to split  the single particle levels, which will cause a fragmentation of the 
collective isovector pair vibrational states (see Sections \ref{sec:PV} and \ref{sec:transfer}).
{The spin-orbit splitting between the single particle levels reduces the level density near the Fermi level, which attenuates 
both isovector and isoscalar pair correlations. This corresponds to a move inwards in  radial direction in Fig. \ref{fig:NPA450f4}. 
In addition, it attenuates the isoscalar pairing matrix element, which corresponds to a move towards smaller $\gamma$ in angle. }

The two $\ell$-shell model discussed above accounts for the large 
spatial overlap of the  nucleons wave function in an $L=0$ state and thus provides the best 
possible scenario to develop isoscalar correlations.
The spin-orbit splitting $v_{ls}$ will increase the energy to form the $S=1$ pairs and will then
 favor  $J=1$ pairs.  A simple estimate gives an effective isoscalar pairing strength, 
$G_{eff} \approx  G_0(1-v_{ls}/2\Delta)$, that becomes zero at $A \approx 20$.  It then
seems interesting to consider the  more realistic case of two $j$- shells, which incorporates
the more appropriate $ jj $ coupling scheme. The difference between the two $\ell$-shell and two $j$-shell models
have been discussed  in terms of the mean field approximation by B\`es {\it et al.} \cite{Bes00},
who found differences in the region of coexisting phases. 

Our approach to study this problem
was to carry out  a Shell Model calculation using a schematic pairing force with two-body matrix elements of the form
\begin{eqnarray}\label{eq:oxbash}
V_{JT}=xG\delta_{J,0}\delta_{T,1} + (1-x)G\delta_{J,1}\delta_{T,0}
\end{eqnarray}  
to model the mixture of the two types of competing interactions, with $x$ playing the same role as in the 
case of a single $\ell$-shell. 

We consider the two spin-orbit partners f$_{7/2}$ and f$_{5/2}$, and study the low-lying spectra of a partially filled f$_{7/2}$
shell  as a function of the splitting $v_{ls}$. In the limit $v_{ls} \approx 0$ we recover the 
results of the $LS$ coupling scheme.  The results for $v_{ls} >> G$ 
agree with those obtained for the f$_{7/2}$ level only.
Fig.~\ref{fig:2jIVIS} is the equivalent to Fig.~\ref{fig:ee-ooIVIS} for the single l- case. As expected
from our initial arguments, one can readily see the effect of the spin-orbit splitting
in depressing the binding in the isoscalar limit. This is due to the fact that an 
important part of the $L=0$ pairing correlations in this channel is missing as the spin-orbit
partner is pushed to higher energies.  This is clearly seen in Fig. \ref{fig:lsIVIS} where  the energy of the ground states 
of the two-body system for both isovector and isoscalar limits are shown as  function of the spin-orbit splitting.

\begin{figure}[tb]
\begin{center}
\epsfig{file=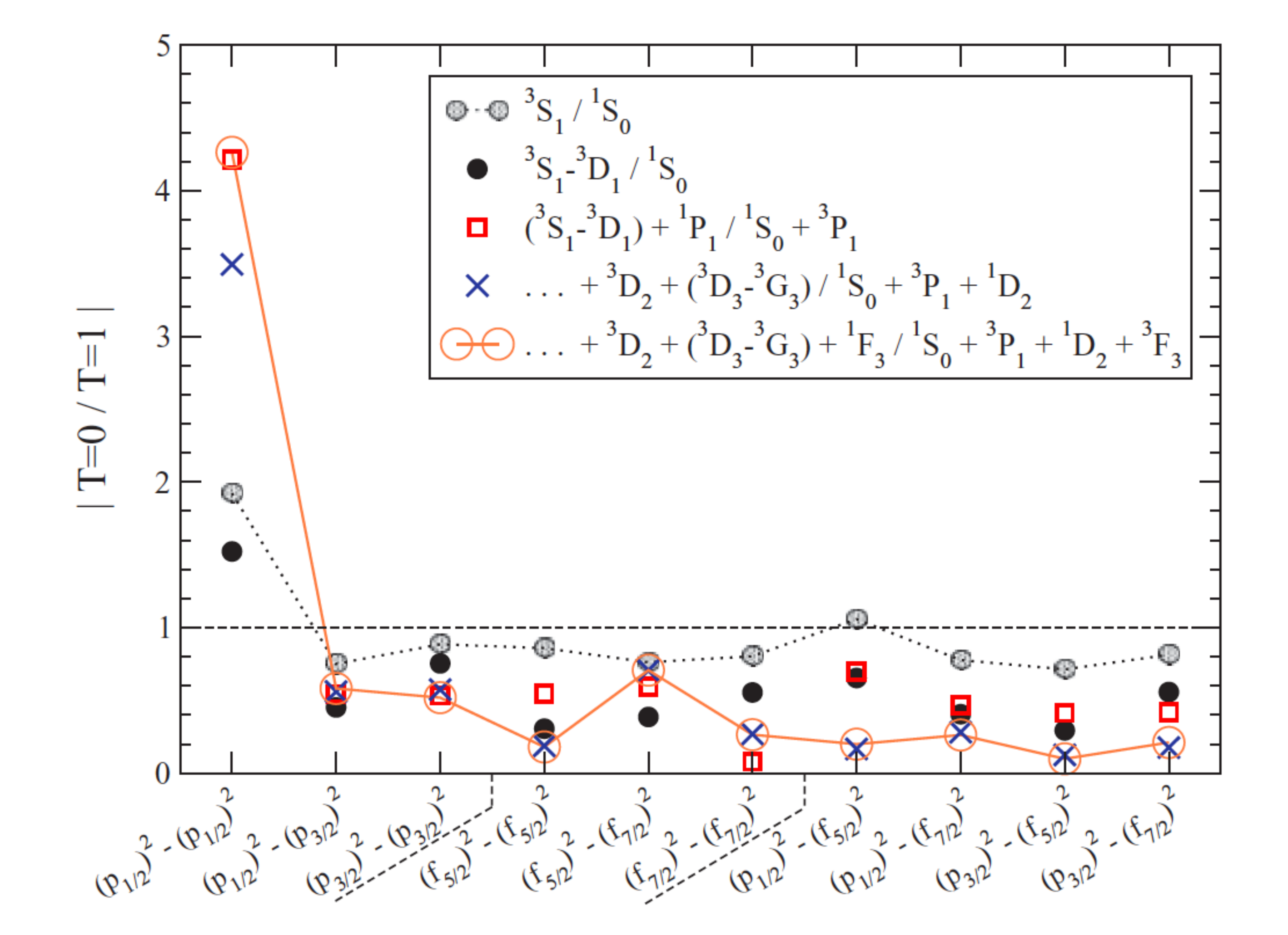,scale=0.4}
\caption{Absolute value of the ratio of the $T =0, ~J = 1$ to the $T = 1,~ J = 0$ pairing matrix elements in the pf shell. The anomalously large ratio for the p$_{1/2}^2$-
p$_{1/2}^2$ matrix element beyond S waves is due to the small $T = 1,~J = 0$ pairing matrix element in this case.  From  \cite{simone}.\label{fig:baroni}}
\end{center}
\end{figure}

A  detailed study of partial-wave contributions of nuclear forces to pairing in nuclei was presented in \cite{simone}. 
The authors found that for  $T = 1,~ J = 0$ pairing, partial waves beyond the standard $^1S_0$ channel play an important role for the pair formation in nuclei. 
The additional contributions are dominated by a repulsive $^3P_1$ partial wave, in particular the spin-triplet
nuclear forces between paired nucleons are influenced by the interplay of spin-orbit partners. 
As shown in Fig. \ref{fig:baroni},  it appears that in-medium nuclear forces favor $T = 1, ~J = 0$ over $T = 0,~ J = 1$ pairing, except in low-$j$
orbitals, casting doubt on the free-space motivation that suggests the formation of deuteron-like $T = 0$ pairs
in $N = Z$ nuclei.  In line with the results  of the simple model above,  this study confirms that the suppression of $T = 0$
 pairing is because the $^3S_1$ strength is distributed over spin-orbit partners
and because of the repulsive contributions of P  and  D waves.

\subsection{Shell Model Calculations}\label{sec:SM}

As discussed in the preceding section for the schematic models, the strengths of the $np$ pair correlations   
of course depend on the strengths of the pertinent effective interactions, which are subject to considerable uncertainty.  
The detailed agreement of the Shell Model results with experiment suggests that the isovector and isoscalar 
paring strength are realistic in the effective interactions used by the Shell Model. In this section we review work in which the authors 
investigated the pair correlation strength predicted by their Shell Model calculations, 
where we focus on the predicted correlation strength of the various pairing modes.
A second goal of the section is to provide some 
information about the various calculations  (Hamiltonian, configuration space, solution method).  The implications of the calculations will then 
be discussed in the subsequent sections.  
We will not address the question of how well the calculations reproduce the spectroscopic data on individual nuclei.  

\subsubsection{Spherical potential}\label{sec:SMs}
Engel {\em et al.} \cite{Engel96,Engel98},  Langanke  {\em et al.} \cite{Langanke95,Langanke96,Langanke97a,Langanke97b},
Dean  {\em et al.} \cite{Dean97},  Poves and 
Martinez-Pinedo \cite{Poves98}, and Martinez-Pinedo {\em et al.} \cite{Martinez99} investigated  
the strength of $np$ pair correlations by analyzing the results of large-scale Shell Model calculations, for which 
there is a certain ambiguity in quantifying the
correlation strength, as none of the authors calculated the pair transfer amplitude, which is the most direct measure.
Refs.  \cite{Engel96,Engel98} used the pair counting operators (\ref{eq:pairnumber}), which   are not  a direct measure of
 the correlations. They count the total  number of pairs, which is different from zero  in the uncorrelated system.  A more direct measure is provided by the
 modified pair counting operators
 \be\label{eq:pairnumberC}
 \Omega\tilde{\cal N}_{T=1\mu}= P^\dagger _\mu P_\mu-\left<0|P^\dagger _\mu P_\mu|0\right>, 
 ~~~ \Omega\tilde{\cal N}_{T=0}=\sum_\mu\left(D^\dagger _\mu D_\mu-\left<0\vert D^\dagger _\mu D_\mu\vert 0\right>\right),
\ee 
where $\left<0|...|0\right>$ denotes the expectation value in absence of the pair correlations. A simple estimate is counting how many pairs of each type
can be placed on a set of $\Omega$ twofold degenerate levels. Assuming $N\ge Z$ one finds, respectively, $N/2, (N-1)/2$ neutron pairs for even or odd $N$, 
$Z/2, (Z-1)/2$ proton pairs for even or odd $Z$,  $Z/2, (Z+1)/2$ $np$ pairs of given $T$ for even or odd $Z$.   
Refs. \cite{Langanke95,Langanke97a,Langanke97b} used the single particle occupation numbers of their Shell Model state
 for estimating  $\left<...\right>_0$.  Refs. \cite{Poves98,Martinez99} calculated the expectation value with a Shell Model state belonging to 
 a Hamiltonian with a quenched pair interaction. Their "switching off" procedure consisted  in subtracting from the KB3 effective interaction
 an isovector interaction of the form (\ref{eq:HIV}) or an isoscalar interaction of the form (\ref{eq:HISls}). The respective coupling strengths $G_{IV}$
 and $G_{IS}$ were determined by averaging  the KB3 matrix elements in a procedure described in \cite{Dufor96}.      
 
 The expectation value of the pair number operator can be expressed by  the sum of the transfer amplitudes by the completeness relations
  \be\label{eq:pairnumberTransfer}
 \langle0\vert\Omega{\cal N}_{T=1\mu}\vert0\rangle= \sum_i\vert\langle 0\vert P^\dagger _\mu\vert i\rangle\vert^2, 
 ~~~  \langle0\vert\Omega{\cal N}_{T=0}\vert0\rangle= \sum_{\mu,i}\vert\langle 0\vert D^\dagger _\mu\vert i\rangle\vert^2, 
\ee   
where the sum runs over all states of the final nucleus. In case of pronounced pair correlations few collectively enhanced amplitudes will
contribute the major part. The contribution of the many other states provides the background of uncorrelated pairs. If there is a substantial condensate present, 
the transfer amplitude between the ground states dominates the sum.

\begin{figure}[t]
\begin{center}
\begin{minipage}[t]{8 cm}
\epsfig{file=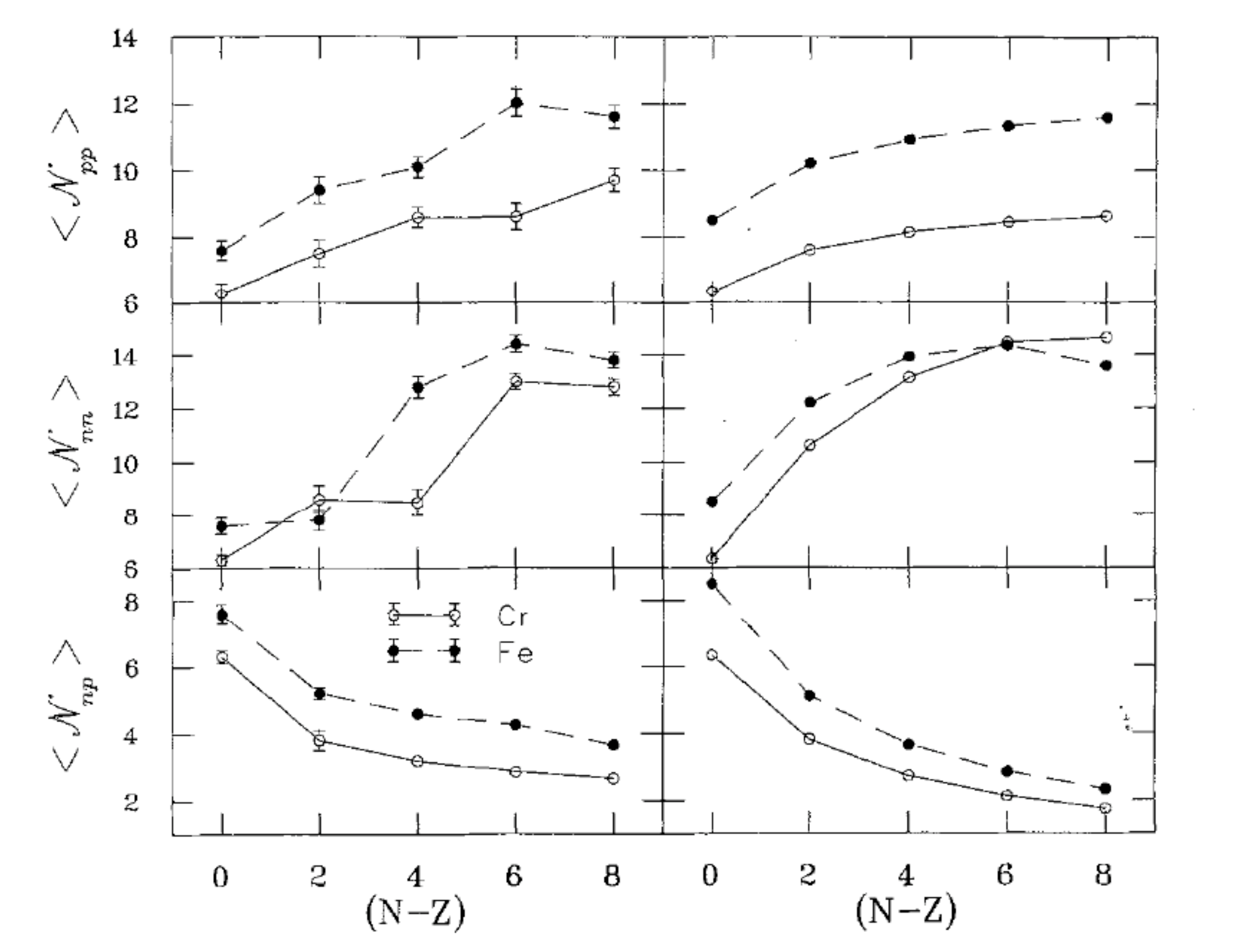,scale=0.25}
\caption{The pair number operators $\Omega {\cal N}_{T=1\mu}$  for the Fe and Cr isotopes, as a function of $N-Z=2T_z$.
On the left are the SMMC results, and on the right the single shell  (O(5)) results with $\Omega=10$ and scaled by a factor of 0.5. 
Note that $<...>$ denote expectation values of the operators (\ref{eq:pairnumber}) multiplied by $\Omega$.
\label{fig:PLB389f1}  From \cite{Engel96}.}
\end{minipage}
\hspace*{0.5cm}
\begin{minipage}[t]{8 cm}
\epsfig{file=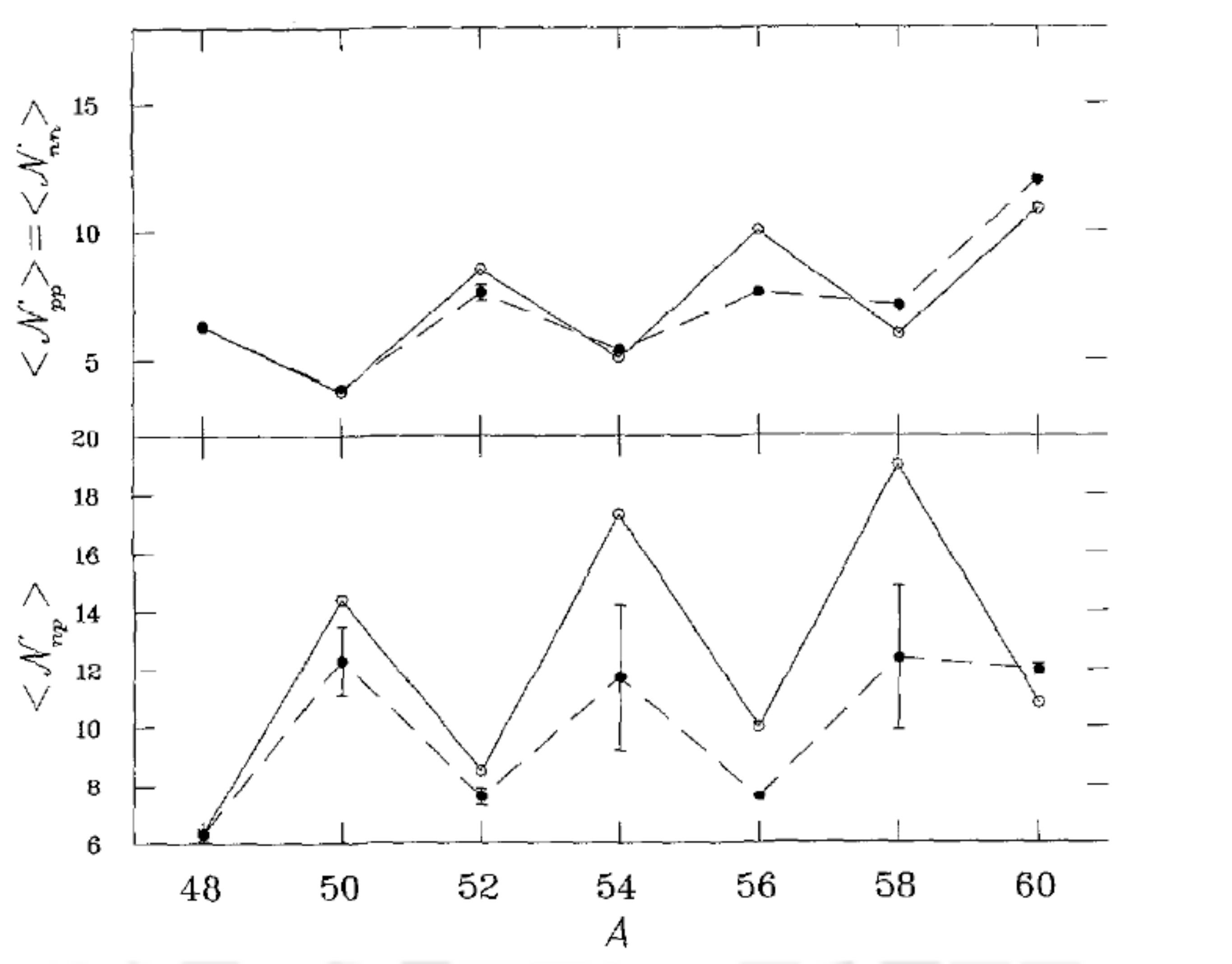,scale=0.3}
\caption{  The pair number operators $\Omega {\cal N}_{T=1\mu}$ (\ref{eq:pairnumber}) for $N=Z$ nuclei. The SMMC results are connected by 
dashed lines and the single-shell results ($\Omega=10$, scaled by 0.5) by full lines. 
Note that $<...>$ denote expectation values of the operators (\ref{eq:pairnumber}) multiplied by $\Omega$.
 From \cite{Engel96}.\label{fig:PLB389f2}}
\end{minipage}
\end{center}
\end{figure}

Engel {\em et al.}  investigated the relative strength of isovector pair correlations  by means of the Shell Model Monte Carlo (SMMC) method performed with the
 modified Kuo-Brown interaction KB3 within the pf model space. 
  Figs. \ref{fig:PLB389f1} and  \ref{fig:PLB389f2} show the expectation values of the
 pair number operators (\ref{eq:pairnumber}). 
   The one-shell model    \ref{sec:2jIVISD0} (O(5))  reproduces reasonably well  the SMMC results when
 scaled  by a factor of 0.5. In the subsequent paper \cite{Engel98}, the authors compared SMMC calculations for 
  the Pairing plus Quadrupole-Quadrupole (PQQ) Hamiltonian
  \footnote{A variant of the PQQ Hamiltonian 
 (\ref{eq:PQQ}) without quadrupole pairing $G_M=0$ and replacing $P^\dagger_\mu$ by $A^{1\mu\dagger}_{00}$.}  with the results of the two-shell model 
 of section \ref{sec:IV2shell}.  They considered $\Omega_1=4$ $\Omega_2=6$  split by $\epsilon=10G$, which is a realistic value for the 
 spin-orbit splitting between, respectively, the  f$_{7/2}$ shell and the rest of the pf shell.  Taking into account the spin-orbit splitting
   improved the agreement between the SMMC and the simple model and,  in particular,  no scaling factor was no longer required (c.f. Fig. 2 of \cite{Engel98}).  The value of the splitting for the considered nuclei 
 corresponds to $y=G/G_c$ somewhat larger than one.  As seen in Fig.    \ref{fig:NPA153f2}, the transfer amplitude $(001)\rightarrow (111)$
 is reduced by about a factor of  1.4, which corresponds to a reduction of the pair number operators by about a factor of 2 (c.f section \ref{sec:sigmf}). 
 Figs. \ref{fig:NPA153f1} and \ref{fig:NPA153f2} show that the  decrease of $y=G/G_c$ essentially only reduces the scale, as long as it stays above the critical value 
 of 1. This indicates that the relative values are given by the ratios of isorotation. We will elaborate on this point in section \ref{sec:IVcoll}.

 \begin{figure}[tb]
\begin{center}
\epsfig{file=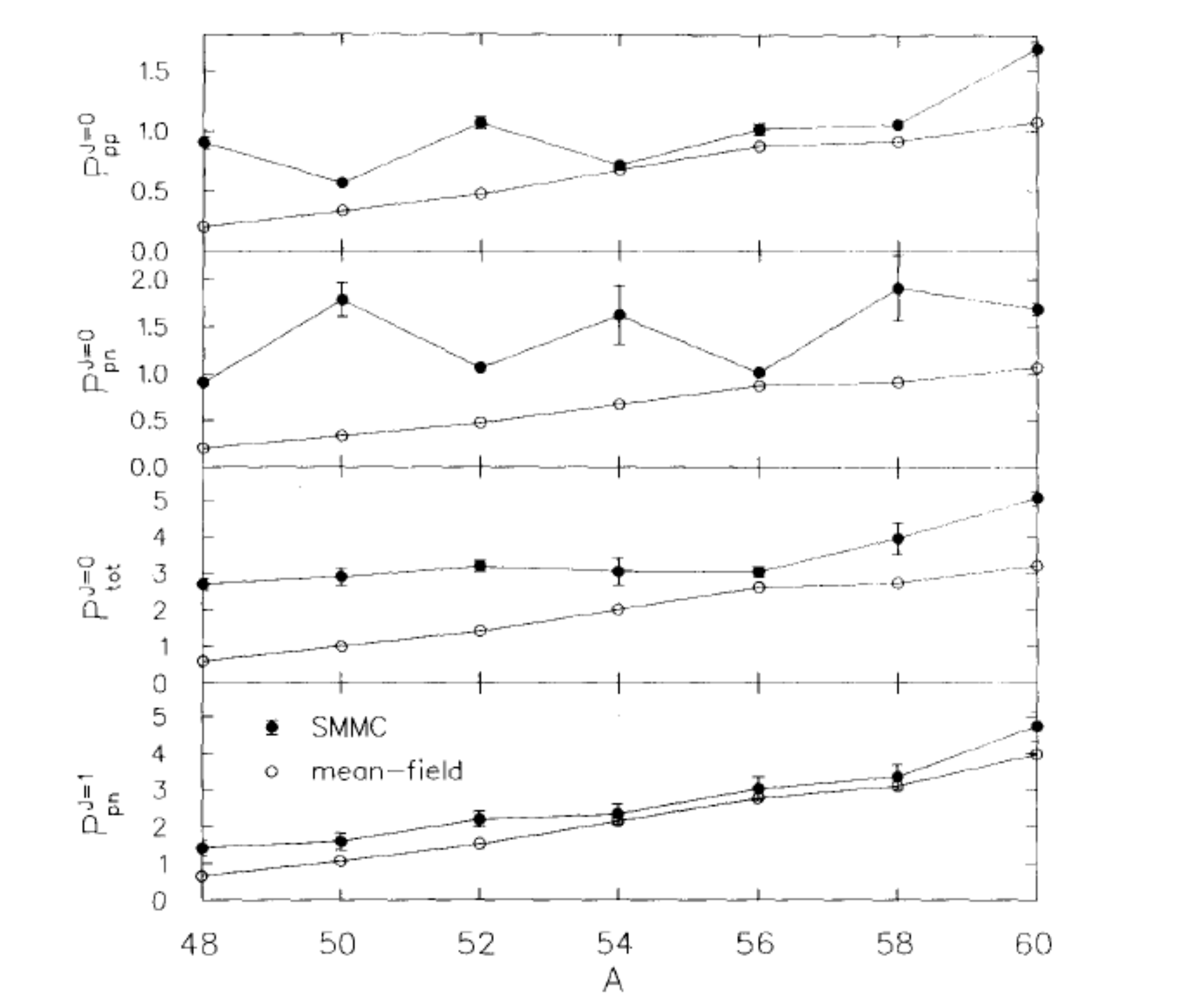,scale=0.5}
\caption{ Pairing correlations $P^J$ in the ground state of the $N = Z$ nuclei with masses A = 48-60. The full
circles show the SMMC results, while the open circles are the mean-field values. $P^{J=0}_{tot}$ gives the sum of the
three isovector pair correlations in the $J = 0$ channel. Note that $P^{J=0}_{pp}= P^{J=0}_{nn}$ for $N = Z$ nuclei.
  From \cite{Langanke97a}.\label{fig:NPA631f1}}
\end{center}
\end{figure}

Langanke  {\em et al.} \cite{Langanke95,Langanke96,Langanke97a,Langanke97b} investigated the pair correlations 
by means of the Shell Model Monte Carlo (SMMC) method performed with the
 modified Kuo-Brown interaction KB3 and the PQQ interaction within the fp model space. The results for the relative strength 
 of  the three isovector components measured by the pair number operators are qualitatively the same as shown in 
 Figs. \ref{fig:PLB389f1} and  \ref{fig:PLB389f2}. In addition, they also estimated the strength
of the isoscalar strength.  Fig. \ref{fig:NPA631f1} shows parts of their results. The values of $P^J$ are somewhat differently defined than the expectation
values of the pair counting operators in Eq. (\ref{eq:pairnumberC}), but the authors state that they obtained similar results by using them ("BCS pair operators"
in their terminology).  The "mean field" values in the figure correspond to $\left<...\right>_0$, and the correlation strength is the difference between the two
calculations.   The isovector terms show the discussed staggering between even-even and odd-odd $N=Z$ nuclei.
 The isoscalar  $J=1$ correlations are weaker than the isovector ones ($J=0$). However, they are significant. This has also been found  for
 $^{74}$Rb \cite{Dean97}. The isoscalar correlations become gradually weaker with the neutron excess $N-Z$.

{Poves and Martinez-Pinedo  \cite{Poves98} and Martinez-Pinedo {\em et al.} \cite{Martinez99}  also calculated the properties of the $A=$46, 48, and 50 isobar chains
by diagonalizing the KB3 interaction within the fp shell. Compared to the SSMC approach,  their use of the conventional Shell Model technique allowed them a 
state to state comparison with experiment and to analyze the wave functions.  Using their definition of correlation strength, they found that the isovector strength is
about 2 - 3 times stronger than the isoscalar strength as long as $T\leq 2$. The isoscalar strength is strongly reduced for  $T=3$  and zero for $T\geq 4$.  

{Volya and Zelevinsky \cite{Volya03} studied the appearance of an isoscalar pair condensate in $^{24}$Mg. The carried out Shell Model calculations 
in the sd shell using the Brown-Wildenthal interaction \cite{Brown88}, which very well reproduces the spectroscopic data on the sd shell nuclei.
They multiplied the matrix elements generating the isovector and isoscalar 
pair correlations by two scaling factors, respectively. They analyzed the Shell Model states by calculating  the Invariant Correlation Entropy (ICE),
which is a measure of how sensitive the state is to changes of the control parameters. The  maxima of ICE indicate the critical cross-over from the 
unpaired to the paired regime (see discussion in Section \ref{sec:schematic}). The analysis indicated that the  ground state of $^{24}$Mg is located   
in the cross-over region between the isovector pair vibrational and pair rotational regimes  on the side of the condensate. The isoscalar pair correlations turned out to be
weak. The authors  had to increase the  isoscalar pair matrix   elements by a factor of three compared to the realistic interaction in order to cross over
into the regime with an isoscalar condensate.}

Stoicheva {\em et al.} \cite{Stoicheva06} diagonalized  the FDP6 effective interaction  within the sdfp space. Qi {\em et al.} \cite{Qi11} performed standard shell
model calculations based on the effective interaction of Lisetzkiy {\em et al.} \cite{Lisetskiy04} within the f$_{5/2}$,
p$_{3/2}$, p$_{1/2}$, g$_{9/2}$ model space. Coraggio {\em et al} \cite{Coraggio12} diagonalized a realistic low-momentum
two-body effective interaction derived from the CD-Bonn nucleon-nucleon potential in the g$_{9/2}$ model space.  Zerguine and Van Isacker \cite{Zerguine11}
investigated several effective interactions within the  g$_{9/2}$ model space. The  results of these studies, which aimed at  rotational
excitations, will be discussed in Section \ref{sec:rotation}.

Hasegawa and Kaneko {\em et al.} \cite{Kaneko99,Hasegawa99,Kaneko99b,Hasegawa00,Hasegawa01,Kaneko01,Kaneko02,Hasegawa04,
Hasegawa04b,Kaneko04,Hasegawa05,Hasegawa05a,Hasegawa05b,Hasegawa05c,Hasegawa07,Kaneko08,Kaneko10}
systematically studied $N\approx Z$ nuclei in the regions $44\leq A\leq 74$ and $88\leq A\leq 96$.
They diagonalized  the Extended Paring plus Quadrupole - Quadrupole (EPQQ) Hamiltonian in various model spaces with manageable dimensions constructed from
the spfg orbitals. EPQQ is an extension of the Paring plus Quadrupole-Quadrupole (PQQ) Hamiltonian     
 \be\label{eq:PQQ}
H_{PQQ}=\sum_j\varepsilon_j\hat N(j)-G_M\sum_{\mu=-1}^1 P^\dagger _{\mu} P_{\mu}-
 G_Q\sum_{\nu=-2,\mu=-1}^{2,1} P^\dagger _{2\nu\mu} P_{2\nu\mu}-\frac{\kappa}{2}\sum_{\nu=-2}^2Q^\dagger _\nu Q_\nu,
  \ee
which is defined by the $J=0$, $T=1$  monopole pair operators $P^\dagger _\mu$  introduced in Eq. ({\ref{eq:P}) ,\\
the $J=2$, $T=1$  quadrupole pair operators (cf. Eq. (\ref{eq:A}))
\be
P^\dagger _{2\nu\mu}=\sum_{j,j'}\left<j|| r^2Y_2||j'\right>A^{1\mu\dagger}_{2\mu},
\ee
and the isoscalar quadrupole operators
\be
Q^\dagger _\nu=\sum_{jm,j'm'}\left<jm\vert r^2Y_{2\nu} \vert j'm'\right>\left(n^\dagger _{jm}n_{j'm'}+p^\dagger _{jm}p_{j'm'}\right).
\ee
The EPQQ \cite{Hasegawa99} takes an additional  isoscalar interaction into account.  It is composed of $T=0, J=odd$ pairs, which contribute with equal weight,     
\be\label{eq:EPQQ}
H_{EPQQ}=H_{PQQ}-k_0\sum_{J=odd,\nu}A^{00\dagger}_{J\nu}A^{00}_{J\nu}.
\ee
They diagonalized the interaction within the model spaces 
\mbox{p$_{3/2}$, p$_{1/2}$, f$_{7/2}$} for  $44\leq A\leq 56$,\\ \mbox{d$_{5/2}$, p$_{1/2}$,f$_{5/2}$, g$_{9/2}$} for $56\leq A\leq 74$,
and \mbox{p$_{3/2}$, p$_{1/2}$,f$_{5/2}$, g$_{9/2}$} for  $88\leq A\leq 94$.
 In the course of their extended studies, the authors  refined their Hamiltonian by adding an Octupole-Octupole interaction and an isoscalar monopole term. 
 They achieved a good description of the spectroscopy of the $N\approx Z$ nuclei.   
{ In Ref. \cite{Kaneko99} they calculated the expectation values of the isovector pair number operators, which qualitatively agree with Figs. \ref{fig:PLB389f1}
 and   \ref{fig:PLB389f2}. However, the authors pointed out that their special isoscalar interaction (\ref{eq:EPQQ}) does not generate isoscalar pair correlations.
 In this respect it has the character of a diagonal matrix, which just modifies the energies but not the states. This is at variance with the above discussed results
 of other authors, who use a different interaction. It may be taken as some warning that a good
 description of the spectroscopic data  does not guarantee that the pair correlations are correctly accounted for. 
 The  results of these authors  will be further discussed in Sections \ref{sec:Wigner} and \ref{sec:rotation}.}

 Aroua {\it et al.} \cite{Aroua03} studied high-lying 1$^+$ states in $^{8,10}_{4}$Be$_{4,6}$ and $^{20,22}_{10}$Ne$_{10,12}$, which are excited  
 from the ground state by the $\Delta T=1$ orbital $M1$ operator ($\propto (\vec l_\nu-\vec l_\pi)$). They  diagonalized the PQQ Hamiltonian (\ref{eq:PQQ})
 with $G_Q=0$ within the configuration space of the p$_{3/2}$ and p$_{1/2}$ shells.  
 Excitations of this type in $N>Z$ nuclides are interpreted as the "scissors mode" of deformed nuclei or as "mixed symmetry" excitations 
 of spherical nuclei (see the review article \cite{Heyde10}). The summed $M1$ strength of these excitations increases with the degree
 of collectivity of the quadrupole modes (isoscalar and isovector) in these nuclides.    Aroua {\it et al.} point out for $N\approx Z$ that a) 
 $M1$ strength is zero when the  isovector $np$ pair correlations  are absent, and b)  it approaches a constant small  value when the coupling strength 
 of the QQ interaction $\kappa \rightarrow 0$. It would be interesting to include an isoscalar interaction with realistic strength
 in order to see if the $M1$ strength contains complementary information about the competition between isoscalar and isovector $np$ pair correlations.

\subsubsection{Deformed potential}\label{sec:SMd}

Frauendorf, Rowley,  Sheikh, and Wyss \cite{Frauendorf94,Frauendorf99a,Frauendorf99b,Frauendorf00,Sheikh00} diagonalized the Hamiltonian
\be\label{eq:defjshell}
H_{QP}=-4\kappa q_{20}-G\delta(\vec r_1-\vec r_2)-\omega j_x
\ee
within the configuration space of protons and neutrons in an f$_{7/2}$ shell. The quadrupole term $4\kappa q_{20}$ induces the energy splitting of the
$j$-shell by the axial average potential, where their choice of  $\kappa$ corresponded  to a deformation of $\beta$=0.16. 
The pair correlations are generated by the $\delta$-interaction, which fixes the relative strength of the 
isovector and isoscalar interaction. The angular momentum dependence of the spectrum was
investigated by means of the cranking term $\omega j_x$, which will be discussed in Section \ref{sec:rotation}. 
The authors compared the spectra obtained with the full interaction $G\delta$ with the  ones obtained with the isovector
model interaction represented by  the $T=1$, $J=0$ multipole of the same interaction $G\delta$. 
 The results turned out to be very similar, which indicates that the isovector pair correlations determine the structure of the rotational 
 spectra. The re-coupling from $\ell s$-orbitals to $j$-orbitals, which is a consequence of the spin-orbit potential, substantially attenuates the 
 isoscalar pair correlations (see Fig. \ref{fig:2jIVIS}). 
 {This work focused on the influence of the pair correlations on the rotational response and will be discussed 
 in Section \ref{sec:rotation}.}

 Bentley and Frauendorf \cite{Bentley13} studied the competition between isovector and isoscalar pairing starting from a set of single particle energies  
$\epsilon_k$ in a  Nilsson potential,    at a  deformation  calculated by means of the Micro-Macro (Strutinsky) method, and  diagonalized the 
Hamiltonian in a basis generated by   all configurations within the seven   single particle states closest to the Fermi level. They used an isospin invariant
model Hamiltonian
\begin{eqnarray}
\label{eqn:IVIS}
 H_P=\sum_{k} \epsilon_{k} \hat{N}_k  - G_{IV}
\sum_{kk', \tau} \hat{P}^\dagger _{k,\tau}  \hat{P}_{k',\tau}-G_{IS}\sum_{kk'}D^\dagger _kD_k'+\frac{\kappa}{2}\vec T\cdot\vec T,\\
\hat{N}_{k}=\hat{p}^\dagger _{k}\hat{p}_{{k}}+\hat{p}^\dagger _{\bar k}\hat{p}_{\bar{k}}+
\hat{n}^\dagger _{k}\hat{n}_{{k}}+\hat{n}^\dagger _{\bar k}\hat{n}_{\bar{k}},\\
 \hat{P}^\dagger _{k,0}=\frac{1}{\sqrt{2}}\Big(\hat{n}^\dagger _{k}\hat{p}^\dagger _{\bar{k}}+\hat{p}
^\dagger _{k}\hat{n}^\dagger _{\bar{k}}\Big),~~
 \hat{P}^\dagger _{k,-1}=\hat{p}^\dagger _{k}\hat{p}^\dagger _{\bar{k}},  \textrm{ and }
\hat{P}^\dagger _{k,1}=\hat{n}^\dagger _{k}\hat{n}^\dagger _{\bar{k}},\\
D^\dagger _k=\frac{1}{\sqrt{2}}\Big(\hat{n}^\dagger _{k}\hat{p}^\dagger _{\bar{k}}-\hat{p}^\dagger _{k}\hat{n}^\dagger _{\bar{k}}\Big),\label{eq:Dk}
\end{eqnarray}
where $\hat{p}^\dagger _{k}$ and  $\hat{n}^\dagger _{k}$ create a proton and a neutron,
respectively, on the level $k$, and $\bar k$ denotes the time reversed state of $k$. The isoscalar interaction involves $np$ pairs in time
reversed deformed orbitals. We will refer to this kind of $np$ correlations as $n\bar p$ pairing (c.f. Section \ref{sec:MF}).
\footnote{In keeping with theoretical references in the literature, here and where necessary for clarity, we explicitly introduce the forms $n\bar n$, $p\bar p$ and  $n\bar p$ to indicate pairs in time reversal orbits.
This is to distinguish from  isoscalar $np$ pairs in the same orbits.  However, throughout most of the article, the generic forms  $nn, pp$ and $np$ are used. }
The model parameter $G_{IV}$ and $C=\kappa/2$ were adjusted to the energy differences of $N=Z$ odd-odd nuclei relative to their $N=Z$ even-even neighbors
and to the energy difference between the $T=0$ and $T=1$ states in the odd-odd $N=Z$ nuclei. Investigating different ratios  between the
isovector and isoscalar interaction strength, they  found that the experimental data could be equally well reproduced for  $0\leq G_{IS}/G_{IV} \leq 0.5$.
For larger ratios the agreement deteriorated, and for $G_{IS}\geq G_{IV}$ no reasonable fit was possible.
The investigation focused on the Wigner energy. The relevant results will be discussed in Section \ref{sec:Wigner} that is devoted to it.

\subsection{Significance of the mean field solution} \label{sec:sigmf}    
 This section is devoted to the interpretation of the mean field results for isovector pairing. 
 The fact that the isovector pair field spontaneously breaks isospin symmetry leads us to consider the collective rotation in isospace 
 of an intrinsic state represented by the mean field solution.  It seems to us that this concept is often overlooked in  the literature, and therefore should be 
 discussed in some detail here.   
Refs. \cite{Camiz65,Camiz66,Ginocchio68,Frauendorf99a,Frauendorf00,Frauendorf01,Satula01,Glowacz04} studied the consequences of  spontaneous symmetry breaking 
by the isovector condensate 
in the general frame of the HFB mean field approach. The essential arguments can be most simply conveyed for  the isovector
pair Hamiltonian (\ref{eq:HIV}).   In analogy  to the deformed nuclear potential, the pair field  
$\Delta_\mu$, which  is the expectation value $\left<GP^\dagger _\mu\right>$   with respect to the HFB vacuum state, represents the condensate.  The pair field 
is a vector in isospace, which breaks the isotropy. Since the Hamiltonian is isospin invariant, all directions $\vec\Delta$ have the same mean field energy
\cite{Camiz65,Camiz66}. 
Ginocchio and Weneser \cite{Ginocchio68} pointed out that the different directions of $\vec\Delta$, which  correspond to different values of 
mean value of $\left<T_z\right>$, represent the members of the isospin multiplet $T$.  If $\vec\Delta$ is parallel to the y-axis, there are only $nn$  and $pp$ pairs 
and $\left<T_z\right>$ takes its maximal value,  if parallel to the z-axis, there are only $np$ pairs
 and $\left<T_z\right>=0$.  Considering only the mean values is not sufficient.   
  The isovector pair field breaks the isotropy  in isospace   like the the deformed 
nuclear potential breaks the  spatial isotropy, which gives rise to rotational bands and collective vibrational excitations. This far reaching analogy
suggests that one should  employ the familiar concept of "collective"  motion of the pair field  $\vec\Delta$  and "intrinsic" excitations, which are
constructed as  configurations of quasiparticles that belong to the mean field Hamiltonian.  

\subsubsection{Collective motion of the isovector pair field}\label{sec:IVcoll}
 The mean field approximation for the isovector pair Hamiltonian (\ref{eq:HIV}) takes the form
\be\label{eq:HIVmf} 
 h_{mf}=\sum_j\varepsilon_j\hat N(j)-G\sum_\mu\left(\Delta_\mu  P^\dagger _\mu +\Delta_\mu^* P_\mu\right)-\lambda_p\hat Z-\lambda_n\hat N,
 ~~\hat Z=\sum_j\hat N_p(j),~~\hat N=\sum_j\hat N_p(j),
  \ee
where $\lambda_{p,n}$ control the particle numbers such that $\left<\sum_j\hat N_{p,n}(j)\right>=Z,N$.
 If there is a strong isovector condensate one can replace the operators by their expectation values.  
The relative strength of the three types of isovector pair correlations is essentially of rotational nature
 Fig. \ref{fig:NPA645f2}  illustrates  the concept for  the special case $N=Z$.  Choosing the z-direction (only $np$ pairing) as the
  intrinsic z-axis, the pair field has the projection $K_T=0$ and the magnitude
 $p=\Delta/G$. The pair field in the laboratory system is then 
\be\label{eq:Pcoll}
 P^\dagger _\mu=D^1_{\mu 0}(\Phi,\Theta)p.   
 \ee
 The $T=0$ state is isotropic, the numbers of the three types of correlated pair are 
 equal to $\Omega\tilde {\cal N}_\mu =p^2/3$. The $T=1$ states of the axial   isorotor Hamiltonian are $D^{1*}_{M_T0}(\Phi,\Theta)$ (not normalized). 
 The odd-odd $N=Z$ nuclei ($T=1, M_T=0$) have strong $np$  and weaker $nn$ and $pp$ correlations because $d^1_{00}=\cos(\Theta)$. 
 The even-even $N+1=Z-1$ nuclei ($T=1,M_T=1$) have strong $nn$, $pp$ and weaker $np$ correlations because  $d^1_{\pm10}=\mp\sin(\Theta)/2$. 
Evaluating the expectation values, one finds $\Omega\tilde {\cal N}_{pp} =\Omega\tilde {\cal N}_{pp}=p^21/5,~p^22/5$ and $\Omega\tilde {\cal N}_{pn}=p^23/5,~p^22/5$
for $T=1,M_T=0$ and $T=1,M_T=1$, respectively. The ratios between the fractions are not far from the ones for the single shell model (c.f. Fig.  \ref{fig:NeqZ}), which fairly well account 
for the ratios of the SMMC (c.f. Fig.\ref{fig:PLB389f2}).

Dussel {\em et al.} \cite{Dussel71,Dussel72} generalized the concept of the rotating isovector  pair field to a dynamic pair field. They introduced a collective 
Hamiltonian in analogy to the Bohr Hamiltonian for the quadrupole shape degree of freedom, which allowed them to describe both the rotational and the vibrational
regimes and  the cross-over between them. The behavior of the energies and transition matrix elements  of the discussed two-shell model \cite{Dussel70,Dussel72}
were explained in detail by deriving the potential for the pair degrees of freedom from the mean field solution, where  Ref.  \cite{Dussel70} 
studied  the approximate solutions of the pair vibrational type and the isorotational type and Ref.  \cite{Dussel72} found
numerical solutions for full collective Hamiltonian. 
 For example, the reduction of the matrix element for the $(111)\rightarrow(002)$
state in Fig. \ref{fig:NPA153f2} is understood as a change of the $(002)$ state from a two-phonon pairing vibration to an
 intrinsic $\Delta$ vibration when crossing over to the isorotational regime $y >>1$.
{To our knowledge, these types of full solutions  have not been compared with data from realistic nuclei.   }

\begin{figure}[tb]
\begin{center}
\begin{minipage}[t]{8 cm}
\epsfig{file=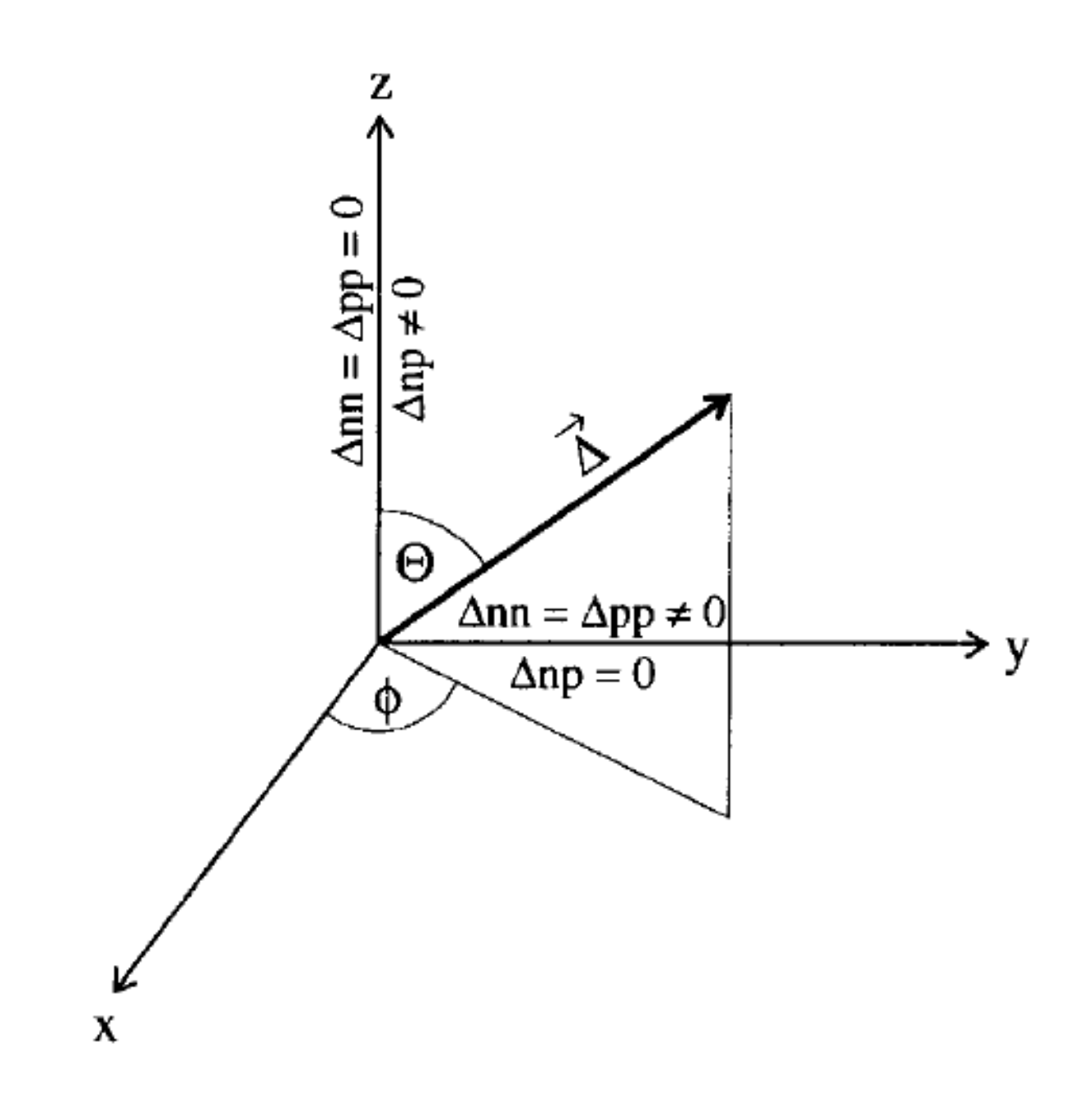,scale=0.5}
\caption{The family of mean field solutions (\ref{eq:HIVmf}) belonging to the  the isovector pair Hamiltonian (\ref{eq:HIV}). All orientations
of $\vec \Delta$ have the same energy. The fractions of $pp, nn$ , and $np$ pairs change with orientation. The special cases of no $np$ pairing 
and pure $np$ pairing  are indicated. The absolute value  $|\vec\Delta|=\sqrt{2}\Delta_{nn}=\sqrt{2}\Delta_{nn}=\Delta_{pn}$. 
  \label{fig:NPA645f2} From \cite{Frauendorf99a}.}
\end{minipage}
\hspace*{0.5cm}
\begin{minipage}[t]{8 cm}
\epsfig{file=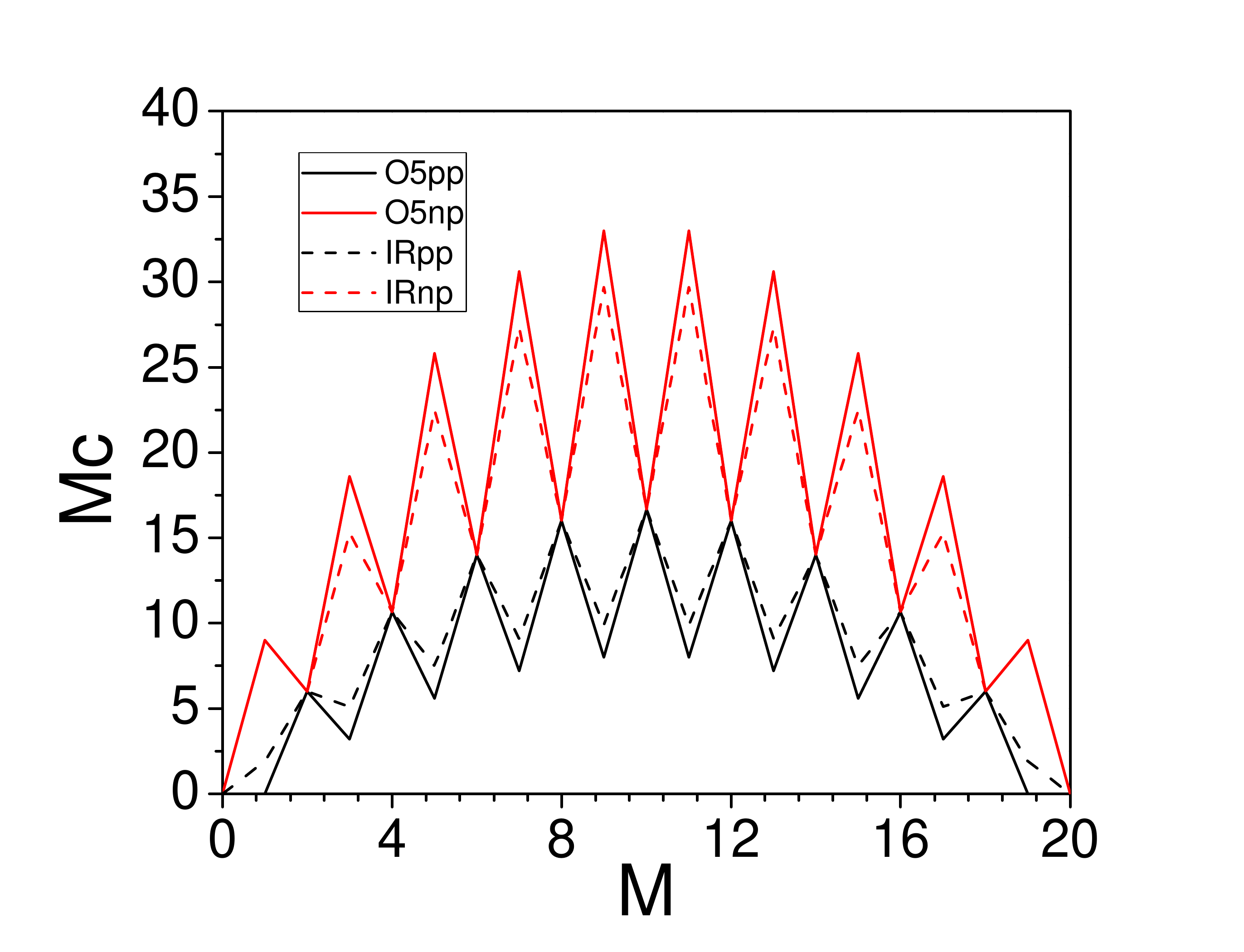,scale=0.33}
\caption{  The number of correlated isovector  pairs $Mc=\Omega \tilde{\cal N}_{T=1\mu}$ (\ref{eq:pairnumberC}) in $N=Z$ nuclei for $M$ nucleon pairs
in a $j$-shell of degeneracy $\Omega=10$. The expressions for the O(5) taken  from \cite{Engel96} are denoted by O5, and the expressions for the isospin rotor 
discussed in the text are denoted by IR.\label{fig:NeqZ}}
\end{minipage}
\end{center}
\end{figure}

\subsubsection{Intrinsic configurations}
 Frauendorf and Sheikh \cite{Frauendorf99a,Frauendorf00,Frauendorf01} further explored  the analogy with collective and intrinsic states of the shape degrees of freedom.
They suggested that the quasiparticle configurations that belong to the  mean field    
Hamiltonian (\ref{eq:HIVmf}) are  intrinsic states upon each of which an isorotational
band is built.  As in the case of spatial rotation, $E(T)=E_{int}+T(T+1)/2\theta$. The spectrum of
intrinsic excitation does not depend on the direction of $\vec\Delta$.  They suggested using the direction of the y-axis as the most convenient choice, 
because the $np$ pair field is zero, which naturally extends  the familiar description  of pairing in $N>>Z$ nuclei.
  The moment of inertia $\theta$ can be calculated by means of isocranking (see  \ref{sec:isocrank}) or can be taken from the
experimental symmetry energy. A detailed discussion of constructing the quasiparticle configurations is given in \cite{Frauendorf99a}.
The symmetries of the intrinsic states and their consequence for spatial and isotopic rotation are discussed in \cite{Frauendorf00,Frauendorf01}. 
As will be discussed in section \ref{sec:rotation}, this approach provides a description of  the excitation spectra in 
nuclides with $N\approx Z$, which is as good as for nuclides with $N > Z$.

\subsubsection{Isocranking}\label{sec:isocrank}

The appearance of isorotational sequences is a necessary consequence of the presence of an isovector pair condensate. However, isorotational
bands also emerge as a consequence of breaking isorotational invariance by the nuclear potential.  
The separation between intrinsic and isorotational degrees of freedom is only approximate and will deteriorate with increasing $T$. As
for  spatial rotation, the modification of the intrinsic quasiparticle states by isorotation can be  described in a semiclassical way by the cranking
approach, which amounts to finding the mean field solution under the constrain that the  projection of $T$ is fixed. The natural choice is the 
z-axis because only states with $T$ not much larger than $T_z$ are of relevance.  With this choice, isocranking corresponds to the standard 
particle number constraint of
mean field Hamiltonian (\ref{eq:HIVmf}), because $-\lambda_p\hat Z-\lambda_n\hat N=-\lambda\hat A-\omega T_z$ with
$\lambda=(\lambda_p+\lambda_n)/2$ and $\omega=\lambda_n-\lambda_p$. It is important to note that as soon as $\omega$ is finite, the
isovector pair field will be $\Delta_p\not=\Delta_n\not=0$ and $\Delta_{np}=0$. For $\omega=0$ all directions of $\vec\Delta$ are solutions, and 
 {the direction found may depend on the details of how the calculation is implemented}. This however does not mean that the $np$ correlations are absent. They emerge via the quantal fluctuations of the direction of the pair field, which are neglected in the cranking approach.

\begin{figure}[tb]
\begin{center}
\begin{minipage}[t]{8 cm}
\epsfig{file=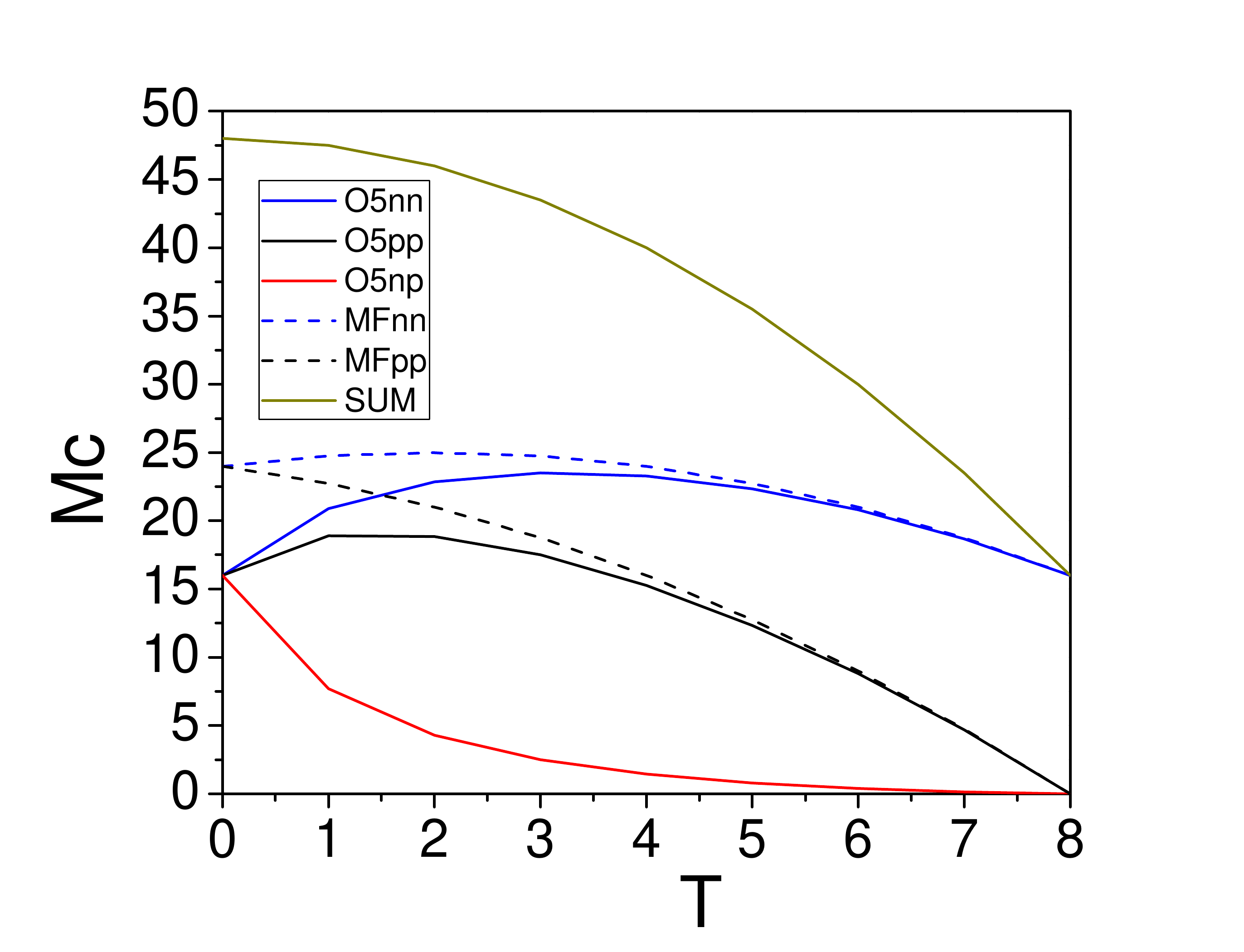,scale=0.3}
\caption{ The number of correlated isovector  pairs $Mc=\Omega \tilde{\cal N}_{T=1\mu}$ (\ref{eq:pairnumberC}) in the isobaric chain with $M=8$ nucleon pairs
in a $j$-shell of degeneracy $\Omega=10$. The expressions for the SO(5) taken  from \cite{Engel96} are labelled by O5.
The expressions for the mean field values  discussed in the text are labelled by MF. SUM denotes the sum of the components, which is the same
for SO(5) and MF. \label{fig:N8T}}
\end{minipage}
\hspace*{0.5cm}
\begin{minipage}[t]{8 cm}
\hspace*{-0.9cm}
\epsfig{file=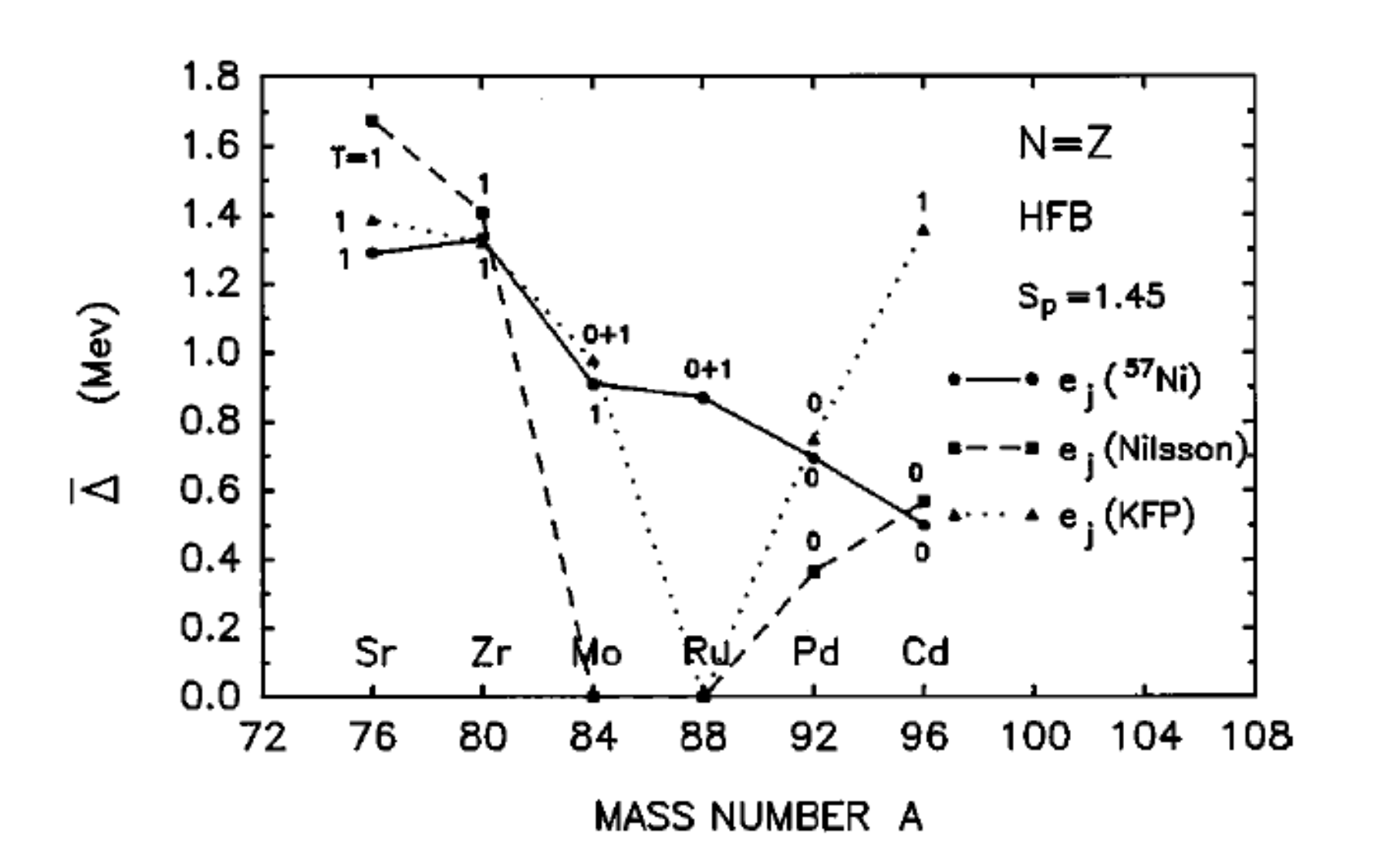,scale=0.55}
\caption{The average pair gap  $\bar\Delta$ for $N=Z$ nuclei obtained by HFB calculations using the KB3 effective interaction
and three different sets of spherical single particle levels labelled by $^{57}$Ni, Nilsson, KFP (see text). The numbers indicate the isospin of the pair field,
where 0+1 means that the isovector and isoscalar pair fields coexist. 
The $T=0$ field is of $np$ type.  Only for $A=44$ it is $p\bar n$.  
 From \cite{Goodman99}.
 \label{fig:PRC60f14}}
\end{minipage}
\end{center}
\end{figure}

Fig. \ref{fig:N8T} illustrates the situation for the one $j$-shell model. The mean field expressions    
 \be\label{eq:pairnumberMF}
 \Omega\tilde{\cal N}_{pp,nn}^{MF}= \left<BCS|P^\dagger _{\mp} P_{\mp}|BCS\right>-\left<0|P^\dagger _{\mp} P_{\mp}|0\right>=\Omega(M\mp T)(1-(M\mp T)/2\Omega)
\ee 
are given in  textbooks. The mean field correlation energy $-G\Omega\left (\tilde{\cal N}_{pp}^{MF}+  \tilde{\cal N}_{nn}^{MF}\right)$ is equal to the
exact correlation energy  $-G\Omega\sum_\mu\tilde{\cal N}_{\mu}$. Hence, the mean field approximation gives the correct correlation energy even though
it has no $np$ component. Isorotation of the pair field redistributes the correlation strength in moving it from the like particle fractions into the $np$
fraction.   Fig. \ref{fig:NeqZ} shows that the redistribution is well accounted for by an axial isorotor if $T$=0, 1.   For larger $T$ values the assumption 
of axiality is no longer a good approximation. The wave functions for the triaxial rotor, which depend on the relative values of the moments of inertia,
are not available. 

Satu\l a and Wyss \cite{Satula01,Satula01a} analyzed  cranking about the x-axis in isospace, which due to isospin invariance is equivalent to cranking about the z-axis,
however connects less directly with the standard mean field treatment of pair correlations by means of the Hamiltonian (\ref{eq:HIVmf}).  
G\l owacz {\em et al.} \cite{Glowacz04} considered cranking about an axis of arbitrary direction in isospace, and worked out in detail the mean field solutions
for different orientations of the pair field.

\subsubsection{ QRPA corrections}\label{sec:QRPA}

The one $j$-shell model captures the main features of the general situation. In the case of strong isovector correlations, the standard mean field solution
without the $np$ pair field already  gives  the total correlation energy. In realistic nuclei with a finite level spacing there is an additional
contribution from the zero point fluctuations of the pair field about its mean values. 
Due to the small number of valence particles/holes in the medium mass nuclei of relevance, one expects strong fluctuations of the pair fields around their mean values. 
These fluctuations enhance the pair correlation strength substantially. Hence the mean field values should be considered as a lower 
limit of the actual correlation strength.
The fluctuations were studied  by Ginocchio and Weneser \cite{Ginocchio68} and 
Neergard \cite{Neergard02,Neergard03,Neergard09} in the framework of the Quasiparticle Random Phase Approximation (QRPA). 
The main correction to the ground state energy was found to be generated by the QRPA spurious mode $\propto T_+$, 
which appears as a consequence of breaking the isospin symmetry by the isovector mean pair field. Its contribution changes the ground state energy 
from the isocranking value    $\left<T_z\right>^2/2\theta$ to  $\left<T_z\right>(\left<T_z\right>+1)/2\theta$. This correction is taken into account
by using the constrain $\left<T_z\right>=\sqrt{T(T+1)}$ in the isocranking prescription. Neergard also studied the contributions of the other QRPA modes, 
and obtained a $T(T+X)/2$ dependence of the ground state energies on isospin, with $X$ slightly larger than 1, the value for rigid isorotation. 
 %This  explains why the contributions from the zero point fluctuations result in a nearly constant shift of the ground state energy as obtained for the 
 %two-shell model in Fig. \ref{fig:PRC55f78}.  
 
 {
 In order to calculate physical quantities that are sensitive to the relative fractions of the three types of pair correlations one 
 needs the know the collective wave function of the pair field. QRPA should  give reliable results for large enough isospin. 
 For $T=0,1$ it gives a reliable estimate of the total fluctuation-induced correlation
 energy, however it may become problematic for estimating  physical quantities that are sensitive to differences  between the three types of pair fields.  
The comparison with the large-scale Shell Model calculations (cf. Sect. \ref{sec:SM}) and the two $j$-shell
 model (cf. Sec. \ref{sec:IV2shell}) shows that the ratios between the fractions are relatively well reproduced by the SO(5) ratios of the one $j$-shell model if there is a strong
 isovector condensate. One may combine the estimate for the total correlation strength with these well known ratios for a reliable estimate. }

\begin{table}
\begin{center}
\epsfig{file=GoodT4.pdf,scale=0.08}
\end{center}
\caption{The calculations for $N>Z$ nuclei by Wolter {\em et al.} \cite{Wolter71}. From \cite{Goodman79}.}
\end{table}
\begin{table}
\begin{center}
\epsfig{file=GoodT5.pdf,scale=0.08,angle=180}
\end{center}
\caption{The calculations for $N=Z$ ($^{44}$Ti, $^{48}$Cr) and $N>Z$ nuclei  by Wolter {\em et al.} \cite{Wolter71}. From \cite{Goodman79}}
\end{table}
\begin{table}
\begin{center}
\epsfig{file=GoodT6.pdf,scale=0.08}
\end{center}
\caption{The table summarizes calculations for $N=Z$ nuclei by Goodman {\em et al.} taken from various papers. From \cite{Goodman79}}
\end{table}

\begin{figure}[tb]
\begin{center}
\begin{minipage}[t]{8 cm}
\hspace*{-0.8cm}
\epsfig{file=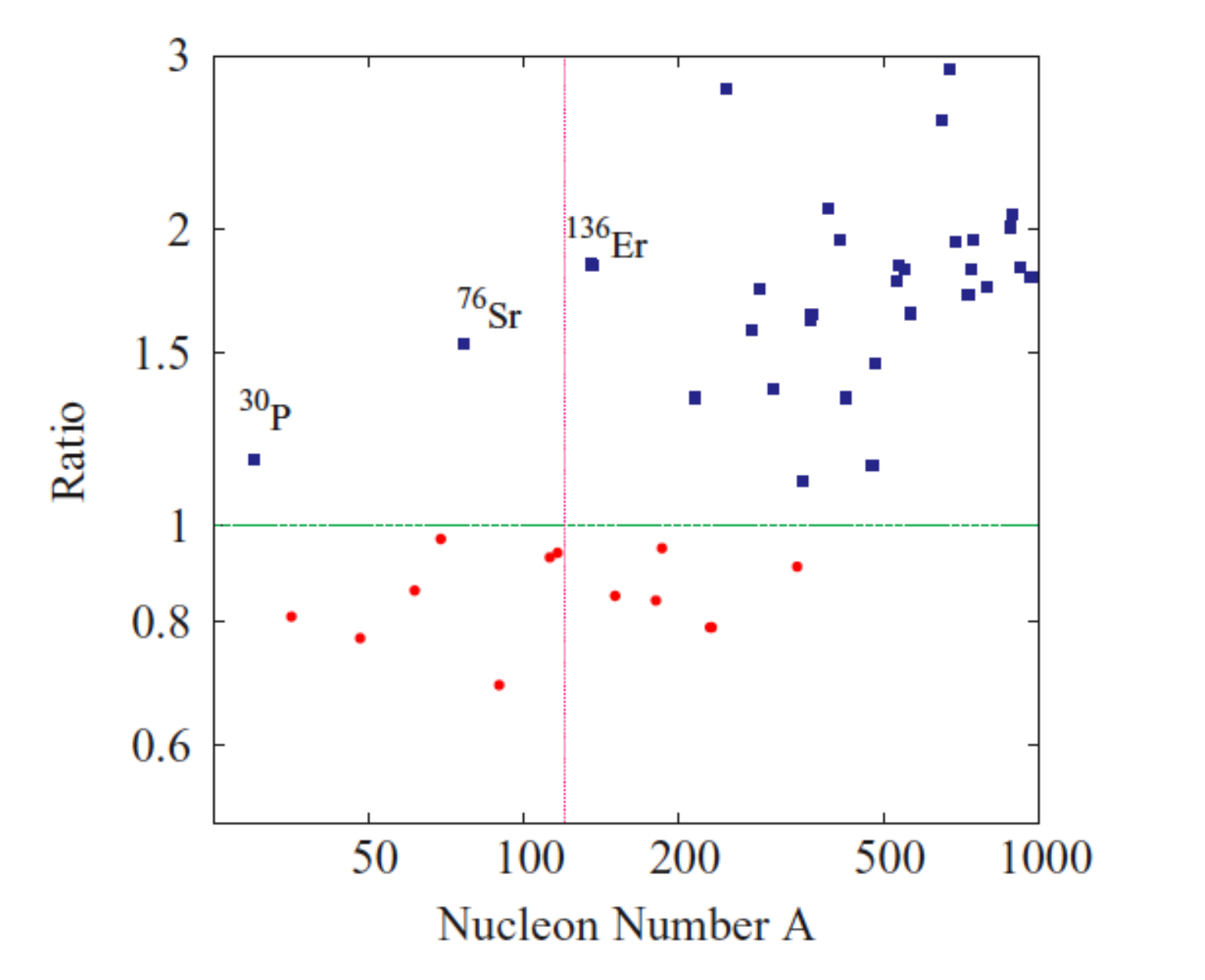,scale=0.55}
\caption{ The ratio of the isoscalar  and isovector pair correlation energies for spherical $N=Z$
nuclei assuming a ratio 3:2 for the respective interaction strengths and a Woods-Saxon 
spin-orbit potential. From \cite{Bertsch10}.   \label{fig:PRC81f2}}
\end{minipage}
\hspace*{0.5cm}
\begin{minipage}[t]{8 cm}
\hspace*{-0.9cm}
\epsfig{file=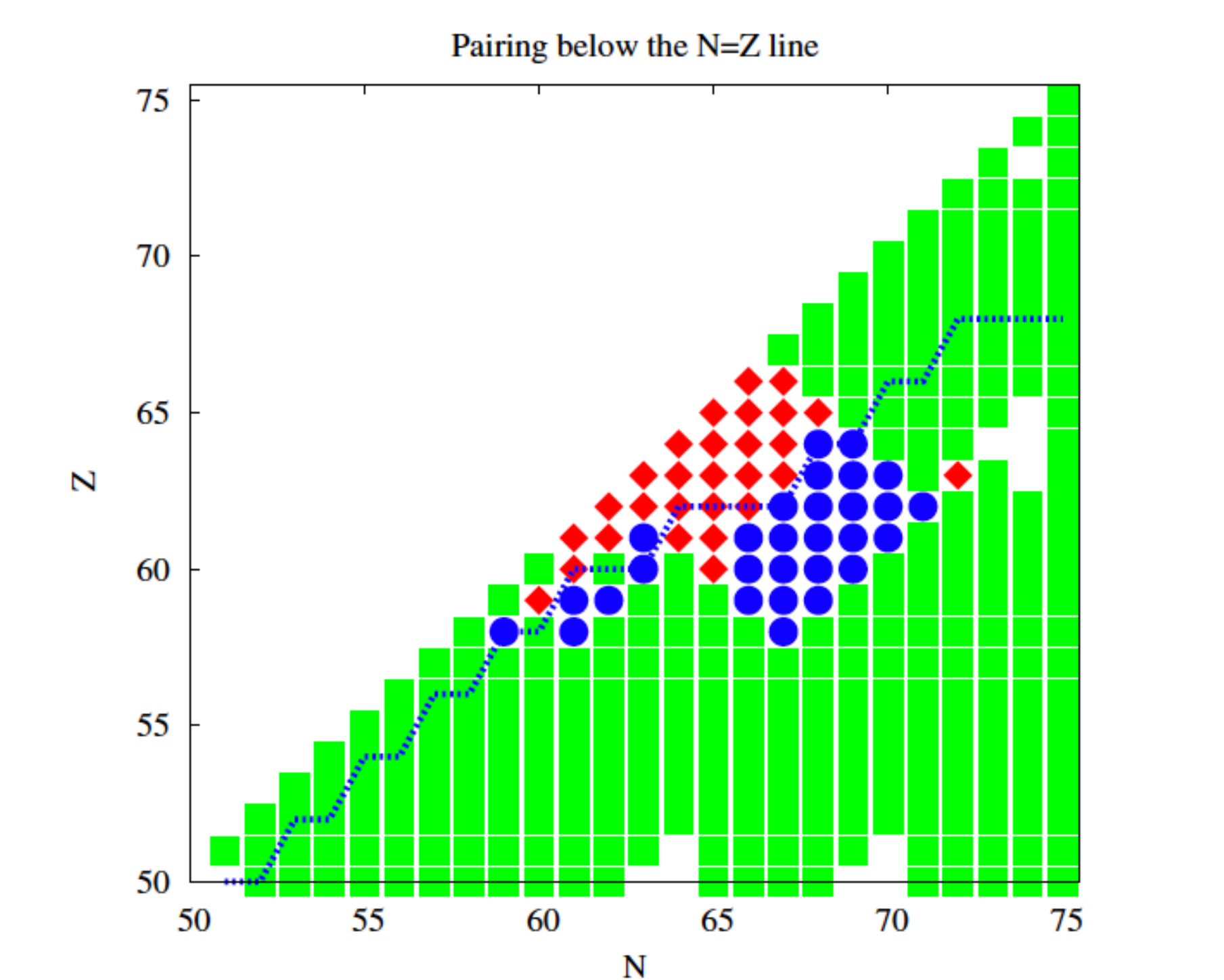,scale=0.5}
\caption{Competition between isoscalar and isovector pairing in  heavy, very proton-rich spherical nuclei,
 assuming a ratio 3:2 for the respective interaction strengths and a Woods-Saxon 
spin-orbit potential.  Green: Isovector condensate, red: isocalar condensate, blue: coexistence of both condensates.  
 From \cite{Gezerlis11}.\label{fig:PRL109f1}}
\end{minipage}
\end{center}
\end{figure}

\subsection{Mean field calculations}\label{sec:MF}
{Neutron-Proton pairing has been extensively studied in the framework of the Hartree-Fock-Boguljubov (HFB) theory. At the early stage, 
it was the method of choice because it was computationally feasible.  Nowadays, when beyond-mean field approaches have become numerically tractable,
it has retained its value. The HFB solution  represents the natural extension of the macroscopic superconducting/superfluid state when the number of
active particles becomes small. In this sense the presence or absence of a HFB pair field constitutes a demarkation line between the regimes with or without a pair condensate.
The beyond-mean field corrections induce fluctuations of the pair field, which wash out this border line. As pointed out before, we think that this is the appropriate way to
address the question: "Is there $np$ pairing?"   In the Subsection \ref{sec:HFBresults} we review the results of  HFB calculations that were based on a
realistic nuclear potential and pair interaction.   In particular the role of the spin-orbit coupling in suppressing the isoscalar pairing will be discussed. 
The HFB solutions are the starting point of various beyond-mean field approaches to be discussed in Section \ref{sec:SP}. }

Goswami \cite{Goswami64}, Camiz, Covello and Jean\cite{Camiz65,Camiz66}, and Ginocchio and Weneser \cite{Ginocchio68} extended the HFB theory by including the isovector 
$np$ pair field. Goswami and Kisslinger \cite{Goswami65} worked out the HFB equations for the case of  a pure isoscalar pair field. Chen and Goswami
\cite{Chen67} and Goodman, Struble, and Goswami \cite{Goodman68} extended the theory to coexisting $T=0$ and $T=1$ pair fields with the restriction to $N=Z$ nuclei.    
Wolter, Faessler and Sauer \cite{Wolter70,Wolter71} developed the full HFB scheme without any restriction on the pair fields and the $N$ and $Z$ values.
This pioneering and the subsequent work through 1979 have  been thoroughly reviewed by Goodman  \cite{Goodman79}.

In order to keep this review sufficiently self-contained, we start with Subsection \ref{sec:HFBtheory} that introduces  the HFB theory applied to the case of $np$
 pairing.
We follow Wolter, Faessler, and Sauer \cite{Wolter71}, who considered the isovector  and deuteron-like isoscalar pairs  discussed for the schematic models above. 
Goodman {\em et al.}  \cite{Goodman70}  extended the theory to include isoscalar pairs with the proton and neutron in identical orbitals, which they called $\alpha \alpha$ 
or $ np $ pairs in order to distinguish them
from the deuteron-like pairs, which they called $\alpha \bar \alpha$ or $ n \bar p$ pairs.   As it will be discussed in Section \ref{sec:SpinAligned}, we do not consider correlations of the $ np$ ($\alpha \alpha$)  type
as candidates for an isoscalar pair condensate.  For this reason we restrict the presentation of the HFB theory to the case of $n \bar p$ pairing. The more general version  is well
described in Goodman's paper  \cite{Goodman99}.    
Subsection \ref{sec:HFBsch} connects the general HFB scheme with the schematic models discussed in Section \ref{sec:schematic}. 

\subsubsection{HFB Theory}\label{sec:HFBtheory}

The pair fields that are taken into account by  the HFB theory are determined by the Bogoliubov transformation, which relates the nucleon creation operators $c^\dagger_k$
to the quasiparticle creation operators $a^\dagger_\alpha$,
\be\label{eq:Bog1}
a^\dagger_\alpha=\sum _{k>0} \left( A^*_{\alpha k} c^\dagger_k-B^*_{\alpha k} c_{\bar k}   \right).
\ee
Proton and  neutron operators are not explicitly distinguished. The index $k$  contains the isospin projections as well. Time reversal symmetry of the single particle orbitals
is assumed.  The summation  $k>0$ runs  over one of each time reversed orbitals (for example $m>0$ for orbitals in a spherical potential).
 The other orbital, denoted by $\bar k$, is explicitly taken into account.   Introducing column vectors $c^\dagger\equiv \left(c^\dagger_k\right)$, $\bar c\equiv \left(c_{\bar k}\right)$,
 $a^\dagger\equiv \left(a^\dagger_\alpha\right)$, and $\bar a\equiv \left(a_{\bar \alpha}\right)$, the Bogoliubov transformation can be written in compact form  
  by means of the transformation matrix $X$,
 \be\label{eq:Bog2}
\left( \begin{array}{c}a^\dagger \\ \bar a \end{array} \right)=
 \left( \begin{array}{cc} A^* & -B^* \\
B^* & A^* \end{array} \right)
\left( \begin{array}{c} c^\dagger \\ \bar c \end{array} \right)\equiv
 X\left( \begin{array}{c} c^\dagger \\ \bar c \end{array} \right).
  \ee
The HFB equation, which determine $X$, are given by
 \be\label{eq:HFBeq}
{\cal{H}}_{mf} \left( \begin{array}{c} A^* _\alpha\\ B^*_\alpha \end{array} \right)=
  \left( \begin{array}{cc} h & -\Delta \\
-\Delta & -h \end{array} \right)
\left( \begin{array}{c} A^* _\alpha\\ B^*_\alpha \end{array} \right)=
 E_\alpha \left( \begin{array}{c} A^* _\alpha\\ B^*_\alpha \end{array} \right),
  \ee  
where the quasiparticle amplitudes are arranged in column vectors  $A^*_\alpha=\left(A^*_{\alpha k}\right)$ and $B^*_\alpha=\left(B^*_{\alpha k}\right)$.
The HF Hamiltonian $h$ and the HF potential $\Gamma$ are defined as 
\be\label{eq:HFpot}
h_{kl}=T_{kl}-\lambda_p\delta_{kl}-\lambda_n\delta_{kl}+\Gamma_{kl}, ~~~\Gamma_{kl}=\sum_{m,n}V_{km,ln}\rho_{mn},
\ee
where $T$ is the kinetic energy, $\lambda_p$ and $\lambda_n$ are the chemical potentials  controlling $\langle Z\rangle$ and $\langle N\rangle$, $V$ is the antisymmetrized 
effective nucleon-nucleon interaction, and $\rho$ is the density matrix
\be\label{eq:rho}
\rho_{kl}=\langle{c^\dagger_l c_k}\rangle,~~~\rho=B^\dagger B.
\ee
The pair potential $\Delta$  and the pairing tensor $\kappa$  (pair field) are given by
\be\label{eq:kappa}
\Delta_{k\bar l}=\sum_{m>0,n>0}V_{k\bar l,\bar m n}\kappa_{n\bar m},~~~\kappa_{k\bar l}=\langle{c_k c_{\bar l}}\rangle,~~~\kappa=-A^\dagger B=-B^\dagger A
\ee
and the total energy  by
\be\label{eq:EHFB}
E=\sum_{kl}\left(h_{kl}\rho_{kl}-\frac{1}{2}\Gamma_{kl}\rho_{kl}+\frac{1}{2}\Delta_{k\bar l}\kappa_{k\bar l}\right).
\ee 
The Bogoliubov matrix can be decomposed into a product of three subsequent transformations, $X=  RBD$. The first, $D$, transforms to  the canonical 
single particle basis, which diagonalizes density matrix $\rho$ and the pairing tensor $\kappa$,
\be\label{eq:canonical}
b^\dagger_\mu=\sum_k D_{\mu k}c^\dagger_k,~~~\rho_{\mu\mu'}=\rho_{\mu}\delta_{\mu\mu'},~~~\kappa_{\bar \mu\mu'}=\kappa_{\mu}\delta_{\mu\mu'}.
\ee
The $B$ transformation, which is a special Bogoliubov transformation as used in the BCS theory, consists of 2x2 blocks for each $\mu$,
\be\label{eq:Bog3}
\left( \begin{array}{c}b^\dagger_\mu \\  b_{\bar\mu} \end{array} \right)=
 \left( \begin{array}{cc} u_\mu& -v_\mu \\
v_\mu & u_\mu \end{array} \right)
\left( \begin{array}{c} d^\dagger_\mu \\ d_{\bar \mu} \end{array} \right).
\ee
 The final transformation $R$ diagonalizes the mean field Hamiltonian ${\cal H}_{mf}$,
\be\label{eq:Bog4}
a^\dagger_\alpha=\sum_\mu R_{\alpha \mu }b^\dagger_\mu,~~~{\cal H}_{mf}=\delta_{\alpha\alpha'}\left( \begin{array}{cc} E_\alpha& 0 \\
0& -E_\alpha \end{array} \right).
\ee
The quasiparticle vacuum takes the simple BCS form when expressed by the creation operators of the canonical  basis,
\be\label{eq:vacuum}
\vert 0\rangle=\prod_\mu\left( u_\mu+v_\mu d^\dagger_\mu d^\dagger_{\bar \mu}\right)\vert 00\rangle,~~~a_\alpha\vert 0\rangle=0,~~~a_{\bar\alpha}\vert 0\rangle=0.
\ee
Finally, we specify the isospin of the pair fields,
\be\label{eq:A}
A^{TM_T \dagger}_{k\bar k'}=\sum_{\tau\tau'}\left<\frac{1}{ 2}\tau \frac{1}{ 2} \tau'\vert TM_T\right>
c^\dagger_{k,\tau}c^\dagger_{\bar k',\tau'},~~~\kappa^{TM_T}_{k\bar l}=\langle A^{TM_T \dagger}_{k\bar  l}\rangle,~~~
\Delta_{k\bar l}^{TM_T}=\sum_{m>0,n>0}V_{k\bar l,\bar m n}^T\kappa^{TM_T}_{n\bar m}.
\ee

The HFB equations in this most general form allow  coexistence of all $T,M_T$ fields. In solving, only special combinations are needed.
\renewcommand{\labelenumi}{\roman{enumi}.}
\begin{enumerate}
\item Pure isoscalar pair field $\Delta^0$, which is the  "isopairing" introduced in Ref. \cite{Goswami65}.
\item Pure isovector pair fields  $\Delta^{1\mu}$, which was introduced in Refs. \cite{Goswami64,Camiz65,Camiz66}. 
For $N\not=Z$ the solution has always $\Delta^{10}=0$ , no $np$ pair field. As discussed, for $N=Z$ isobaric invariance 
leads to infinite many equivalent solutions that are related by rotation in isospace. Two of them are $\Delta^{10}$, $\Delta^{1\pm 1}=0$,
pure isovector $np$ pairing, and  $\Delta^{10}=0$, $\Delta^{1 1}=\Delta^{1-1}$, no $np$ pair field.
\item  Coexistence of $\Delta^{1 1}$, $\Delta^{1-1}$, $\Delta^0$, where $\Delta^{10}=0$,  for all combinations of $N$and $Z$. 
\item  Coexistence of $\Delta^{1 0}$, $\Delta^{0}$ for $N=Z$, which is the "generalized pairing theory" of Refs. \cite{Chen67,Goodman68}. 
Since $\Delta^0$ is isorotational invariant, this solution represents only one of the infinite many generated by rotation in isospace. One of these  is
 $\Delta^{1 1}=\Delta^{1-1}$, $\Delta^0$, where $\Delta^{10}=0$, which is a special case  of iii) for $N=Z$.
\end{enumerate}
Case i) is realized with a real field.  There are infinite many solution generated by the gauge rotation $\exp(-i\phi \hat A)$, which multiplies  the field  with
the phase $\exp(-2i\phi)$. Case ii) is realized with two real fields. This solution corresponds to the isovector field lying in the x-y plane, where the angle with the x-axis
determines the ratio $\Delta^{1 1}/\Delta^{1-1}$. There are additional solutions by rotation about the z-axis in isospace and gauge rotation. For cases iii) and iv) one has to 
invoke complex pair fields. One solution is realized by real isovector fields and the  isoscalar field imaginary. As for case ii), there are additional solutions by rotation 
about the z-axis in isospace and gauge rotation.  
The only remaining type of solution is the possiblity that the isovector field of $N\not=Z$ nuclei has a z-component as well, i.e. all three isovector components are finite. All calculations so far
have found     $\Delta^{10}=0$ for $N\not=Z$.  Assuming that this is a general property of nuclear pairing, it is sufficient to consider case iv), which includes  i), ii), and iv) as special cases.
Then the Bogoliubov transformation to the quasiparticles has the following form \cite{Wolter71}
\bea\label{eq:Bog5}
a^\dagger_{\alpha 1}=\sum _{k>0} \left( A_{\alpha 1, pk} c^\dagger_{pk}-iA_{\alpha 1, nk} c^\dagger_{nk}-B_{\alpha1, pk} c_{p\bar k}+i B_{\alpha1, nk} c_{n\bar k}  \right),\\
a^\dagger_{\alpha 2}=\sum _{k>0} \left( -iA_{\alpha 2, pk} c^\dagger_{pk}+A_{\alpha 2, nk} c^\dagger_{nk}+iB_{\alpha2, pk} c_{p\bar k}-B_{\alpha2, nk} c_{n\bar k} \right),
\eea
where the coefficients are real. This reduces the dimension of the HFB problem. The numerical solutions are either found in the traditional way 
by iterating Eqs. (\ref{eq:HFBeq}-\ref{eq:kappa}) or by numerically minimizing the HFB energy (\ref{eq:EHFB}) with respect the coefficients $A$ and $B$ by means of the Steepest Gradient Method
\cite{Bertsch10,Gezerlis11}. 

\subsubsection{Schematic models}\label{sec:HFBsch}
For completeness we apply the general HFB scheme to the above discussed schematic models: \\

{\it Isovector pairing }\\
The model is given by the Hamiltonian (\ref{eq:HIV})
One can consider the proton and neutron parts separately.  This leads to the standard BCS  scheme for the protons,
\bea
V^1_{pk,p\bar l, p\bar m ,n}=-G\delta _{kl}\delta_{mn},~~~\kappa_{k\bar l}=\delta_{kl}u_kv_k,\\ \label{eq:BCS1}
E_{pk}=\sqrt{(\varepsilon_{pk}-\lambda_p)^2+\Delta_p^2},~~~
\begin{array}{c} u_k \\ v_k \end{array}=\frac{1}{\sqrt{2}}\left(1\pm \frac{\varepsilon_{pk}-\lambda_p}{E_{pk}}\right)\label{eq:BCS2}.
\eea 
The pair field $\Delta_p$ and the chemical potential $\lambda_p$ are determined by the gap equation and the particle number equation, respectively,
\be
\frac{1}{G}=\sum_{pk>0}\frac{1}{E_{pk}},~~~Z=2\sum_{pk>0}v^2_k.\label{eq:BCS3}
\ee 
The neutron scheme is analogous. 

For the considered case of two shells of degeneracy  $\Omega=j+1/2$, $\varepsilon_{1,2}=\pm D/2$, and $Z=2\Omega$,
one has $\lambda=0$. The gap equation becomes
\be
\frac{1}{G}=\Omega\frac{1}{E},~~~E=\sqrt{\left(\frac{D}{2}\right)^2+\Delta^2},~~~\Rightarrow~~~\Delta=\frac{1}{2}\sqrt{(2G\Omega)^2-D^2}.
\ee

{\it Isoscalar pairing}\\
The model is given by the Hamiltonian (\ref{eq:HISls}). It  suffices  to consider one of the spin projections, i. e. $S_z=0$. For $N=Z$ the
the HFB equations take the same form as Eqs. (\ref{eq:BCS1}, \ref{eq:BCS3}) if one drops the p (for proton) in the labels for the single particle states, which now
denote the identical proton and neutron states. \\ 

{\it Isoscalar and isovector pairing for a $\ell$-shell}\cite{Engel97}\\
The model is given by combining the Hamiltonians (\ref{eq:HISls}) and (\ref{eq:HIV}) and considering one $\ell$-shell of degeneracy $\Omega$. The structure of the HFB
density matrix and pairing tensor is described in    \cite{Engel97}. In accordance with the considerations in the preceding section,
the HFB solutions are of type iii). The two densities $\rho_n=N/2\Omega$ and $\rho_p=Z/2\Omega$ are fixed by the particle numbers. The energy is  expressed
in terms of these two densities and 
the three pair fields $\kappa^0$, $\kappa^{11}$. and $\kappa^{1-1}$, which are found by   numerically minimizing the energy. The minimization is constraint by 
additional  relations,  which account for the unitarity of the Bogoliubov transformation. \\

{\it Isoscalar pairing for a $j$-shell }\cite{Goswami65}\\
The spin-orbit potential enforces  $jj$ coupling. Therefore it is  relevant to consider
isoscalar pair correlations in an axially deformed $j$-shell described by the   Hamiltonian
\be\label{eq:HISjj}
 H=\sum_m\varepsilon_m \hat N(m)-G'\sum_\mu D^\dagger _\mu D_\mu,
  \ee
which is defined by the $J=1$, $T=0$ pair operators 
\be\label{eq:Djj}
D^\dagger _\mu=\frac{1}{\sqrt{2}}\sum_{m}\sqrt{j+1/2}\left<jm,j\mu-m\vert1\mu\right>\left(n^\dagger _{jm}p^\dagger _{j\mu-m}-p^\dagger _{jm}n^\dagger _{j\mu-m}\right)=\sum_{j}\sqrt{j+1/2}A^{00\dagger}_{1\mu}(j,j)\ee 
and the number operators
\be
\hat N(m)=\sum_m\left(p^\dagger _{jm}p_{jm}+n^\dagger _{jm}n_{jm}\right),
\ee
where $p^\dagger _{jm}$ creates a proton  and $n^\dagger _{jm}$ a neutron with angular momentum $j$. 
Consider the case $\mu=0$ and abbreviate $\Delta^{00}_{1\mu}=\Delta$. 
Using 
$\sqrt{j+1/2}\left<jm,j-m\vert10\right>\propto (-)^{j-m}m$,
the HFB equations take the form
\bea
V^0_{k,p\bar l, \bar m ,n}=-G\delta _{kl}\delta_{mn}(-)^{k+m}km,~~~\kappa_{k\bar l}=\delta_{kl}u_kv_k,\\ \label{eq:BCS3}
E_{k}=\sqrt{(\varepsilon_k-\lambda)^2+(k\Delta)^2},~~~
\begin{array}{c} u_k \\ v_k \end{array}=\frac{1}{\sqrt{2}}\left(1\pm \frac{\varepsilon_k-\lambda}{E_{k}}\right)\label{eq:BCS4}.
\eea
The pair field $\Delta$ and the chemical potential $\lambda$ are determined by the gap equation and the particle number equation, respectively,
\be
\frac{1}{G}=\sum_{k>0}\frac{k}{E_{k}},~~~A=4\sum_{k>0}v^2_k.\label{eq:BCS3}
\ee 
The pair potential $\Delta_k=k\Delta$ depends strongly on the angular momentum projection  projection $k$. For the case of a spherical, degenerate,
half filled  $j$-shell, the quasiparticle energies become $E_k=k\Delta$.  The energy of the lowest two-quasiparticle state will be reduced by the factor $\sim 2/j$
compared to the average two-quasiparticle energy of $j\Delta$.

 \subsubsection{Results}\label{sec:HFBresults}
The nuclei with $A<50$ and $N, ~Z ~even$ have been studied in framework of the HFB approximation starting from realistic effective nucleon-nucleon
interactions. These early studies assumed that the $T=1$ and $T=0$
pair modes are mutually exclusive, i.e. the HFB calculations were carried out allowing only either isovector or isoscalar pair fields being present. 
The early work was thoroughly reviewed by Goodman in Ref. \cite{Goodman79}. 
The results for even-even nuclides are summarized in Tab. 1-3, taken from his review.  Most $N=Z$ nuclei  have an $n \bar p$ isoscalar pair field in the ground state.
 They have excited 
 HFB solutions, which belong to different shape and have isovector pair fields or do not show pair correlations. The nuclei with $N>Z$ have isovector pair fields.
 As expected from the preceding discussions, the HFB calculations give the like-particle pair fields for nuclei with $T_z>0$ and for excited $T=1$ solutions in $N=Z$ nuclei.
 One exceptions is  $^{24}$Mg.  In the ground state, it  has correlations of the $np$ type and a triaxial shape. Only slightly higher, there is a prolate $n \bar p$ solution. 
 The other exception is $^{44}$Ti,  which has an unpaired solution lightly below an isoscalar $n \bar p$ solution. Summarizing, all even-even $N=Z$  nuclei have 
 an isoscalar pair field in the ground state or in a state  only few 100 KeV above it.

After a break of two decades, advances of the experimental techniques brought heavier $N=Z$ nuclei   
within the reach of measurements, and initiated  mean field studies of the nuclei with $A$ up to 100 alongside with the above discussed Shell Model studies. 
Goodman \cite{Goodman98,Goodman99,Goodman01} carried out HFB calculations for $N=Z$ nuclei with $76\leq A\leq 96$ based on the KB3 interaction used for SMMC.
He took into account  the possibility of  coexisting   
  $n\bar p$, isovector and $n \bar p$ isoscalar pair fields and, in addition, their coexistence with $np$ isoscalar pair correlations between particles in identical orbitals.
The generalization of the HFB equations compared to Section \ref{sec:HFBtheory} and  Ref. \cite{Goodman79} are described in Ref. \cite{Goodman99}.
The results are summarized in Fig. \ref{fig:PRC60f14}. 
 The displayed average pair gap $\bar \Delta$  is a measure of total strength of the pair field, where its type is indicated. 
 ($\bar \Delta$ is the average of the diagonal matrix elements of the pair potential in the 
 canonical single particle  basis (see Ref. \cite{Goodman99}).
 The results depend somewhat on the choice between three different sets of single particle levels. On the whole, there is a development from isovector pairing 
 for $A=$76, 80 to $np$ pair correlation for $A=$92, and 96 with coexistence of the two types for $A=$84, 86. Combining this with the early results for the lighter nuclides, 
 one concludes that the HFB calculations based on realistic effective interactions indicate for the ground states of  even-even $N=Z$ nuclei $p\bar n$ isocalar pairing
 below $A=$52, a change to isovector pairing between  $A$=52 and 76 (no calculations) and a further change to $np$ correlations around $A=$ 84, 88, which 
 are fully established for $A=$92, 96. No comparable HFB calculations for odd-odd $N=Z$ nuclei have come to our attention. Such studies should provide essential
 clues to what extend the HFB results for the even-even $N=Z$ can be taken as evidence for the existence of an isoscalar pair condensate.

Goodman studied the  rotational properties of $^{80}$Zr by including  a cranking term $-\omega j_x$ and
 allowing  for time odd components in the mean field \cite{Goodman01}. Terasaki {\em et al.} \cite{Terasaki98} carried out an analogous study of 
 $^{48}$Cr based on the Skyrme energy density functional combined with a zero-range interaction accounting for the pairing.  
 Both  investigations will be discussed in Section \ref{sec:rotation}.

%{\bf include Bertsch {\em et al.} \cite{Bertsch10b}}

Bertsch and Luo \cite{Bertsch10} investigated the competition between isoscalar ($n \bar p$) and isovector pairing in spherical $N=Z$ nuclides on a large mass range. 
They used a spherical Woods-Saxon potential and a zero-range interaction. 
The strength in the isoscalar channel was taken  3/2 times the  strength in the  isovector channel, which  is roughly the ratio 
between the channels for realistic effective interactions. Only   nucleon numbers that correspond to half-filled $j$-shells were considered.
Although allowing for coexistence, their solutions turned out to be pure isovector or pure isoscalar pair fields. Fig. \ref{fig:PRC81f2} shows the ratios of the isoscalar to isovector pair correlation 
energies. The systematics  reveal the competition between the stronger  isoscalar interaction strength and the spin-orbit splitting. The latter disfavors 
the isoscalar pair correlations because the strong pair matrix elements between a proton and a neutron in states with the same orbital angular momentum $l$, Eqs.
(\ref{eq:Dls1},\ref{eq:Dls0},\ref{eq:Dls-1}), is fragmented into matrix elements  within the multiplets $j=l\pm1/2$ and between   $j=l+1/2$ and    $j=l-1/2$
of the two the spin-orbit partners.
The energy separation between the spin-orbit partners attenuates the isoscalar pair correlations. 
The systematics is explained by realizing that the spin-orbit coupling  increases proportional to the spin angular momentum $j$ of the orbitals.
Below $A=40$ the spin-orbit splitting is small, and the isosclar correlations prevail. Above $A=40$ orbitals with progressively larger $j$  dominate the region 
around the Fermi level, which attenuates the isoscalar correlations, and the isovector correlations take over.  Above $A=300$ the fraction of low-$j$ orbtitals 
becomes larger than the one of high-$j$ orbitals, because the high-$j$ orbitals are localized in the surface region, which decreases compared to 
the interior region. As a consequence  the isoscalar correlations return.  There are exceptions from the general trend.  
In case of the nuclides $^{76}$Sr and $^{136}$Er there is a local accumulation of low-$j$ orbitals near the Fermi level, which favors  isoscalar pairing.
The multi-$j $-shell calculations give the strong reduction of the lowest two-quasiparticle energy below two times the average 
value of the gap (see Fig. 1 of \cite{Bertsch10}), which  characterizes    isoscalar pairing in single $j$-shell (see Section \ref{sec:HFBsch}).  
However, this reduction of the pair gap may be removed by nuclear deformation.  The HFB calculations of Ref. \cite{Goodman99}, which 
 allow for deformation result in  isovector pairing for the ground state of  $^{76}$Sr.  Experimentally, the long rotational band is evidence for 
 large deformation and there is no evidence  for a low-energy two-quasiparticle excitation. 
 
\begin{figure}[tb]
\begin{center}
\epsfig{file=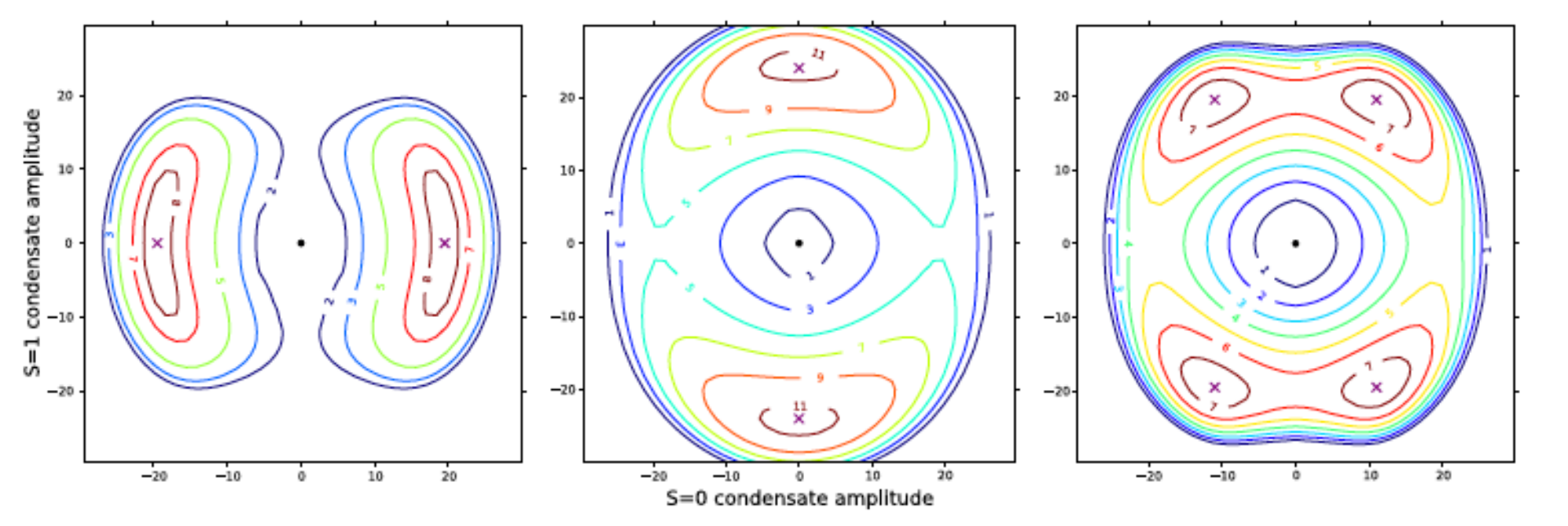,scale=0.9}
\caption{Contour plots of the correlation energy in three different $A =132$ nuclei as a function of the amplitudes of
of the isoscalar pair field $\kappa^0$ (S=1 axis) and the isovector fields $\kappa_n=\kappa_p$ (S=0 axis). 
Left panel: $^{132}_{60}$Nd$_{72}$
with dominating isovector pairing; middle panel: $^{132}_{66}$Dy$_{66}$ with dominating isoscalar pairing;
right panel: $^{132}_{64}$Gd$_{68}$ with a mixture of both pairing types. The numbers
show correlation energies in MeV. In all three cases, the maximum is marked by an X. 
 From \cite{Gezerlis11} 
 \label{fig:IVISmap}}
\end{center}
\end{figure}

Gezerlis, Bertsch, and Luo \cite{Gezerlis11} extended the study to 
nuclei with $N>Z$. Fig. \ref{fig:PRC81f2} shows their results for the  region around  $^{136}$Er, where  there is an island of isoscalar pairing near the $N=Z$ line. 
It is bordered by a region of coexistence of isoscalar and isovector pair fields. As illustrated in Fig. \ref{fig:IVISmap}, the character of the pair correlations changes
gradually with increasing $N-Z$ from isoscalar to isovector via a coexistence region. This agrees with the schematic $\ell$-shell model discussed in Section \ref{sec:2jIVISD0}.
As demonstrated by Fig. \ref{fig:PRC55f11}, for $N>Z$ the isovector and isoscalar fields coexists in the region $G_{IS}>G_{IV}$, where the isovector field is
gradually quenched with increasing $G_{IS}$.   The softness of  the potential energy surface in Fig.   \ref{fig:PRC55f11} suggests that for solutions with a pure isovector
field  the isoscalar field should be present  as a soft collective vibration.  One would expect that this softness with respect to the two pair modes is a general feature.
This is consistent with the  results  the Shell Model calculations discussed in Section \ref{sec:SM}, which indicate the presence of noticeable  isoscalar pair correlations
on top of the dominating isovector correlations (see Fig. \ref{fig:NPA631f1}).

The mean field studies show a scenario of delicate balance between isoscalar and isovector pair fields. The
$N=Z$ nuclei below about mass 40 have an isoscalar condensate in the ground state. Nuclides with  $N>Z$ have an isovector ($p\bar p$, $n\bar n$) condensate.
 The  question whether there is a mixed phase  has not been addressed in the existing calculations for these nuclei.  
 Above about mass 40  the ground states of $N=Z$ nuclei have an isovector condensate, because the progressive spin-orbit splitting
 between the orbitals disfavors the isoscalar correlations. This general trend is modulated by shell structure.  In the nuclides around $^{76}$Sr and $^{136}$Er,\
 there is  an accumulation of low-$j$ orbitals 
 near the Fermi surface. The smaller spin-orbit splitting of the low-$j$ orbitals let the isoscalar field win over the isovector field in the
 calculations of Refs. \cite{Bertsch10,Gezerlis11}, which assume a spherical shape. However, nuclear deformation may tip the balance in favor of isovector
 pairing.  In general, one expects 
that the channel which does not develop a condensate will still be present in the form of strong dynamical correlation.  
The strongest isoscalar correlations, be they static,  mixed, or only dynamic,  are expected in the regions around  $^{76}$Sr and $^{136}$Er.

\subsection{Mean field approaches with partial symmetry restoration}\label{sec:SP}

Chen, M\"uhter and Faessler \cite{Chen78} and Raduta {\it et al.} \cite{Raduta00,Raduta01,Raduta00a,Raduta03,Raduta12} worked out the techniques for
projecting the mean field solution onto good particle number and isospin. Ref. \cite{Raduta12} suggested an interesting new approach, which replaces the 
standard projection integrals over the angles in isospace and gauge space by  recursion relations, which may present a new efficient way for numerical 
evaluation of energies and matrix elements.  This refers not only to the pairing equations but also to analytical expressions for $np$ transfer
reactions and double beta decay transition matrix element.
The authors applied their methods to one $j$-shell and two $j$-shells with the isovector interaction (\ref{eq:HIV}) and
demonstrated that variation of the mean field after projection reproduced the exact energies. In particular, Ref. \cite{Raduta03}
reproduced the energy relations between the single and double analog states in the Ti isotopes, which are observed in experiment.

Petrovici, Schmid, and Faessler \cite{Petrovici96,Petrovici99,Petrovici00,Petrovici02a,Petrovici02b,Petrovici03} studied the competition between isoscalar and 
isovector pairing in the framework of their "Complex
Excited Vampir" variational approach.   The trial wave function is a  deformed quasiparticle configuration which is projected on good
particle numbers, $N$, $Z$, and good angular momentum. The quasiparticles are defined by a complex valued Bogoliubov transformation, which 
ensures the possibility of coexisting isoscalar and isovector pair fields (c.f. \cite{Goodman79},\cite{Schmidt87}).
The coefficients of the transformation are found by minimizing the energy  of the symmetry-projected trial wave function. 
A number of excited states is found in the same way, and the Hamiltonian is diagonalized within this non-orthogonal basis.
Two shells above $N=Z=20$ are used as configuration space. The Hamiltonian is a renormalized G-matrix derived from
the Bonn-A interaction. It is modified by phenomenological correction terms of Gaussian type, which control the strength of the paring interaction.
The isovector corrections are not isospin invariant, however the deviations are moderate. 
The composition of the pair correlations in the nuclides $^{66}$As, $^{62,64}$Ge, $^{74}$Rb, and $^{72,74}$Kr
were studied in \cite{Petrovici99}.
The information about the pair fields of the lowest basis states is presented in the form of "average pair gaps", which are the trace of the pair potential divided by
 the dimension of the singe particle basis.  The pair fields have a finite isovector $np$ component. This is in contrast to pure mean field calculations based on an
 isospin invariant Hamiltonian. As discussed,  the corresponding mean field solution for $N>Z$ has always a zero  isovector $np$ component. 
 It is not clear whether the isospin dependent part of the Hamiltonian or the restoration of good particle number generate the isovector $np$ component.
 The relative strength of the three types isovector pair correlations will be redistributed by restoration of (approximately) good isospin. Coexisting, an isoscalar pair
  field is found, which has an average pair gap of about 1/3 of the mean value isovector components. A major focus of the studies was the coexistence of
  prolate and oblate shapes in $N\approx Z$ nuclei with $56\leq A \leq 84$. Their investigations of the rotational response will be 
  discussed in Section \ref{sec:rotation}
  
 In a series of papers, Chasman \cite{Chasman02, Chasman03a, Chasman03b, Chasman10} investigated the pair correlations by means of a version of the 
 model Hamiltonian (\ref{eqn:IVIS}).
 At variance with \cite{Bentley13}, he did not include the  term  $\kappa \vec T\cdot\vec T/2$, assumed  $G_V=G_S$, 
 and multiplied all non-diagonal matrix elements  by a factor of 0.5. 
 Instead of diagonalization, he optimized the trial wave function 
 \bea\label{eq:chasman}
 \langle \Psi \vert H_P\vert\Psi \rangle\rightarrow min,~~~~~~~~~~~~~~~~~~~~~~~~~~~~\Psi=P(N,Z,Q)\times~~~~~~~~~~~~~~~~~~~~~~~~~~\nonumber\\
 \prod_k\left[1+U(1,k)P^\dagger _{1,k}+U(2,k)P^\dagger _{-1,k}+U(3,k)P^\dagger _{0,k}+
 U(4,k)iD^\dagger _{k}+U(5,k)\left(p^\dagger _{k}p^\dagger _{\bar k}n^\dagger _{k}n^\dagger _{\bar k}\right)\right]
 \eea
 with respect to the amplitudes $U(\nu,k)$, where $P(N,Z,Q)$ projects on exact particle numbers $Z$ and $N$ and the parity $Q$ of the number of 
 the $T=0$  pairs $D^\dagger _{k}$.   He constructed excited states by blocking single particle levels, which correspond to the quasisparticle excitations in the mean
 field approximation. In addition, he included collective pair modes in the framework of the Generator Coordinate Method.  The Hamiltonian $H_P$ was
 diagonalized  within the discrete set    
$\Psi_i$ of the type (\ref{eq:chasman}), which were generated by modifying artificially the interaction strength of the various terms in $H_P$.   
Similar to the Vampir  solutions of Ref. \cite{Petrovici99}, he found the isovector $np$ correlations comparable with the like-particle ones.
The isoscalar correlations, which are of the $n\bar p$ type, were found to be as strong as the isovector ones, 
which is probably a consequence of the ad hoc assumption $G_{IV}=G_{IS}$.
More details will be given in connection with the Wigner energy in Sec. \ref{sec:Wigner}. 

Satu\l a and Wyss \cite{Satula97, Satula00, Satula01, Satula01a} studied the competition between the isovector pair modes  and the isoscalar pair mode  of the $np$ type.
They used a model Hamiltonian of the form  (\ref{eqn:IVIS}) without the  term  $\kappa \vec T\cdot\vec T/2$. 
 The single particle energies were calculated by means of a deformed Woods-Saxon potential.
At variance with Eq. (\ref{eq:Dk}), the isoscalar interaction was a product  pair operators of the $np$ type $D_k^\dagger =n^\dagger _kp^\dagger _k$. This type of correlations is favored by 
rotation, because the angular momenta of the proton and neutron add up. The consequences for the rotational response were studied by adding a cranking term 
$-\omega j_x$ to  (\ref{eqn:IVIS}). The many body states were mean field solutions generated by a complex valued Bogoliubov transformation, subject to
the standard self-consistency conditions, the standard  constraints on the expectation values of $N$ and $Z$, and   the Lipkin-Nogami (LN)  constraints on the particle number 
fluctuations. In the case of $N=Z$ nuclei,  the standard mean field solution without the LN correction corresponds to a pure isovector field for $G_{IS}/G_{IV}<1$ and to a pure isoscalar field  for  $G_{IS}/G_{IV}>1$. The LN correction shifts the transition to the ratio of 1.1 and it leads to coexistence of the $T=0$ and $T=1$ fields 
above the critical ratio. Isospin was restored in an approximate way by the isocranking approximation with the constraint $\left<T_x\right>=\sqrt{T(T+1} $ (see Sections \ref{sec:isocrank} and \ref{sec:QRPA}). 
The results of Satu\l a and Wyss for the binding energies and the moments of inertia for spatial rotation will be discussed in sections 
\ref{sec:Wigner} and \ref{sec:rotation}, respectively. 

Sun {\em et al.} \cite{Hara99,Velaquez01,Sun01,Palit01,Sun04} used the Projected Shell Model (PSM) to study the rotational excitations of $N\approx Z$ nuclei. 
Their approach is based on the PQQ Hamiltonian (cf. Eq. (\ref{eq:PQQ})) without the $np$ terms of the monopole
and quadrupole pair interactions.
They start with the mean field solution, which is the standard deformed BCS vacuum. The pair gaps $\Delta_n$ and $\Delta_P$ are taken from experimental systematics, and  the quadrupole deformation is treated as a parameter, which is adjusted  the experiment. The ratio $G_Q/G_M$ is taken as  0.16-0.20. The three 
coupling constants are determined by the self-consistency conditions for the mean field. A set of states is generated by exciting few quasiparticles, which together with   
the vacuum state are  projected onto good angular momentum $I$. The Hamiltonian is diagonalized within this non-orthogonal basis. Concerning 
pairing, this approach remains on the mean field level. The authors aimed at the description  of the rotational behavior, which will be discussed in Section \ref{sec:rotation}.

\section{Binding energies}\label{sec:Wigner}

\subsection{Phenomenology}\label{sec:Wphen}

%Start AOM 11/4/2013

Guided by studies of  $nn$ and $pp$ pairing,  it is not surprising that the search for experimental signatures  of $pn$ pairing
has been centered around effects in binding energy differences.    We first follow a
phenomenological approach as introduced early by Janecke \cite{JAN}, Zeldes and Liran \cite{Zeldes}, and more recently by
Vogel \cite{Vogel} and Macchiavelli {\it et al.}\cite{aom1}.
%\subsection{Qualitative Pair Blocking Arguments}
Let us start by considering the simple case depicted in Fig.~\ref{fig:IVISBE1}. 
For a situation where the ground state of a nucleus of mass A can be considered
to be a condensate of nucleon pairs, adding further pairs will result in a
smooth change in binding energy if these pairs  can ``take part'' in the 
correlations already present in nucleus A. 
The gain in energy is proportional to the
number of pairs. If the additional nucleon pairs are such that they
cannot take part in the correlation, then there is no gain in
energy. In fact, these ``unpaired'' nucleons serve to reduce the pairing
energy by removing those nucleon orbitals from the pairing
correlations. This is the phenomenon of pair blocking:  All else 
being equal, the reduced pairing energy is reflected in a reduced binding energy. 
By studying the change of the observed binding energies caused by adding
 various nucleon pairs  (e.g. $nn$, $pp$, or $np$) it is not only possible
to determine whether that nucleus contains strong pairing correlations but 
also to determine the type of nucleon pairs contributing to this correlation.

\begin{figure}[tb]
\begin{center}
%\begin{minipage}[t]{8 cm}
\epsfig{file=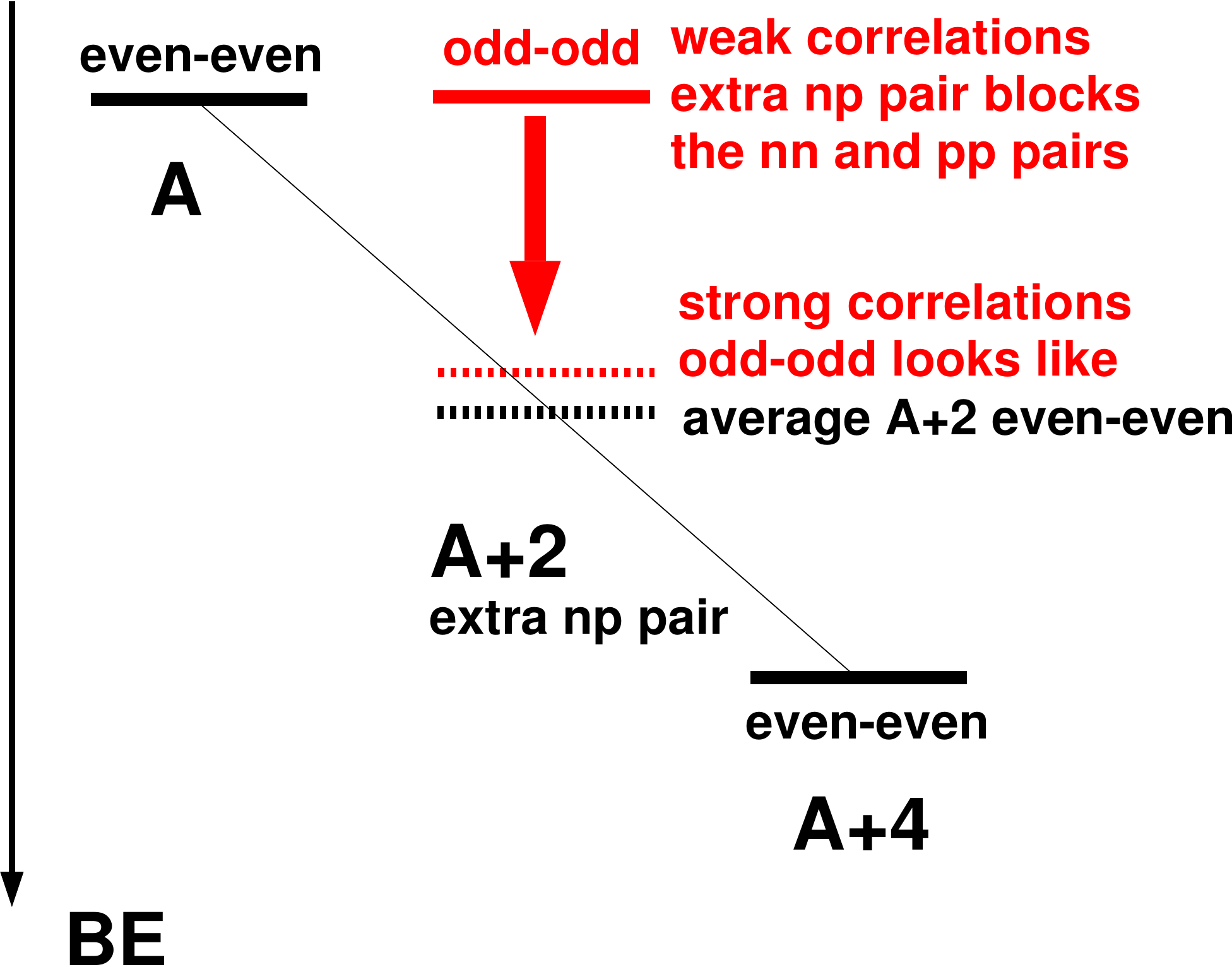,scale=0.5}
%\end{minipage}
%\begin{minipage}[t]{16.5 cm}
\caption{Schematic diagram showing the relative energies of the ground states of two even-even nuclei along the $N=Z$ line, and the expected position of the an odd-odd nucleus  with
strong and weak isoscalar pairing correlations.\label{fig:IVISBE1}}
%\end{minipage}
\end{center}
\end{figure}

\begin{figure}[tb]
\begin{center}
%\begin{minipage}[t]{8 cm}
\epsfig{file=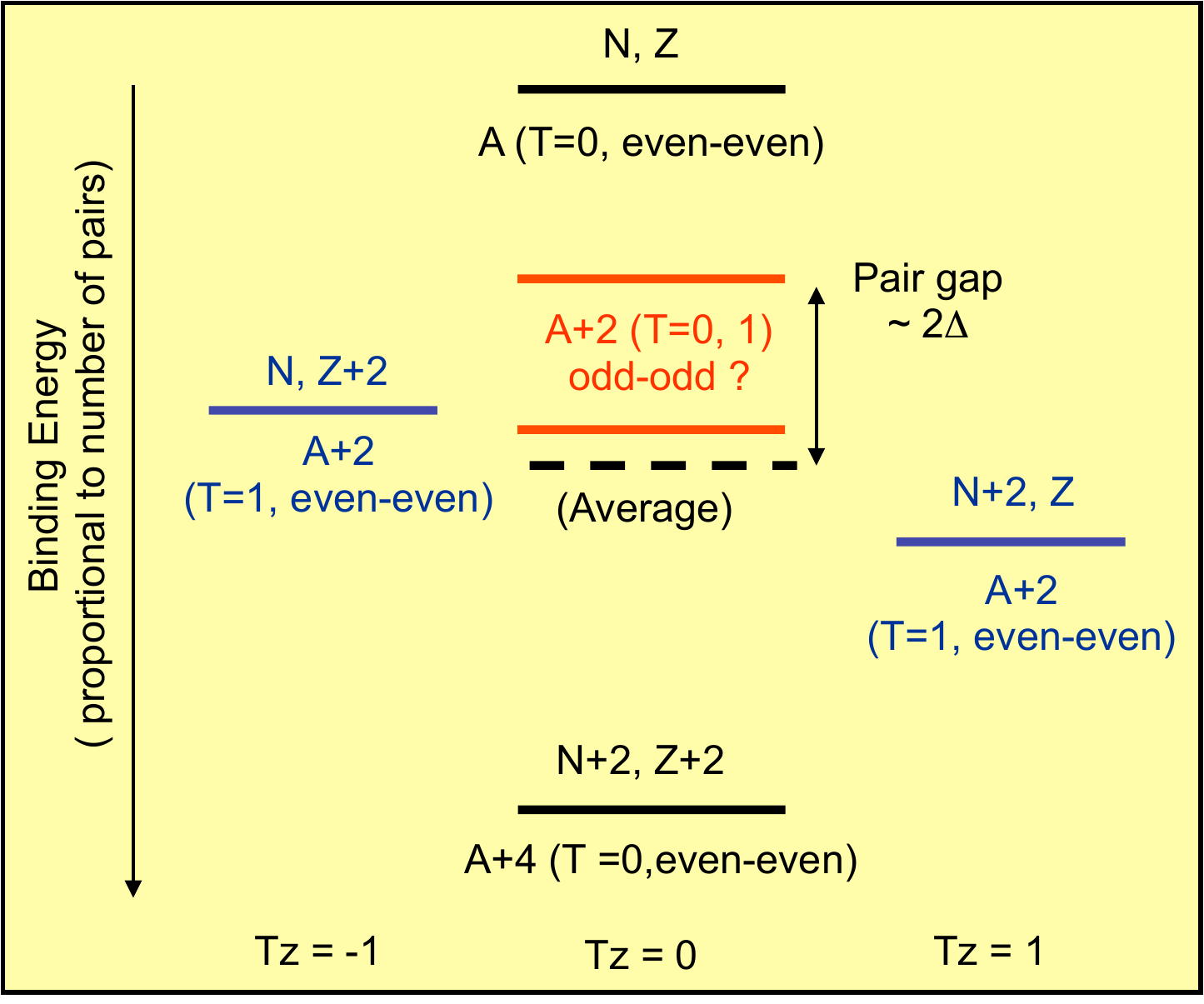,scale=0.5}
%\end{minipage}
%\begin{minipage}[t]{16.5 cm}
\caption{The nuclei involved in the evaluation of the "average" even-even binding energy
 used in Eq. (\ref{eq:BE1}), which takes into account the different isospins.\label{fig:IVISBE}}
%\end{minipage}
\end{center}
\end{figure}

Away from $N=Z$, pairing is dominated by separate $nn$
and $pp$ correlations. The addition of a neutron and a proton to an
even-even nucleus with mass $A$ leads to the well known reduction in
the binding energy of the odd-odd nucleus from that given by the smooth
average of the neighboring even-even nuclei. See  
the solid line labeled odd-odd in Fig.~\ref{fig:IVISBE1}.
The difference in binding energy 
\be\label{eq:BE1}
BE_{<ee>}-BE_{oo} \approx \Delta_p + \Delta_n \approx 2\Delta
\ee
provides a measure of the pair gap, $\Delta$, for both protons and
neutrons: $BE_{<ee>}$ is the average for even-even nuclei
with ($N$, $Z$) and ($N$+2, $Z$+2), and $BE_{oo}$ is for the nucleus ($N$+1, $Z$+1).
For the example being discussed, the $pp$ and $nn$ pairs couple to spin $J=0$
and $T=1$; that is, the even-even nucleus has total spin $J=0$ and seniority
$\nu = 0$.  The state with seniority $\nu = 2$ (a broken pair) will 
be less bound than the ground state by the amount of $2\Delta$.

Let us now restrict ourselves to $N=Z$ nuclei and ask what happens when
we add an $np$ pair to the even-even nucleus $A$. In this case
the near degeneracy of the neutron and proton orbitals near the Fermi surface
allows the possibility for correlated $np$ pairs which
can couple to isospin $T=0$ or $T=1$. In the following we consider the effect
of both isospin couplings on the binding energy of an even-even $N=Z$ core with
 (i) pure $T=0$ pairing, (ii) pure $T=1$ pairing, and (iii) mixed $T=0$  and $T=1$
pairing.  We note that at $N=Z$,  the application of Eq. (\ref{eq:BE1}) is based on combinations of the  nuclei shown in Fig. \ref{fig:IVISBE},  in order 
to define the average even-even reference according to the isospin.

\begin{itemize} 

\item For pure $T=0$ pairing (i.e.~a ``condensate'' of
deuteron-like $T=0$ pairs), the $T=0$ state of the odd-odd $N=Z$ nucleus
with $A+2$ is expected to have a binding energy equal to that
of the average of its even-even neighbors (dashed line labeled odd-odd state in
Fig.~\ref{fig:IVISBE1}, see also Section \ref{sec:ISsym}).  The extra $np$ pair, which is coupled to $T=0$, can take part
in the correlations already present in the even-even core, and  the
difference in binding energy given in eq. (\ref{eq:BE1}) is zero.  On the other hand, if
the $np$ pair were coupled to $T=1$, then it could not
contribute to the "condensate" of $T=0$ pairs, which are assumed to be present in
the even-even core, and its presence would block pairing correlations 
leading to a pair gap of $2\Delta$, as given in Eq. (\ref{eq:BE1}).

\item If only $T=1$ pairing were present, then the
addition of the $T=0$ quasi-deuteron blocks the $T=1$ pairs in the
even-even condensate in the same way as any other odd-odd nucleus and
eq. (\ref{eq:BE1}) now gives the isovector pair gap. Adding a $T=1$ $np$ pair would
not disrupt the pair correlations since it can ``blend-in'' with the existing
$T=1$ pairs, and the $T=1$ state in the odd-odd nucleus would show 
zero binding energy difference from Eq. (\ref{eq:BE1}).

\item When both $T=1$ and $T=0$ pairing correlations are equally favored
(SU(4) limit) the binding energy difference in Eq. (\ref{eq:BE1})
would be essentially zero for both the $T=1$ and $T=0$ states. This is
because in both cases, whether $T=1$ or $T=0$, the core already contains
pairing correlations of that character. There is no odd-odd even-even staggering.

\end{itemize}

A direct consequence of these simple arguments is that the observed
gaps given by Eq. (\ref{eq:BE1}) are a maximum when one type of correlation
dominates; mixing in, for example, $T=0$ pairing correlations to a core
which has a $T=1$ condensate will reduce the observed pair gap, not
increase it. From the charge independence of the nuclear force, the
$T=1$ pair gap at $N=Z$ cannot be greater than that found in any other
nucleus away from $N=Z$.

In light of the blocking arguments presented above, it appears
possible to probe the nature of the ground state pairing correlations in
$N=Z$ nuclei by comparing the experimental binding energies 
in a manner consistent with Eq. (\ref{eq:BE1}).
In the top left of Fig.~\ref{fig:sumB} we show the binding energy differences between the
ground state of an 
even-even $N=Z$ nucleus and
the lowest $T=0$ states  in the neighboring odd-odd $N=Z$ nuclei. 
 The extra binding of the
even-even nuclei is clearly seen  and follows remarkably
well the known $1/A^{1/2}$  dependence (solid blue curve)\cite{BM}.
It  is then  natural to associate the observed binding energy
differences between the lowest $T=0$ states in $N=Z$ nuclei with the
full $T=1$ pair gap.
The $T=0$ states in odd-odd $N=Z$ nuclei then behave
like those in any other odd-odd nucleus where the extra $n$ and $p$
block the $T=1$ pairing to the same degree as any ``standard''
two quasiparticle state.  
Therefore, the data in Fig.~\ref{fig:sumB} suggest that $T=0$
pairs do not contribute significantly to the pair correlation energy
in $N=Z$ nuclei \cite{aom1}. 

\begin{figure}
\centerline{\psfig{file=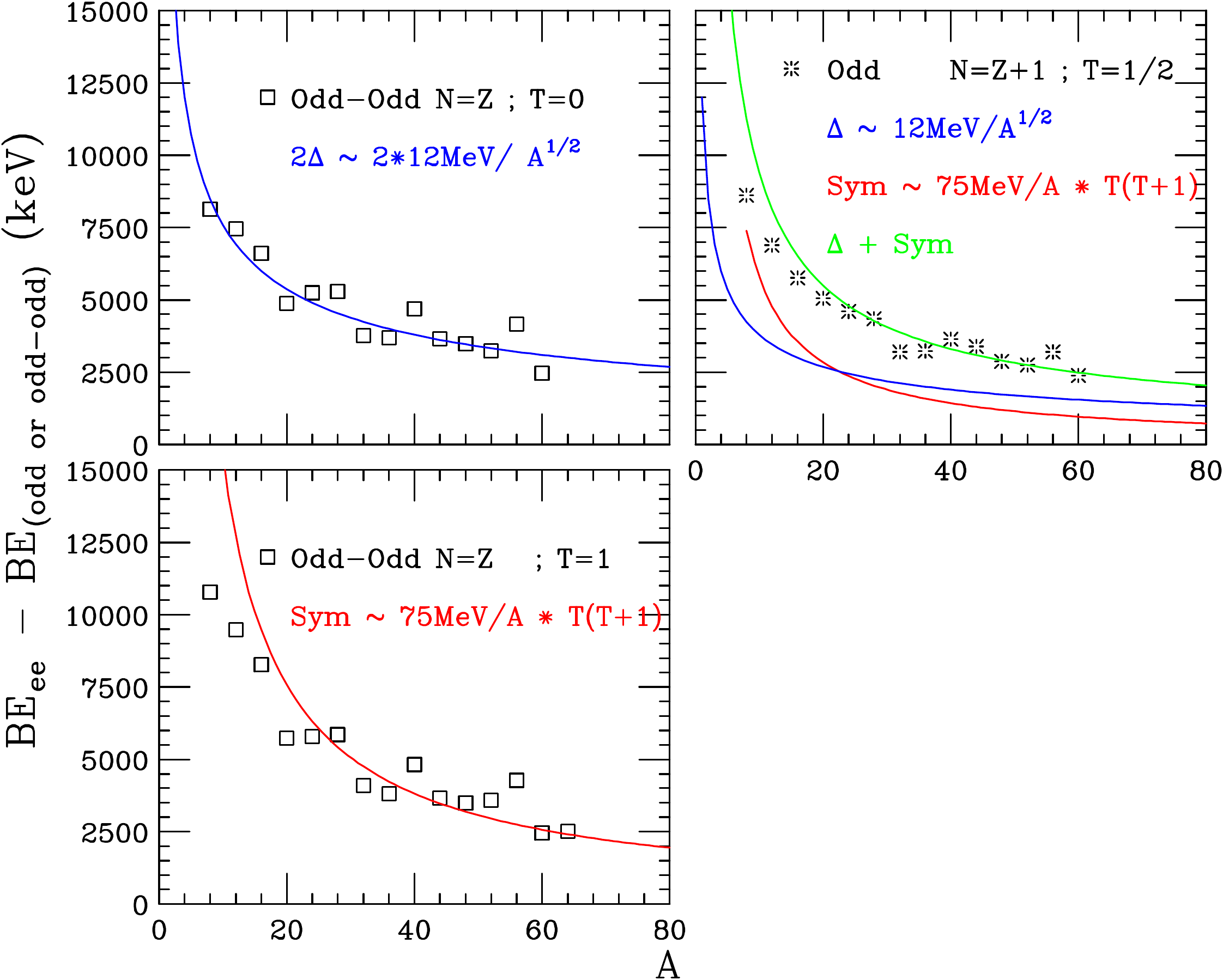,scale=0.6, angle=90}}
\caption{ Summary of binding energy differences for nuclei along the $N=Z$ line.  }
\label{fig:sumB}
\end{figure}

Let us now turn our attention to the $T=1$ states shown at the bottom left of Fig.~\ref{fig:sumB}. 
Because the even-even reference core has isospin $T=0$, one has to correct Eq. (\ref{eq:BE1})
for the known isospin
contribution to the smooth binding energy.
The $T=1$ states in odd-odd nuclei must include a bulk symmetry energy term  \cite{JAN,BM, Duflo}
\be\label{eq:Esym}
 E_S(T)=\frac{75MeV}{A}T(T+1),
 \ee
  and Eq. (\ref{eq:BE1}) should be modified to
\begin{eqnarray}\label{eq:BE2}
BE_{ee}-BE_{oo} \approx 2\Delta_{T=0} + \frac{75MeV}{A}1(1+1).
\end{eqnarray}  
In applying Eq. (\ref{eq:BE2}) to the data, we see that all of the binding energy
 difference is accounted for by the symmetry term (solid red line), and  there is no much room 
 to associate with an isoscalar pairing gap (i.e. $2\Delta_{T=0} \approx 0$).
The $T=1$ state in the odd-odd 
$N=Z$ nucleus appears as bound as the neighboring even-even $N=Z$ ground state, 
and just as the addition of an $nn$ or $pp$ pair to an even-even nucleus
does not block pair correlations, neither does the addition of an $np$ $T=1$ 
pair in $N=Z$ nuclei. The near degeneracy (after correcting for Coulomb effects) of the 
$T=1$ isobaric analogs
with $T_z = -1, 0, 1$ clearly shows that it does
not matter whether a $pp$, $np$, or $nn$ $T=1$ pair is added to the 
$N=Z$ core -- neither results in a blocked state.   This proves the existence of $T=1$ pairing correlations in $N=Z$ \cite{aom1,Vogel},
which in view of the charge-independence of the nuclear force is not surprising. 

Finally the single-particle $T=1/2$ states in the odd systems, 
(top right of  Fig.~\ref{fig:sumB}), complete the picture elaborated above. The binding energy difference now consists of two parts
of approximately equal magnitude
\begin{eqnarray}
BE_{ee}-BE_{o} \approx \Delta_{T=1} + \frac{75MeV}{A}\frac{1}{2}(\frac{1}{2} +1)
\end{eqnarray}  
one related to the isovector pairing (blue line) and the other to the symmetry term (red line), that reproduce the data remarkably well (green line).

Within this phenomenologically based picture,  the most striking result is perhaps the natural explanation of the competition between  $T=0$ and  $T=1$ 
as ground states  in the odd-odd nuclei \cite{Zeldes,aom1} shown in Fig.~\ref{fig:deltaoo}.  
Due to the symmetry energy,  the  $0^+$ states should always be excited  (red line).  
It is only because of the $T=1$ pairing correlations that they get pushed down by $\approx 2\Delta_{T=1}$ (blue line) and become the ground states (green line).  
Near doubly-closed shells, $^{40}$Ca and $^{56}$Ni, where static pair correlations are quenched, the  gain in energy
is not large enough to revert the ground states to  $T=1$. It is interesting to speculate that a similar situation may be encountered again near $^{100}$Sn.
As summarized in Table \ref{tab:summaryoo}, the ground state energies of  nuclei near N=Z  can be estimated by the phenomenological analysis 
 in terms of only two ingredients, namely the isovector gap $\Delta_{T=1}$ and the symmetry term $E_S(T)$.

\begin{figure}
\centerline{\psfig{file=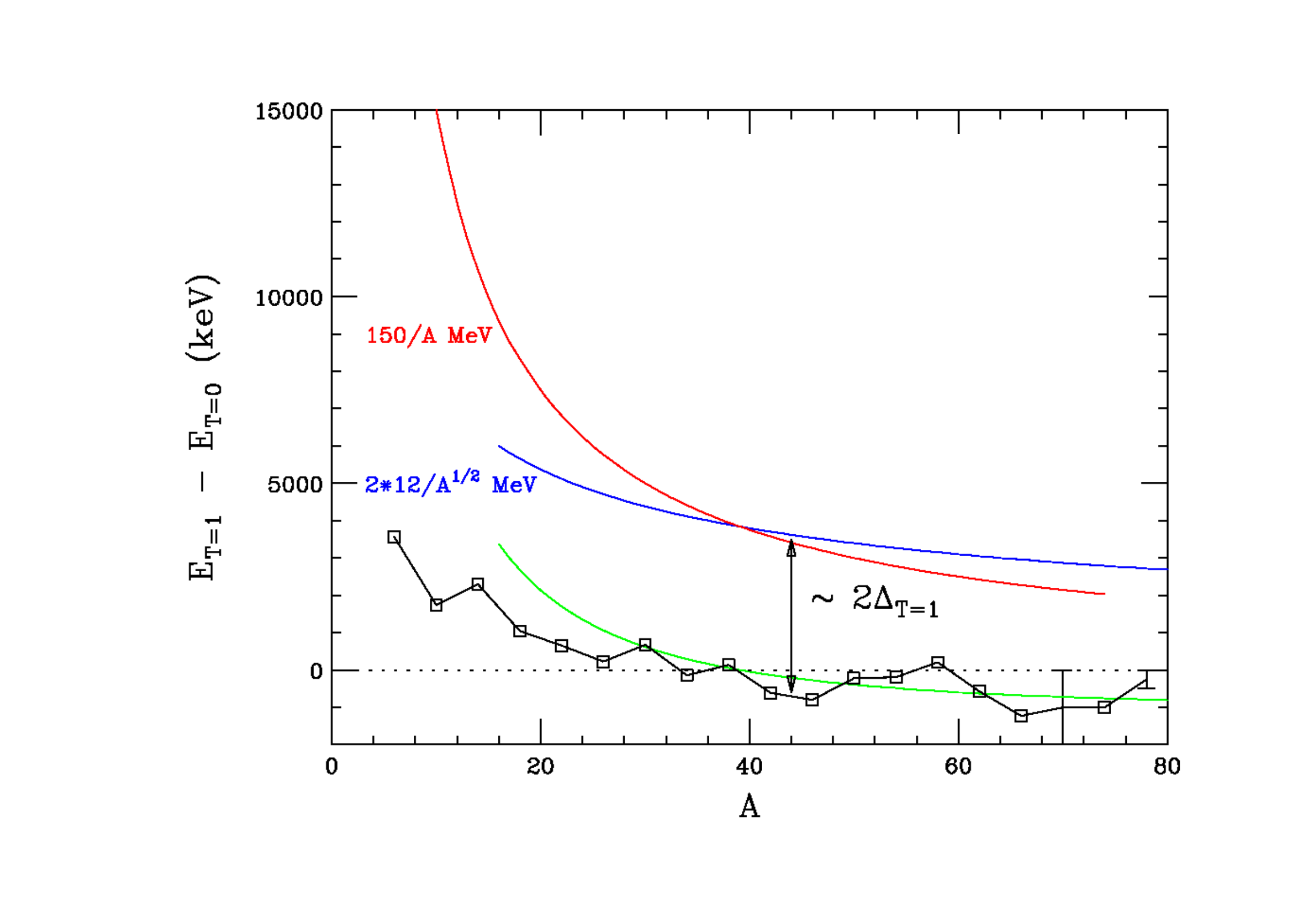,width=10cm}}
\caption{ Energy difference between the lowest $T=0$ and $T=1$ states in odd-odd $N=Z$ nuclei.  }
\label{fig:deltaoo}
\end{figure}

% qualitative arguments discussed  above are borne out by the results of 
%the simple models introduced in subsection \ref{sec:IV2shell}.

At this point it is important to explicitly mention the so-called  "Wigner Energy", which refers to the
"apparent" enhancement of the binding energy near $N=Z$.
 In the framework of Wigner's super multiplet theory \cite{Wigner37}  (SU(4), see Section \ref{sec:1jIVIS}) the energy increases as $T(T+4)/2\theta$.
 The term $2T/\theta=\vert N-Z\vert/\theta$ pushes up  the $N>Z$ nuclei relative to the symmetry energy parametrized by  Weizs\"acker's expression  $(N-Z)^2/8\theta$.   
 It has become custom to call any additional binding  energy proportional to $\vert N-Z\vert$   
 "Wigner Energy", which is often parametrized as
\be
\label{eq:Wigner} 
E_W(A)=W(A)\vert N-Z\vert +d(A)\delta_{NZ}\pi_{np},  ~~~\pi_{np}=\frac{1}{4}\left(1-(-)^N\right)) \left(1-(-)^Z\right)).    
\ee
Myers and Swiatecki \cite{Myers82} derived this expression by counting the number of $np$ pairs in spatial identical orbitals and assigning an
extra binding energy to these pairs. They called the sum "congruence energy". 
Brenner {\em et al.} \cite{Brenner90} analyzed binding energies of sd-shell nuclei by means of the double differences
\be\label{eq:Vpn}
\delta V_{pn}(Z,N)=\frac{1}{4}\left(E(Z+1,N+1)-E(Z+1,N-1)-E(Z-1,N+1)+E(Z-1,N-1)\right),
\ee
which they interpreted as the interaction between the last proton and neutron. For  $N=Z$ they found strong negative values of about twice
the values of the $N\not=Z$ neighbors. As we discussed above, it is  important to consider the proper reference that takes into account the isospin of the
relevant states.   The "filter" (\ref{eq:Vpn}) is just another way to expose the linear Wigner term in the binding energies, and as such does not tell much about the
existence or not of a $T=0$ condensate. 
Satu\l a {\em et al.} \cite{Satula97b} suggested certain combinations of $\delta V_{pn}(Z,N)$, 
which allowed them to "filter out" the two parameters $W(A)$ and $d(A)$ when the other terms in the mass formula
depend sufficiently weakly on $A$.  Fig. \ref{fig:PLB407f1} shows that the average trend is well reproduced by $W(A)\approx 47 MeV/A$ and $d(A)\approx W(A)$,
if the lowest $T=0$ state in the odd-odd $N=Z$ are taken in the filtering procedure. Of course, this again reflects the discussed fact that the
binding energies are approximately $\propto T(T+1)$.  
\begin{table}[t]
\begin{center}
\epsfig{file=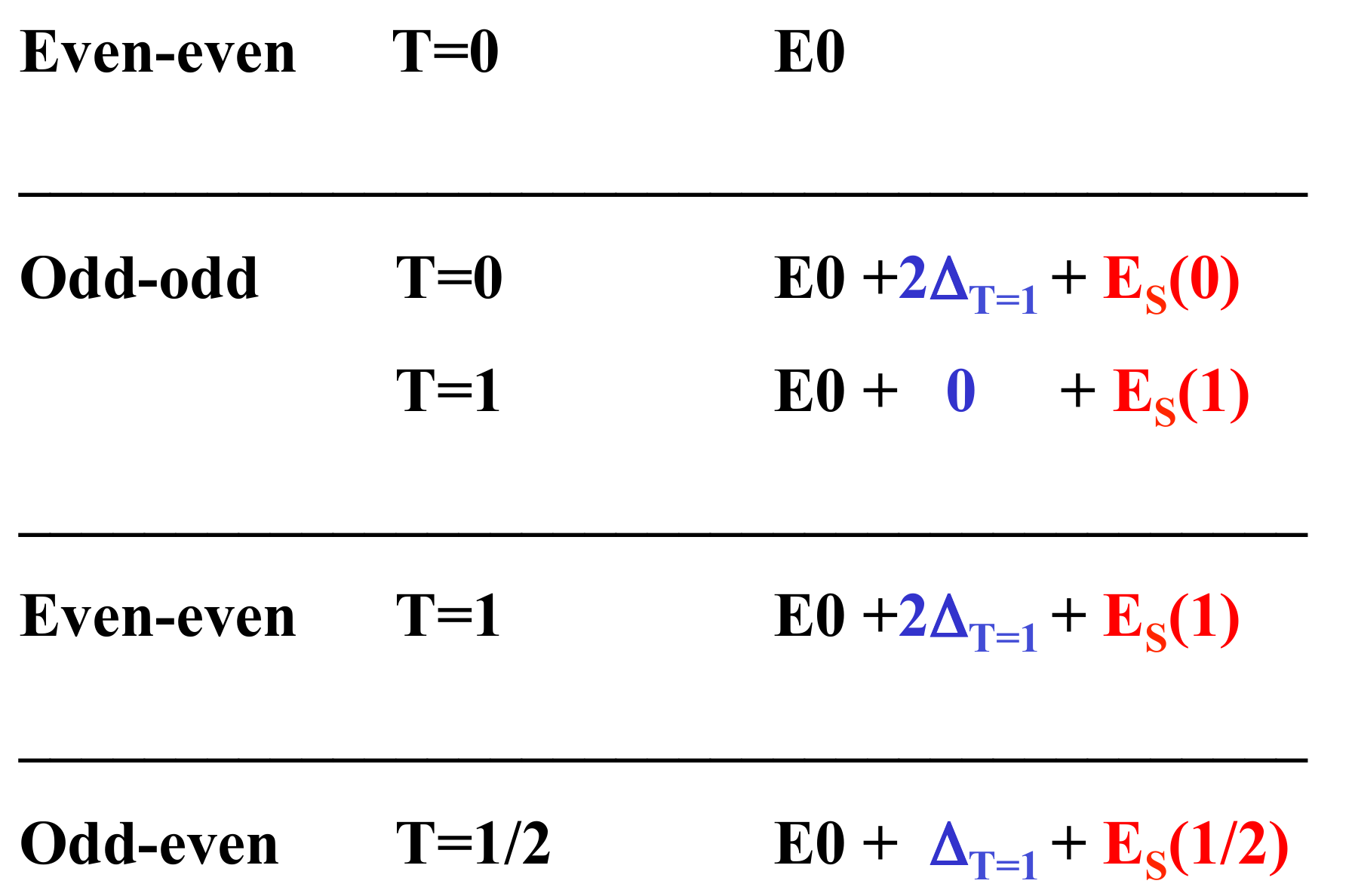,scale=0.5}
\end{center}
\caption{Expected ground-state energies based on the phenomenological analysis discussed in the text.}
\label{tab:summaryoo}
\end{table}

\begin{figure}[tb]
\begin{center}
\begin{minipage}[t]{8 cm}
\hspace*{-0.8cm}
\epsfig{file=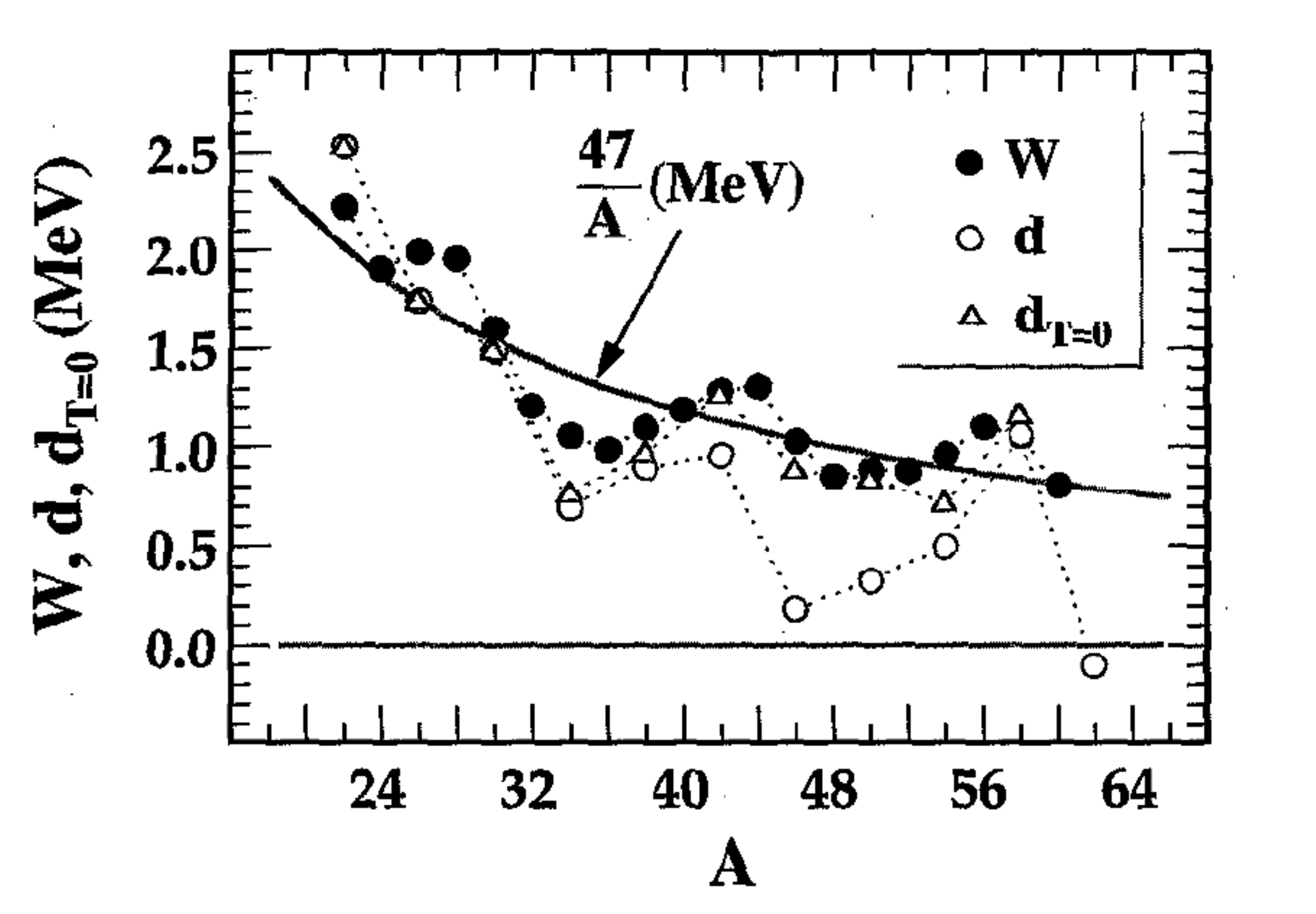,scale=0.35}
\caption{ The experimental values of the two contributions  $W(A)$ (full circles) and $d(A)$ 
(open triangles) to the Wigner energy (\protect \ref{eq:Wigner}). The values are obtained by means of 
combinations of differences ("filters") between the binding energies tabulated in the 1995 mass evaluation.
 From \cite{Satula97b}, where the filter expressions are given. The  open circles were obtained using the binding energies of the lowest
 $T=1$ states in the odd-odd $N=Z$ nuclei and the open triangle by using the lowest $T=0$ states.\label{fig:PLB407f1}}
\end{minipage}
\hspace*{0.5cm}
\begin{minipage}[t]{8 cm}
\hspace*{-0.9cm}
\epsfig{file=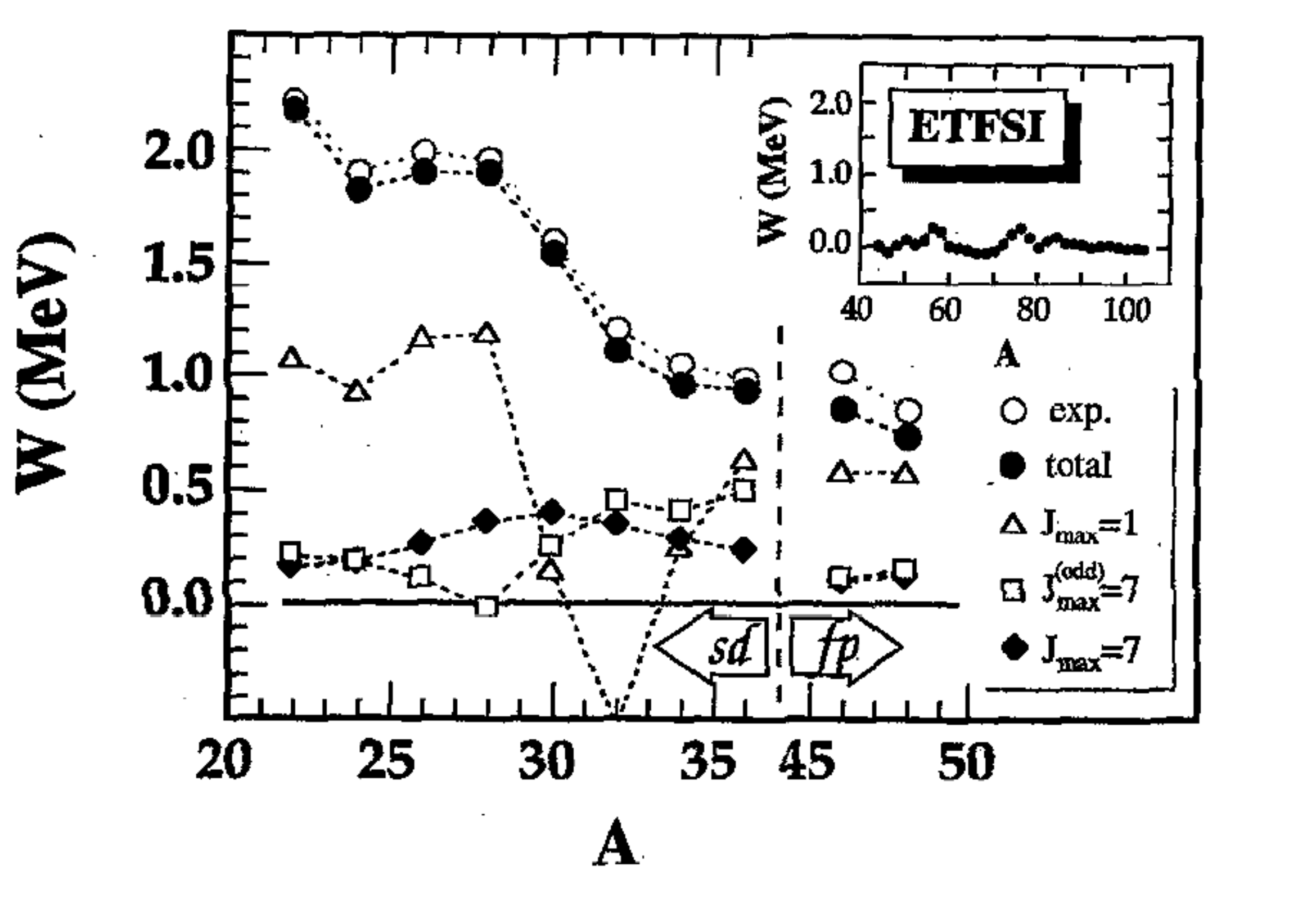,scale=0.35}
\caption{The Wigner term $W(A)$ extracted from the binding energies calculated calculated by diagonalizing the USB and KB3 interactions
in the sd- and fp-shells, respectively, as compared with the experimental values shown in Fig. \ref{fig:PLB407f1}. The results of calculations with
interaction matrix elements removed are labelled as $J_{max}=1$: all $T=0, J=1$ removed,  $J_{max}=7$: all $T=0$ removed, 
and $J_{max}^{odd}=7$: all $T=0,J=1,3,5,7$ removed. The insert shows the values extracted from the ETFSI mass formula.
 From \cite{Satula97b}.\label{fig:PLB407f3}}
\end{minipage}
\end{center}
\end{figure}

\subsection{Microscopic calculations}
The goal establishing a relation between the binding energies and  $np$ pairing has initiated a number of theoretical  studies, the methods of which were reviewed in Sections
\ref{sec:SMs}, \ref{sec:SMd}, \ref{sec:QRPA}, \ref{sec:MF}, and \ref{sec:SP}.  Depending on the approach, the authors concluded that the systematics of the binding energies
can be explained as a consequence of the isovector pair correlations only (including the $np$ component required  by isospin invariance) or that the  isoscalar
pair interaction and $T=0$ pair correlations play a significant role. In the following we discuss the different results from a common point of view.

\subsubsection{Isovector scenario}\label{sec:WIV}
Frauendorf and Sheikh \cite{Frauendorf99a} first pointed out that the $T(T+1)/2\theta$ 
 dependence ($T\approx T_z=\vert N-Z\vert/2$) of the strong interaction part of the binding energies is expected to emerge as a consequence of strongly breaking
the isospin symmetry by the isovector pair field.   In the subsequent paper \cite{Frauendorf 00} they drew attention to the fact that the systematic displacement of the 
lowest $T=0$ state in odd-odd nuclei relative to they even-even neighbors observed for $N=Z$ nuclides indicates that a substantial isoscalar pair field 
cannot be present (c.f. Sec. \ref{sec:2jIVISD1} and \ref{sec:Wphen}). They also noted that standard self-consistent theories involve the isocranking procedure automatically, and the coefficient
of the symmetry energy $C_{sym}\left<T_z\right>^2/2$ provides the isorotational moment of inertia $\theta=1/C_s$.   Focusing on spatial rotational structures, they did not calculate
${\cal J }$ but took the experimental value for describing the relative energies of $T=0$ and $T=1$ rotational bands (cf. Section \ref{sec:rotation}).

In his QRPA studies (cf. Section \ref{sec:QRPA}), Neergard \cite{Neergard02,Neergard03,Neergard09} used a fixed set of single particle levels in a deformed potential.
The isocranking procedure for a fixed levels gives a value of $\theta$ that is about a factor of 2 too large, because  it accounts only for the so-called "kinetic" part of the 
symmetry energy, which is generated by the asymmetric occupation of the levels according to the Pauli Principle.   An equally large "interaction" part arises from the 
change of the single particle levels with $N-Z$.  Self-consistent mean field theories provide the isospin dependence average potential and the correct symmetry coefficient.
To account for the deficiency, Neergard introduced     the "symmetry interaction" $\kappa \vec T\cdot\vec T/2$ and carried out the standard HFB + QRPA procedure for the
Hamiltonian (\ref{eqn:IVIS}) with $G_{IS}$=0. The symmetry interaction generates a constant up and down shift of proton and neutron levels levels, respectively. The 
strength $C$ was adjusted to reproduce the experimental symmetry energy and the $G_{IV}$  was chosen to enforce for the HFB solution 
standard values $\Delta_p=\Delta_n=12 MeVA^{-1/2}$.
He could reproduce well the binding energies of the isobaric chains with $A$=48, 56, 68, 80, 100 (see Figs. 3 and 4 in Ref. \cite{Neergard09}). 
 
 Bentley and Frauendorf \cite{Bentley13} calculated the binding energies of even-A nuclei with $T\leq 5$ 
by numerically diagonalizing the Hamiltonian (\ref{eqn:IVIS}), which avoids the problems that are caused by the small amplitude approximation of the QRPA
(see Section \ref{sec:SMd}).
They determined the equilibrium deformation individually for each isobaric chain for $30\leq A \leq 100$ by means of the Micro-Macro method in order to 
properly account for the changing level density near the Fermi surface.  Several fixed ratios $G_{IS}/G_{IV}$ were investigated. 
 For a detailed comparison with experiment, the Coulomb energy was removed from the data in such a way that mirror pairs (same $|T_z|$) appear at the same energy.

 The resulting energies were written in the form  $E(A,T)=E_{int}(A)+T(T+X)/2\theta$. The relative energies of the 
 considered  nuclei with $T\leq 5$ can be re-expressed in terms of energy staggering $2\Delta= E_{int}(A)-\left(E_{int}(A-1)+E_{int}(A+1)\right)/2 $
 between the even-even and odd-odd $N=Z$ nuclei, the isorotational moment of inertia $\theta$, the "Wigner X", and the energy difference
  $E(A,T=1)-E(A,T=0)$ of the odd-odd $N=Z$ nuclei.  Fig. \ref{fig:PRC88f9} displays  the results for the six-level calculations. 
  Seven-level calculation were carried out for $G_{IS}=0$ only, which additionally included the $T$=1, 3, 5 chains.  They were nearly identical with
  the six-level results after renormalization of the interaction strength, the values of which are quoted in Fig. \ref{fig:PRC88f9}. 
The strengths of isovector 
and  the symmetry interactions were parametrized by smooth A-dependences $\propto A^{-3/4}$ and $A^{-1}$, respectively.  The coefficients
were adjusted to the experimental values of $2\Delta$ and $E(A,T=1)-E(A,T=0)$.

 Fig. \ref{fig:PRC88f9} demonstrates that the model Hamiltonian (\ref{eqn:IVIS}) describes all four quantities in a consistent way. 
 The authors attribute the slight systematic overestimation of $1/\theta$ to the small number of single particle levels.   
Together with Neergard's work, the results 
  place  the phenomenology discussed in Section \ref{sec:Wphen} on a microscopic basis. The local fluctuations, which
 reflect the changes of the level density near the Fermi surface, are reasonably well reproduced. Some deviations are expected, because there 
 are uncertainties of the single particle energies. As seen in Fig. \ref{fig:PRC88f9}, the presence of moderate isoscalar pair correlations of the $p\bar n$ type changes the results
 only insignificantly as long as $G_{IS}/G_{IV}< 0.5$. For  $G_{IS}/G_{IV}> 1$ a consistent description of the data was not possible. The authors
interpret this as an  signal for the appearance of a isoscalar pair field, which is accompanied by quenching the even-even odd-odd 
energy staggering $2\Delta$.  The results substantiate the suggestion of Ref. \cite{Frauendorf99a} 
 that  the Wigner term $T/2\theta$ originates from the spontaneously broken isospin symmetry. As a new point, they note that
symmetry is broken by the mean field in two ways, by the isovector pair field in the particle-particle channel and the by isospin dependent part of the
average potential in the particle-hole channel, where both contribute  to about equal parts.   
  
 \begin{figure}[tb]
\begin{center}
\epsfig{file=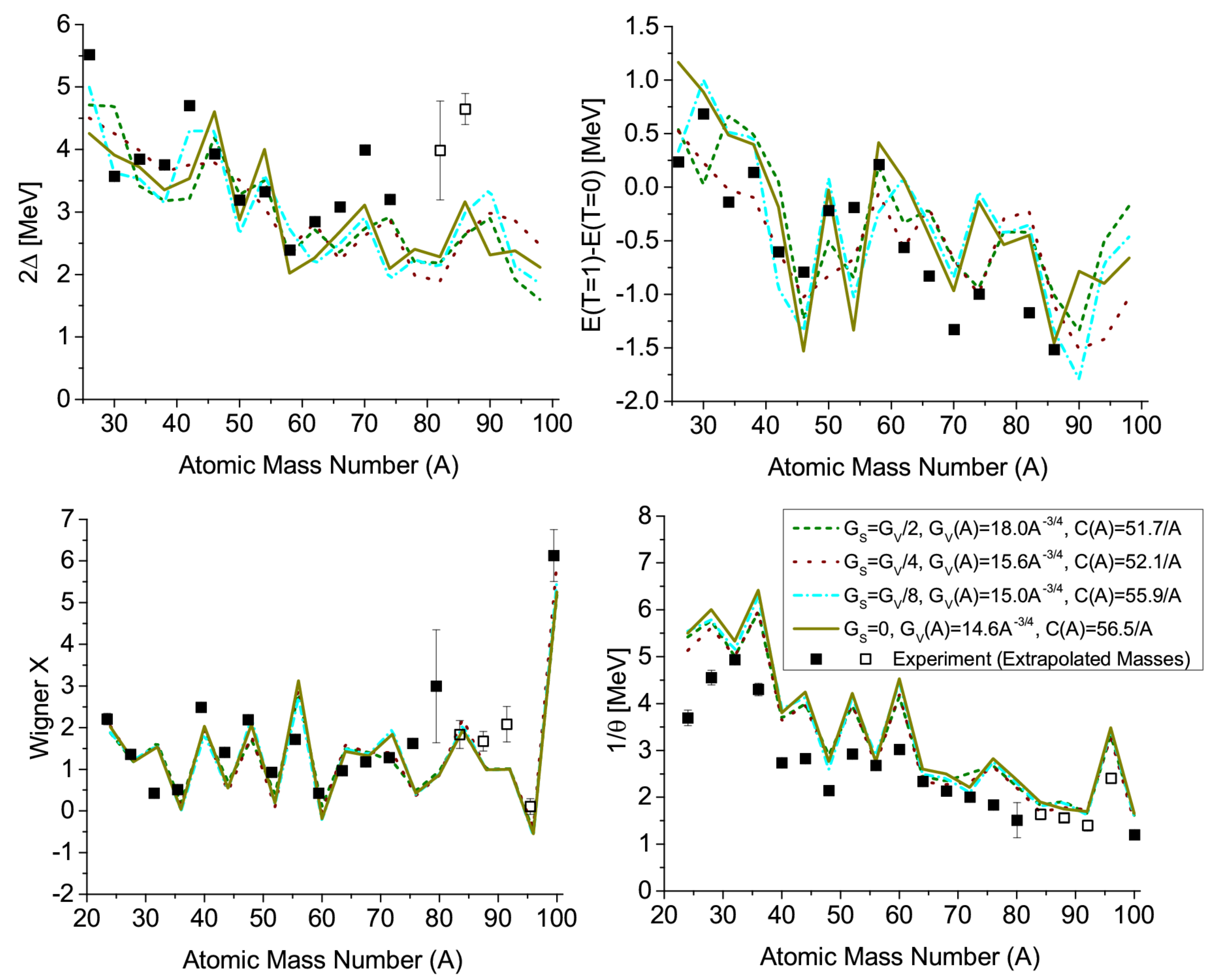,scale=0.35}
\caption{ The even-even odd-odd mass difference $2\Delta$ of $N=Z$ nuclei, the energy difference between the lowest $T=1$ and $T=0$ states
in odd-odd $N=Z$ nuclei, the Wigner X, and the isorotational moment of inertia $\theta$ for $N\approx Z$ nuclei calculated by diagonalizing  the
Hamiltonian (\ref{eqn:IVIS}) compared with the values derived form the experimental binding energies. The coupling constants $G_{IV}$, $G_{IS}$,
and $\kappa/2$ are denoted by $G_V$,$G_S$ and $C$, respectively. The average trend of $2\Delta$ is  $24 MeVA^{-1/2}$.
  From \cite{Bentley13}.\label{fig:PRC88f9}}
\end{center}
\end{figure}
Bentley {\it et al.} \cite{Bentley14} demonstrated that QRPA reproduces with good accuracy the  Shell Model results of Ref. \cite{Bentley13}
as long as the nucleus is sufficiently away from the phase transition. They suggested an interpolation for the cross-over region based on the 
QRPA solutions above and below the transition, which very well approximates the Shell Model results. The method was 
incorporated to  the Micro-Macro scheme for calculating ground state binding energies, where  three major single particle 
shells were taken into account. As anticipated in Ref. \cite{Bentley13}, the systematic overestimate of $1/\theta$ seen in Fig. \ref{fig:PRC88f9} was a consequence of the small number 
of single particle levels (6 or 7) to which the Shell Model diagonalization had to be restricted,  and it was removed in the extended QRPA single particle  space. 
The authors  calculated 
the binding energies without explicitly introducing a Wigner term, where the  accuracy was comparable with other approaches to large-scale
mass calculations. The method holds promise to avoid the need of introducing   a phenomenological Wigner term into 
self-consistent mean field approaches to nuclear binding energies.   HFB+QRPA calculations within such self-consistent 
framework could provide a completely microscopic description of the binding energies of nuclei near the  $N=Z$ line.

\subsubsection{Isoscalar versus isovector pairing} \label{sec:WIS}

Brenner {\em et al.} \cite{Brenner90}   could reproduce the presence of the $N=Z$ spikes 
in the $\delta V_{pn}$ values (\ref{eq:Vpn})  by diagonalizing a surface $\delta$- interaction within the sd-shell. 
The spikes disappeared when the $T=0$ part of the interaction was removed. 
 Diagonalizing  the realistic USD interaction,
 they could quantitatively reproduce the experimental  spikes of $\delta V_{pn}(Z,N)$ at $N=Z$. 
 Satu\l a {\em et al.} \cite{Satula97b} extended this kind of investigation to pf-shell nuclei, for which they diagonalized the KB3 interaction. 
 They used combinations of $\delta V_{pn}(Z,N)$   to  filter out the two terms $W(A)$ and $d(A)$ in Eq. (\ref{eq:Wigner}) from the experimental and calculated energies.
 The results are displayed in Figs. \ref{fig:PLB407f1} and \ref{fig:PLB407f3}. The Shell Model calculations accurately reproduce the Wigner term $W(A)$.
 The Thomas-Fermi approach ETFSI shown in the inset does not provide it, as other mean field approaches fail to do. Removing the $T=0$ matrix elements 
 of the interaction drastically reduces the size of $W(A)$, which quantifies the findings of  Brenner {\em et al.} \cite{Brenner90}. The authors state that they 
 confirm the conclusion of Ref. \cite{Brenner90} that the bulk part of the Wigner energy comes from $T=0$ $np$ interaction.
  Hasegawa and Kaneko  \cite{Hasegawa05b} carried out a similar analysis of their Shell Model results. 
 They related  combinations of $\delta V_{pn}(Z,N)$ to various parts of their interaction and concluded that both  isoscalar and  isovector 
 terms of their interaction enhance the binding entergies of $N=Z$ nuclei. 
 Although the  work in Refs. \cite{Brenner90,Satula97b,Hasegawa05b} demonstrated that the presence of the $T=0$  part in the interaction is necessary for the 
  enhancement of  the binding energies of $N=Z$ nuclei, it did not establish the mechanism how this comes about. In particular, the 
 work did not demonstrate that   $T=0$ pair correlation generate the Wigner term.

 The relation between the isoscalar interaction and the Wigner energy was clarified by Bentley {\it et al.} \cite{Bentley14}, who carried out 
 a Shell Model calculation  for $A=48$ within the configuration space of  the f$_{7/2}$ shell.
They used the interaction ``Model~I'' of Zamick and Robinson~\cite{ZR02}
with a normalization to zero of the largest matrix element
(two-nucleon angular momentum $J=6$) so as to make the total
interaction attractive. Like Satu\l a~\textit{et al.}~\cite{Satula97b}, they
switched off successively the interactions in individual $J$ channels. However they did not 
restrict their analysis to the linear Wigner term.  Like in Ref. \cite{Bentley13}, they parametrized the calculated ground state energies
in the form $E(A,T)=E_{int}(A)+T(T+X)/2\theta$. The values for $1/2\theta$ and $X$ are shown in Table~\ref{table:1f7/2}.
As observed  by Satu\l a~\textit{et al.} \cite{Satula97b},  the Wigner term  $X/2\theta$ (=$2W$ of Fig. \ref{fig:PLB407f3}) decreases when the
isoscalar interactions $J=1,3,5,7$ are switched off successively and
very much so when the $J=7$ interaction is switched off finally. It is
seen, however, that this is due \emph{not} to a decrease of $X$, which
actually increases, but to a decrease of the symmetry energy
coefficient $1/2\theta$. The isoscalar terms of the  interaction thus
contribute significantly to the \emph{entire} symmetry energy and not
just to its Wigner term. The reduction of the Wigner energy when the
isoscalar interactions are switched off is only a side effect of this
general reduction of the symmetry energy. In our view, there is no obvious relation to 
the presence of isoscalar pair correlations.

\begin{table}
   \caption{\label{table:1f7/2}
The symmetry energy coefficient $1/2\theta$, the Wigner coefficient $X/2\theta$, and the Wigner $X$ parameter   for the $A=48$ isobars.
The values are derived from the binding energies obtained by a Shell Model calculation within the  f$_{7/2}$ shell. The table
lists the values obtained by progressively removing the all matrix elements of a given $J$.     
From \cite{Bentley14}.}
%\begin{ruledtabular}
\begin{center}
\begin{tabular}{ccccccccccc}
\multicolumn{8}{c}
  {\parbox{.30\columnwidth}{\noindent$J$ of included matrix elements}}
&\parbox{.15\columnwidth}{~~~$1/2\theta$(MeV)}
&\parbox{.15\columnwidth}{~~~$X/2\theta$(MeV)}
&\parbox{.15\columnwidth}{~~~~~~~~$X$}\\[6pt]
\hline
0&1&2&3&4&5&6&7&1.21&1.85&1.31\\
0& &2&3&4&5&6&7&1.12&1.18&1.07\\
0& &2& &4&5&6&7&0.93&1.14&1.23\\
0& &2& &4& &6&7&0.69&1.21&1.75\\
0& &2& &4& &6& &0.12&0.21&1.71\\
0& &2& &4& & & &0.12&0.21&1.71\\
0& &2& & & & & &0.22&0.30&1.41\\
0& & & & & & & &0.40&0.40&1.00
\end{tabular}
\end{center}
%\end{ruledtabular}
\end{table}

Poves and Martinez-Pinedo \cite{Poves98} diagonalized  the KB3 interaction and
very well reproduced the  binding energies  of several pf-shell nuclei. They   
addressed the origin of the Wigner term by calculating its value for the full KB3 interaction and for the modified KB3 interaction
from which its isovector $L=0$, $=0$ part or  isoscalar $L=0$ $S=1$ part was removed. As discussed in Section \ref{sec:SMs}, they quenched the respective 
pair correlations in this way. The value of Wigner term did not very much change. They concluded that the Wigner term is not related to 
the strength of the pair correlations. This result is not inconsistent with the findings of Refs. \cite{Brenner90,Satula97b} , because these authors only 
demonstrated a connection with the $T=0$ matrix elements of the residual interaction, however not with the presence of substantial isoscalar pair correlations
in the wave function. On the other hand, it concurs with the interpretation of Ref. \cite{Bentley13} that the Wigner term appears as the consequence of 
breaking the isospin symmetry by the isovector pair field and the average potential. The symmetry breaking leads to the $T(T+1)/2\theta$ dependence of  energy 
irrespective of  how the symmetry is broken, provided it is strong. Switching off the isovector field does not change much because there is still the
symmetry breaking by the average potential. The isorotational moment of inertia is not  very sensitive to the pair correlations and so it is the Wigner term $T/2\theta$.   

Poves and Martinez-Pinedo discuss in detail the relative position of the lowest  $T=0$ and $T=1$ states in odd-odd $N=Z$ nuclei. It is the result of a balance between
the repulsive isovector monopole part of the KB3 interaction, which pushes up the $T=1$ state relative to the $T=0$ state, and the isovector pairing part, which pulls down the 
$T=1$ state relative to the $T=0$ state.  The monopole part has a smooth dependence on $A$   whereas the isovector pair correlation energy
grows linearly with the shell degeneracy (see Section \ref{sec:IV2shell}), which tips the balance as follows.
 The ground state of $^{46}$V is $T=1$ because the Fermi level is located in 
the well degenerate 1f$_{7/2}$ shell. The ground state of $^{58}$Cu is $T=0$, because the Fermi levels located in 
the barely  degenerate 2p$_{3/2}$ shell. All odd-odd $N=Z$ nuclides with   the Fermi level located in 
the well degenerate 1g$_{9/2}$ shell have $T=1$ as ground states. This balance is qualitatively the same as the one  discussed
in Section \ref{sec:Wphen}, because the isovector monopole interaction generates the interaction part of the symmetry energy.
The authors note an additional weight in the balance. The small spin-orbit splitting between the 2p$_{1/2}$ and 2p$_{3/2}$ levels
enhances the isoscalar pair correlations in the case of $^{58}$Cu, which results in extra binding of its $T=0$ level.
The latter effect was studied in detail by Bertsch and Luo \cite{Bertsch10} (see Section \ref{sec:MF}.) 
    
Using the filters of Ref. \cite{Satula97b}, Chasman \cite{Chasman03b} (see Section \ref{sec:SP}) derived $W(A)=45.5MeV/A$ from his calculations   
assuming  equally spaced single particle levels, which agrees very well
with the average trend in experiment $W(A)=47MeV/A$ (cf. Fig. \ref{fig:PLB407f1}). The result was obtained assuming that 
the isovector and isoscalar coupling strengths are
equal $G_{IV}=G_{IS}$. For this combination, the author predicts that the $T=0$ and $T=1$ states in the odd-odd $N=Z$ nuclei have the 
same energy.  His value of $d=0.44MeV$ does not particularly well correlate with 
the experimental values in Fig. \ref{fig:PLB407f1}, because  it is  to be compared with the values 
derived from the both the $T=0$ and $T=1$ states in the odd-odd system, which are predicted to be the same. 
Chasman makes a strong point that the Wigner energy essentially originates from the diagonal matrix elements of the pairing interactions,
because the pair correlation energy, which is generated by the non-diagonal matrix elements, turned out to be nearly constant as a function 
of $\vert N-Z\vert$. However, his approach, which remains on the mean field level concerning isospin conservation, misses an essential part of the 
total correlation energy. As discussed in Section \ref{sec:sigmf}, the correlations  that restore good isospin  generate a term $\propto T$, which 
enhances the binding of $N=Z$ nuclei. In fact, the work by Bentley, Frauendorf, and Neergard \cite{Bentley13,Neergard09} (see above) demonstrates
that the  Wigner term can be explained by these correlations. In our view, the question of the relative importance diagonal and non-diagonal
matrix elements of the pairing interaction needs to be re-addressed in the frame  of an isospin conserving approach. 
We expect that restoring isospin within Chasman's approach will 
enlarge $W(A)$ beyond the experimental value.

\begin{figure}[tb]
\begin{center}
\begin{minipage}[t]{8 cm}
\hspace*{-0.8cm}
\epsfig{file=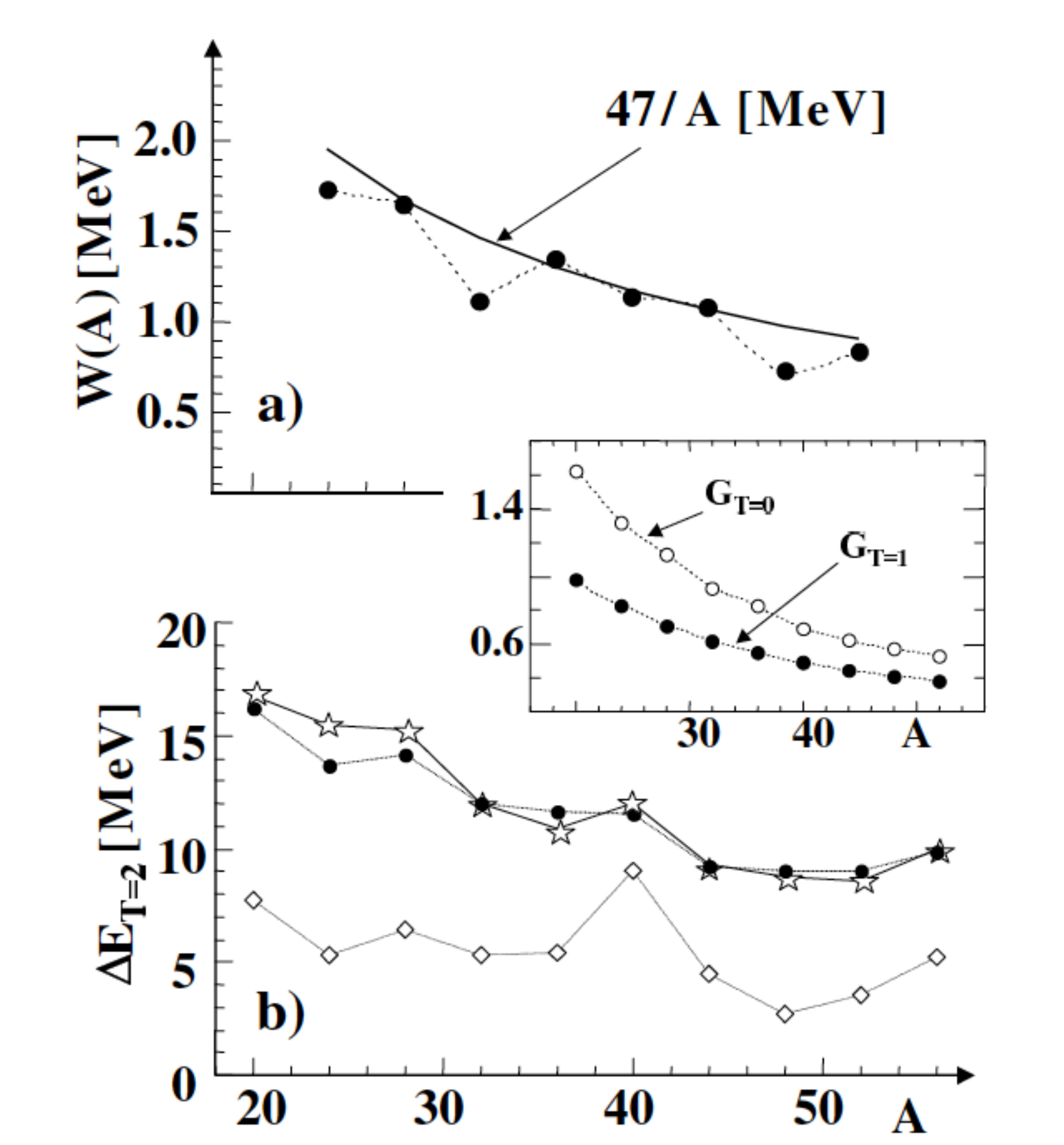,scale=0.5}
\caption{Upper panel: Calculated Wigner term $W(A)$ compared with the average experimental values from Fig. \ref{fig:PLB407f1}. 
Lower panel: Calculated (stars) and experimental (dots) excitation energies of the $T=2$ states in even-even $N=Z$ nuclides. 
Inset: Values of $G_{IV}=G_{T=1}$
and $G_{IS}=G_{T=0}$ used in the calculation. Diamonds show a calculation with $G_{IS}=0$.
From  \cite{Satula00b}, \label{fig:PRL86f3}}
\end{minipage}
\hspace*{0.5cm}
\begin{minipage}[t]{8 cm}
\hspace*{-0.9cm}
\epsfig{file=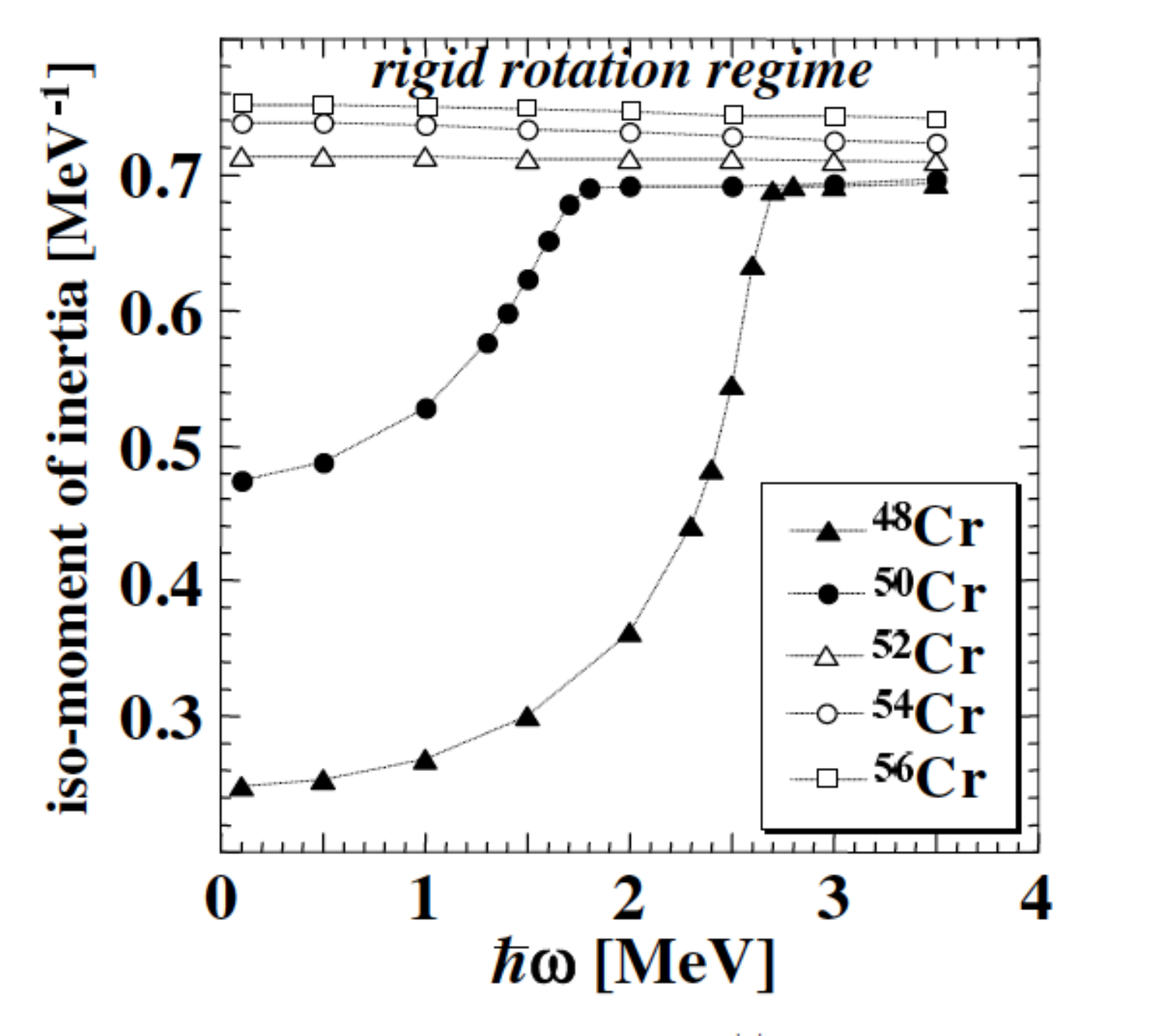,scale=0.55}
\caption{Isorotational moment of inertia $\theta=\left<T_x\right>/\hbar \omega$ as function of isorotational  $ \omega$
for the Cr isotopes. The value of $\left<T_x\right>$ is related to the total isospin $T$ by 
$\sqrt{\left<T_x\right>^2+\left<T_z\right>^2}=\sqrt{T(T+1)}$.
The isoscalar pair correlation are absent in the rigid rotation regime.  From  \cite{Satula00b}.\label{fig:PRL86f5}}
\end{minipage}
\end{center}
\end{figure}

 Satu\l a and Wyss \cite{Satula00b,Satula01} calculated the Wigner term $W(A)$ in the framework of their approach described in Section \ref{sec:SP}.
  They used single particle energies 
 from a slightly deformed Woods Saxon potential ($\beta=0.05$), which did not depend on isospin. The isovector interaction strength $G_{IV}$ was determined 
 by means of the average gap method from  
 the experimental the even-odd mass differences. The coupling  constant $G_{IS}$ for isoscalar spin-aligned $np$ pairs was  adjusted to reproduce the experimental values of $W(A)$. As seen in 
 Fig. \ref{fig:PRL86f3}, good agreement was achieved for $G_{IS}>G_{IV}$, the actual values of which are shown in the inset. The
 excitation energies of $T=2$ states of even-even $N=Z$ nuclei  were also very well reproduced. Interpreting the $T=0$ states in odd-odd $N=Z$ nuclei as
 two quasiparticle excitations, they  obtained a rather good description of the energy difference between the lowest $T=1$ and $T=0$ states in 
 odd-odd $N=Z$ nuclei (Fig. 4 Ref. \cite{Satula01}). 
 
Satu\l a and Wyss concluded that the isosccalar $pn$  pair correlations are the main source for the Wigner energy,
 which is in contrast to the conclusions of Sections \ref{sec:Wphen}, \ref{sec:WIV}, and Ref. \cite{Poves98}. 
 In our view, their use of a set of single particle energies which do not depend on isospin is problematic. 
As discussed above, the isospin dependence of the nuclear potential generates about one half of the 
symmetry energy (the interaction part), i.e. it reduces the isorotational moment of inertia $\theta$ by a factor of about two. Neglecting the isospin dependence, will result in 
a value of $\theta$ that is a factor of two too large for pure isovector pairing. The large value of $\theta$ is reflected by the small values of the excitation energies $E(T=2)$
shown in Fig. \ref{fig:PRL86f3} for the case $G_{IS}=0$.  By introducing strong isoscalar pairing, Satula and Wyss reduce the value of $\theta$  to be consistent with experiment.
Fig. \ref{fig:PRL86f5} illustrates the situation. The isoscalar pair correlations are absent in the rigid rotation regime, where  $\theta\approx 0.7 MeV^{-1}$,
  which is to be compared with the experimental value of $\theta=0.45 MeV^{-1}$ for $^{48}$Ca from Fig. \ref{fig:PRC88f9}. Switching on the isoscalar pairing reduces 
  $\theta$ to $0.25MeV^{-1}$ for the $T=0$ ground state of $^{48}$Ca. Increasing the isorotational frequency $\hbar \omega$ (i.e. $T$) 
  suppresses the isoscalar correlations by blocking. They are quenched at $\hbar \omega=2.5 MeV$, which correspond to $T\approx 2$. 
We expect that taking into account the isospin dependence of the nuclear potential
in the framework of Satula and Wyss' approach will reduce $\theta$ remove the necessity to introduce strong isoscalar pair correlations. 
  
As  discussed in Section \ref{sec:Wphen} and  pointed out in Refs. \cite{Frauendorf00,Frauendorf01} the presence of a strong
 isoscalar pair field breaks the $N$ and $Z$ number parity symmetry, 
 which has the consequence that the quasiparticle vacuum represents the ground states of both even-even and odd-odd $N=Z$ nuclei.
This general consideration is born out by the exactly soluble models in Sections \ref{sec:2jIS}-\ref{sec:2jIVISD1} (see Fig. \ref{fig:ee-ooIVIS}). Within their framework,
 Satu\l a and Wyss postulate that the ground state of odd-odd $N=Z$ has two quasiparticle nature (see Fig.  4 of Ref. \cite{Satula01}). We consider such 
 assignment as a problem of their approximate approach, because a complete many body treatment of their Hamiltonian will provide another $T=0$ state that is
 located much lower.

Summarizing it seems that the $T$ dependence of the ground state binding energies does not provide evidence for an isoscalar pair condensate.
 The scenario that invokes only isovector pairing accounts very well for the data.  We consider the attempts to explain the difference between experimental 
 energies and the results obtained in a pure isovector scenario by the presence of strong isoscalar pair correlations as problematic, because the isovector 
 baseline used in these calculations  missed  ingredients that contribute to the binding energy in essential way. Comparing Shell Model results with different
 strength of the relevant pair interaction have not revealed a causal relation between the appearance a term proportional to $\vert N-Z\vert$ in the binding energy and the 
 presence of isoscalar pair correlations.

\section{Rotational response}\label{sec:rotation} 

It has been early recognized that the pair fields respond differently to rotation, i. e. the increase of angular momentum. The $J=0$ pairs, which dominate
the isovector correlations, generate the $T=0,I=0$ ground states of the even-even $N=Z$ nuclei. In order to increase the
angular momentum, the pair correlations must be partially lifted, which costs energy. Hence, the isovector pairing increases the excitation
energy of states with finite $I$ (decreases the moment of inertia in case of regular rotational bands), and it will be destroyed 
 when enough pairs are broken with increasing $I$. Pairs with $J\not=0$ can gradually align to build up angular momentum,
 which will decrease the excitation
energy of states with finite $I$ (increase the moment of inertia). Their  presence is expected to decrease slower
 or even to be enhanced with increasing $I$ (up to a certain value when the pairs are fully aligned).  
Isovector pairs  have even values of $J$. The $J=2$ pairs (quadrupole pairing), which are known to be significant in accounting for the 
correct rotational energy in  $N\gg Z$,  nuclei will play a similar role in $N\approx Z$ nuclides. Quadrupole pairing
reflects the nuclear deformation, which induces a deformation of the pair field.        
The isoscalar pairs have odd values of  $J$.
 The pairs with maximal spin $J=2j$ are particularly favored, because
  the maximal overlap of their wave functions provides a large gain in energy (see Fig. \ref{fig:SchTru}). 
 However, the number of pairs that may correlate decreases with $J$, because the interaction, which conserves angular momentum,
 cannot correlate pairs with different angular momentum projection. The number is maximal for $J=0$ pairs and one for $J=2j$ pairs.
 Hence one must distinguish between the contribution of the isoscalar interaction to the energy, which may be just a diagonal term,
 and the generation of isoscalar pair correlations.   
 For $J=1$  there are still many pairs that may correlate.  Since their alignment gain of angular momentum  is 
 relatively small, it is not clear from the outset whether
 the $J=1$ correlations will increase or decrease the rotational energy.
 As in the case of the $T=1$ pairs, one expects that nuclear deformation induces a certain admixture of $J=3$ pairs. 
 
 The different response promises a possibility to asses the relative importance of isovector and isoscalar  pair correlations. 
Systematic investigations of the rotational response have been carried out assuming that there is only isovector pairing.
In Section \ref{sec:sigmf} we discussed the isorotational approximation, which allows one to avoid the  explicit calculation of
the isovector $np$ field.    In  Section \ref{sec:IVRestore} we elaborate on this approach and demonstrate its accuracy for 
a rotating 
deformed $j$-shell. In Section \ref{sec:IVrot} the studies within this approximation are reviewed.   
In Section \ref{sec:IVISrot} we discuss calculations based on realistic 
effective interactions, which are expected to quantitatively account for the different rotational resonse to isovector and isoscalar pairing.
The relation between isoscalar interaction and isoscalar pair correlations will be analyzed in more detail.
In particular the recently suggested existence of a "spin-aligned phase" \cite{Cederwall11} is
discussed thoroughly.
Section \ref{sec:ISsym} discusses the symmetry aspects of isoscalar pair correlations. 
In Section \ref{sec:ISrot} possible  evidence in the rotational spectra  for the presence isoscalar pair correlations  is reviewed.

%\begin{figure}[tb]
%\epsfysize=9.0cm
%\begin{center}
%\begin{minipage}[t]{8 cm}
%\epsfig{file=cartoon.pdf,scale=0.8}
%\epsfig{file=Nature.pdf,scale=0.6}%\end{minipage}
%\begin{minipage}[t]{16.5 cm}
%\caption{The energy gain from the short range attractive  interaction is maximal when
%the anguar momenta $j$ of two nucleons are anti aligned ($J=0$, $T=1$) or aligned ($J=2j$, $T=0$, labelled by $J>0$).
% From \cite{Cederwall11}.\label{fig:Nature469f1}}
%\end{minipage}
%\end{center}
%\end{figure}

\begin{figure}[tb]
\begin{center}
\begin{minipage}[t]{8 cm}
\epsfig{file=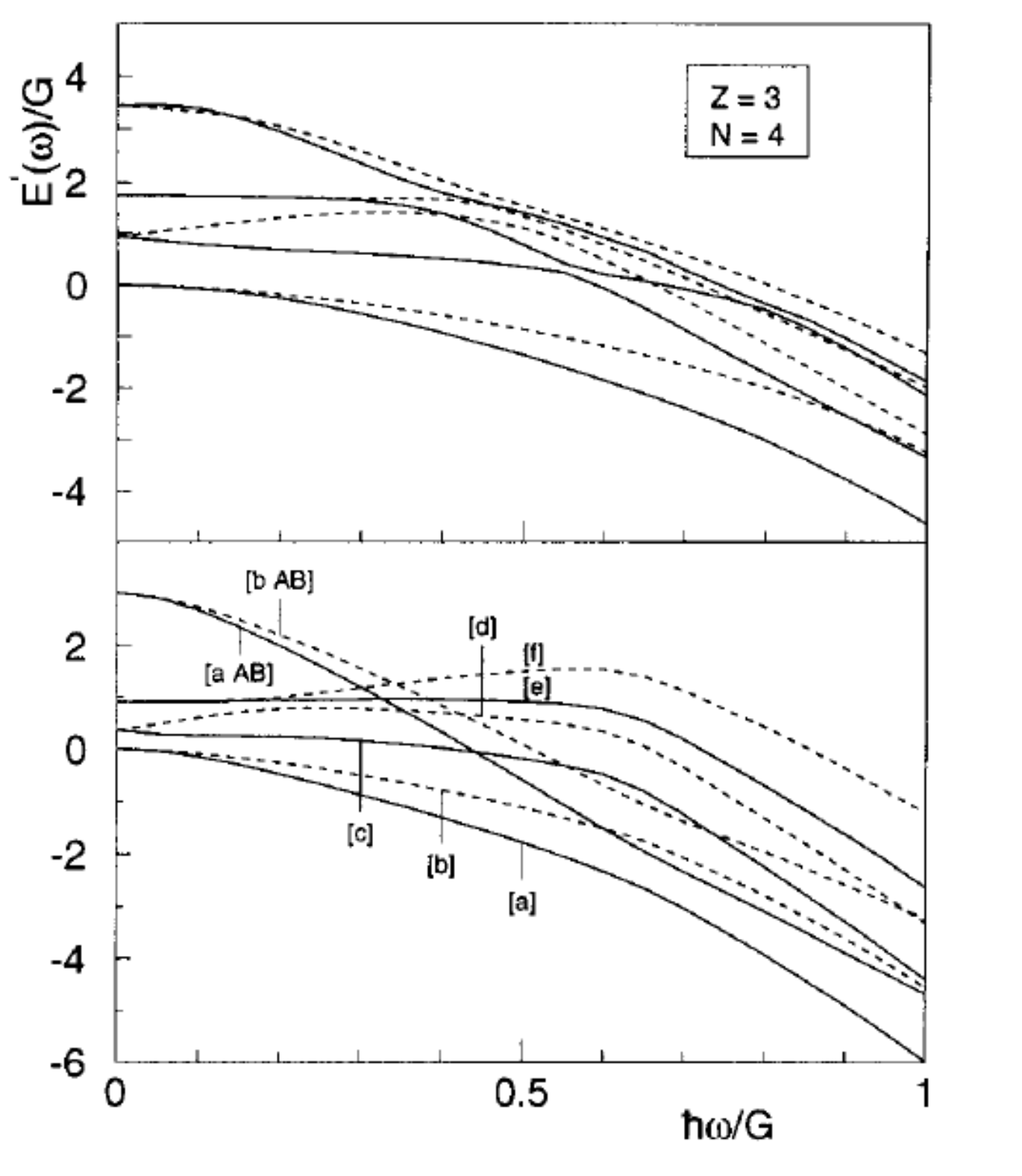,scale=0.59}
\end{minipage}
\hspace*{0.5cm}
\begin{minipage}[t]{8 cm}
\epsfig{file=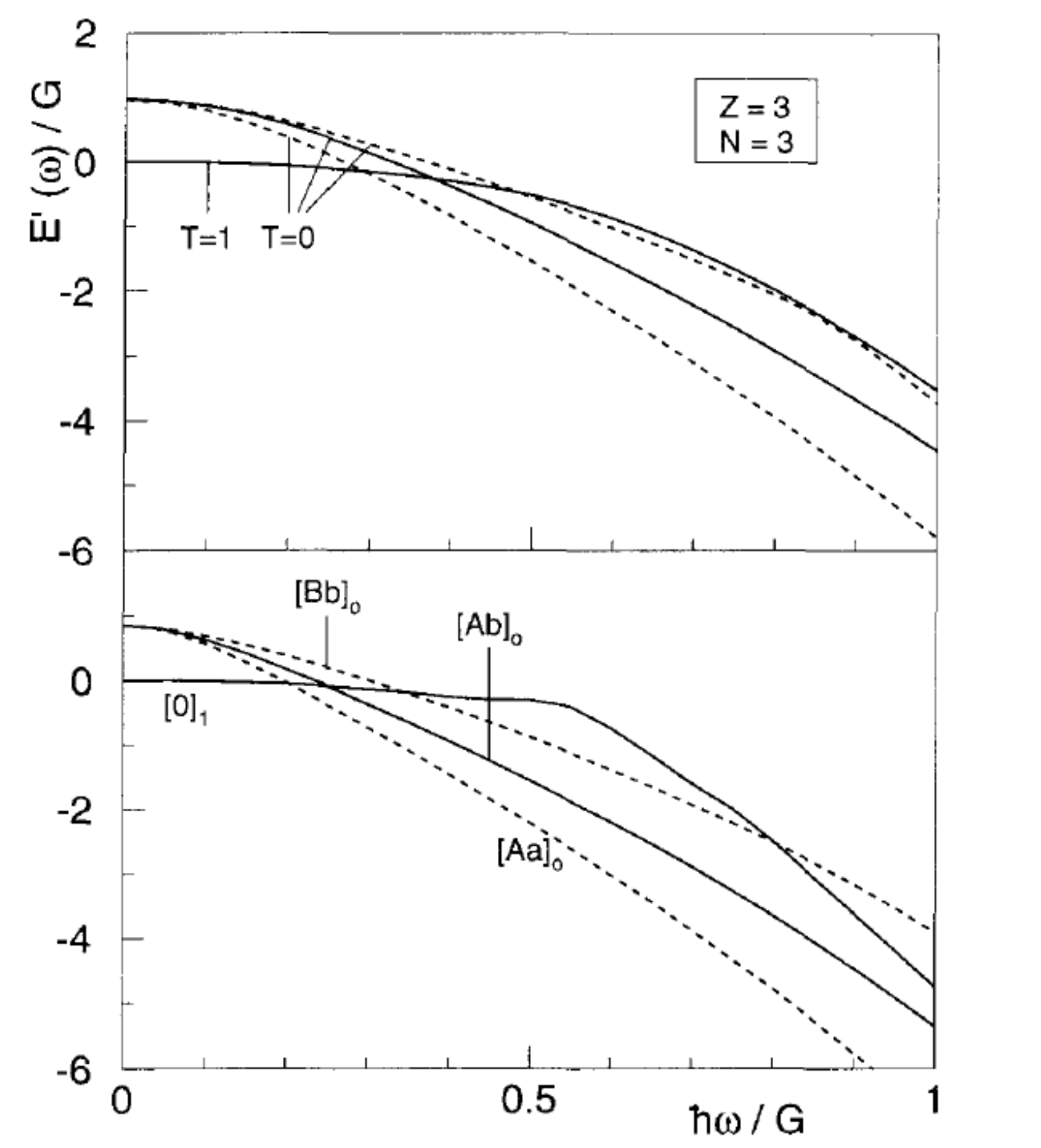,scale=0.5}
\end{minipage}
\caption{Upper panels: Exact Routhians for 3 protons and 4 neutrons (left) and 3 protons and 3 neutrons (right)
in a deformed f$_{7/2}$ obtained by diagonalization 
of the two-body Routhian (\ref{eq:defjshell}).  Lower panels: Isorotational CSM approximation (\ref{eq:AdRou}) 
of the results in the upper panels. The various configurations are indicated as $C=[ab...,AB...]_T$, where the excited protons are labelled $a,~b,~...$
and the excited quasineutrons  $A,~B,~...$. From  \cite{Frauendorf99a}. \label{fig:NPA645f4f5}}
\end{center}
\end{figure}

\subsection{Approximate isospin restoration}\label{sec:IVRestore}
Frauendorf and Sheikh \cite{Frauendorf99a} studied how the spectrum depends on the spatial rotational frequency $\omega$ 
for the model system of protons and neutrons in a deformed f$_{7/2}$ shell interacting via a $\delta$-force.
The rotational response was described by means of the cranking approximation (see Section \ref{sec:SMd}).
They obtained the exact energies in the rotating frame of reference (Routhians) by diagonalization of the two-body Routhian (\ref{eq:defjshell})
and and compared them with the corresponding values in Boglujubov mean field approximation. The pair field turned out to be purely isovector.
As already discussed in Section \ref{sec:sigmf}}, they noted that all orientations of the isovector are equivalent and that the y-direction is particularly 
convenient, because the $np$ component is zero. 

They discussed in detail how to combine intrinsic configurations of quasiparticle
 in the conventional like-particle pair field with the isorotational mode to construct states of good isospin. 
Applying the isorotational approximation,
the Routhian is simply given by the sum
\be\label{eq:AdRou}
E'(T,\omega,C)=E'_0(\omega,C)+T(T+1)/2\theta,
\ee
where $E_0(\omega,C)$ is the Routhian of the  configuration $C$ quasiparticles belonging to the deformed mean field, which rotates
with the angular velocity $\omega$. With the choice of the y-axis for the orientation of the isovector pair field, the configuration is determined by 
separately allocating  the excited quasiprotons  and quasineutrons. Fig. \ref{fig:NPA645f4f5} 
compares the exact Routhians with the ones obtained by this procedure. The mean field Routhian was calculated in 
Cranked Shell Model (CSM) approximation \cite{BF79}, which
keeps the isovector pair field constant equal to its  value at $\omega=0$.  The isorational moment of inertia $\theta$ was obtained by means of the isocranking
method. The fair agreement demonstrates that presence of a strong isovector pair field leads to a simple decription of the rotational spectra of $N\approx Z$
nuclei in terms  configurations that combine quasiprotons with quasineutrons. 

As discussed in detail in Ref. \cite{Frauendorf99a}, isospin  conservation imposes rules how to combine quasineutron and quasiproton excitations,
which exclude certain combinations. By these rules, the two degenerate configurations   
in $N=Z$ nuclei obtained by exchanging quasiprotons by quasineutrons will combine into two mixed configuration with $T=0$ at lower and $T=1$
at higher energy. In the framework of their deformed $j$-shell model (see Section \ref{sec:SMd}), Frauendorf {\em et al.} \cite{Frauendorf94}, studied the splitting between 
the $T=0$ and $T=1$ s-bands. These states  represent an equal-weight mixture of the two configurations with a broken pair of quasineutrons or neutrons, respectively, which align
their spins with the rotational axis. In the case of $N=Z$, the $T=1$ state is shifted  to substantially higher energy by the isovector part of the interaction.
In accordance, only one back-bend, which reflects 50\% $pp$ and 50\% $nn$ alignment, is observed in rotational bands of even-even $N=Z$ nuclei.  This is in contrast to
$N\not=Z$ nuclei, where both the neutron- and proton-s bands are observed at close energies.   In this sense the splitting between the $T=1$ and $T=0$ states
originating from two degenerate mean field configurations generated by exchange of protons with neutrons can be considered as evidence for isovector
$np$ pair correlations that restore the isospin symmetry.   

 \begin{figure}[tb]
\begin{center}
\epsfig{file=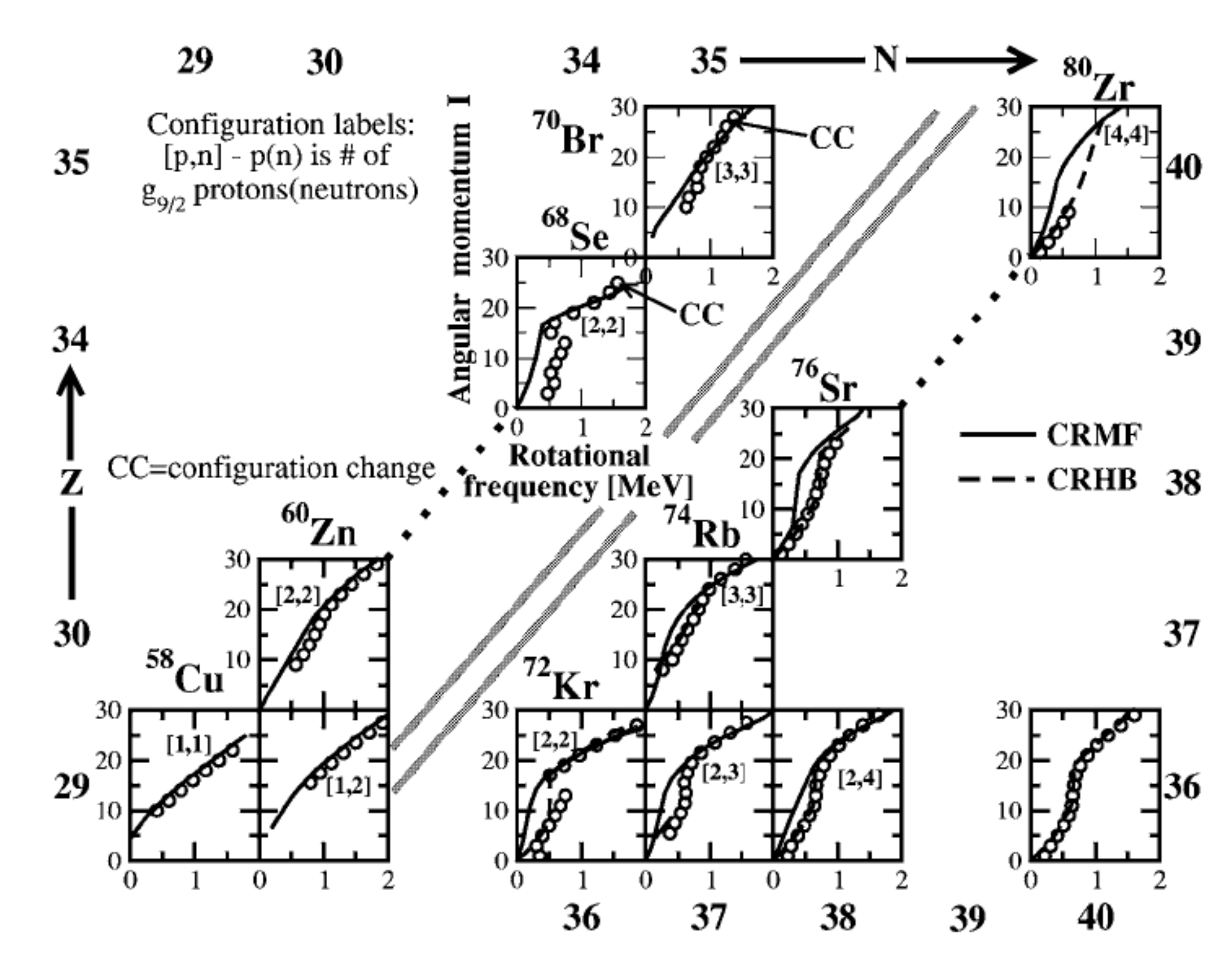,scale=0.8}
\caption{ Angular momentum $J$ (denoted by$I$) as function of the rotational frequency $\hbar \omega=(E(I)-E(I-2))/2$. The circles show the experimental values. The  CRHB calculation
include the isovector pairing on the mean field level with the Lipkin-Nogami correction. The CRMF calculations assume zero pairing. The configuration label $C$=[p,n] 
indicates the number of respective protons and neutrons, respectively, in the g$_{9/2}$ shell. The configuration of the other quasiparticles is omitted.
  From \cite{Afanasjev07}.\label{fig:IJMPE16f6}}
\end{center}
\end{figure}

\subsection{Pure isovector pairing}\label{sec:IVrot}
Frauendorf and Sheikh \cite{Frauendorf99a} used the CSM \cite{BF79} based on the realistic deformed Nilsson potential and a value of $\theta$ derived from the
 experimental binding energies to describe the rotational bands in $^{72}_{36}$Kr$_{36}$ \cite{deAngelis97}, $^{73}_{36}$Kr$_{37}$ \cite{Freund93},
  and $^{74}_{37}$Rb$_{37}$ \cite{Rudolph96}.  The calculations accounted reasonably well the available data. 
  Before this work, Satu\l a and Wyss  \cite{Satula94} analyzed the rotational bands in $^{74}$Kr, $^{78}$Sr and $^{82}$Zr \cite{Rudolph97} and $^{95}$Rb \cite{Gross97}
  in the framework of the Total Routhian Surfaces (TRS) approach, which optimizes the shape of the nucleus for each angular momentum. 
  The authors of the papers \cite{Rudolph97,Gross97} were surprised about the excellent agreement of the calculations with the data, because 
  the proton - neutron pairing should be important for these $T=1$ and $1/2$ nuclei, which is not taken into account in the TRS. 
  Frauendorf and Sheikh \cite{Frauendorf99a}
  clarified the issue. Within the isorotational approximation (\ref{eq:AdRou}), the relative energies  of states of the same isospin  
  are given by their intrinsic values $E_0(\omega,C)$. They can be calculated for the y-direction of the isovector field, which does not
  have a $np$ component. Their TRS calculation for $^{74}$Ru  well reproduced the data ( compare Figs. 5 and 11 
  of  \cite{Frauendorf99a}).
      
  The simplicity of  (\ref{eq:AdRou}), and the fact that sophisticated Cranking codes without $np$ pairing were at 
 hand for calculating   $E_0(\omega,C)$  made the isorotational approximation the standard tool for describing the rotational response of   $N\approx Z$
 nuclei. After the year 2000, the   rotational response   of $N\approx Z$ nuclei was investigated in a number of experimental papers
 \cite{Fisher01, Marginean02,Kelsall01,Kelsall02,OLeary03,Kelsall04,
 Davies07,Andreoiu07,Valiente08,Johnston08,Davies10,Cederwall11}.  Afanasjev , Frauendorf and Ragnarsson
 \cite{Afanasjev04,Afanasjev05}, %\cite{ Kelsall01,Kelsall02,OLeary03,Kelsall04,Davies07,Andreoiu07,Valiente08,Johnston08,Davies10}%
  interpreted the rotational bands 
 of the nuclides with $A=58-80$ in a systematic way, where $E_0(\omega,C)$ was calculated by means of cranked relativistic   Hartree+Boguljubov (CRHB) \cite{Afanasjev00}.
 For quantitative agreement the suppression
 of the isovector pair field by the rotation-induced  alignment of quasiparticles with the rotational axis (cf. e.g Frauendorf in \cite{50yearsBCS}) had to be taken into account,
 which was facilitated by means of the Cranked Relativistic Mean Field (CRMF) theory \cite{Afanasjev96} or the Cranked Nilsson-Strutinsky (CNS) approach \cite{Afanasjev99}.  
The rotational spectra and the reduced transition probabilities $B(E2)$ of $N=Z,~Z+1$ nuclei were reproduced  as  accurately as for nuclei with substantial neutron excess. 
  Detailed  discussions and systematic comparisons with experiment can be found in Refs. \cite{Afanasjev04,Afanasjev05} and Afanasjev in \cite{50yearsBCS}. 
   Fig. \ref{fig:IJMPE16f6}  displays a summary, which  compares  the calculated angular momentum as function of the rotational frequency,  
   $ J(\hbar \omega)$, with the experimental values. These functions, which expose deviations of the 
 calculations from experiment  better than the energies themselves,  are systematically  well reproduced by the calculations.  
 The summary Fig.  4 of Afanasjev in the review \cite{50yearsBCS} shows that the calculated  transition quadrupole moments
   also well accounted for the experiment.

Sun {\em et al.} \cite{Hara99,Velaquez01,Sun01,Palit01,Sun04} used the Projected Shell Model (PSM, see Section \ref{sec:SP}) 
to study the rotational excitations in $N\approx Z$ nuclei.
Their PSM only includes like-particle isovector pairing, i. e.  it provides a description of the relative energies for states of the same isospin. Figs. \ref{fig:PRL83f1} and
\ref{fig:NPA686f3} show examples of their calculations, which very well agree with experiment.  The results for $^{47}$V, $^{47}$Cr, $^{49}$Cr, $^{49}$Mn,
and  $^{51}$Mn \cite{Velaquez01} and  $^{72,74,76}$Kr, $^{76,78}$Sr, $^{80,82}$Zr, $^{84,86}$ and $^{88,90}$Ru \cite{Sun01,Palit01,Sun04} are of comparable quality.
Other quantities as transition quadrupole moments are also very well accounted for. In particular, the observation of a sharp crossing for $Z=36, N=36,38$ and a very washed out for
$Z=,38,N=38,40$ could be reproduced (Fig. 2 of Ref. \cite{Sun04}). It was interpreted as oscillations of the interaction between the crossing bands when the g$_{9/2}$ shell
is progessively filled, which is a  the well known feature of the isovector pair fleld (cf. e.g. \cite{Frauendorf01}).

Langanke {\em et al.} \cite{Langanke03} carried out SMMC calculations based on the purely isovector   Pairing+QQ Hamiltonian (isospin invariant version)
 the fp-gds model space for  $N\approx Z$ nuclei around $A=80$. They could very well reproduce the energies and 
 reduced transition probabilities $B(E2, 2^+ \rightarrow 0^+ )$ and demonstrate
how the progressive occupation of the g$_{9/2}$ shell induces  a transition from spherical to deformed shape.  
 
Based on their experiments,  de Angelis {\em et al.} \cite{deAngelis97}, Fisher {\em et al.} \cite{Fisher01}, and M\u{a}rginean {\em et al.} 
\cite{Marginean02} (and previous work cited therein) suggested a delay
of the crossing between the g-band and s-band in the $N=Z$=36, 38,40,42, 44 nuclides as compared to the $N =Z+2$  isotopes. 
These results initiated several theoretical studies.

Kaneko and Zhang \cite{Kaneko98} carried out CSM calculations. They could generate an upshift  of the crossing frequency by adding a 
$np$ component to the like-particle isovector
pair field, which they interpreted as evidence for additional  $np$ pairing in $N=Z$ nuclei. As we discussed in Section \ref{sec:sigmf}, their modification
amounts to a reorientation of the isovector field and an increase of its absolute value. The reorientation does not change the energies because of isospin invariance.
Increasing the absolute value shifts up the crossing frequency. They could have obtained identical results by increasing  the like-particle pair fields only. 

Frauendorf and Sheikh \cite{Frauendorf99b} calculated the band crossing frequency  for different nucleon combinations in a deformed rotating f$_{7/2}$ shell (see Sect. \ref{sec:SMd}).
 For fixed prolate deformation they identified two trends (see Figs. 4 and 5 of \cite{Frauendorf99b}). (i) With increasing number of nucleons, the crossing moves up and 
 becomes more gradual. This  had been known for nuclei with $N>>Z$ for long and is well understood as caused by the increasing coupling of the $j$-shell states to 
 the deformed potential. (ii)  The crossing frequency increases with a decrease of $\vert N-Z\vert$ with the maximal value at symmetric filling.
 By switching off the $T=0$ part of the $\delta$-interaction, they demonstrated that the shift is caused by the $T=1$ part of the interaction. 
    It is remarkable that the trend involves both even and odd $Z$ (or $N$) for fixed even $N$ (or $Z$). 
 The shift  reflects the additional binding energy of the ground state caused
 by the isovector $np$ pair correlations. 
 
 Sheikh and Wyss \cite{Sheikh00} demonstrated in the framework of the same model that 
an increase of the isoscalar $J=1$ term of the interaction also shifts up the crossing frequency. 
Sun \cite{Marginean02} showed that an increase of
the strength of the QQ interaction (cf. Section \ref{sec:SP}) increases the crossing frequency too, which is expected, because  
a stronger QQ force enlarges the deformation. 

The later experiment 
by O\' ~Leary {\em et al.} \cite{OLeary03} identified another rotational band in $^{72}$Kr, which crosses at about the same frequency as in $^{74}$Kr and fits into the systematics of band crossing shown in
Fig. \ref{fig:IJMPE16f6} (see also Fig. 2 of Ref. \cite{Sun04}). Davies {\em et al.}\cite{Davies07} demonstrated that the smooth alignment  in $^{76}$Sr
is observed at the frequency predicted by the CRHB calculations assuming pure isovector pairing.
 In the heavier $N=Z$ nuclei  the angular momentum of the observed yrast levels is too low to decide if there is  a delay of the band crossing or not.

In summary, calculations based on the isorotational approximation and the assumption of pure isovector pair correlations describe the rotational response of
$N\approx Z$ nuclei as well as they do it for $N \gg Z$ nuclei. In particular,  the crossing frequency between a band and another
containing two rotationally aligned  g$_{9/2}$ quasiparticles is well accounted for.

 \begin{figure}[tb]
 \begin{center}
\begin{minipage}[t]{8 cm}
\hspace*{-0.9cm}
\epsfig{file=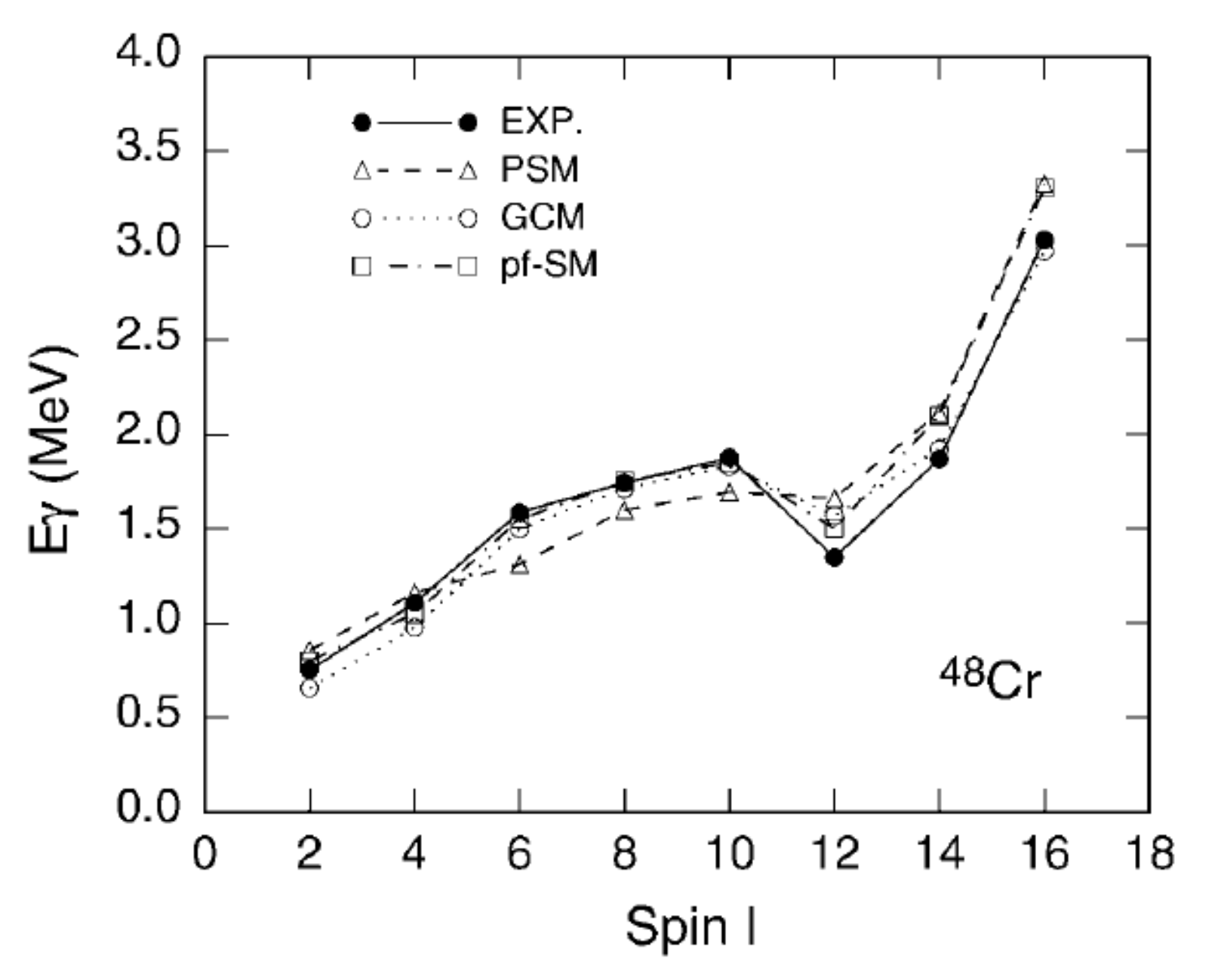,scale=0.55}
\caption{   Transition energy between the yrast levels of $^{48}$Cr as function of the angular momentum $J=I$.  
The experimental values are compared three calculations:  PSM, GCM (a modification of the PSM, cf. Section
\ref{sec:SP}) and the spherical Shell Model calculations \cite{Poves98} (cf. Section \ref{sec:IVISrot}).
From  \cite{Hara99}.\label{fig:PRL83f1}}
\end{minipage}
\hspace*{0.5cm}
\begin{minipage}[t]{8 cm}
\hspace*{-0.8cm}
\epsfig{file=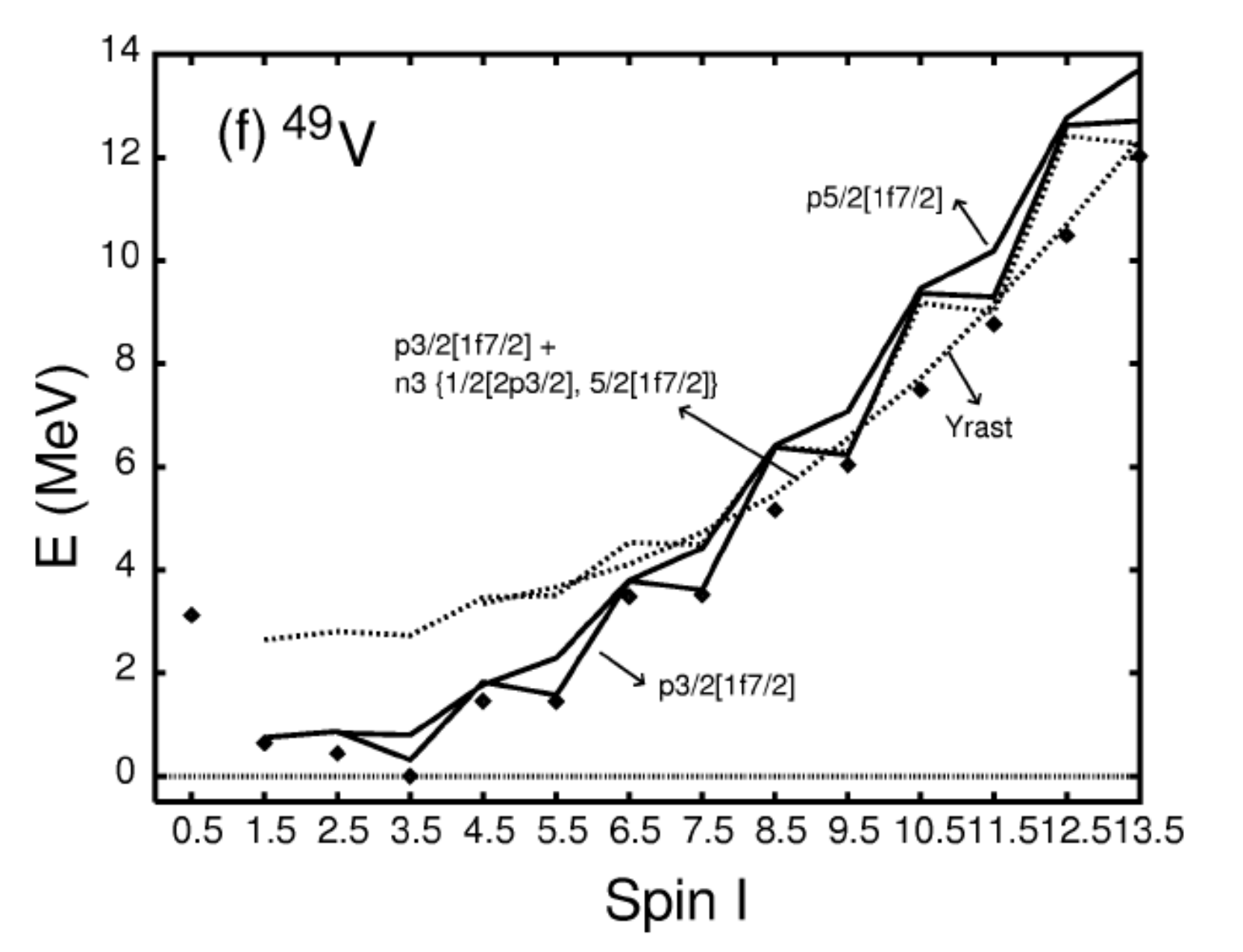,scale=0.55}
\caption{ Experimental yrast energies (diamonds)  of $^{49}$V as functions of the angular momentum $J=I$. 
 compared with the lowest configurations from PSM calculations (full lines one and dashed line
 three quasiparticles). The  configuration
$C$ of the major component in the mixed wave function is indicated for some results. The label p3/2[1f7/2] denotes a
quasiproton from the 1f$_{7/2}$ shell with projection 3/2 on the symmetry axis. Other labels analogous. 
From  \cite{Velaquez01}. \label{fig:NPA686f3}}
\end{minipage}
\end{center}
\end{figure} 
\begin{figure}[tb]
\begin{center}
\begin{minipage}[t]{8 cm}
\hspace*{-0.9cm}
\epsfig{file=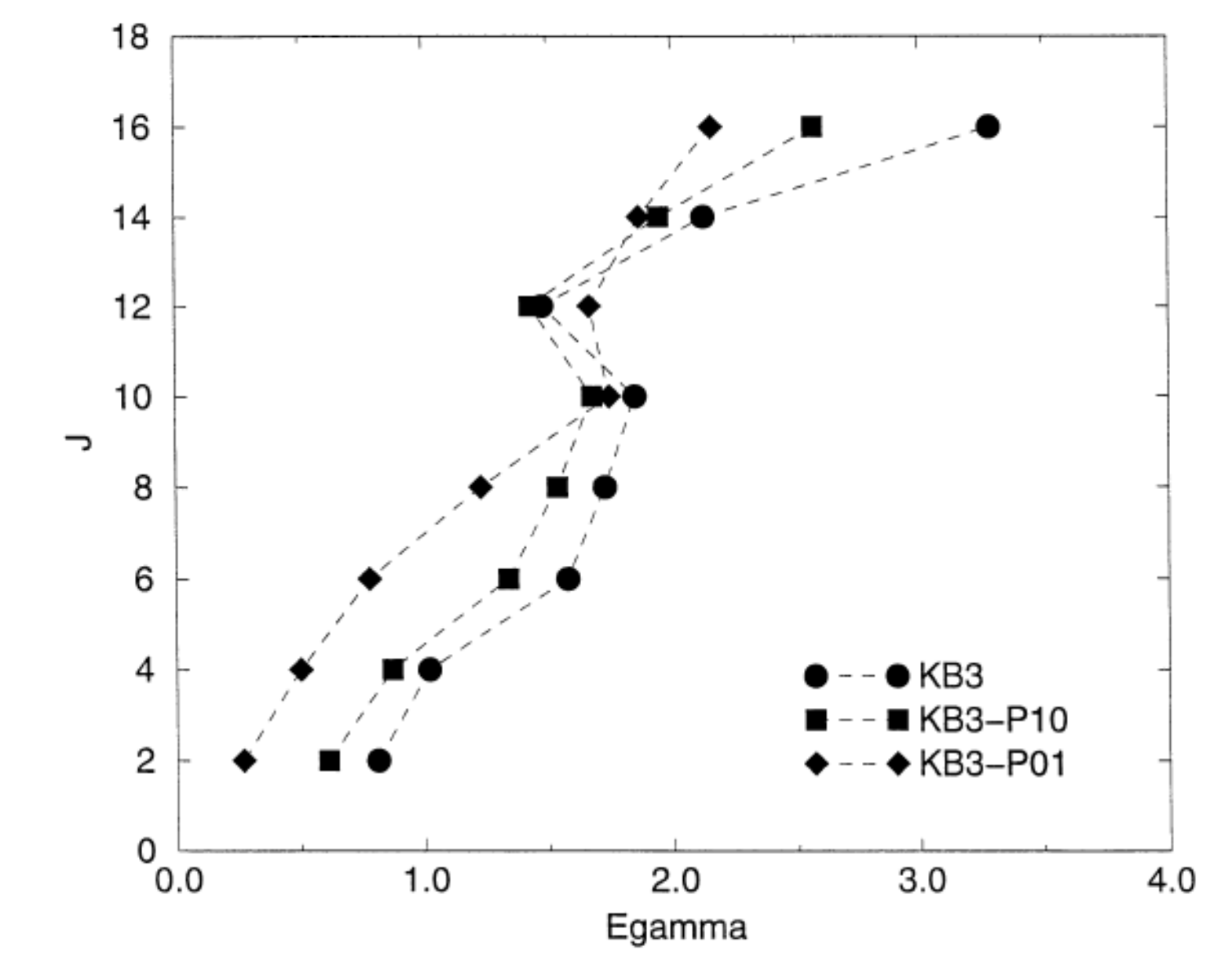,scale=0.55}
\caption{ Angular momentum $J$ as function of the transition energy between the yrast levels of $^{48}$Cr.
KB3 shows the results obtained with the complete effective interaction.
KB3-P01 was obtained by switching off the isovector pairing and KB3-P10 the isoscalar
$J=1$ pairing.  From  \cite{Poves98}.\label{fig:PLB430f5}}
\end{minipage}
\hspace*{0.5cm}
\begin{minipage}[t]{8 cm}
\hspace*{-0.8cm}
\epsfig{file=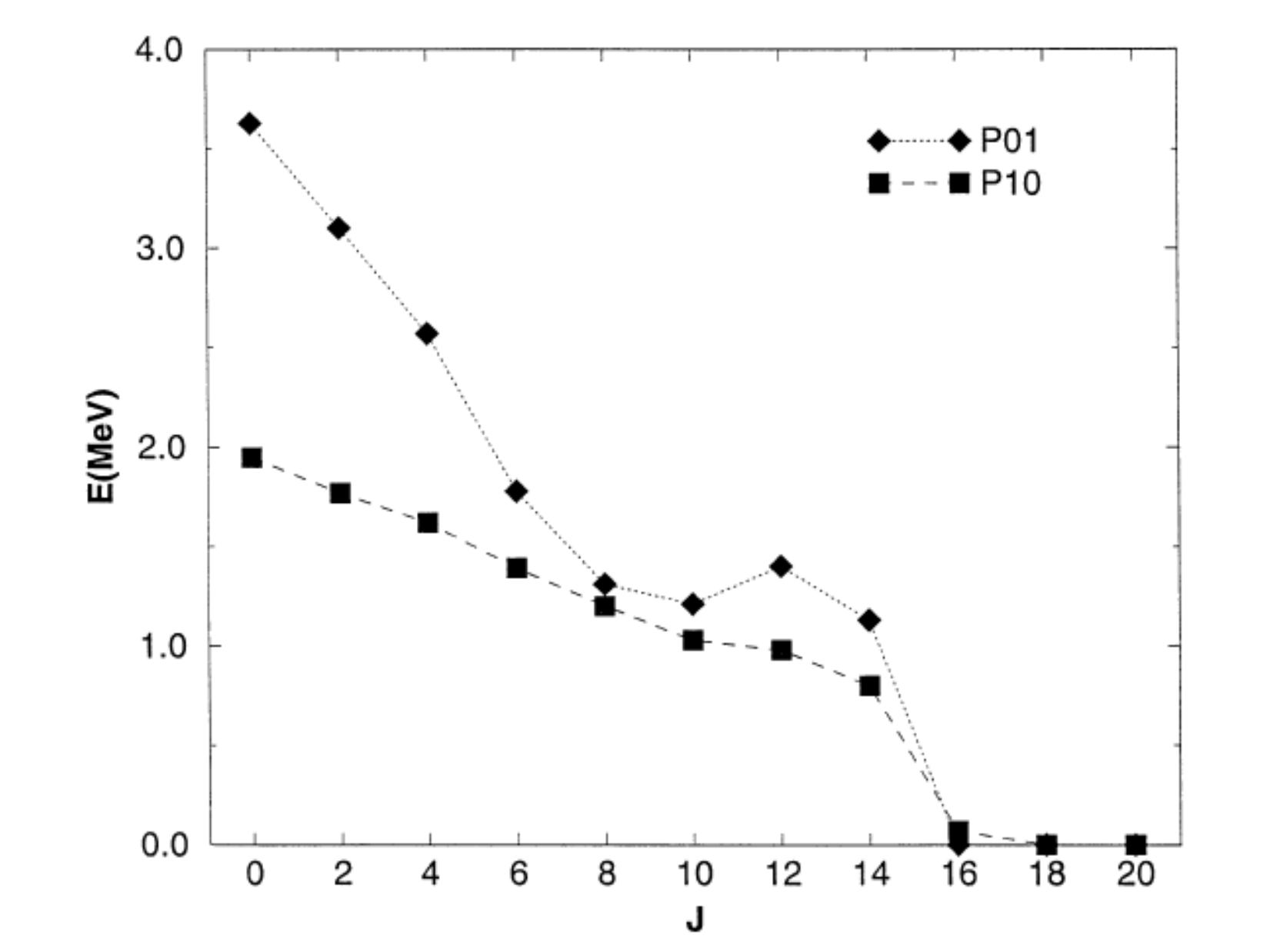,scale=0.55}
\caption{Isovector pair correlation energy P01 and $J=1$ isoscalar pair correlation energy P10 as functions
of the angular momentum $J$ for $^{48}$Cr. See Section \ref{sec:SMs} for the definition of the correlation energies.
From  \cite{Poves98}. \label{fig:PLB430f4}}
\end{minipage}
\end{center}
\end{figure} 

 \subsection{Isovector and isoscalar pairing}\label{sec:IVISrot}
 \subsubsection{Shell Model, mean field, and combined studies }
 
Poves and Martinez-Pinedo \cite{Poves98} calculated the yrast levels of  several $N\approx Z$ nuclei in the lower part of the fp shell (See Section \ref{sec:SMs}).
Fig. \ref{fig:PLB430f5} shows the energy differences between the yrast levels of $^{48}$Cr, which are twice the rotational frequency $\hbar\omega$.
The calculated energies are very close to experiment. The authors extracted pair correlation energies by "switching off" the isovector paring and $J=1$ isoscalar pairing.
 (Their "switching off" procedure is explained in Section \ref{sec:SMs}). Fig. \ref{fig:PLB430f4} shows that the isovector correlation energy is about twice the
 isoscalar one in the ground state.  It rapidly decreases with $J$. The isoscalar correlation fall off at a slower rate. Both disappear at $J=16$,  which is the maximal
angular momentum that can be generated by 4 protons and 4 neutrons f$_{7/2}$ shell (band termination). No correlations can bulid up, because there is only one state.
Fig. \ref{fig:PLB430f5} shows also the calculations with  the two types of pairing switched off. As expected, isovector pairing strongly reduces the moment of inertia
${\cal J}=J/\omega$.  The $J=1$ isoscalar correlations cause a reduction as well, which is however substantially smaller. This is understood as follows. 
Fighting against the correlation of nucleons
to $J=0$ pairs requires most energy when generating angular momentum.  Fighting against the $J=1$ correlations costs less  because the correlation energy
is smaller and angular momentum is generated by aligning the spins of the pairs. Nevertheless, it is more costly than generating angular momentum in the absence
of these correlations. A reduction of the moment of inertia is the consequence. There is  little promise to use this shift as a measure for 
the $T=0$ correlation strength. As discussed below, $T=0$ correlations with large $J$ are present. They increase the angular momentum
and tend to outweigh the shift. As seen in Fig. \ref{fig:PRL83f1}, the PSM calculations of Ref. \cite{Hara99}, which disregard isoscalar correlations, 
reproduce the experiment nearly as well as the spherical Shell Model calculation of Ref. \cite{Poves98}.

Dean {\em et al.} \cite{Dean97}  studied the rotational behavior of $^{74}$Rb \cite{Rudolph96,OLeary03} in the framework of the SMMC described in Section \ref{sec:SMs}.
They added a cranking term $-\omega j_x$ to the Hamiltonian, which  ensured that the mean value of the angular momentum was finite.
Their SMMC calculations provided only low- temperature canonical averages, which  agreed with
experiment within the statistical uncertainties. In particular, their mean values of $T^2$ reflected  the transition  from  $T=1$  to  $T=0$ around  $J=5$ observed
in the  yrast sequence of $^{74}$Rb by Rudolph {\em et al.} \cite{Rudolph96}.  They calculated mean  values of the  modified pair number 
operators (\ref{eq:pairnumberC}) as function of the rotational frequency $\omega$.  The $np$ component of the isovector correlations  dominates  in the ground
state, which is  characteristic for a $T=1, M_T=0$ state (cf. Section \ref{sec:IVcoll}). The strong isovector 
correlations  are rapidly quenched with increasing $\omega$ by breaking a $J=0$ isovector pair
composed of a  g$_{9/2}$ proton and a  g$_{9/2}$ neutron, which align of their spins forming a $J=9$ pair.  Remarkably, the alignment induces noticeable $J=9$ isoscalar 
pair correlations, which are absent in the ground state. They remain constant up to the value of 15 of the angular momentum. 
As discussed, rotation favors the formation of $T=0$ pairs with large $J$. The 
correlation between the $J=9$ may be related  to the presence of   a pair field of the $\alpha\alpha$ type found in CHFB calculations
 of Refs. \cite{Satula00,Goodman01} discussed below. Note, within a spherical $j$-shell, the pairs with $J=2j$ have different 
spin projections $-J\leq M_J \leq J$, between  which the interaction matrix elements are zero. That is, the correlations must be a consequence of 
some deformation, which breaks angular momentum conservation on the mean field level. A significant $J=1$ isoscalar correlation of $P_{corr}=1.9\pm 0.6$ is reported,
which has also been found in SMMC calculations for other $N=Z$ nuclei \cite{Langanke96,Langanke97a} (see Fig. \ref{fig:NPA631f1}).

 Hasegawa {\em et al.}  systematically investigated the rotational spectra of $N\approx Z$ nuclei in the lower part of the fp shell \cite{Hasegawa00,Hasegawa05a,Hasegawa05b}
 and of the nuclides $N=Z=$34,35,36,37,38 \cite{Hasegawa07} in the framework of their
 Shell Model version (See Section \ref{sec:SMs}).   Overall,  the calculations well describe  the experimental spectra,
 in particular, the onset of deformation  and the shape coexistence phenomena.  
 Ref. \cite{Hasegawa00} shows a figure for $^{48}$Cr similar to Fig.  \ref{fig:PLB430f5}, which includes calculations with  modified interactions.
 Their full EPQQ interaction reproduces the experiment with comparable accuracy. Switching off the $T=0$ part of EPQQ, i.e. using PQQ, did not  change the 
 function in any significant way, which is consistent with the above discussed weak influence of the $T=0$ interaction on the moment of inertia that  Poves and Martinez-Pinedo observed. 
The relative position of the $T=1$ and $T=0$ bands in the odd-odd $N=Z$ was well accounted for. The expected dependence on the $T=0$ part of the interaction was not quantified 
in these studies.

Goodman \cite{Goodman01} investigated the rotational response of $^{80}$Zr in the framework of CHFB approach (see Section \ref{sec:MF}) using
the pair potentials as a measure the correlation strengths \mbox($[\sum_{M_JM_T}\vert \Delta_{JM_JTM_T}\vert^2]^{1/2}$)},
and found two coexisting  mean field solutions. 
 One solution has  a pure isovector pair potential.  The $J=0$ component dominates. A weak $J=2$ and a very small $J=4$ component
 reflect the deformation of the mean field. The calculated moment of inertia is not far from the experimental value.
 The second solution has  a pure isoscalar pair field.    The dominant components are $J=9$ and $J=5$, which are
 about a factor 4 weaker than the isovector $J=0$ component  of the other solution. They do not change much with increasing  the angular moment, whereas
 the $J=1$ and 3 components grow from zero to comparable values at $I=16$. The isoscalar solution lies 0.65 MeV above the isovector one at zero spin.  
It becomes yrast above $I=4$, because it has a larger moment of inertia. Goodman explains the difference between moments of inertia as follows. 
Generating angular momentum
in the isovector phase, one has to partially lift the $J=0$ pair correlations,  which is costly energy-wise. 
In the isoscalar  phase,  most angular momentum  is generated by gradually aligning the pairs that carry substantial angular momentum, which costs less energy.  
The predicted band crossing is not seen in experiment. The moment of inertia of $^{80}$Zr stays nearly constant up to $I=10$, which is consistent with the calculations that assume pure isovector pairing (see Fig. \ref{fig:IJMPE16f6} and Ref. \cite{Sun04}). Presumably, Goodman's interaction favors the isoscalar solution too much.

Terasaki {\em et al.} \cite{Terasaki98} carried out CHFB calculations for $^{48}$Cr based on the Skyrme interaction combined with  surface $\delta$-interactions
generating  isovector and isoscalar  pair correlations. Below $I=16$, they found an isovector yrast solution and an excited isoscalar solution  and above 
$I=16$ a deformed pure isoscalar solution.    
Sheikh and Wyss \cite{Sheikh00} also found  a pure isovector and a pure isoscalar CHFB solutions in the framework of their deformed f$_{7/2}$ model, which are similar 
to Goodman's solutions.  They demonstrated that the exact results obtained by numerical diagonalization  have  the character of a mixing of the two CHFB solutions.
They calculated the expectation values of the pair counting operators ${\cal N}_{JT\mu}$, Eq. (\ref{eq:pairnumberJT}), for various values of the angular momentum 
$J$ of the pairs as functions of the rotational frequency. 
The isovector part shows the expected drop when a pair of f$_{7/2}$ nucleons aligns with the rotational axis.  The total isoscalar value does not change with frequency.
However there is a shift of strength from $J=1$ to $J=7$, which reflects the  alignment of the nucleons with the rotational axis. Since the authors do not calculate 
uncorrelated reference values, one cannot asses the correlation strength.  
% Sheikh and Wyss only report two distinct CHFB solutions with isovector
% and isoscalar character whereas Goodman also found a third solution with mixed character.

Petrovici, Schmid, and Faessler \cite{Petrovici96,Petrovici99,Petrovici00,Petrovici02a,Petrovici02b,Petrovici03} calculated the rotational spectra of
$N\approx Z$ nuclei in the framework of their "Excited Vampir" approach, which does not conserve isospin.
 As far as one can tell from the level schemes shown, the calculated energies of even-spin positive parity states 
reproduce  the experimental  ones well in the case of even-even nuclei.  In particular, the 
 coexistence between prolate and oblate shapes is accounted for. 
  As  discussed in  Section \ref{sec:IVRestore}, models that assign a direction to the 
 isovector pair field are capable of describing the relative energies of states with the same isospin, which is the case.
  In Ref. \cite{Petrovici99}, the authors  provide extensive information about the expectation values of  operators counting pairs ${\cal N}_{JT\mu}$ (cf. Eq. (\ref{eq:pairnumberJT})),
  in the $^{72,74,76,78}$Kr isotopes and $^{74}$Rb.
 In the ground state the $T=1$ pairs are  distributed over $J$ with a maximum at $J=4$, which reflects the deformation of the mean field solutions
 used in Vampir.   With increasing angular momentum $I$ of the yrast states the number of $T=1, J=0$ and
  $T=0, J=1$ pairs decreases and  the number of $J=8$ and 9 pairs increases. 
 However it is difficult to estimate the correlation strength  from the results, because there is no
 reference to the uncorrelated system.
 
Stoicheva {\em et al.}  \cite{Stoicheva06} investigated  terminating states in the framework of the Shell Model. These  states
have a simple structure. All valence nucleons align their spins in accordance with the Pauli Principle.  
Their calculations reproduce the observed energy differences between the aligned configurations d$_{3/2}^{-1}$f$_{7/2}^{n+1}$ and  f$^{n}_{7/2}$
in $N>Z$ nuclei around mass 44 but  underestimate it by 500 keV for the $N=Z$ nuclides. Mean field calculations based on the Skyrme
interaction combined with approximate restoration of isospin along the   lines described in Section \ref{sec:IVRestore} or by isospin 
projection \cite{Satula10} gave  results very similar to the Shell Model. The authors suggest that missing isoscalar components of 
the interaction may be responsible for the discrepancy. 

Summarizing, the calculations do not reveal a clear signal from the  $J=1$ deuteron-like pair correlations in the rotational response.
They indicate the emergence of pairs with large angular momentum, which are favored by the isoscalar interaction. Their role  will be discussed in the next two subsections. 

\begin{figure}[tb]
 \begin{center}
\begin{minipage}[t]{8 cm}
%\hspace*{-0.9cm}
\epsfig{file=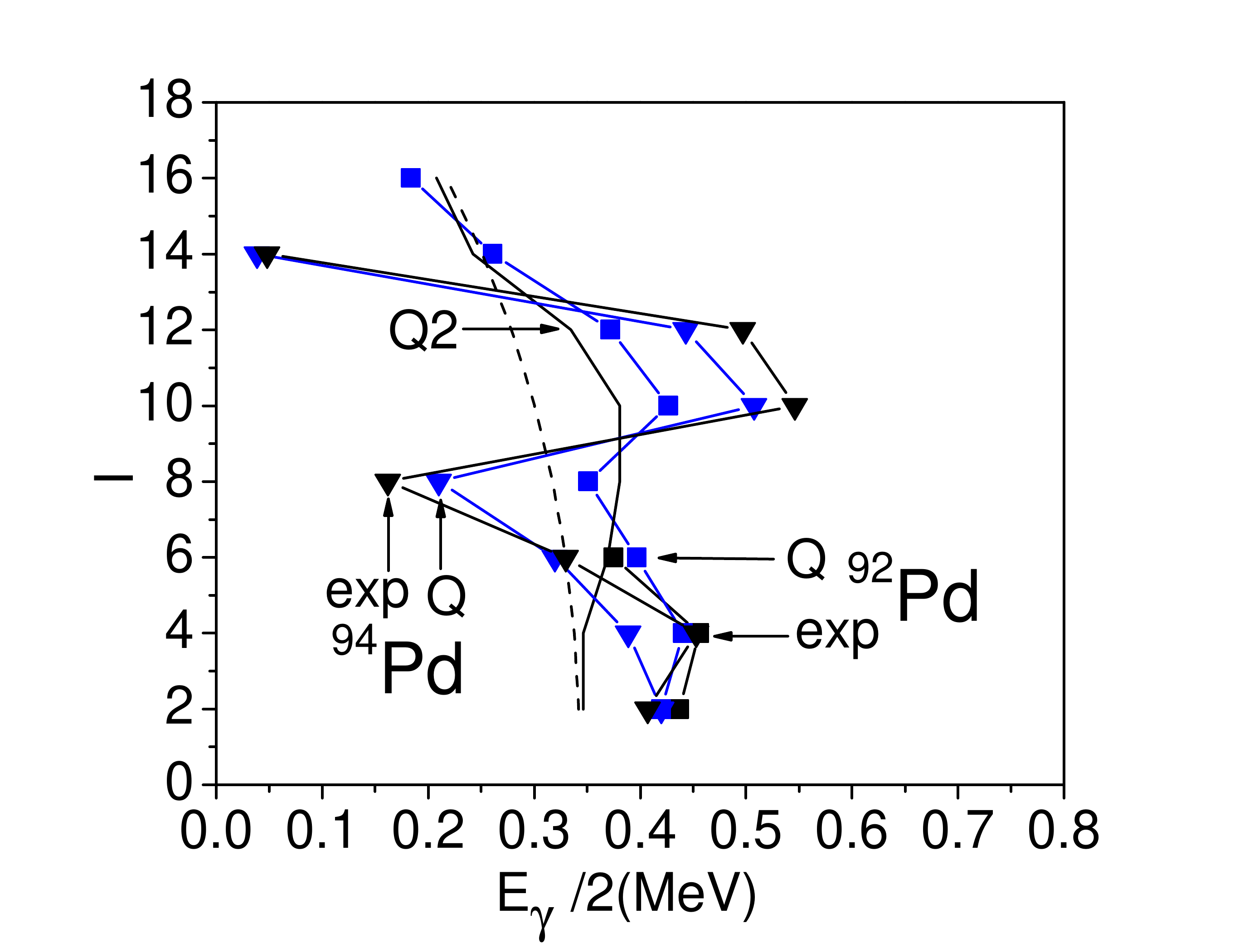,scale=0.3}
\caption{ Transition energy $E_\gamma$ between the yrast levels of
$^{92,94}$Pd calculated by Qi {\em  et al.} (Q)  compared to experiment (exp).
Q2 shows the calculation with doubled strength of the isoscalar interaction.  The dashed
line shows the  estimate using  the AMR coupling scheme.
Data from  \cite{Qi11} and \cite{privcom}.\label{fig:Iom92Pd94Pd}}
\end{minipage}
\hspace*{0.5cm}
\begin{minipage}[t]{8 cm}
\epsfig{file=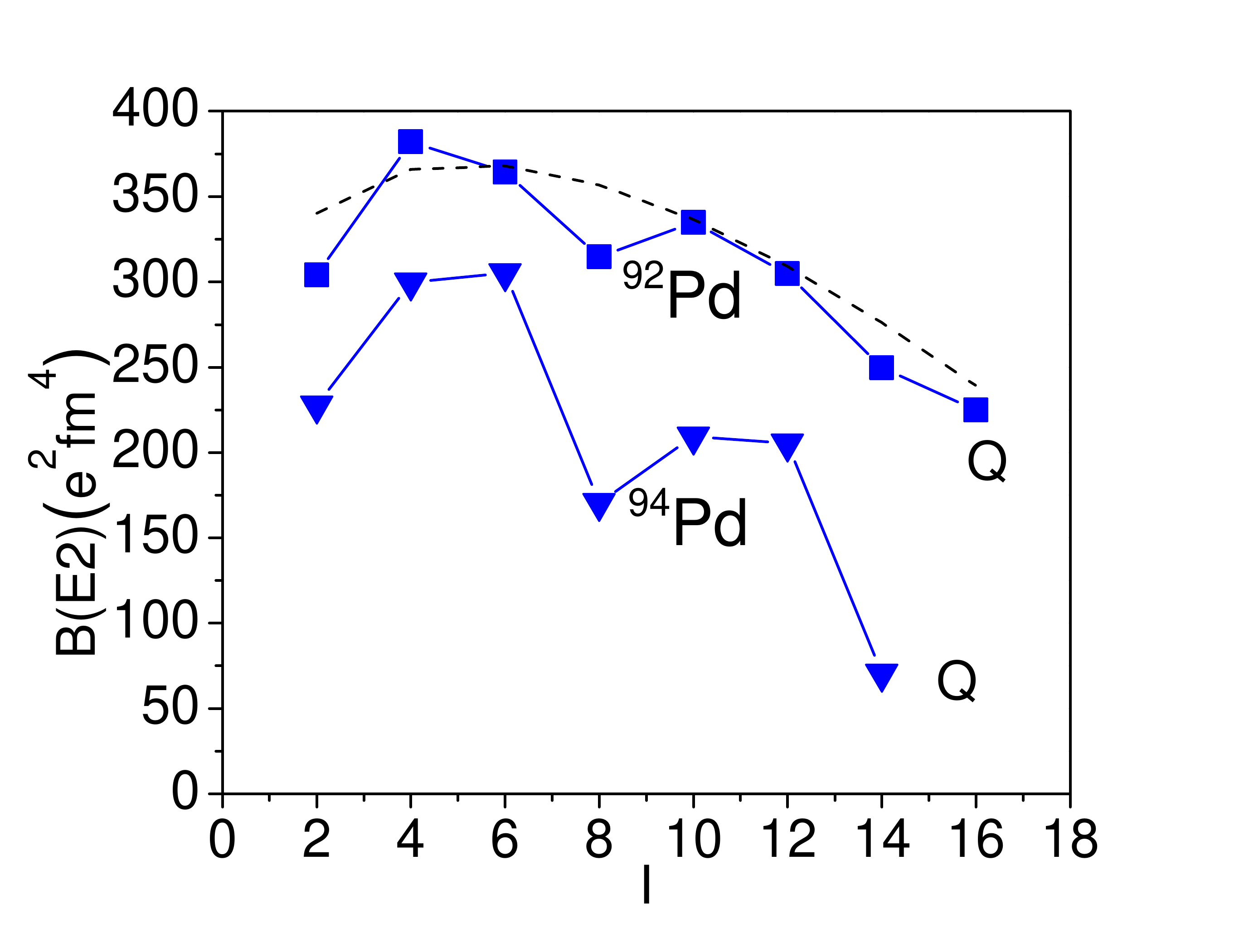,scale=0.3}
\caption{ Reduced transition probability  between the yrast levels of
$^{92,94}$Pd calculated by Qi {\em et al.} (Q).
The dashed line shows the  estimate using the AMR coupling scheme, where $B(E2, I\rightarrow I-2)=$
\mbox{$4/15\pi\langle I~0~2~0\vert I-2~0\rangle^2Q_t(I)^2$}.  
 Data from \cite{Qi11} and \cite{privcom}. \label{fig:BE292Pd94Pd}}
\end{minipage}
\end{center}
\end{figure} 
\begin{figure}[tb]
\begin{center}
\begin{minipage}[t]{8 cm}
\epsfig{file=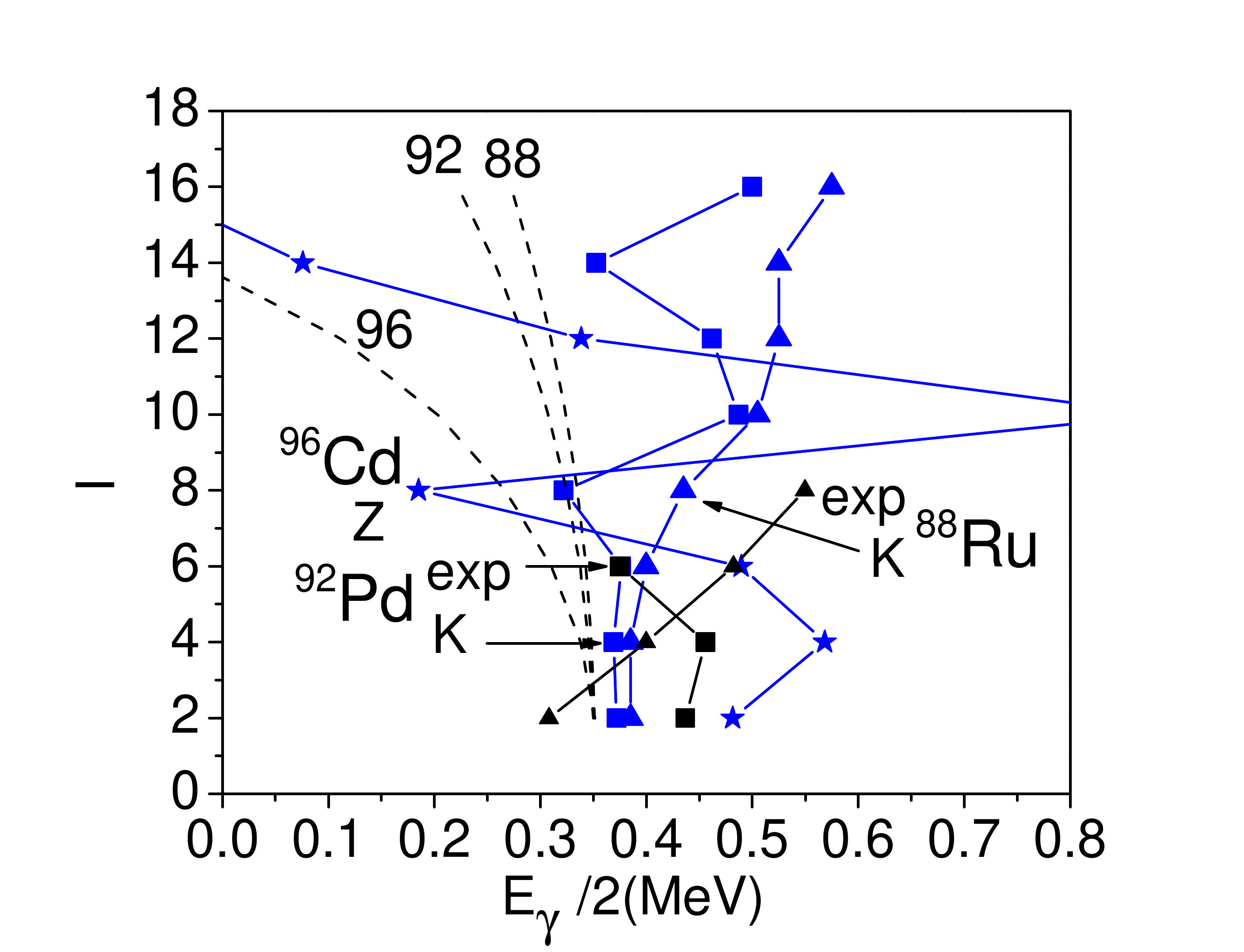,scale=0.3}
\caption{ Transition energy $E_\gamma$ between the yrast levels of
$^{88}$Ru and $^{92}$Pd calculated by  Kaneko {\em et al.} (K) and 
$^{96}$Cd calculated by Zerguine and Van Isacker  (Z) compared to experiment (exp).
 The dashed
lines show the  estimates using the AMR coupling scheme.
 Data from  \cite{Hasegawa04,Kaneko08,Zerguine11} and \cite{privcom}.\label{fig:Iom88Ru92Pd96Cd}}
\end{minipage}
\hspace*{0.2cm}
\begin{minipage}[t]{8cm}
\hspace*{-0.5cm}
\epsfig{file=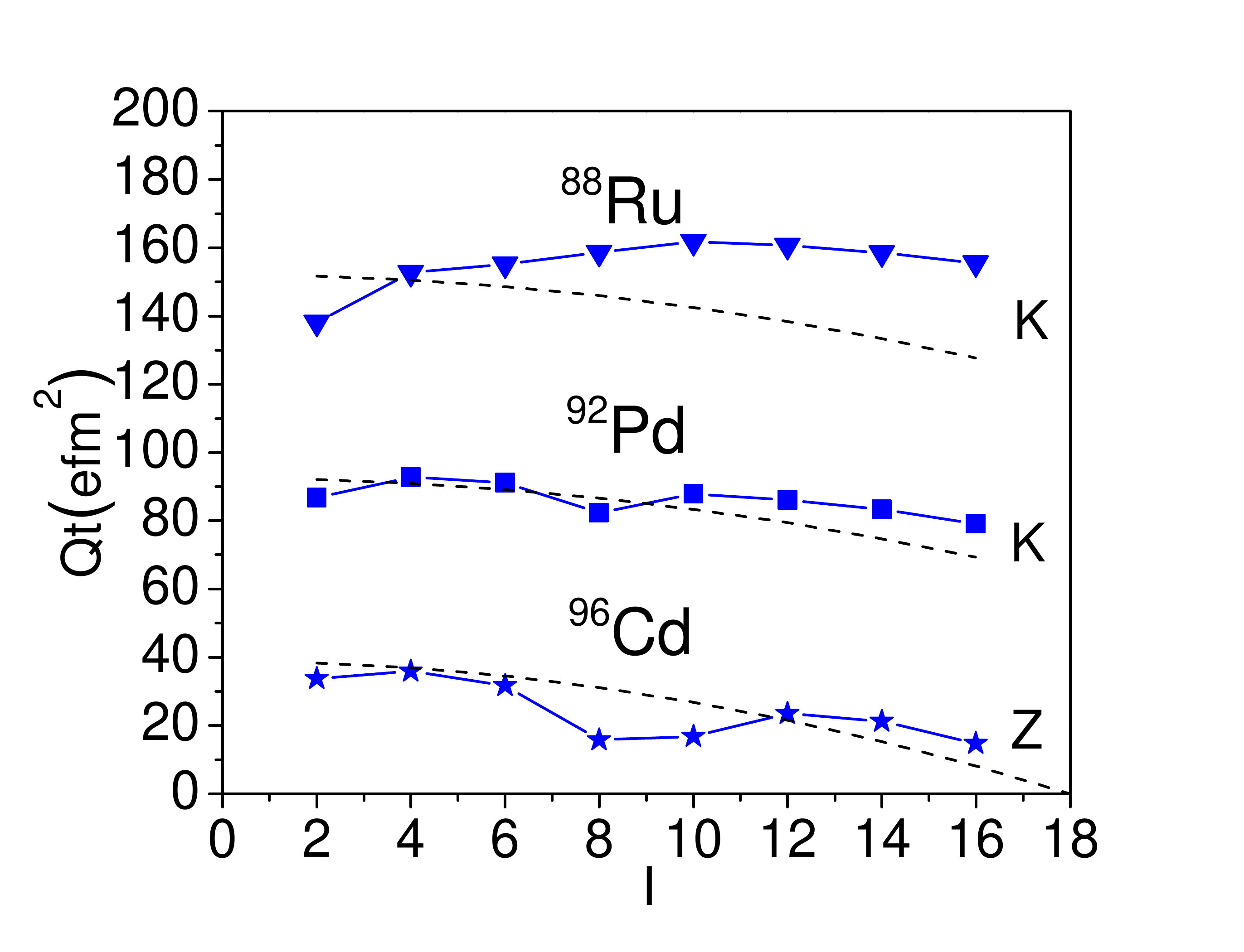,scale=0.3}
\caption{ Transition quadrupole moment between the yrast levels of
$^{88}$Ru and $^{92}$Pd calculated by  Kaneko {\em et al.} (K) and 
$^{96}$Cd calculated by Zerguine and Van Isacker  (Z).
The dashed lines show the  estimate using the AMR coupling scheme.  
  Data from  \cite{Hasegawa04,Kaneko08,Zerguine11} and \cite{privcom}. \label{fig:BE288Ru92Pd96Cd}}
\end{minipage}
\end{center}
\end{figure}

\subsubsection{The spin-aligned coupling scheme}\label{sec:SpinAligned}

Cederwall {\em et al.} \cite{Cederwall11}, who  measured the yrast  levels of $^{92}$Pd for the first time, suggested the existence of a 
"spin-aligned phase" based on an analysis of the Shell Model results of Ref. \cite{Qi11}. In our view, this "new pairing phase"  
represents a mechanism for generating angular momentum that has been suggested before under different names (Stretch Scheme, Magnetic/
Antimagnetic Rotation), which does not incorporate correlated pairs as an essential ingredient. The following discussion is based on
the results obtained  in the framework of the conventional Shell Model (see section \ref{sec:SMs}) by various authors.
Qi {\em et al.} \cite{Qi11}  and Kaneko {\em et al.} \cite{Kaneko08} studied
 $^{92,94}$Pd and $^{96}$Cd,  Hasegawa {\em et al.} \cite{Hasegawa04} studied $^{88,90}$Ru, Zerguine and Van Isacker 
 \cite{Zerguine11} studied $^{92}$Pd and $^{96}$Cd,  and Coraggio {\it et al.} \cite{Coraggio12} studied $^{92,96}$Pd and $^{96,98}$Cd.
 Their results were used to produce  Figs. \ref{fig:Iom92Pd94Pd} - \ref{fig:BE288Ru92Pd96Cd}, which show the angular momentum $I$
as the function of the transition energy $E_\gamma=E(I)-E(I-2)$  and the reduced transition probabilities $B(E2, I \rightarrow I-2)$ or the respective
transition quadrupole moments $Q_t(I)$ between the yrast levels.  

 Cederwall {\em et al.}
pointed out that the transition energies  are approximately
independent of $I$ like for a sequence of vibrational excitations, whereas 
 the  reduced transition probabilities $B(E2)$ are nearly independent of $I$ in contrast to the standard collective vibrational model, which predicts a 
linear increase of $B(E2)$ with spin (see Figs. \ref{fig:Iom92Pd94Pd} and \ref{fig:BE292Pd94Pd}).
 Analyzing the wave functions  (see \cite{Qi11,Qi10,Qi11b} for details), they found that
 the yrast states  to good approximation are composed of four $J=9$ $np$  pairs, which are coupled to the total angular momentum. 
The authors interpreted their findings  as follows. The nucleus $^{92}$Pd can be considered to be a manifestation of a $np$ paired phase corresponding 
to an isoscalar spin-aligned coupling scheme. In  the ground state  these $J=9$ $np$ pairs are pairwise anti-aligned coupled. 
With increasing angular momentum, they re-couple by gradually aligning their spins.

This "spin-aligned phase" is closely related the Stretch Scheme suggested by Danos and Gillet \cite{Danos67} to explain nuclear rotational bands and the
 concept Magnetic Rotation (MR) introduced by Frauendorf \cite{Frauendorf93} (see \cite{Frauendorf01,Clark00}) for further development) to explain regular rotational bands in near-spherical nuclei.  
Danos and Gillet  assumed that half of the valence nucleons are coupled to maximal spin according to the Pauli principle and the other half is coupled in the same way.
These two groups  are then coupled to the total angular momentum, 
which increases as  the two groups  align their spins.
The mutual overlap between the orbitals  from the two groups decreases when they close. The resulting increase of energy, 
 the rotational energy,  was qualitatively accounted for.
 In the case of MR, the angular momentum is generated by the same  mechanism.
A group of high-$j$ valence particles and a group of high-$j$ valence holes form the two "blades of a pair of shears",  which are 
assumed to interact via the slightly deformed nuclear potential.  
Regular rotational sequences, the "shears bands",  are obtained by "closing the shears". 
The  nearly quadratic increase of the energy with $I$ and the very small $B(E2)$ values
were calculated   in quantitative agreement with the  experiment \cite{Frauendorf01,Clark00}. 
  Antimagnetic Rotation (AMR)  \cite{Frauendorf01}  represents a modification of MR for which the
  particles and holes arrange in a symmetric way, such that there is no magnetic radiation and the spin increases in units of two along the band.
  The following discussion uses the AMR scheme.
    
 Figs.   \ref{fig:Iom92Pd94Pd} and \ref{fig:BE292Pd94Pd} include an application of the AMR shears mechanism to $^{92}$Pd. 
 Each blade is composed of two g$_{9/2}$ hole pairs. Their interaction
 is assumed to be  of the Schiffer-True  form \cite{Schiffer76} (see Fig. \ref{fig:SchTru}).  Approximating its  isoscalar part  by $-A+B\sin \theta_{12}$, 
 where $\theta_{12}$ is the angle between the blades, 
 and applying the expressions from Ref. \cite{Clark00}, one obtains  the energy $E(I)$ and
 the transition quadrupole moment $Q_t(I)$. The scale of both quantities is a free parameter, which is 
 adjusted to the Shell Model calculations.
 The dashed lines carry Cederwall's {\it et al.} signature of the spin-aligned phase:  Both  the transition energies $E_\gamma$ and reduced probabilities
 $B(E2)$ do not change much with the angular momentum. As seen in the figures, both quantities slightly decrease with $I$ in accordance with the 
 Shell Model calculations. Figs.  \ref{fig:Iom88Ru92Pd96Cd} and \ref{fig:BE288Ru92Pd96Cd} compares the AMR scheme with 
 the Shell Model calculations for $N=Z$ nuclei. For $^{96}$Cd, $^{92}$Pd, and $^{88}$Ru, the blades have the length  $J=18$, 32, and 40,
 corresponding to 2, 3, and 4  g$_{9/2}$ hole pairs, respectively. The AMR coupling scheme qualitatively accounts for the  Shell Model results
  for $^{96}$Cd and  $^{92}$Pd but it does not  for  $^{88}$Ru
 and even less for the experimental $E_\gamma$ values from this nucleus. The discrepancy  indicates that
with a larger number of valence holes 
the deformed  average potential will become more important. As a consequence,  the  interaction
between the blades will change from the Schiffer-True form towards the $-B\cos^2\theta_{12}$ form, which gives the familiar linear
increase of $E_\gamma$ with $I$ of rotational bands (see Fig. 12 in \cite{Clark00}). 

 Cederwall {\em et al.} suggest that  the spin-aligned coupling scheme appears in the $N=Z$  as a transition from the isovector pair phase 
  to new phase characterized by 
 strong $J=9,T=0$ pair correlations.  In justifying, they point out that the $N>Z$ isotopes show the signature of the isovector seniority scheme.
 Fig. \ref{fig:Iom92Pd94Pd} illustrates their point for  $^{94}$Pd. The transition energies   decrease rapidly up to $I=8$, jump up, and decrease again,
 which is understood as follows. The first $J=0$ isovector  pair breaks at $I=8$, where the two g$_{9/2}$ holes align their spins. 
This is followed by  the alignment of the second pair.  However, 
 as seen in Fig. \ref{fig:Iom92Pd94Pd}, the signature of pair breaking appears in $^{92}$Pd in a  mitigated form as well.  
This rather points to coexistence of the spin-aligned and the seniority coupling schemes than a phase transition.

 One expects that the interplay between  the coexisting modes is  determined  by the number of available $np$ pairs and the ratio of the isovector and isoscalar 
 components of the interaction.  This is confirmed by the  Shell Model  calculations. Qi {\em et al.} \cite{Qi11} 
 changed   the $T=0$ strength of their interaction.    Taking only $T=1$ matrix elements into account, they 
 obtained a pronounced jump in the function $I(E_\gamma)$, the hallmark of the isovector seniority scheme.
 Enhancing the strength of the isoscalar interaction by a factor of two gave  the results  denoted by Q2 
 in Fig. \ref{fig:Iom92Pd94Pd}. The jump in the function $I(E_\gamma)$, which indicates the breaking of the isovector
pair, is replaced by a bulge (relative to the dashed line), which indicates a gradual reorientation of the two nucleons. 
The  jump in  $I(E_\gamma)$ is progressively washed out with number of isoscalar pairs ($J$=odd and large)
present: As seen in Fig. \ref{fig:Iom92Pd94Pd}, it is more pronounced for $N=Z$  than $N\not=Z$. As seen in Fig. \ref{fig:Iom88Ru92Pd96Cd}, 
it is  most pronounced  in $^{96}$Cd ( 2 pairs), less in $^{92}$Pd (4 pairs) and
least in $^{88}$Ru (6 pairs).   Figs. \ref{fig:BE292Pd94Pd} and \ref{fig:BE288Ru92Pd96Cd} demonstrate that the deformation of the nuclei
increases with the number of holes. It well known from the study of the $N\gg Z$ nuclei that an increase of deformation smoothes out
the sudden break of an isovector pair to a gradual alignment of the angular momenta of the two nucleons  (cf. e. g. \cite{Frauendorf01}).   

 The Shell Model values of the transitional quadruple moments in
Fig. \ref{fig:BE288Ru92Pd96Cd} corroborate the preceding  scenario of a  transition from  the dominance of the shears mechanism in $^{96}$Cd (down sloping)  
to rotation of a soft deformed mean field in $^{88}$Ru (up sloping). As pointed out by Qi {\it et al.}  \cite{Qi11} such a transition is expected 
when the configurations involving f$_{5/2}$, p$_{3,2}$, p$_{1/2}$ become significant. The
calculation of the yrast levels  within the framework of the PSM \cite{Sun04}, which takes these and more single particle orbitals into account,
very well reproduces the experimental $I(E_\gamma)$ function and predicts   a rather abrupt alignment. The PSM  is based on the PQQ Hamiltonian. 
The calculation within the EPQQ \cite{Hasegawa04} , which includes an additional $T=0$ pair interaction (compare Eq. (\ref{eq:PQQ}) with Eq. (\ref{eq:EPQQ}) )
gives a very gradual alignment with $I$ (cf. Fig. \ref{fig:Iom88Ru92Pd96Cd}).  The reason for the discrepancy is not clear.
It may   point to $np$ correlations  beyond the quadrupole deformation of the average potential or just be a consequence of the
larger single particle space used in the PSM calculation \cite{Sun04} because the smoothness of the 
rational alignment depends also sensitively on the position of the chemical potential in the deformed g$_{9/2}$ shell 
(see \cite{Frauendorf01}).  Measuring the yrast levels with $I>8$ would be of great interest.

In our view,  the rotational response of the $N=Z$ nuclei in the upper g$_{9/2}$ shell, as predicted by the Shell Model, 
reveals the familiar scenario of  the quenching  the  isovector pair correlations by rotational alignment of an isovector $np$ pair,
which is combined with the shears mechanism for generating angular momentum.
The isoscalar interaction favors the mutual alignment  of the g$_{9/2}$ holes.  However it
 does not presume the existence of scattering of these   aligned pairs. In following section we address the ambiguity in interpreting
 the Shell Model results, which is a consequence of the small number of valence holes.    

\subsubsection{Isoscalar pair correlations vs. isoscalar interaction}

 Cederwall {\em et al.} based their concept of the spin-aligned phase on the analysis of the Shell Model wave functions.  Their yrast states of $^{92}$Pd
 have an overlap of about 90\% with a spin-aligned state that is composed of four $J=9$ $np$ 
 hole pairs, $A^{00}_{9M_J}\equiv B$, which are coupled as $\left[[B^2]_{18} B^2]_{18} \right]_I\vert CS\rangle$, 
 where $\vert CS\rangle$ represents the completely filled g$_{9/2}$ shell. 
 This large overlap lead them to declare that the nucleus is in a spin-aligned phase.  
Subsequent analyses by Zerguine and Van Isacker \cite{Zerguine11} and Neergaard \cite{Neergard13} revealed a more complex picture.
In the case of $^{96}$Cd, the spin-aligned state is composed of two pairs,  $\vert SA;I\rangle=\left[B^2\right]_I\vert CS\rangle/No(SA;I)$, where $No$ is the 
normalization constant.
Table IV of \cite{Zerguine11} quotes a $\langle SA;0\vert SM;0 \rangle^2$=91\% overlap of the $I=0$ Shell Model ground state $\vert SM;0\rangle$ with this spin-aligned state.
However it also quotes a $\langle IV;0\vert SM;0 \rangle^2$=80\% overlap with the  ground state of isovector pair Hamiltonian (\ref{eq:HIV}),  
\mbox{$\vert IV;0\rangle=\sum_\mu  (-)^\mu P_\mu P_{-\mu}\vert CS\rangle/No(IV;0)$}. 
This is possible, because the spin-aligned  and the seniority-zero states  have a large overlap of $\langle IV;0\vert SA;0 \rangle^2$=52\% among themselves.
Neergaard showed that  with very good accuracy $\vert SM;0\rangle=\alpha \vert SA;0\rangle+\beta\vert IV;0\rangle$. Geometrically, the  overlaps
correspond to angles between the state vectors of $\angle SM,SA=18^\circ$, $\angle SM,IV=26^\circ$, and $\angle SA,IV=44^\circ$=18$^\circ$+26$^\circ$.
 He further demonstrated that
the geometry is very similar for all effective interactions used within the g$_{9/2}$ shell.  

\begin{figure}[tb]
\begin{center}
\begin{minipage}[t]{8 cm}
\hspace*{-0.6cm}
\epsfig{file=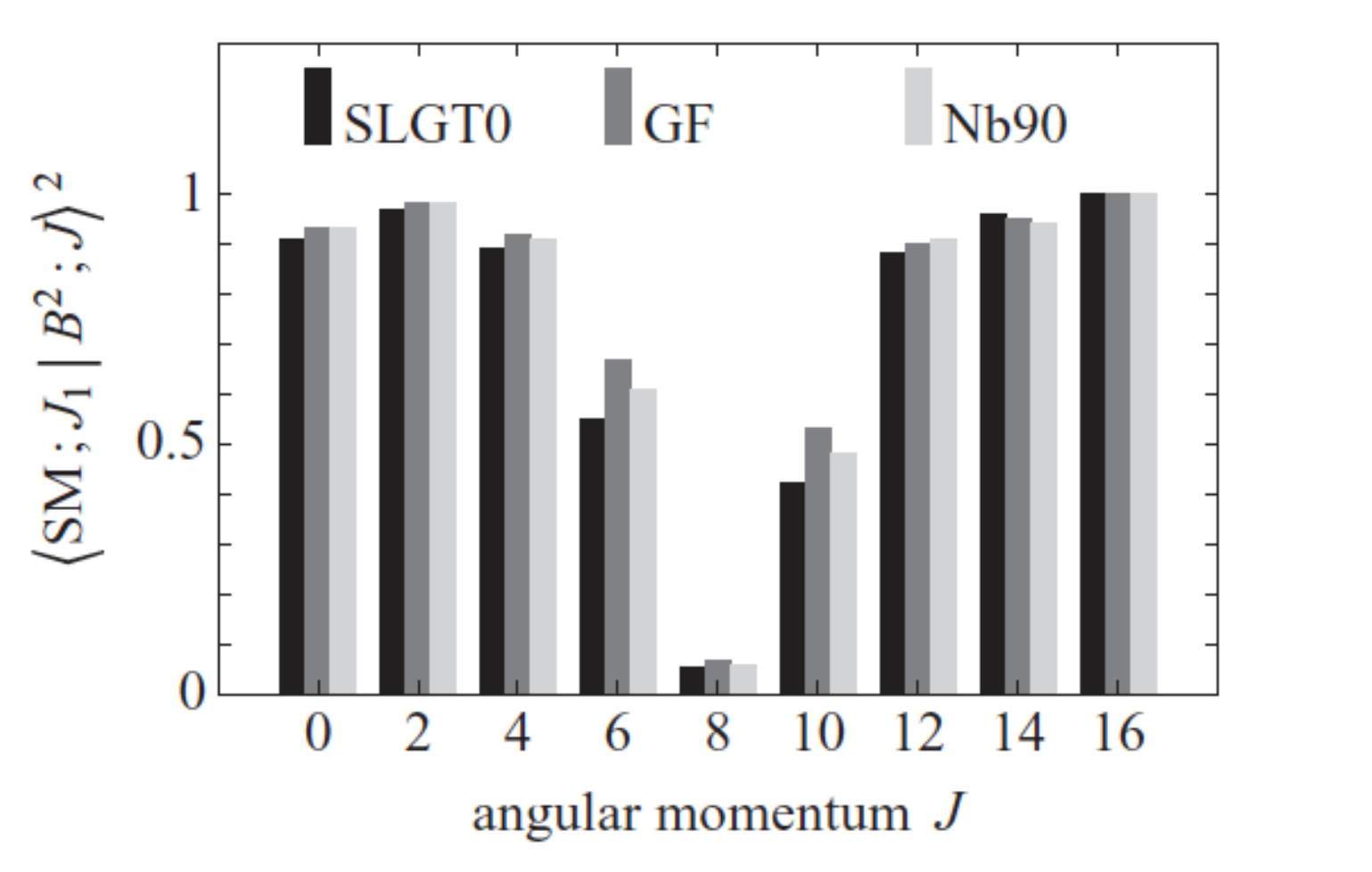,scale=0.6}
\caption{ Overlaps of the $(g_{9/2})^4$ yrast eigenstates in $^{96}$Cd  with the 
 two-pair state $\vert B^2;J=I\rangle$. Zerguine and Van Isacker used three different  interactions interactions,
SLGT0, GF, and NB90, were used in their shell model calculations. 
 From  \cite{Zerguine11}.\label{fig:PRC83f1}}
\end{minipage}
\hspace*{0.2cm}
\begin{minipage}[t]{8cm}
\hspace*{-0.5cm}
\epsfig{file=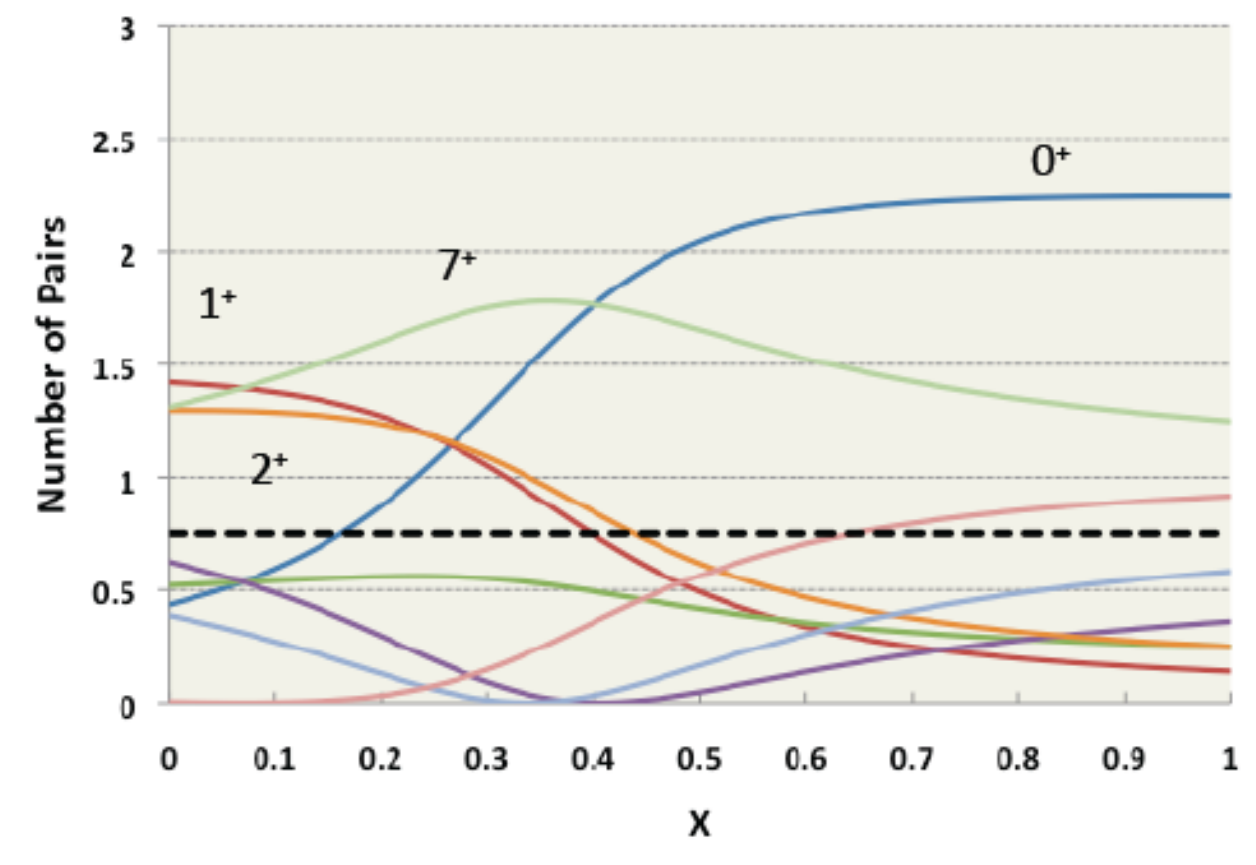,scale=0.65}
\caption{Number of pairs of a given spin in the ground state of $^{44}$Ti as a function of the
relative strength of $T=0$ vs. $T=1$ pairing, measured by the parameter $x$. Pure isoscalar
interaction corresponds to \mbox{$x=0$} and pure isovector \mbox{$x=1$}.
 From Macchiavelli in\cite{50yearsBCS}.\label{fig:AOMpairs}}
\end{minipage}
\end{center}
\end{figure}

This  dualism perpetuates to larger angular momentum.  Fig. \ref{fig:PRC83f1} shows the overlap of the Shell Model wave function 
with the spin-aligned state $I$. For $I=$, 2, 4, 6, and 8, Table IV of \cite{Zerguine11}  quotes, respectively,  the  overlaps of 85\%, 64\%, 70\%, and 83\%  
with the states $\sum_\mu  (-)^\mu P_\mu A^{1-\mu}_{J=I,M_J}\vert CS\rangle/No(IV,I)$, which represent the seniority scheme that couples one $J=0$ isovector
pair with a $J=I$ isovector pair. The $I=8$ state is  almost completely composed of a $J=0$ pair and an aligned $J=8$  pair, which 
is the familiar broken pair coupling of isovector pairing.   For $I=10$, 12, 14, and 16, the overlaps with states that combine 
the isovector pairs $(J_1,J_2)=$(2,8), (6,6), (8,8), and (8,8)
are 58\%, 57\%, 31\% and 100\%, respectively.

 Fig. \ref{fig:AOMpairs} shows the  expectation values of the pair number operators ${\cal N}_{J,T}$  given by Eq. (\ref{eq:pairnumberJT}) 
 for  $^{44}$Ti with two protons and two neutrons in the 
 spherical f$_{7/2}$ interacting by $S=0,T=1$ and $S=1,T=0$ pairing forces (see Section \ref{sec:2jIVISD0}), which is
 analogous to the two proton and two neutron holes in the  g$_{9/2}$ shell.  
 The numbers of  $J=0$ and $J=1$ pairs show the change from enhancement due to correlations to the uncorrelated
 values, which was discussed in Section \ref{sec:2jIVISD0} (cf. Fig. \ref{fig:PRC55f11}). 
 However, the number of isoscalar pairs with the maximal spin of $J=7$ does not change much with the balance between isovector and isoscalar
 interaction strength. Moreover, Zamick {\em et al.} \cite{Zamick05} quote the values 2.0 for a {\em pure} 
 Quadrupole-Quadrupole interaction and   1.9 for an effective interaction
 adjusted to experimental energies, which are to be compared with 0.7 for the non-interacting system. 
 It seems that the number of spin-aligned $T=0$ pairs is enhanced by
 any attractive $np$ interaction.   

\begin{figure}[tb]
\epsfysize=9.0cm
\begin{center}
\begin{minipage}[t]{8 cm}
\epsfig{file=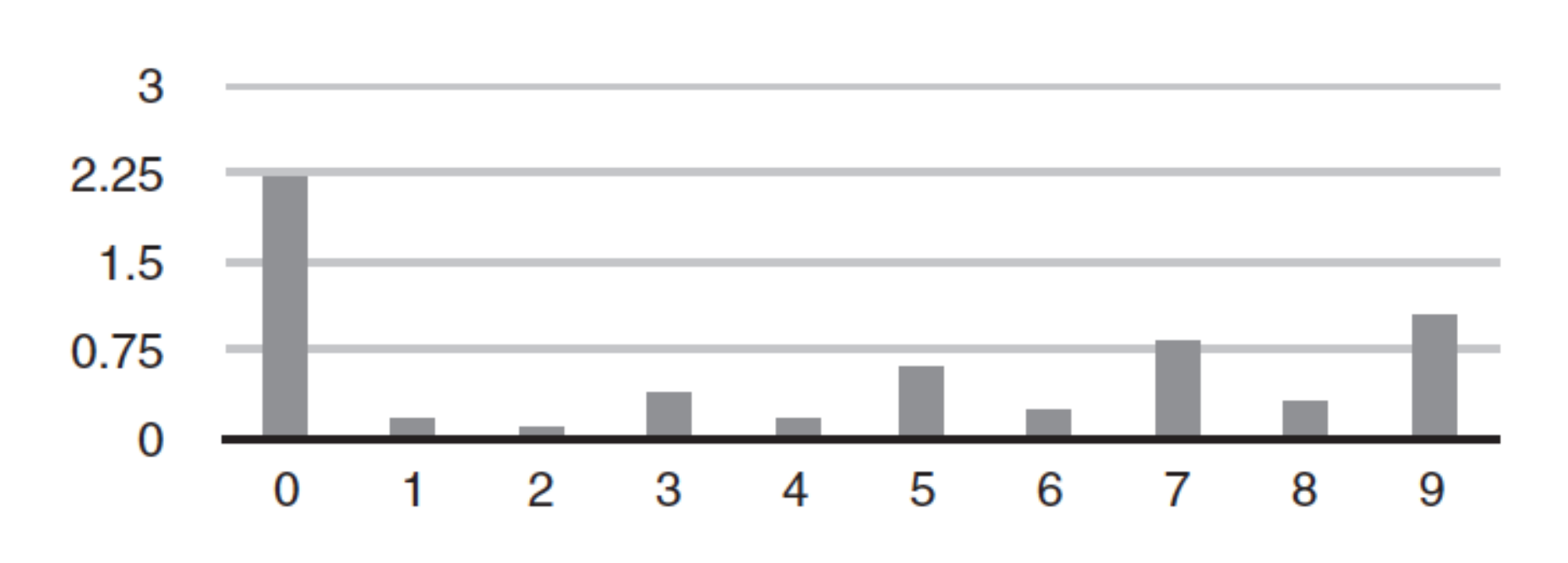,scale=0.45}
\end{minipage}
\hspace*{0.2cm}
\begin{minipage}[t]{8cm}
%\epsfig{file=PRC88f3.pdf,scale=0.43}
%\hspace*{0.4cm}
\epsfig{file=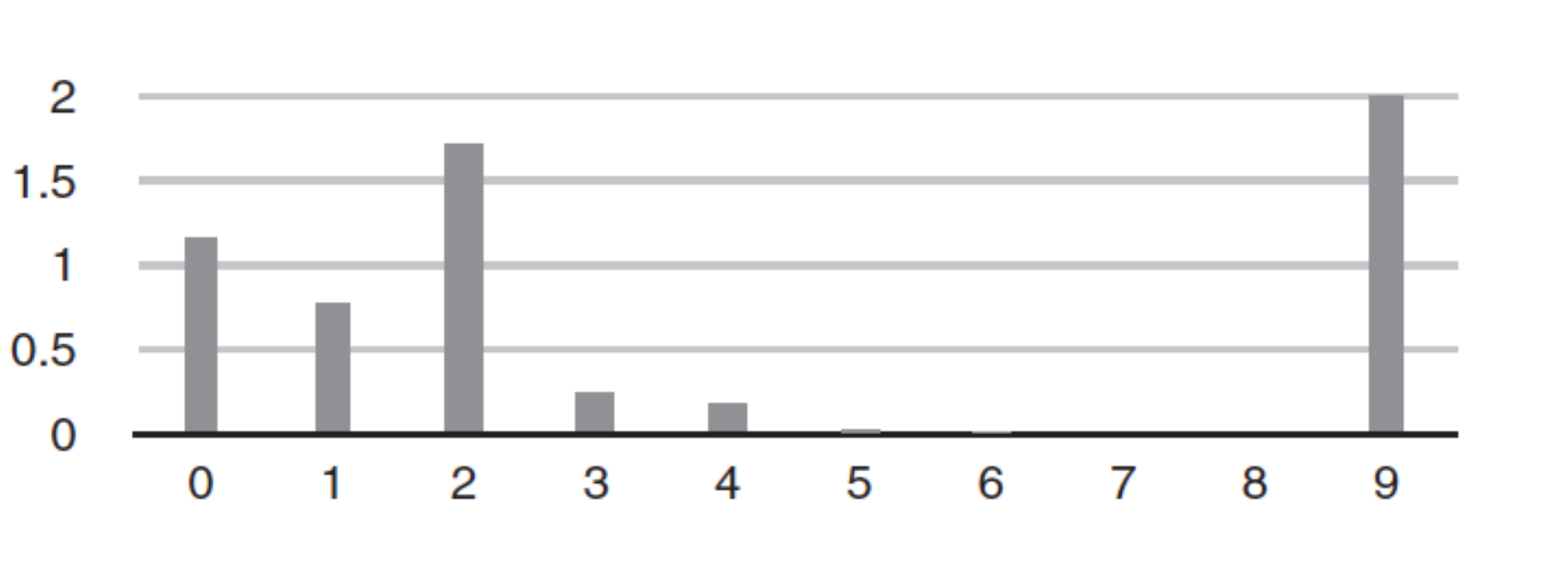,scale=0.45}
\end{minipage}
\caption{ Expectation values of the pair number operators ${\cal N}_{J,T}$ for two neutron and two proton
holes in the g$_{9/2}$ shell. The  $y$ -axis displays ${\cal N}_{J,T}$ and the $x$-axis the pair angular momentum $J$ .
 The left panel shows ${\cal N}_{J,T}$ for the isovector pairing ground state $\vert IV; 0\rangle$ and the 
right panel for the spin-aligned ground state   $\vert SA; 0\rangle$.  From \cite{Neergard13}.\label{fig:PRC88f2f4}}
\end{center}
\end{figure}

 The reason for the insensitivity is illustrated by Fig. \ref{fig:PRC88f2f4}. The spin aligned state $\vert SA;I\rangle$ has about two $J=9$ pairs
 and  the isovector pairing ground state $\vert IV;I\rangle$ has about one pair. Accounting for Neergard's observation that  the Shell Model ground states
 for various interactions are
 well described by the linear  combination $\vert SM;0\rangle=\alpha \vert SA;0\rangle+\beta\vert IV;0\rangle$ (see Table I of \cite{Neergard13}), 
 one expects the number of $J=9$ pairs  to be between one and two. 
 The substantial fraction of $J=9$ pairs in the isovector pairing ground state  appears counterintuitive. It is a consequence of the  antisymmetrization 
 of the wave function, which limits the appropriateness of simple concepts as the spin-aligned scheme or the pair condensate of macroscopic 
 super conductors. Another aspect of this limitation is the non orthogonality of the states  $\vert SA;I\rangle$ and  $\vert IV;I\rangle$.
 The smaller the system the more severe the limitations become.  The overlap $\langle IV;0\vert SA;0 \rangle^2$ 
 changes from 30\% for a k$_{15/2}$ shell to 100\% for a p$_{3/2}$ shell (see Fig. 5 in Ref. \cite{Neergard13}).   
 
 The $J=0$ pairs are a coherent superposition of $j+1/2$ components with $m_1=-m_2$. Several such pairs can be present in a $j$-shell, which form the $T=1$
 condensate. The interaction between the different components generates the correlation energy (see Section \ref{sec:2jIV}).   
 Because of the stretched coupling, the $J=2j, M_J$ pairs are composed of one large component with $m_1=m_2=M_J/2$ and small admixtures of the 
 other components $m_1+m_2=M_J$.  Only one such $JM_J$ pair can be present, because of the Pauli principle. The interaction does not correlate pairs with
 different $M_J$ values, because it conserves angular momentum. Therefore the aligned pairs do not qualify for pair correlations in our  sense.  The $J=1,M_J=0,\pm1$ 
 pairs have still enough components $m_1+m_2=M_J$ to generate pair correlations. Generally, the number of coherently contributing components 
 decreases as  $j-J/2$, which  limits the phase space for building  pair correlations. The deformation of nuclei to some extent removes these limitations
 imposed by the Pauli principle and angular momentum conservation.

    There is a suggestive similarity between the linear combination  of the $J=0$ isovector and spin-aliged isoscalar components
    in the Shell Model states and the coexistence of the isovector and   
  the isoscalar CHFB solutions reported by several authors. As already mentioned,  Goodman \cite{Goodman01} found for $^{70}$Zr
  in addition to the isovector solution an excited  isoscalar solution of the spin-aligned type.  
  The pair field of the CHFB solution is a reasonable indicator of pair correlations.
 As discussed  above,  a spin-aligned pair field is only possible if the nuclear potential is deformed,
because angular momentum conservation forbids scattering between pairs  $JM_J$ and $JM_J'\not=M_J$.  
Goodman {\it et al.} \cite{Goodman68} pointed out that the appearance of a spin aligned pair field 
in the ground state of $^{24}$Mg is coupled with the triaxial shape of this state.
 Goodman \cite{Goodman99,Goodman01} reported  isoscalar pair fields of the $np$ type for the ground states 
 of $^{92}$Pd and $^{96}$Cd  and   a  weak $J=9$ pair field, which coexists with $J=5$ and $J=3$ fields of comparable strength,
 for finite angular momentum in  $^{70}$Zr.
   Likewise, Sheikh and Wyss \cite{Sheikh00}, 
suggested that their Shell Model solutions for the rotating deformed f$_{7/2}$ shell (see \ref{sec:SMd}) have the character 
of a mix between the  isovector and isoscalar    CHFB solutions found for the same Hamiltonian. 
 Satu\l a and Wyss'  CHFB (+LN correction) calculations (Ref. \cite{Satula00}, cf. Section \ref{sec:SP}),
 suggest the possibility of an excited  superdeformed band in $^{88}$Ru, 
 composed of spin-aligned $T=0$ pairs, which coexists with the isovector yrast solution.
 Depending on the case, the spin-aligned CHFB solution may become yrast with increasing angular momentum.
  It seems that the above discussed  mixed structure of the Shell Model wave functions is reflected by the coexistence of the 
two CHFB solutions on the mean field level.
The question of orthogonality of the competing CHFB solutions has  not been  addressed so far. 
The role played by this spin-aligned pair field and its relation to the structure of the Shell Model states needs further clarification. 
  
  \subsection{Symmetry considerations}\label{sec:ISsym}

 The presence of a strong pair field breaks particle number symmetries in a specific way, which is reflected by the appearance of a pair rotational band.
 Essential aspects  of this symmetry breaking were already discussed in Sections \ref{sec:sigmf}.and \ref{sec:Wphen}.
 Frauendorf and Sheikh  \cite{Frauendorf00} analyzed the consequences  of the symmetry of the  pair fields for finite angular momentuml.  
   The  direction of the  $J=0,T=1$  isovector field in isospace can be chosen such that
 there is no $np$ component. The like-particle components are invariant with respect to  $\exp(-i\hat N\pi)$ and  $\exp(-i\hat Z\pi)$ (rotations in gauge space by $\pi$),
 which has the consequence that the pair rotational bands are sequences of nuclei with $\Delta N=\pm2$ and $\Delta Z=\pm2$.   That is, the isovector field conserves
 the  number parity of $N$ and $Z$ separately. The isorotational bands  are the sequences 
 $N-Z=2T_z=0$, 4, 8, ... or 2, 6, 10, ... in the case of even $A$  or 1, 5, 9, ... or 3, 7, 11, ... in the case of odd $A$.
 In addition there are the gauge rotational bands with $A-A_0=$0, ... ,-8, -4, 0, 4, 8, ... and fixed $N-Z$.   
  Angular momentum adds a new element to the symmetries. The isovector field contains even spin pairs $J=$0, 2, 
  the components of which are invariant with respect to a spatial rotation
 by $\pi$ about the rotational axis $\exp(-i\hat J_x\pi)$. This implies that the signature quantum number $\alpha=I + even~ number$,
  is the same within a isorotational or gauge rotational sequence
 ($I$=0, 2, 4,.. for even-even nuclei).  See e.g. Ref. \cite{Frauendorf01}
 for a detailed discussion of the symmetries of rotating nuclei and the signature quantum number in particular.
   
 The isoscalar pair field conserves the isospin. It changes sign under the separate gauge rotations $\exp(-i\hat N\pi)$ and  $\exp(-i\hat Z\pi)$, which implies that
 only  the number parity of $A$ is conserved. This lower symmetry is manifest by gauge rotational bands with  with $A-A_0=$ ... ,-4, -2, 0, 2, 4, ... and fixed $N-Z$. In particular,
 the presence of a sufficiently strong (compared with the single particle level distance) isoscalar pair potential has the consequence that the energies of lowest $T=0$ states
 in even-even and odd-odd $N=Z$ change in a smooth way with $A$  (in contrast to experiment as discussed in Section \ref{sec:Wphen}). The spatial rotation 
 $\exp(-i\hat J_x\pi)$ gives -1 in the case of the isoscalar field, because it is  
 composed of odd-spin  pairs. Then the isoscalar field is invariant with respect to the combined operation  $\exp(-i\hat J_x\pi)\exp(-i\hat N\pi)$,
  which implies that $N+I=\gamma+even~number$, were $\gamma$ is a fixed quantum number called gaugeplex by Frauendorf and Sheikh. 
  Within the deformed f$_{7/2}$ shell, they carried out Shell Model calculations for $Z=N=3,4$ switching off the $T=1$ part of the $\delta$ interaction
  (cf. Section \ref{sec:SMd}). The low-energy  part
  of the resulting two spectra becomes very similar for $\hbar \omega/G>0.7$, if one swaps even and odd $I$ (see Figs. 8 and 9 in Ref. \cite{Frauendorf00}). 
 %Rotation about a principal axis  can be  expected for  moderately deformed $N\approx Z$ nuclei at values of $I$ of a few units larger than the ground state value.
 As seen in Fig. \ref{fig:IJMPE16f6}, the experimental spectra of $^{68}$Se, $^{70}$Br,  $^{72}$Kr, $^{74}$Rb, and $^{86}$Sr do not contain evidence for such similarity. 
 The same holds for the Shell Model calculations for $^{88}$Ru, $^{90}$Rh, $^{92}$Pd, $^{94}$Ag, and $^{96}$Cd, which base on one and 
 the same effective interaction  \cite{Hasegawa04,Kaneko08}. This speaks against the existence of a strong isoscalar pair field.
 
  \subsection{Experimental evidence from rotational spectra}\label{sec:ISrot}
 
The  rotational spectra of $N\approx Z$ nuclei have been described in a systematic way assuming pure isovector pairing, provided that the $np$ component, which
is required for good isospin, has been taken into account. The latter generates a shift by $T(T+1)/2/\theta$, which is seen in experiment. In particular, it has the consequence
that only one band, the  $T=0$ combination, is observed, instead of the two close bands that could be generated by exchanging protons with neutrons  if there was 
no $np$ interaction. Examples are the odd-spin yrast sequence in odd-odd and the s-bands in even-even $N=Z$ nuclei.
With this provision, the rotational spectra and transition probabilities of $N=Z$ nuclei are reproduced with the same accuracy as    for  $N\not=Z$ nuclides.
The spectra confirm the rapid quenching of the isovector pair correlations with angular momentum, which is predicted by both  the Shell Model 
and the various mean field approaches that partially restore the symmetries.     

Deviations of experimental spectra from  pure isovector calculations have been repeatedly  suggested as possible evidence for the existence of an
isoscalar pair field. In our opinion, such evidence is not significant. The calculated moments of inertia are close to experiment, 
both in the paired and unpaired regimes. This is consistent with Shell Model calculations, which find moderate isoscalar correlations
that do not change the moments of inertia enough to be significant. Calculated band crossing frequencies also agree with experiment within 
the uncertainty margin of the applied approaches. The early evidence for a delay of the crossing between the g- and s bands (first backbend) 
in $N=Z$ nuclei has been disproved by more recent experiments.  Moreover some delay
is also predicted within an isovector scenario, which is  also within the general margin of uncertainty. The spin-aligned regime suggested by the
Shell Model studies may represent  certain isoscalar correlations, the nature of which has still to be understood. Experimental evidence
does not exist, because the relevant data on energies and $B(E2)$ values for $N=Z\geq 44$ have yet to be measured.    
        
\begin{figure}
\centerline{\psfig{file=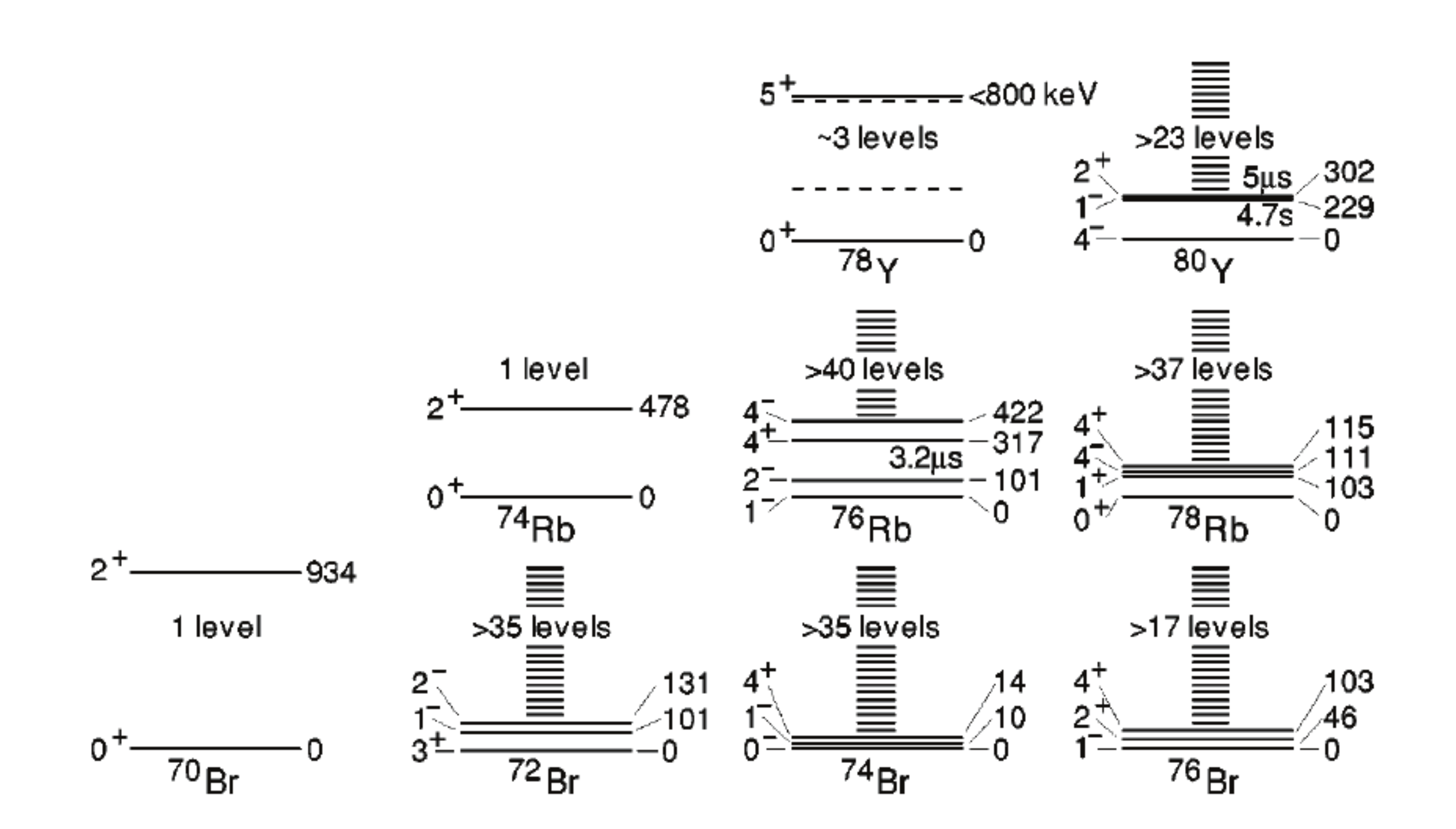,width=12cm}}
\caption{ Summary of observed levels  in Br , Rb and Y  isotopes near N=Z.  The low level density at $N=Z$ is apparent. 
From  \cite{Jenkins}}
\label{fig:levelsBr}
\end{figure}

\begin{figure}
\centerline{\psfig{file=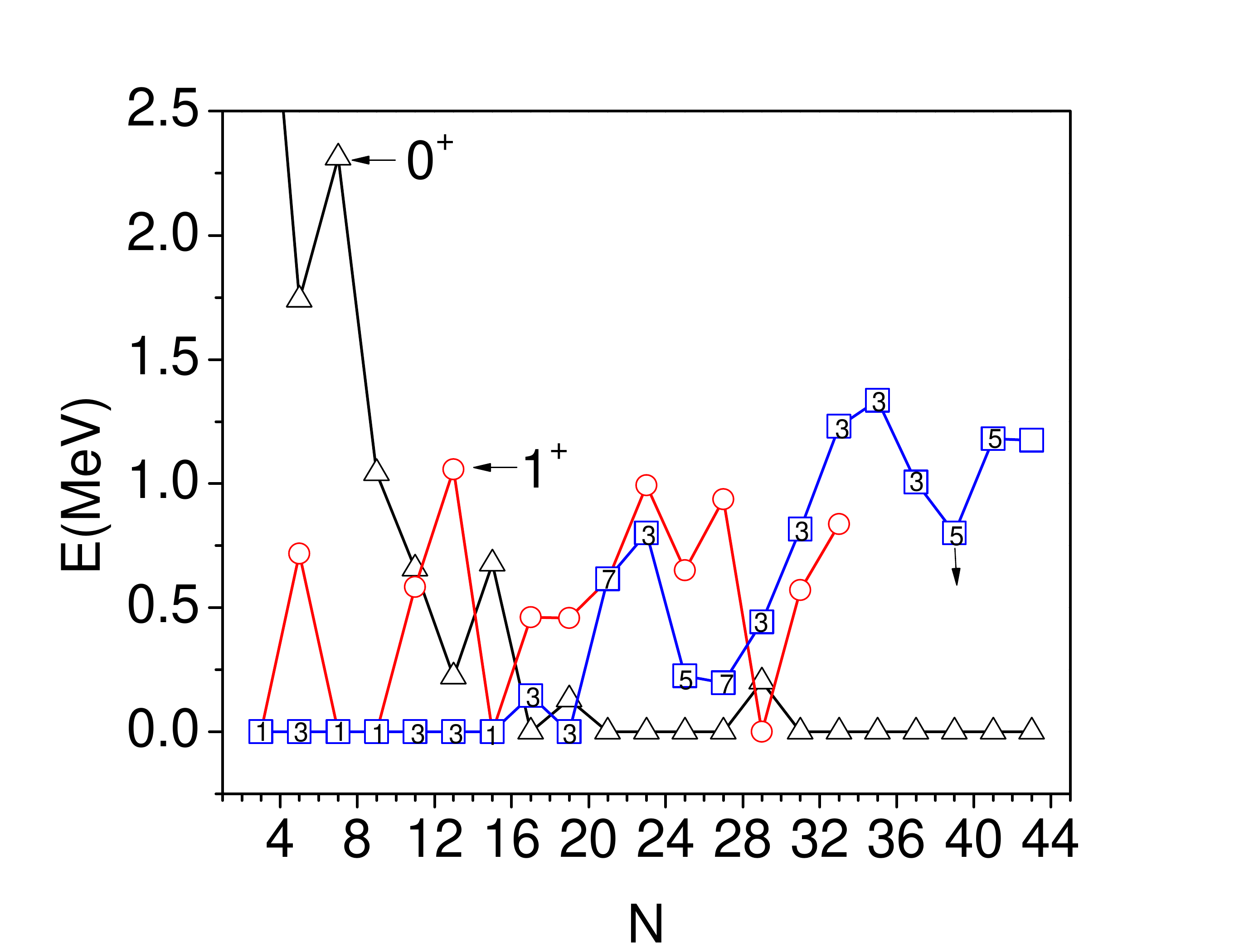,width=12cm}}
\caption{Low-lying  states in $N=Z$ odd-odd nuclei. Shown are the lowest $T=1$, $0^+$ states, the lowest $T=0$, $1^+$
states, and the lowest $T=0$ states, for which $I^+$ is indicated by the numbers in the squares. Only a limit of the energy of the $5^+$ state in 
$^{78}$Y is reported, and $I^\pi$ of the lowest $T=0$ state in $^{82}$Tc is not known. Data from the ENSDF data base and  \cite{Jenkins,Paddy}.\label{fig:oo}}
\label{fig4}
\end{figure}

\section{Low-lying states of odd-odd self-conjugate nuclei }\label{sec:lowoo}

%This pattern is not seen in experiment. The $T=0$ ground states of odd-odd nuclei $N=Z$ nuclei from the p- and sd-shells are about 24 MeV$A^{-1/2}$ above
%the average of their even-even neighbors (c.f. Figs. \ref{fig:IVISBE1} and \ref{fig:sumB} ).  
%Only three of these odd-odd $N=Z$ nuclei have $1+$ ground states ($^{14}$N, $^{18}$F, and $^{30}$P).
%The ground state spin and parity values of all othe $N=Z$ nuclei from  p- and sd- shell nuclei accounted for by assuming that the proton and the neutron occupy the %same Nilsson
%orbital with angular momentum projection $\Omega$ on the symmetry axis, i.e. $I^\pi=2\Omega^\dagger $.   
% Proton-neutron pairs with aligned angular momentum are favored over the deuteron-like pairs (see Section \ref{sec:lowoo}).    

Recent experimental work has  focused on the detailed spectroscopy of low lying levels in 
odd-odd $N=Z$ nuclei.   Fig.~\ref{fig:levelsBr} illustrates a general observation.
The known levels in Br , Rb and Y  isotopes near $N=Z$ are shown.  One readily notices that
the odd-odd $N=Z$  nuclei show low level densities near the ground state. Once again,  this is
consistent with the picture presented in section 2.1. The odd-odd ground states at $N=Z$  are dominated by
the $T=1$ pairing correlation and  the isovector gap $2\Delta_1$ is 
clearly recognized as in the even-even isobaric analogues. 
The high density of $ T=0 $ states appears higher in energy only at $N=Z$,  where to make a $T=0$ $np$ pair
requires breaking the gap. Fig. \ref{fig:oo} demonstrates that this is a general phenomenon. There is a systematic 
appearance of a gap between the $T=1$, $0^+$ ground state and the onset of the  $T=0$ states 
above $A=40$. {Going below $A=34$, the ground states have $T=0$, and  the $T=1$, 0$^+$ state becomes progressively disfavored. 
This might be taken (and has been) as an indication that the isoscalar pair correlations become more important than the isovector ones, which would be
consistent with the early HFB calculations (see Section \ref{sec:HFBresults}). However as discussed  in Section \ref{sec:Wphen},
the  energy difference between  the lowest $T=0$ and $T=1$ states in odd-odd $N=Z$ nuclei can be well understood in terms of the
competition between the isovector pair correlation energy and the symmetry energy.   In contrast to Fig.~\ref{fig:levelsBr},  the level density of the odd-odd $N=Z$ nuclei
with $A<30$ is comparable with the one the odd-odd $N>Z$ isobars, which speaks in favor of a two quasiparticle character too. }

 As seen in Fig. \ref{fig:oo}, the spin of the lowest $T=0$ state is systematically equal to $I=2j$, where
$j$ is  the ground state spin of the odd-A neighbors.\footnote{In  few cases $j$ is the ground state spin of only
the two heavier (lighter) nuclides. Then a state with $I=2j$ of the $j$ in the lighter (heavier) neighbors is  found at close energy.}
Obviously the strong attraction  between the odd nucleons in identical orbitals brings these orbital  down most. The $T=0$, 1$^+$
states are always higher, except when the odd-A neighbors have a $I=1/2$ ground state, and spin-aligned coupling gives
1$^+$. This indicates that the $J=1$ part of the isoscalar interaction acts less effective than the $J=2j$ part, which is a general feature of the
Schiffer-True interaction shown in Fig. \ref{fig:SchTru}. The expression for the interaction Eq. (41.2) in Ref. \cite{Molinari75}  gives
an energy difference of $E(2j)-E(1)$=0.8 MeV and 0.3 MeV for $A=20$ and 60 respectively.  These differences  are of the order of the distances
between the two states in Fig. \ref{fig:oo}.     Thus, the data is consistent with assuming that the relative position is determined by
the diagonal matrix element of the interaction and does not provide clear evidence for additional correlation energy of the $J=1$ isoscalar
type. This is consistent with Chasman's finding \cite{Chasman02,Chasman03a,Chasman03b } that the diagonal matrix elements of the $T=0$ 
interaction control the energy of the lowest states in the odd-odd $N=Z$ nuclei. 

Lisetzkiy {\em et al.} \cite{Lisetzkiy99} studied the M1 transitions between the 1$^+$ and 0$^+$ states. They  assumed that these states represent 
a pair of a proton and a neutron in the ground state  j-orbital of the odd-A neighbors, which are coupled to $J=1$  and $J=0$, respectively.   
The reduced transition probabilities are  very well reproduced if the $g$ - factors are taken from the experimental magnetic moments 
of the odd-A neighbors. In particular, the strong difference between orbitals with $j=l\pm 1/2$ is accounted for. For $j=l+1/2$ the spin and orbital
contributions to the transitional magnetic moment add up in a constructive way, which results in a large $B(M1)$ value. The mechanism is the 
same as the one that leads to the strong M1 transitions of between the members of the shears bands representing 
Magnetic Rotation \cite{Frauendorf93,Frauendorf01,Clark00}( see Subsection \ref{sec:SpinAligned}).
 For $j=l-1/2$ the different terms in the sum largely cancel each other and
the resulting $B(M1)$ values are small.
Using the same scheme, Ronen and Shlomo \cite{Ronen98} showed that the magnetic moments
of the  lowest $T=0$ states can be well described by assuming the stretched coupling $I=2j$. 
The success of simply  coupling the quasiproton and quasineutron magnetic moments from the odd-A neighbors to describe the magnetic properties
of the odd-odd $N=Z$ nuclei  suggests that  the $J=2j$ and  $J=1$ states have  two-quasiparticle character. 
For collective states the pronounced differences between g-factors of the quasiparticles, which are found enhanced in 
two-quasiparticle  states, would be washed out by the superposition of a  large number of components with different magnetic properties.  
 
A final example refers to the study of metastable state decays in the proton dripline nuclei $^{82}$Nb and $^{86}$Tc \cite{Paddy}.
Isomeric-decay spectroscopy has been studied using the RISING array at GSI in Germany, following the projectile fragmentation of a 
$^{107}$Ag beam.   In line with the previous results, the level structure  of  these $T_z = 0$ odd-odd nuclei at  low excitation shows a 
preference for $T = 1 $ states,  associated with the existence of a $T = 1$ $np$ pairing gap.  Comparison with PSM
calculations \cite{Sun04} (cf. Section \ref{sec:SP}) support this observation. 
 
{In conclusion,  the spectroscopic information on the low-lying states of odd-odd self-conjugate nuclei does not provide evidence
for the existence of strong isoscalar pair correlations. The properties of lowest $T=0$ states are well accounted for by considering them 
as two quasiparticle states generated from an isovector pairing vacuum.  The spin-aligned coupling  of the quasi proton and quasineutron 
is energetically preferred. However it is not clear whether this is caused just by the diagonal matrix element of the isoscalar interaction, or if additional 
pair correlations are involved.  }

\begin{figure}[tb]
\epsfysize=9.0cm
\begin{center}
\begin{minipage}[t]{8 cm}
\epsfig{file=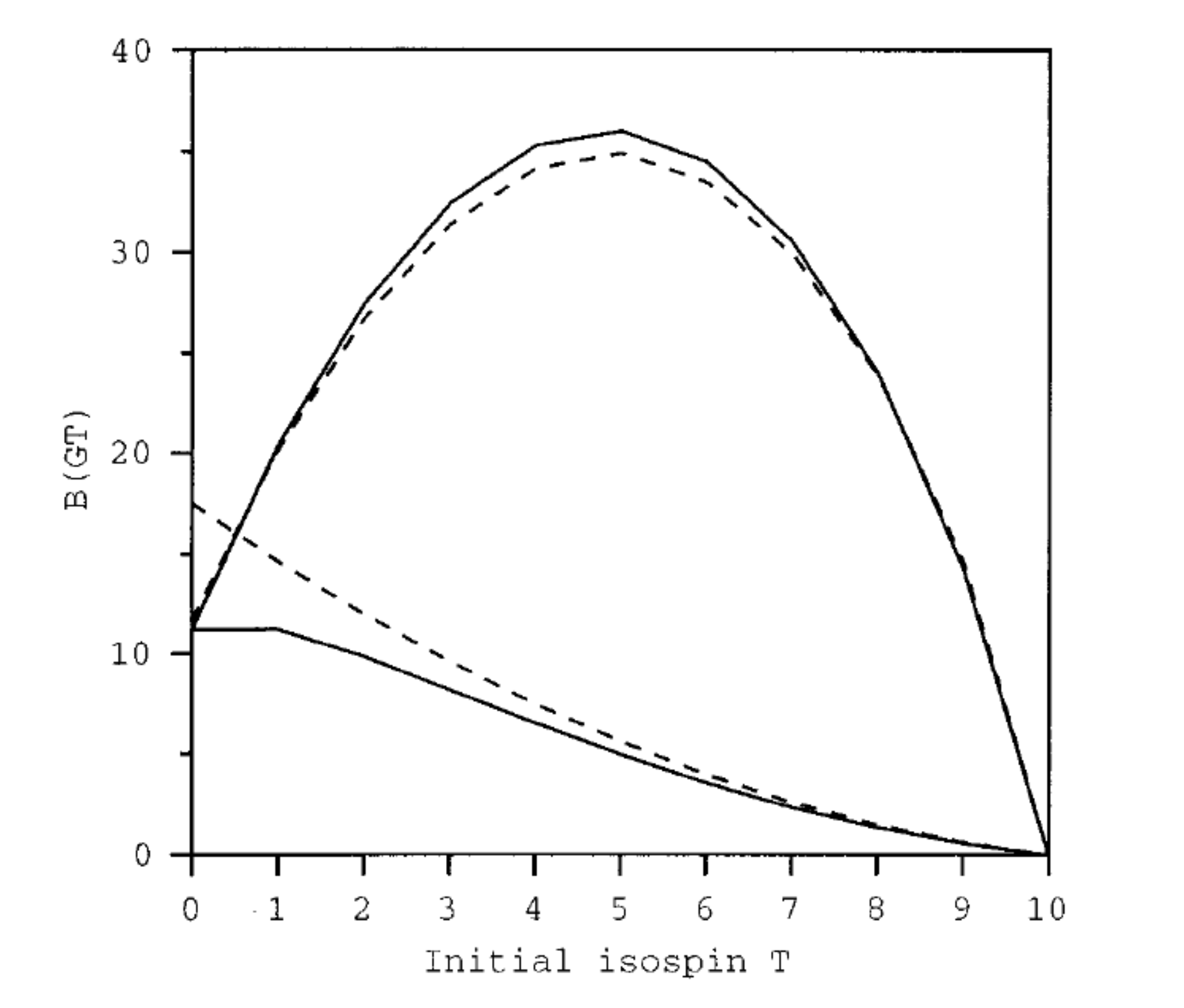,scale=0.40}
\caption{The Gamow-Teller $\beta^+$ strength $B(GT)$ for $\Omega$=12, $N$=10
as a function of the isospin $T$ in the two limiting phases.
Solid lines represent the exact solutions and dashed lines the generalized
BCS solutions in the isovector (bottom two lines) and isoscalar
(top two lines) phases. From \cite{Engel97}.\label{fig:PRC55f9}}
\end{minipage}
\hspace*{0.2cm}
\begin{minipage}[t]{8cm}
%\epsfig{file=PRC88f3.pdf,scale=0.43}
%\hspace*{0.4cm}
\epsfig{file=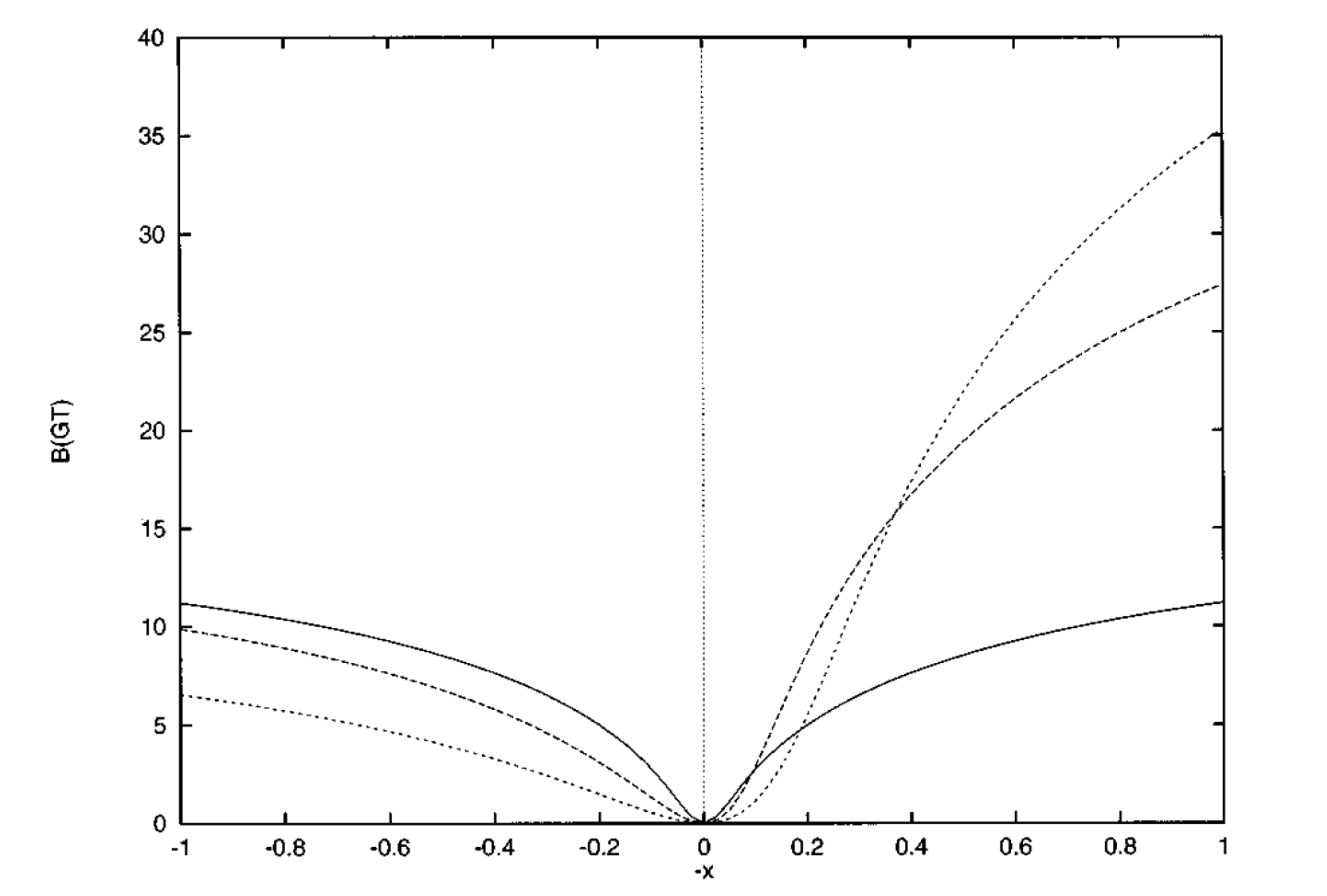,scale=0.23}
\caption{ The Gamow-Teller $\beta^+$ strength $B(GT)$ vs the Hamiltonian
parameter -x, for $\Omega$=12, $N$=10. The
solid curve corresponds to $T$=0, the long dashes to $T$=2, and the
short dashes to $T$=4. 
From \cite{Engel97}.\label{fig:PRC55f4}}
\end{minipage}
\end{center}
\end{figure}

\section{Gamow-Teller $\beta$-Decay}\label{sec:GT}

Qualitative relations between the Gamow-Teller $\beta$-decay strength  $B(GT)$ and the pair correlations have been established by
 Engel {\it et al.}  \cite{Engel97}, and Engel, Vogel and Zirnbauer \cite{Vogel86,Engel88}. They studied the 
 the degenerate case $D=0$ of the schematic Hamiltonian discussed in Section \ref{sec:1jIVIS}, for which exact solutions 
 are available. The summed Gamov-Teller $\beta^+$ strength can be expressed by the elements of the  the SO(8) basis. 
 Fig. \ref{fig:PRC55f9} shows that for pure isovector pairing  $B(GT)$ decreases with $T=M_T$, because the neutrons block  an increasing 
 number of states into which the protons can decay (Pauli blocking). For pure isoscalar pairing $B(GT)$ behaves in an unexpected way.
 It first {\it increases } with $T$ and then decreases. The authors attribute this to the competition between coherent isoscalar  pair
 correlations, which increase $B(GT)$, and Pauli blocking. The counterintuitive increase of $B(GT)$ when moving away from $N=Z$ has not been 
 explained.  The dashed lines demonstrate that these features of $B(GT)$ are caused by the pair condensates.       Fig. \ref{fig:PRC55f4} 
 shows how $B(GT)$ changes as a function of the relative strengths of the isovector and isoscalar interaction. It goes through zero at $x$=0, where
 $G_{IV}=G_{IS}$ and the Hamiltonian obeys the SO(6)  symmetry (equivalent to SU(4)).  The GT matrix element disappears because it represents a ladder operator
 of the group, which is zero when applied to a completely symmetric representation  \cite{Vogel86}.   The mean-field approximation does not
 show this quenching (cf. Fig. 10 of Ref. \cite{Engel97}). For $x=0$, there exist a family of degenerate mean field solutions, which spontaneously break the SO(6) symmetry.
  They are generated  by the rotations of the SO(6) group from one of the mean-field solutions and correspond to different ratios of
  the isovector and isoscalar     pair fields (see for example Table II of Ref. \cite{Bertsch10}).  The exact solution, which is a completely symmetric
   representation of the SO(6) group, is a superposition of all these mean-field solutions with equal weight.   The GT matrix element, which is not a scalar with respect to the 
   SO(6) rotations, averages to zero.   (Note the analogy with the breaking of the  SO(3) isospin symmetry by the isovector pair field and its restoration,
   which were discussed in Section \ref{sec:sigmf}.)
 
 The quenching of the summed GT strength when approaching $x=0$ from the isovector side signals substantial dynamic isoscalar pair correlations, 
 which are the precursor of the isoscalar pair condensate. They reduce the matrix element for the $2\nu\beta^+\beta^+$ decay (double $\beta^+$ decay with emission of two neutrinos) 
 in a significant way. The decay 
 is a second order process  which proceeds via virtual $1^+$ states of the intermediate odd-odd nucleus 
 \mbox{($0^+_{ini}\rightarrow (\beta^++1^+_{int})\rightarrow (\beta^++0^+_{fin}$))}.
  Its strength is determined by the 
 $\beta^+$  matrix elements that connect the final state with the $1^+$ states.  Similar to the summed $\beta^+$ strength, their contribution is quenched near $x=0$,
 where  the decay matrix element goes through zero. (see Fig. \ref{fig:PRC55f6}). The reduction is caused by the strong fluctuations of the pair fields. 
 
 Faessler and Raduta {\it et al.}    \cite{Raduta01,Raduta01a} found an analogue quenching of the  $2\nu\beta^+\beta^+$ Fermi matrix element, which is caused by 
 large fluctuations of the isovector pair field.  They studied the  isovector pair Hamiltonian 
 \be\label{eq:HIVF}
 H=\varepsilon_n\hat N_p(j)+\varepsilon_p\hat N_n(j)-G_{nn} P^\dagger_1 P_1-G_{pp} P^\dagger_{-1} P_{-1}-G_{pn} P^\dagger_0 P_0
 +2\chi\sum_m(-)^{j-m}mp^\dagger_{jm}n_{jm}\sum_m(-)^{j-m}n^\dagger_{jm}p_{jm}
  \ee
  in a $j$-shell with the operators defined in Eqs. (\ref{eq:P}, \ref{eq:NpNn}).  (The last term
  simulates the residual interaction in the particles-hole channel, which is not relevant for our discussion.)
  Exact solutions were found by diagonalizing in the basis of representations of the SO(5) group.
  The $2\nu\beta^+\beta^+$ Fermi matrix element can be expressed in terms of the SO(5) generating operators, and exact values were calculated.    The authors
   considered the pair coupling strengths as independent parameters.  For small $G_{pn}$ the solutions correspond to a $pp +nn$ pair field. At a critical strength 
  of $G_{pn}$, indicated by the break down of the QRPA approximation,  the pair field changes to pure isovector $np$ pairing.  The character of the pair fields is  
  indicated by the energy of the exact solution of the odd-odd intermediate nucleus. 
  As seen in Fig. 1 of Ref. \cite{Raduta01}, $E_{oo}$ is larger than the mean of the neighboring $E_{ee}$ for small $G_{pn}$ which is characteristic for like particle pairing.
  For large $G_{pn}$ (right margin of the figure) $E_{oo}$ approaches the mean of the neighboring $E_{ee}$, which is the signature of the a strong isovector $np$ pair field.
  As seen in Fig. 2 of  Ref. \cite{Raduta01}, the exact $2\nu\beta^+\beta^+$ Fermi  matrix element becomes very small near the instability,  where the large fluctuations of the pair field
  average out the matrix element. The instability occurs at $G_{pn}>>G_{pp},G_{nn}$. In view of the central role of isospin invariance of the isovector pair interaction this
  instability seems not to be relevant to experiment.
    
Engel, Vogel and Zirnbauer \cite{Vogel86,Engel88} used the QRPA method for estimating the Gamow-Teller $\beta^+$ decay strength  the in semi magic nuclei 
$^{94}_{44}$Ru$_{50}$,  $^{96}_{46}$Pd$_{50}$, $^{148}_{66}$Dy$_{82}$, and   $^{152}_{70}$Yb$_{82}$. They started from $pp+nn$ pair field generated by 
a $\delta$-force, the strength of
which was adjusted to the even-odd mass differences in the standard way. In addition they took into account a $\delta$-force $np$ interaction in the particle-hole
channel, the strength of which was adjusted the experimental position of the GT resonance. The third ingredient was a $\delta$-force $np$ pair interaction. The ratio
between its isoscalar and isovector parts was assumed to be 0.6, which is close to 0.66 used by Gezerlis, Bertsch, and Luo \cite{Gezerlis11}(see Section \ref{sec:HFBresults}).
The total strength of the $np$ pair interaction was adjusted to reproduce the experimental Gamow-Teller $\beta^+$ decay strength in the considered nuclei. 
The measured $\beta^+$ GT strengths  could be reproduced with an accuracy of about 10\%. The experimental  lifetimes for the  
$2\nu\beta^+\beta^+$ decays of  $^{130}$Te and $^{82}$Se could be reproduced by a choice of the constant for the $np$ pair interaction that agreed with the value 
determined from the $\beta^+$ decay within 10\%. The authors point out the extreme sensitivity of the  $2\nu\beta^+\beta^+$ matrix element to 
the strength of the $np$ pair interaction,
which in the light of their work appears to be a promising indicator of the strength of the dynamical isoscalar pair correlations.

 It is gratifying that the resulting  constant for isovector  $np$ interaction nearly agrees with  the constants for 
  the $pp$ and $nn$ interactions, i. e. that the pair interaction is isospin invariant    with good accuracy. 
  The final value of the matrix element results from the balance of competing mechanisms \cite{Vogel86}. The $\beta^+$ matrix element in $N>Z$ nuclei is strongly hindered
  by the neutrons blocking the available single particle states into which the proton may decay (Pauli blocking). It is therefore susceptible to correlations. The isovector pair
  correlations enhance it, because they smear out the Fermi surface alleviating the Pauli blocking.  The particle-hole correlation reduce it, because they draw GT strength
  from the low-energy region into the GT resonance. The dynamical  isoscalar $np$ pair correlation reduce it by a comparable amount.  They are strong, because  
  the nuclei are not far below the phase transition  where the isoscalar pair field appears, which is the region somewhat below $x=0$ in  Fig. \ref{fig:PRC55f6}.   
 In the framework   of the QRPA the reduction is caused  by a substantial cancellation of the contributions from the forward amplitudes by the  contributions from
 the backward amplitudes, which are a measure of the dynamical correlations in the ground state. Since the QRPA becomes an unreliable approximation close 
 to the point of instability, the authors used a renormalized version of QRPA (RQRPA) in the actual calculations. They took the good agreementt
  with the exact matrix element in a $j$-shell
  in Fig. \ref{fig:PRC55f6}  as  evidence for the RQRPA being a reliable approximation for the transition region.  
  
  A reliable description of  $2\nu\beta^+\beta^+$ process is of great interest because the nuclear structure that determines its strength is very similar to the one that governs
  the hypothetical $0\nu\beta^+\beta^+$ decay (neutrino-less double $\beta$ decay), the possible existence of which is of central importance for particle physics. 
  The nuclear structure aspects have been recently analyzed by   \v{S}imkovic {\it et al.} \cite{Simkovic08}, where earlier work is cited. The isoscalar and isovector pair 
  correlations have a significant influence on the size of the  $0\nu\beta^+\beta^+$ matrix element.  
\begin{figure}
\centerline{\psfig{file=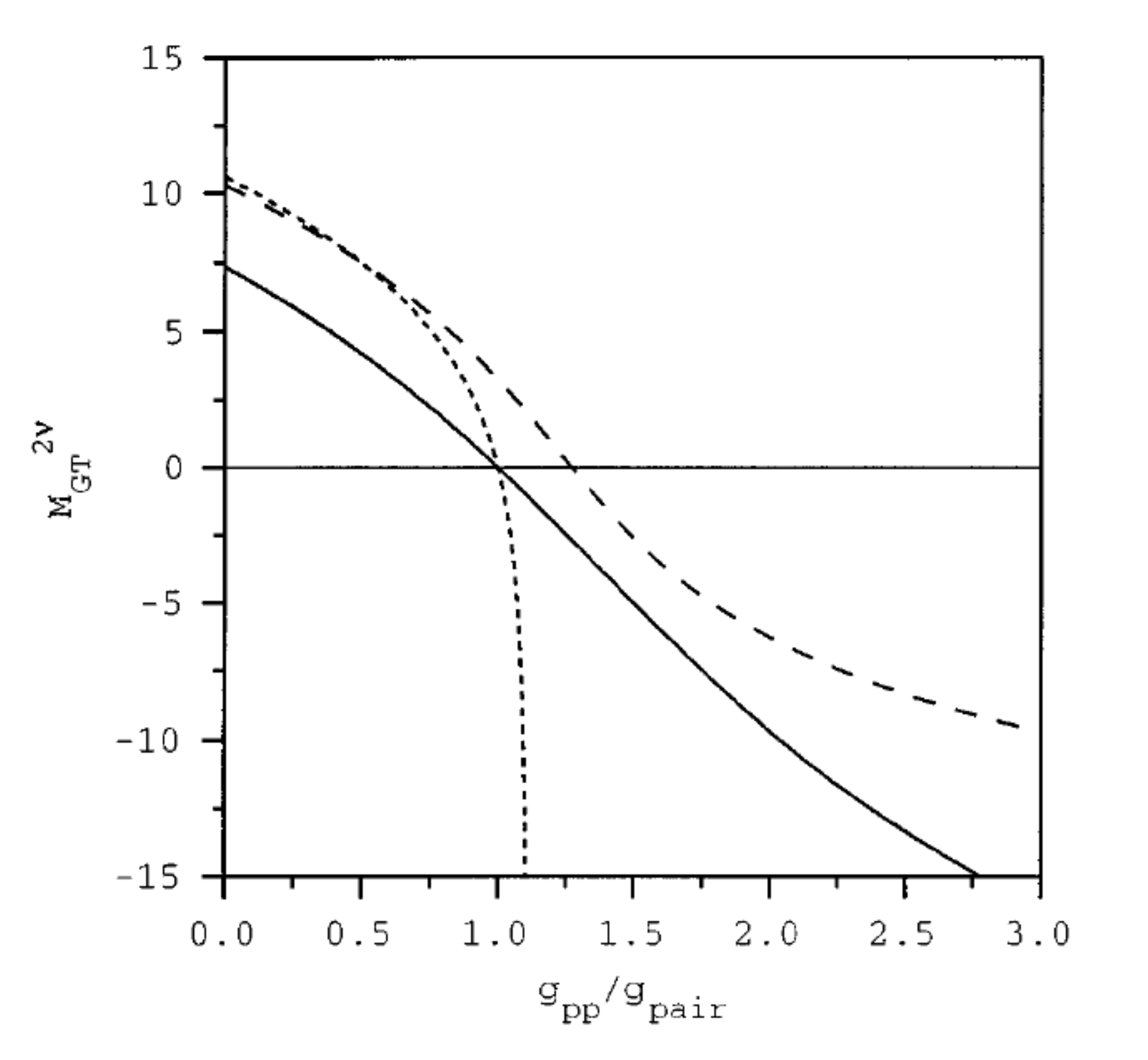,width=10cm}}
\caption{ The $2\nu\beta^+\beta^+$ decay matrix element $M_{GT}^{2\nu}$ vs. $g_{pp} /g_{pair}=(1-x)/(1+x)$. The
exact results are denoted by the solid line, the RQRPA results by
the dashed line, and the QRPA results by the dotted line. Calculated
for $\Omega=12$, $N=12$, $S=0$, $T=4$. From \cite{Engel97}.\label{fig:PRC55f6}}
\label{fig:PRC55f6}
\end{figure}

The  GT $\beta$ transition probabilities have been studied in the framework of the QRPA by several authors. Typically they start from a deformed
mean field Hamiltonian $H_{mf}$ that  includes $pp$ and $nn$ pairing and add separable residual interactions in the particle-hole channel (ph) and 
in the particle-particle channel (pp),
\bea\label{eq:VresQRPAGT}
H=H_{mf}+V^{ph}_{GT}+V^{pp}_{GT}, \\
V^{ph}_{GT}=2\chi^{ph}_{GT}\sum_K(-)^K\beta^+_K\beta^-_{-K},~~~\beta^+_K=\sum_{jm,j'm'}\langle jm\vert\sigma_K\vert j'm'\rangle n^\dagger_{jm}p_{j'm'}\\
V^{pp}_{GT}=-2\kappa^{pp}_{GT}\sum_K(-)^KD^+_KD_{-K},~~~D^+_K=\sum_{jm,j'm'}(-)^{j-m}\langle jm\vert\sigma_K\vert j'm'\rangle n^\dagger_{j'm'}p^\dagger_{j-m}.
\eea  
Here, $\beta^+_K$ are the GT transition operators and $D^+_K$  the isoscalar pair operators (\ref{eq:Dk}), up to a normalization constant.
First the strength of the particle-hole interaction $\chi^{ph}_{GT}$  is adjusted to the position of the GT resonance or derived from the interaction that generates the average potential.
Then  the strength of the particle-particle interaction $\kappa^{ph}_{GT}$ is tuned to reproduce the measured transition rates. The pair interaction influences the GT transitions rates
in a different way than the   above  discussed  $2\nu\beta^+\beta^+$ rates. Due to the finite $Q_\beta$ value, only a relatively small low-energy tail of the GT strength function
contributes to the decay rate. The attractive isoscalar pair interaction pulls GT strength into this region, which increases the transition probability. As discussed above, 
it decreases the total summed GT strength.  The reduction is located at higher energy, in the region of the GT resonance (see Fig. \ref{gt2}).    The sum over the 
intermediate states in the $2\nu\beta^+\beta^+$ decay explores this high-energy region as well, which results in a reduction.  

\begin{figure}
\centerline{\psfig{file=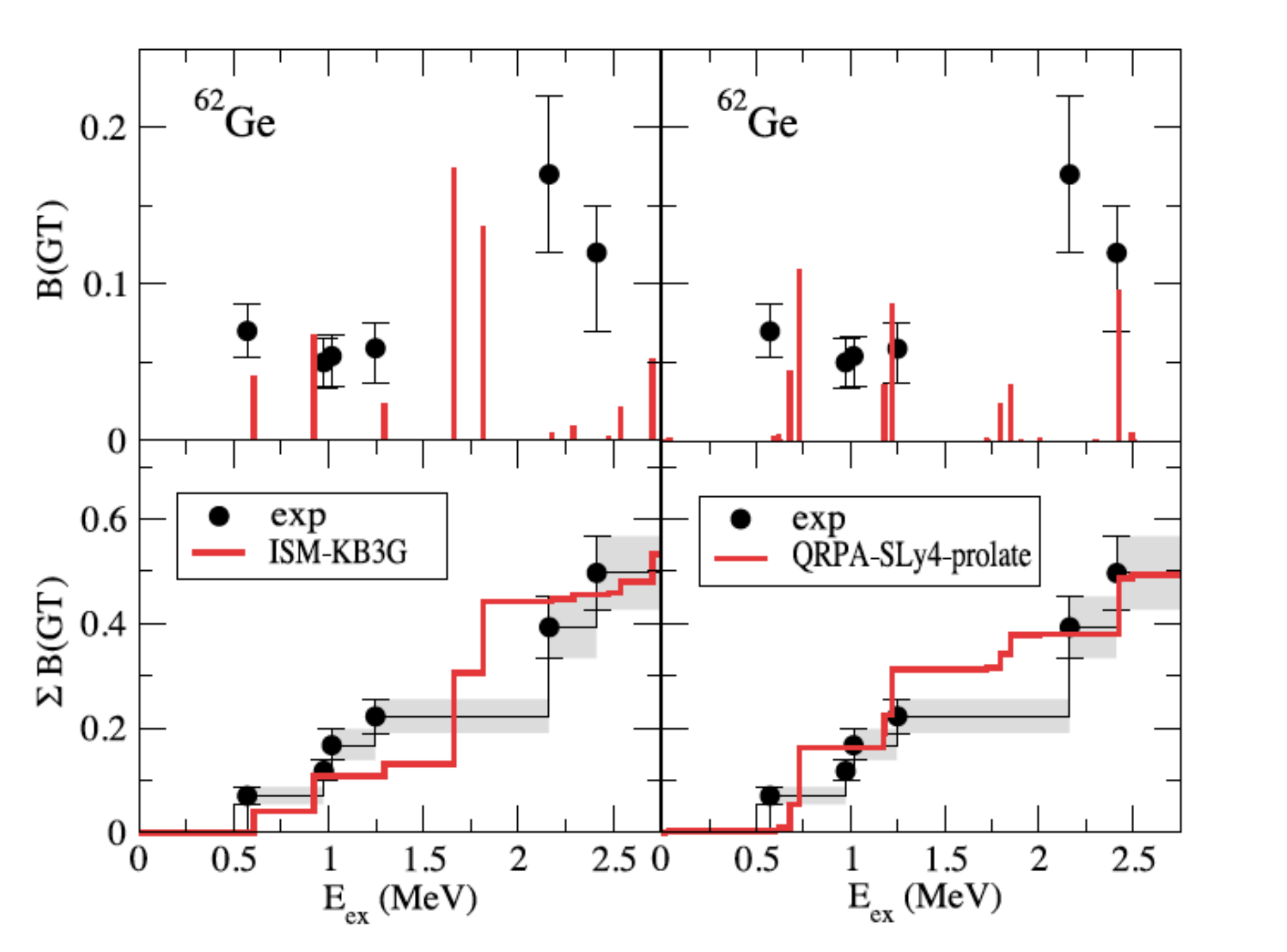,width=14cm}}
\caption{ Experimental (black color) and calculated (red color)
single level $B(GT)$ and accumulated $B(GT)$ values for the $\beta^+$ decay
$^{62}$Ge $\rightarrow$ $^{62}$Ga: left panels with the Shell Model calculations using the
KB3G interaction and right panels with the QRPA approach of Ref.  \cite{Sarriguren09}. 
Experimental uncertainty corridors are indicated in gray. From Ref. \cite{gadea}.}
\label{fig:62Ge}
\end{figure}

\begin{figure}
\centerline{\psfig{file=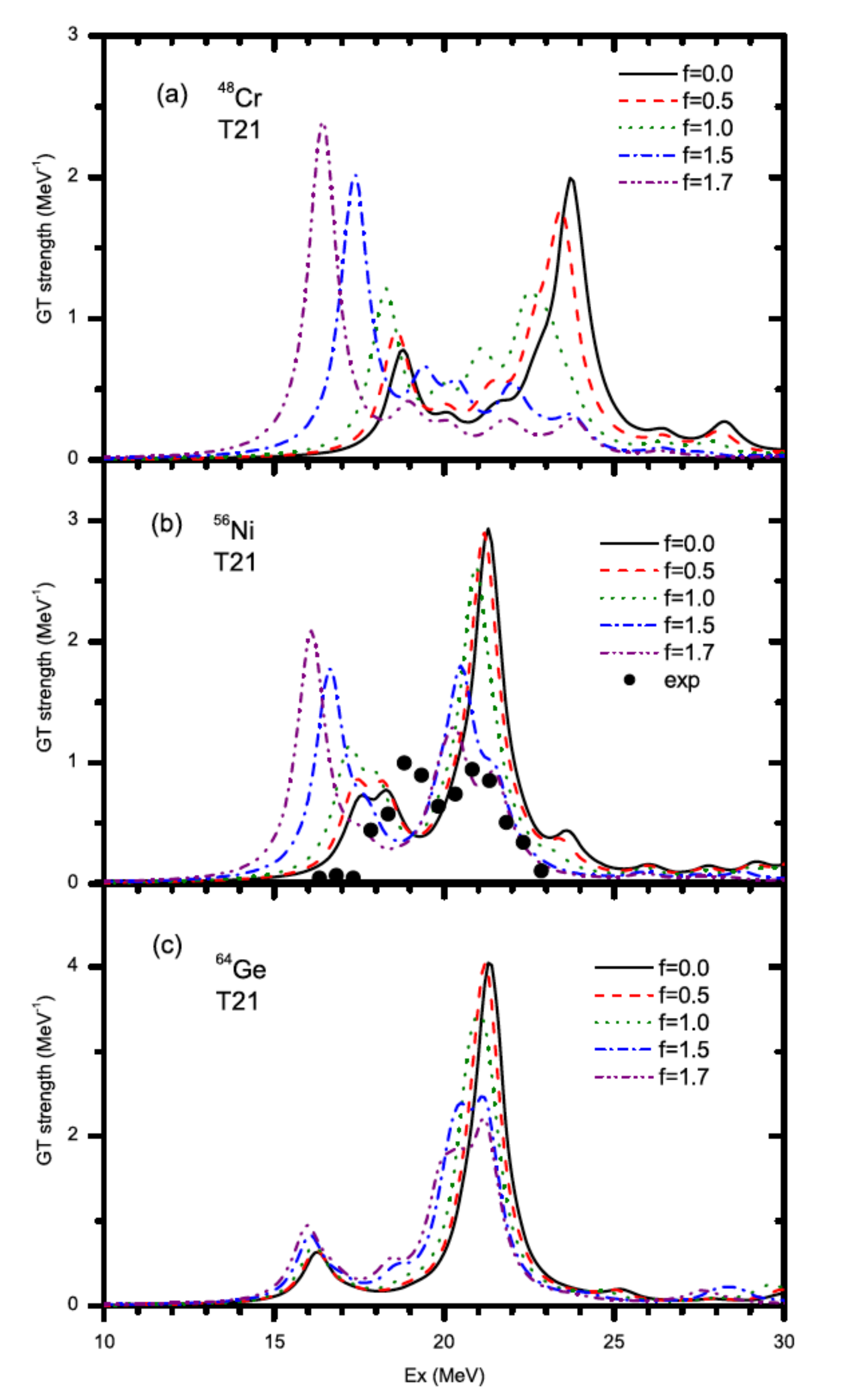,width=10cm}}
\caption{ GT strength in $^{48}$Cr, $^{56}$Ni and $^{64}$Ge by HFB+QRPA
with the Skyrme interaction from Ref. \cite{sasano}.}
\label{gt2}
\end{figure}

Homma {\it et al.}\cite{Homma96} used a modified oscillator potential  and adjusted $\chi^{ph}_{GT}$ to the experimental position of the GT resonance in $^{48}$Ca, $^{90}$Zr, and $^{208}$Pb, which
gave $\chi^{ph}_{GT}=5.2A^{-0.7}$MeV and $\kappa^{pp}_{GT}=0.58A^{-0.7}$MeV. They calculated the lifetimes of all nuclei with $A<150$ for which experimental values were known
and  could reproduce the lifetimes for  90\% of the nuclides within a factor of 10 and for 50\% within a factor of 2. Their isoscalar coupling constant is about one half of the critical value where
the QRPA breaks down.
M\"oller {\it et al.} \cite{Moller97} used a folded Yukawa potential and adopted $\chi^{ph}_{GT}=23A^{-1}$MeV, which is adjusted to the GT resonance in  $^{208}$Pb only.
They did not find it necessary to include a $np$ interaction in the particle-particle channel, i. e.  they assumed$\kappa^{pp}_{GT}=0$. 
They calculated a table 
of lifetimes and further properties that are important for astrophysical applications for all nuclides between  the proton and neutron drip lines. 
The root mean square of the  difference
between the decimal logarithm of the calculated and experimental lifetimes ranges from 3 for $T_{1/2}<1$s to 70 s for 100s $<T_{1/2}<1000$s. 
 Sarriguren {\it et al.} \cite{Sarriguren01} used the  Skyrme energy density functional SG2 from which they derived 
 the coupling constant  $\chi^{ph}_{GT}$ individually for each of the considered nuclei. The SG2 functional reproduces well the position and strength of the GT resonance.
 For nuclei near the $N=Z$ line the coupling constant   $\kappa^{pp}_{GT}=0.07$MeV was adopted, which is 70\% of the critical value. They were able to reproduce the experimental
 lifetimes of $^{64,66,68}$Se,  $^{68,70,72}$Ge, $^{72,74,76}$Kr, and  $^{76,78,80}$Sr within a factor of 10. In a subsequent study, Sarriguren  \cite{Sarriguren09}
 changed to the Sly and Sk3 Skyrme functionals, used the separable interactions (\ref{eq:VresQRPAGT}) with $\chi^{ph}_{GT}$=0.17MeV, and readjusted $\kappa^{pp}_{GT}=0.03$MeV 
 (30\% of the critical value). The calculations very well reproduce the experimental cumulative GT strength as function of the excitation energy for the nuclides
  $^{72,74}$Kr, $^{76}$Sr, and $^{62}$Ge displayed in Fig. \ref{fig:62Ge}. 
Engel {\em et al.} \cite{Engel99} calculated $\beta^-$ GT transition probabilities in neutron-rich nuclei  by combining the Skyrme  HFB with QRPA,
where the residual interaction in the particle-hole channel was fixed by deriving it self-consistently from the Skyrme energy density functional SKO'.
 They used a  finite range $T=0$ pairing  interaction, the 
 strength of which was adjusted to the experimental transition probabilities. Including it  removed   the systematic underestimate of the lifetimes by a factor of 2-5. 
  They pointed out that the resulting strength of the $np$ pair interaction is strongly correlated with strength of the relevant part of the particle-hole 
interaction, which is not very well accounted for by the Skyrme functionals. (The Migdal parameter $g'_0=0.8$ obtained from SKO' is to be compared to the experimental value of 1.8.)
Their strength of the $T=0$ pairing interaction is about one half of the critical value.  

Kaneko and Hasegawa \cite{Kaneko01b}  carried out QRPA calculations based on their EPQQ Hamiltonian (see Section \ref{sec:SMs}). They
found only a  very weak dependence of the summed low-energy GT strength on the parameter $k_0$, which controls  strength of their isoscalar interaction.
They attributed this to the inaccuracies of the  QRPA, because the special choice of their $T=0$ interaction does not induce $np$ pair correlations. 
The reproduced the experimental GT strength of  $^{48,50}_{22}$Ti$_{26,28}$, $^{54,56}_{26}$Fe$_{28,30}$, $^{59}_{27}$Co$_{32}$, 
and $^{58,60,62,64}_{28}$Ni$_{30,32,34,36}$ within a factor of 2 and the $2\nu\beta^+\beta^+$ strength $M_{GT}^{2\nu}$ of $^{76}_{32}$Ge$_{44}$
and $^{82}_{34}$Se$_{48}$ with an accuracy better than 30\%.
Their   calculations should be considered as excluding $T=0$ $np$ pair correlations.

%\begin{figure}
%\centerline{\psfig{file=GT.pdf,width=10cm}}
%\caption{ Expected Gamow-Teller decay in nuclei near N=Z}
%\label{fig4}
%\end{figure}

{ Very recently, Fujita {\it et al.}  \cite{FujitaCa} carried out a detailed study of GT excitations 
for mass number $A= $ 42,46,50, and 54 using the $(^3He,t)$ reaction at the RCNP Facility in Osaka.
Of particular interest to us,  are the results observed for the charge exchange reaction from the gs of $^{42}Ca$ 
to $^{42}Sc$.  The GT strength for this case appears to be concentrated in the lowest $1^+$ state at 0.611 MeV
which the authors interpret as a restoration of the SU(4) symmetry in the form of a low-lying collective GT phonon.  Shell model calculations with the 
Kuo-Brown interaction seem to account for the data rather well.  With only two valence particles outside closed shells, the direct implication of this 
observation on the existence of an isoscalar condensate is not clear and shell model indicators do not signal such a condensate.
The interpretation of this $1^+$ state as an isoscalar phonon, seems at variance with the  
 large $B(M1,1^+  \rightarrow 0^+) $ which is well described by a [$\nu$f$_{7/2}$-$\pi$f$_{7/2}$]$^{1^+}$  configuration \cite{Lisetzkiy99}
and other properties to be discussed in Sections \ref{sec:PV} and \ref{sec:transfer}.  Moreover, low-lying single-particle levels in $^{41}Ca$ and $^{41}Sc$
are consistent with  $jj$- rather than $LS$-coupling.  One cannot but notice in Fig. 48 that in the SU(4)
limit, the single-$l$ model predicts very small values of the total $B(GT)$. }

Grodner {\it et al.}  \cite{gadea} reported the first study of the GT decay of the $^{62}$Ge $T=1$, 0$^+$ 
ground state into excited states of  $^{62}$Ga.  
The experiment was performed at  GSI using the Fragment Separator and the RISING Ge-array 
coupled to an active implantation setup.
The distribution of the cumulative GT strength for the decay of $^{62}$Ge to $^{62}$Ga, which is the
heaviest odd-odd N = Z nucleus investigated via GT decays to date, is shown in Fig. \ref{fig:62Ge}.
 The authors carried out Shell Model calculations using the KB3G effective interaction and QRPA calculations
based on the method of Ref. \cite{Sarriguren09}. As seen in the figure, there is very good agreement 
between experimental data and theoretical calculations.

Iachello \cite{franco}  and  Halse and Barrett \cite{bruce} postulated the existence of $np$ coherent pairs (bosons). 
They  studied their role 
in $\beta $-decay within the framework of the phenomenological IBM4  model and 
pointed out  the enhancement of that transition.  They predicted super allowed GT transitions
with log-ft values less than four. According to these predictions the
Gamow-Teller (GT) $\beta $-decay rates between the ground state of an even-even
$N+2=Z$ nucleus and the lowest $1^+$ state in its odd-odd $N=Z$ daughter nucleus can
provide a fingerprint of $T=0$ pairing.
Contrary to their predictions, 
the measured $ B(GT)= 0.070(0.017) g^2_A/4\pi$ value for the first 1+
state is 40 times smaller than the value of the IBM4 model, which assumes the presence of an
 $T = 0$ pairing condensate.
 As shown in Fig. \ref{fig:PRC55f4},  within the schematic model of one $j$-shell  \cite{Engel97} one also expects that the presence of an isoscalar pair field would 
 strongly enhance the GT strength (compared, for example,  to the Ikeda sum rule).
 Moreover, the increase  of the GT strength with $M_T$, which characterizes the isoscalar regime ($x<0$) (see Figs. \ref{fig:PRC55f9} and \ref{fig:PRC55f4})  is not observed.
  In contrast, the observed decrease is characteristic for  the isovector regime ($x>0$).

The GT transition operator is related the isovector spin part of the magnetic moment by  rotations in isospin space, which change $t_0\rightarrow (t_1\pm t_{-1})/\sqrt{2}$ .
As discussed in the previous Section \ref{sec:lowoo}, the static and transition magnetic moments of the low-lying 1$^+$ states in odd-odd $N=Z$, which are the final states
of the GT decay, are well accounted for by considering them as two-quasiparticle excitations composed of the one-quasiparticle states observed
in the odd-A neighbors. The related GT matrix element should then be determined by the spin parts of the quasiparticles in the odd-A neighbors.
 The measured $B(GT)$ value is below the estimate based on the assumption of a two-quasiparticle structure of the $1^+$ states in $^{62}$Ga. 

The discussed microscopic studies of the GT decay, which presume that there is no isoscalar pair condensate, account
 well for the experimental data. As discussed above,  the absence of an isoscalar pair condensate  is not surprising due to the strong quenching effect of the spin-orbit splitting.
  The data seem to point to the presence of dynamical isoscalar $np$ pair correlations, the strength of which is still
 uncertain. Whereas the   $2\nu\beta^+\beta^+$ data are  accounted for by relatively strong correlations with a $T=0$ pair 
 interaction strength  close to the critical value, the $\beta$ data are better accounted for by weaker or even zero  correlations of this type.
The main source of uncertainty is the  strong correlation between  $np$ interaction in the particle-hole channel, which decreases the
low-energy GT strength, and the $np$ isoscalar pair interaction, which increases it.  
In order to disentangle the two effects, it seems useful  to analyze  GT strength at high-excitation energy.
For example,  Bai et al. \cite{bai} have recently studied GT states in $N = Z$ nuclei with the mass number A from 48 to 64 
by means of QRPA based on a  Skyrme energy density functional and a contact pairing force with an adjustable $T=0$ part.
 An example of their results is displayed in Fig. ~\ref{gt2}. The calculations  demonstrate the  shift of GT strength toward low energy
with increasing strength  of  the $T=0$  force, which is determined  by the factor $f$.  
 However,  in looking at the results for $^{56}$Ni, which are compared to data obtained at RIKEN \cite{sasano},
we believe that the discrepancy between theory and experiment is still too substantial that one can quantify the low-energy enhancement caused by $T=0$ pair correlations. 

The GT strength has been studied in the framework of the Shell Model as well. The various experimental data are well accounted for provided that the 
GT matrix element is attenuated by a quenching factor of 0.75 (see eg. \cite{Langanke95,Zhi13}). The same quenching factor brings the calculated magnetic dipole strength, which is 
closely related to the GT strength by rotation in isospace, into agreement with experiment. This  consistency indicates that the reduction is caused by correlations that are 
beyond the configuration space of the Shell Model  \cite{Caurier05}. The same quenching factor is needed in the discussed QRPA calculations for the GT strength
and in QRPA calculations for the $M1$ strength in order to account for the experimental data. Fig. \ref{fig:62Ge} demonstrates that the SM calculation using the KB3G interaction
in the configuration space of the pf shell very well reproduces the experimental low-energy GT$_+$ strength. Having in mind the discussed sensitivity of this
quantity to the isoscalar pair correlations,  the agreement supports the confidence in this interaction producing a realistic amount of isoscalar pair correlations. 
Petermann {\it et al.} \cite{Petermann07} calculated the GT$_-$ strength using the same Shell Model
 Hamiltonian and configuration space. Since the total GT$_-$ strength $S_-$ is  related to the total GT$_+$ strength $S_+$ by the Ikeda sum rule 
$S_--S_+=3(N-Z)$ one expects $S_-$ to be as sensitive to the isoscalar pair correlations as $S_+$. The authors  found fair agreement with the experimental values of
the accumulated strength $B(GT_-)$ for $^{46}$Ti, $^{50}$Cr and $^{54}$Cr obtained from $(^3He,t)$ charge exchange reactions, which can taken as complementary 
 evidence that their interaction may provide a realistic estimate of the isoscalar pair correlation strength. In particular, the calculations along with the data 
 contradict the concentration of the strength in a single low-lying state of the  daughter nucleus,  which the phenomenological IBM4  model predicts
 as a consequence of an approximate SU(4) symmetry caused by the presumed strong  isoscalar pair correlation  \cite{franco,bruce}. 
 As discussed in Section \ref{sec:SMs}, the analysis of the pair correlation strength generated by this interaction in the pf shell indicates the coexistence of a strong isovector  
 component with a moderate isoscalar one, which is consistent with the above discussed scenario of moderate vibrational isoscalar pair correlations on top 
 of the isovector pair condensate.      

Petrovici {\it et al.} \cite{Petrovici11} calculated the GT strength for the $\beta^+$ decay of the astrophysical relevant nuclides $^{68}_{34}$Se$_{34}$ and $^{72}_{36}$Kr$_{36}$
in the frame work of the Complex  Excited Vampir approach. In the case of Kr,  good agreement with with the data (comparable with the calculations shown in Fig. \ref{fig:62Ge})
 was found. As discussed in Section  \ref{sec:SP}, the analysis of the wave functions of $N\approx Z$ 
  indicated the presence of isoscalar pair correlations with a strength of about 1/3 of the isovector strength.

In summary, it appears that correlations induced by the isoscalar pair interaction in the 
QRPA and Shell Model calculations for the GT strength  do not indicate the presence of  a $T=0$ isoscalar condensate. 
Rather they seem to generate a pair vibrational resonance, which is fragmented among the surrounding particle-hole states. This type of
pairing structure remains to be explored,  by calculating the two-particle transfer strength, for example within the QRPA approach.

\section{Pairing vibrations}\label{sec:PV}
 
Near closed shells, the strength of the pairing force relative to the single-particle level spacing
is expected to be less than the critical value for the existence of a 
pair condensate,  and the pairing field then gives rise to a collective
phonon \cite{bes1} (see Section \ref{sec:schematic}).  

It seems natural to ask whether  $T=0$  collective effects may show as a 
vibrational phonon.  As discussed in \ref{sec:IV2shell},    one can estimate the
frequency of these modes within the framework of a two $\ell$-shell model \cite{dussel},
\be\label{pvfreq}
\hbar\omega_{T} = D\sqrt{ 1 - y_{T}}, ~~~~
y_{T}=2G_{T}\Omega / D=G_{T}/G_{crit}
\ee
in terms of the pairing strenghts $ G_{T}$, the degeneracy $\Omega$,
and the single-particle energy spacing $D$. Clearly, the larger
$y_T$, the stronger the correlations.

\begin{figure}
\centerline{\psfig{file=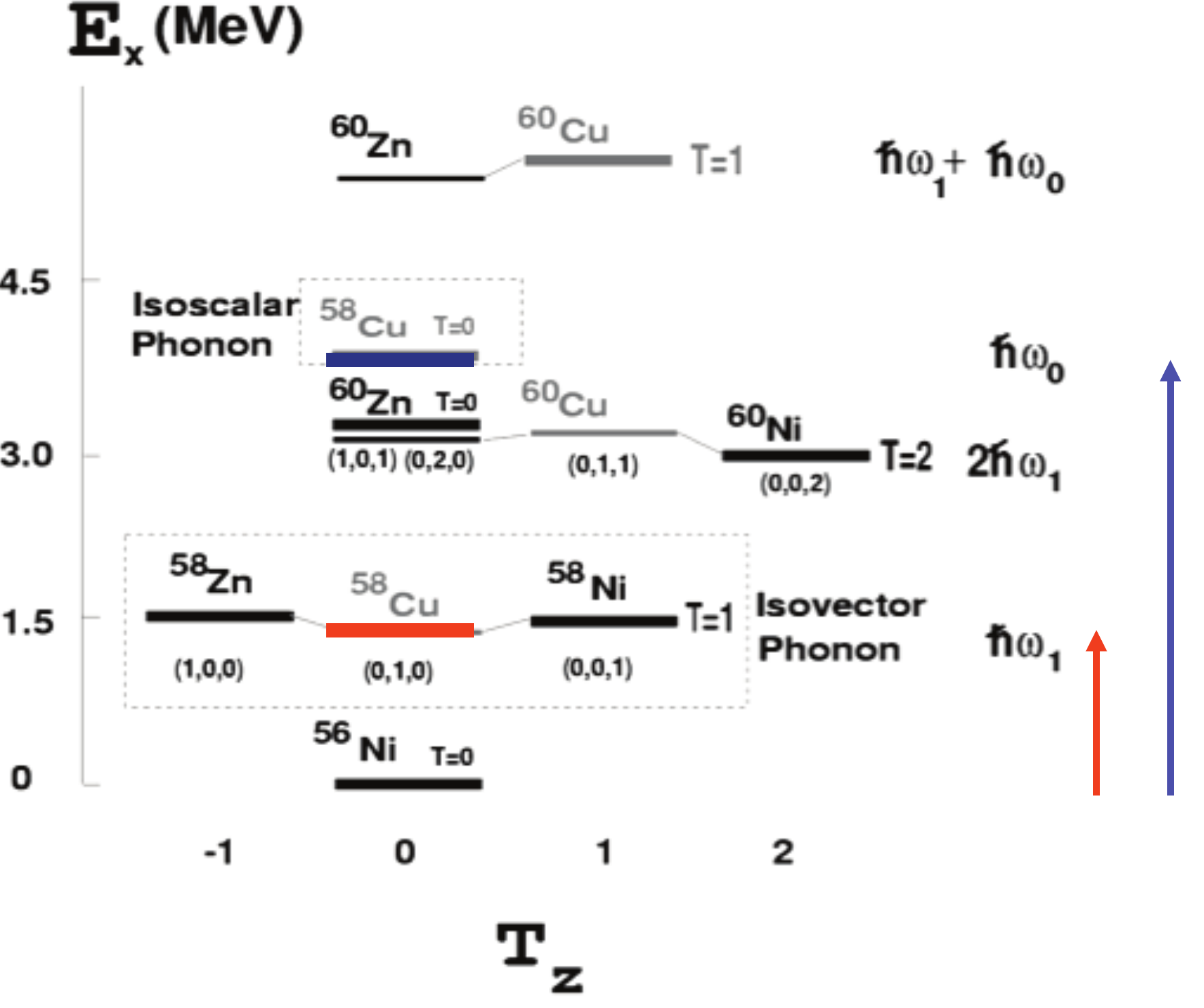,width=14cm}}
\caption{ Experimental
excitation spectrum for addition phonons around $^{56}$Ni. The
excitation energies are derived from experimental binding energies and
the subtraction of average volume, surface, Coulomb, and symmetry
terms. Ground states are represented by a thicker line, and members of
an isobaric multiplet are joined by thin lines. Levels in grey are in
odd-odd nuclei.}
\label{phonon}
\end{figure}

The role of isospin in the pairing vibrations around $^{56}$Ni was initially discussed by Bohr  and Nathan\cite{bohr,nathan}.  
A detailed analysis was presented in the review article by B\`es et al. \cite{bes2}, but the role of $T=0$ 
collective pairing remained unclear.
With this in mind, Macchiavelli {\it et al.}  re-visited the 
excitation spectrum around the doubly closed-shell $N=Z$ nucleus $^{56}$Ni \cite{aom2}.  
They  used $^{56}$Ni as reference and considered  only
even $A$ (even-even or odd-odd) nuclei. The excitation energy of a state of isospin $T$ in a given nucleus $A,Z$  is obtained from
\begin{eqnarray}
E_x(A,T,M_T)=  BE_{exp} - Surf(A) - Coul(A,M_T) -Sym(A,T) - \lambda(A-56) -E_0.
\end{eqnarray}
In this equation $BE_{exp}$ includes both the ground-state binding energy 
and, if applicable, the relative excitation energy of the state under consideration
from the ENSDF database.
They used for the surface energy $Surf(A)=-17.3$MeV$A^{2/3}$,
for the Coulomb term $Coul(A)=-0.65$MeV$(A/2-M_T)^2/A^{1/3}$ and the symmetry energy
$Sym(A,T)=-75$MeV$T(T+1) /A$.  The volume term enters through the coefficient $\lambda = 15.55$MeV which is
adjusted to give the same frequency for the addition and removal phonons\cite{bohr} and, equivalently,
 sets the
Fermi surface in the middle of the $N=Z=28$ shell gap around $^{56}$Ni.
Finally, $E_0$ is chosen such that $E_x(56,0,0)=0$.
The subtraction of average properties (including an average symmetry term) leaves, in principle, only those effects 
associated with pairing and shell structure and can be compared directly with the 
harmonic spectrum discussed above and shown in Fig. \ref{phonon}.  Their method differs slightly from that of Refs. \cite{bes1,bohr}, mainly in the subtraction of
the full symmetry energy term. They thus checked the procedure for nuclei around $^{208}$Pb and obtained for the neutron pairing vibrational phonon
$\hbar\omega \approx 2.3$MeV, in good agreement with the analysis presented in \cite{bm1}.  In our opinion,  this subtraction procedure is the proper
one to generate an excitation spectra whose properties can be,
at least qualitatively, explained by the phonon frequencies in Eq. (\ref{pvfreq}).

The excitation spectrum derived from the experimental data is shown in Fig. \ref{phonon}.
The average energy of the $A=58$ isobaric multiplet gives the frequency
of the $T=1$ phonon at  $\hbar\omega_1 \approx 1.5$MeV.  With a typical 
single-particle level spacing
energy  $D\approx 4-5 $ MeV in this region  we obtain from 
Eq. (\ref{pvfreq}) a qualitative estimate of
$G_{1}/G_{crit} \approx 0.9$ suggesting a strong collective character of the vibration.
Similarly, the lowest $T=0$, $1^+$ state in 
$^{58}$Cu is obtained from $^{56}$Ni by the addition of the 
isoscalar $T= 0$ phonon and, therefore,
defines the frequency for this mode.  It is obvious by simple inspection of the data that 
this excitation is much higher than for $T=1$. In terms of the phonon picture,  
$\hbar\omega_0 \approx 4$MeV,  which translates into 
$G_{0}/G_{crit} < 0.2$ and implies, in contrast to the $T=1$ case, 
weak collective correlations for $T=0$ pairing.

\begin{table}
\begin{center}
\epsfig{file=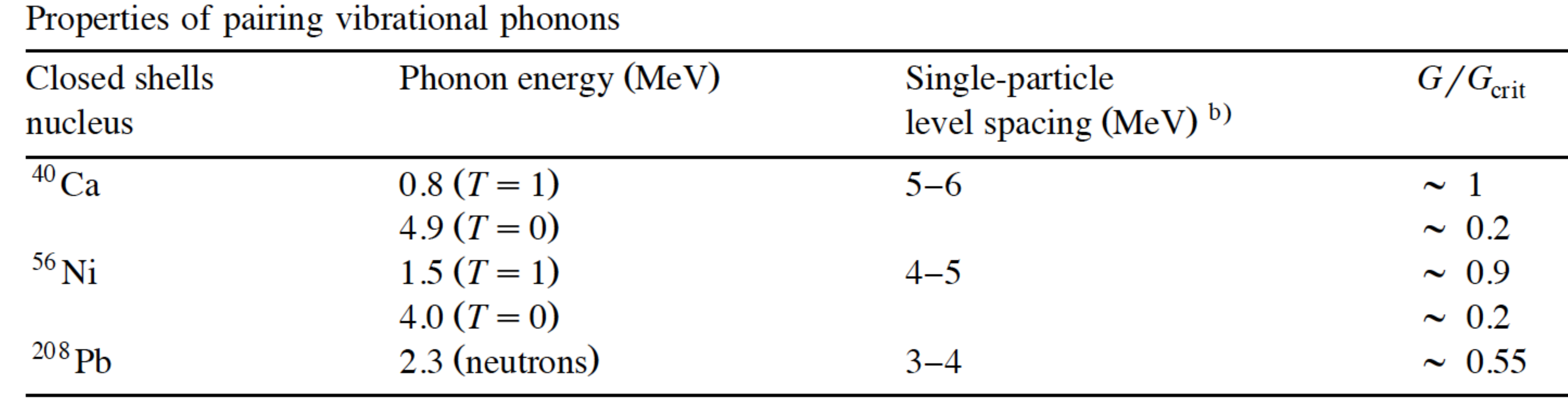,scale=0.5}
\end{center}
\caption{Properties of pairing vibration phonons  from Ref. \cite{aom2}.}
\end{table}

The rather soft character of the isovector addition and removal modes,
having $G_{1} \sim G_{crit}$, suggests that  the
multi-phonon system is the precursor of developing a permanent deformation in gauge 
space and isospace. On the contrary, the isoscalar excitations reflect a weak collectivity , $G_{0}/G_{crit} << 1$, closer to 
single-particle nature.  Based on these observations, one envisions a
coupling scheme whereby the ground state of any even-even nucleus is a
correlated state of $T=1$ pairs maximally aligned in isospace,
i.e. $|A,(T=|M_T|),M_T>$.  In the neighboring odd-odd systems we have two
options: 
{\it i)} the $|A \pm 2,(T=|M_T|+1),M_T>$ state with one extra
$np(T=1)$ pair and the same pairing correlation properties as the
even-even core; and 
{\it ii)} the $|A \pm 2,(T=|M_T|),M_T>$  state, which is equivalent to the
core plus a $np(T=0)$ pair, i.e.~ a $|A,(T=|M_T|),M_T> \otimes ~np(T=0)$. 
 As discussed in Section \ref{sec:lowoo}, the analysis of the magnetic properties of these  $1^+$ states 
 suggests that they have two-quasiparticle character.

 B\`es and Civitarese \cite{BesGT}, have extended the theoretical description of pairing phonons $T=0, J=1$ and $T=1, J=0$ around $^{56}$Ni by including on the same footing
a $T=1, J=1^+$ GT phonon and their couplings.   The experimental information on isospin-spin excitations around $^{58}$Ni was analyzed and found that the proposed
coupling scheme accounts for a sizable amount of the strength associated with isospin-spin excitations, which
include transitions to both one- and two-phonon states.  {It would be interesting to apply this approach to the recent data in Ref. \cite{FujitaCa}.}

{Yoshida\cite{Yoshida14} carried out RPA calculations to describe the pair-addition and pair-removal  modes of the double magic nuclei $^{40}$Ca and $^{56}$Ni. 
The author used the mean field
generated by the SGII Skyrme energy density functional and a zero-range  pairing interaction that depends on the density. 
The isovector strength  of the $\delta$-force was 
fixed by adjusting the HBFB values of $\Delta_n$ and $\Delta_p$ to the experiential even-odd mass differences.   
The ratio between the isoscalar and isovector strengths was treated as a free parameter. With a ratio of about 1.3, energy differences between
the isoscalar and isovector modes come close to the experimental energy differences between the $1^+$ and $0^+$ in the adjacent $N=Z$ odd-odd nuclei.
With this choice of the pairing interaction, the transfer strength to isoscalar mode indicates substantial collectivity, comparable with the collectivity of the
isovector mode. However, as shown in  Fig. \ref{syst} for $^{42}$Sc, the experimental ratio  of the transfer cross section  $\sigma(1^+)/\sigma(0^+)\approx3/4$
corresponds to a ratio of 1/4 for one of the angular momentum  projections  of the $1^+$ state, which is much smaller than  the calculated ratio of 1.2. 
As discussed in Section \ref{sec:Wphen}, the energy difference between the $1^+$ and $0^+$  state is very sensitive to the value of the symmetry energy coefficient.
The calculated mean value of the frequencies of the isovector pair-removal and pair addition modes for $^{40}$Ca is 3.2 $MeV$ is lower than the 
experimental value $B(20,20)-(B(19,19)+B(21,21))/2=4.4 MeV$.  This  may indicate that the symmetry energy coefficient is underestimated, and
 the strength of the isoscalar interaction must be increased to maintain the experimental energy difference between the  $1^+$ and $0^+$  states. }

\section{Two-particle transfer}\label{sec:transfer}

Resorting again on the analogy to the  familiar case of quadrupole shape fluctuations, where an important measure of the 
collective effects is provided by the reduced transition probabilities , $B(E2)$, one can associate a similar role to the transition operator 
$\left<f\Vert P^\dagger \Vert i\right>$  (See Section \ref{sec:theory})  describing a two-particle transfer process. As shown in the discussion of simple models, this transition operator 
could be associated with the "order parameter" describing the phase transition between a normal  and a superfluid state of a macroscopic system.
In fact two-nucleon transfer reactions such as $(t,p) $ and $(p,t)$ reactions, 
provided a unique tool to understand pairing correlations in nuclei\cite{Broglia}.

Schematically the behavior of the two-nucleon transfer cross section between the ground states of nuclei A and (A+2) when pairing correlations are present, will be as follows:
Near closed shells, the system is characterized by an harmonic vibrational pattern and the cross section is proportional to the number of phonons with an enhancement of the order of the single particle degeneracy $\Omega$.  As more phonons are added, there is a transition to a BCS-like ground state containing a pair condensate that breaks gauge symmetry and isotropy in isospace. 
 The cross sections are now rather constant and enhanced by 
$~\Omega^2$,  namely of order $(\Delta/G)^2$.   The change  of the transfer strength from the pair vibrational to the pair rotational  pattern
is discussed in the context of Fig. \ref{fig:NPA153f2} in Section \ref{sec:theory}.

\begin{figure}
\centerline{\psfig{file=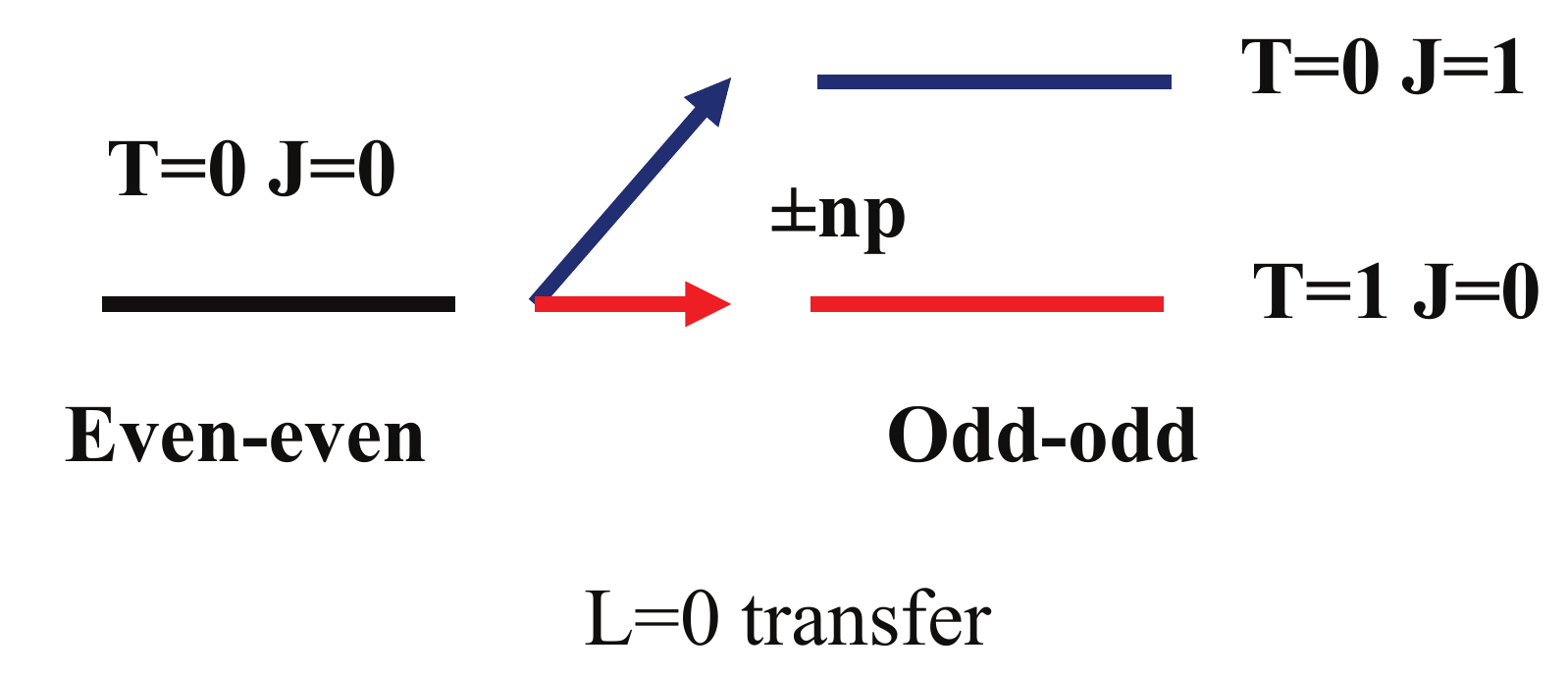,width=8cm}}
\caption{ Schematic diagram depicting the use of  two-particle transfer ($np$) reactions to study $np$ correlations. }
\label{cartoon}
\end{figure}

\begin{figure}[tb]
\epsfysize=9.0cm
\begin{center}
\begin{minipage}[t]{8 cm}
\epsfig{file=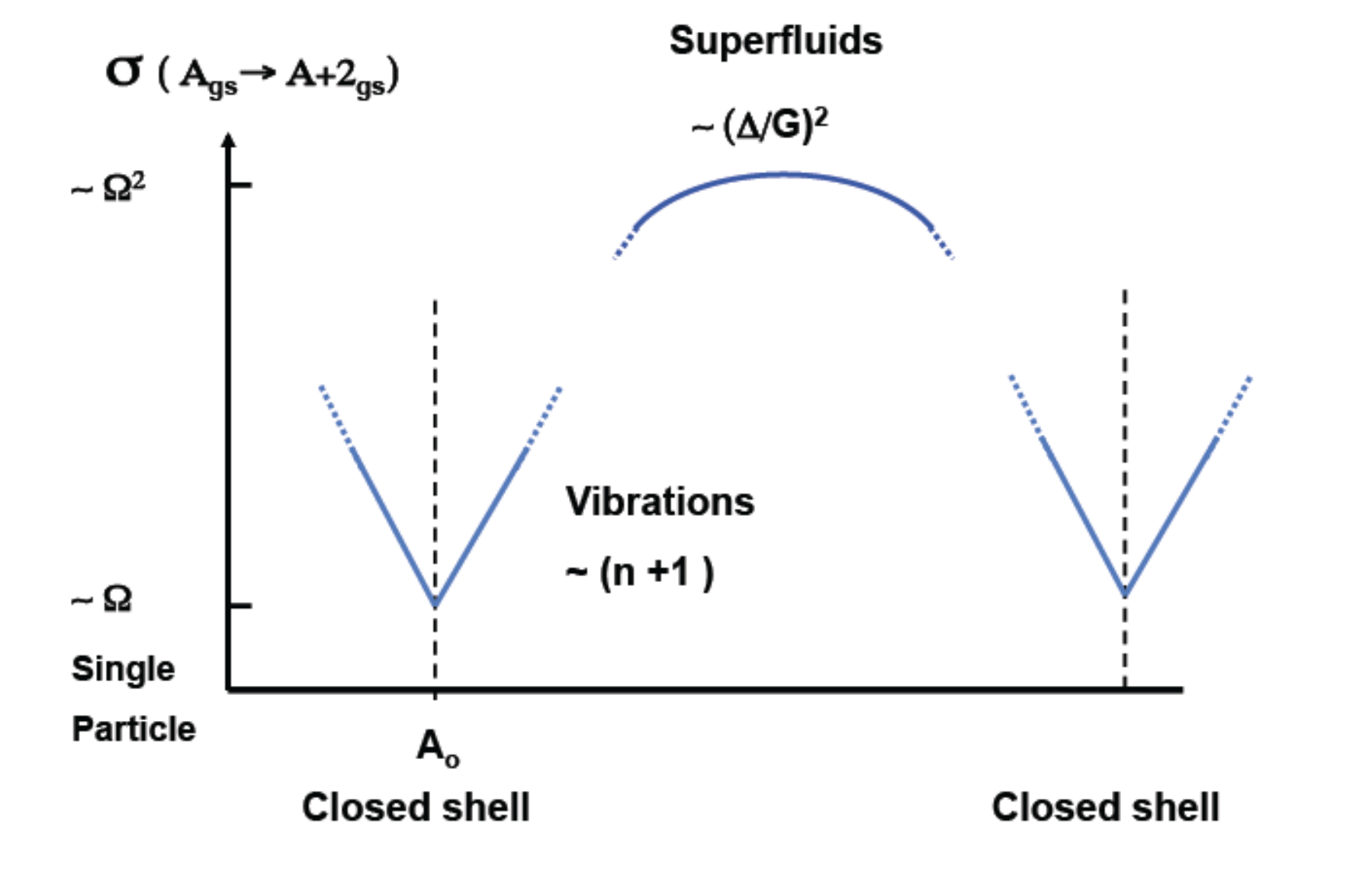,scale=0.46}
\caption{ Schematic diagram depicting the behavior of the 2-nucleon transfer cross-sections. \label{xsec}}
\end{minipage}
\hspace*{0.2cm}
\begin{minipage}[t]{8cm}
\hspace*{-1.5cm}
\epsfig{file=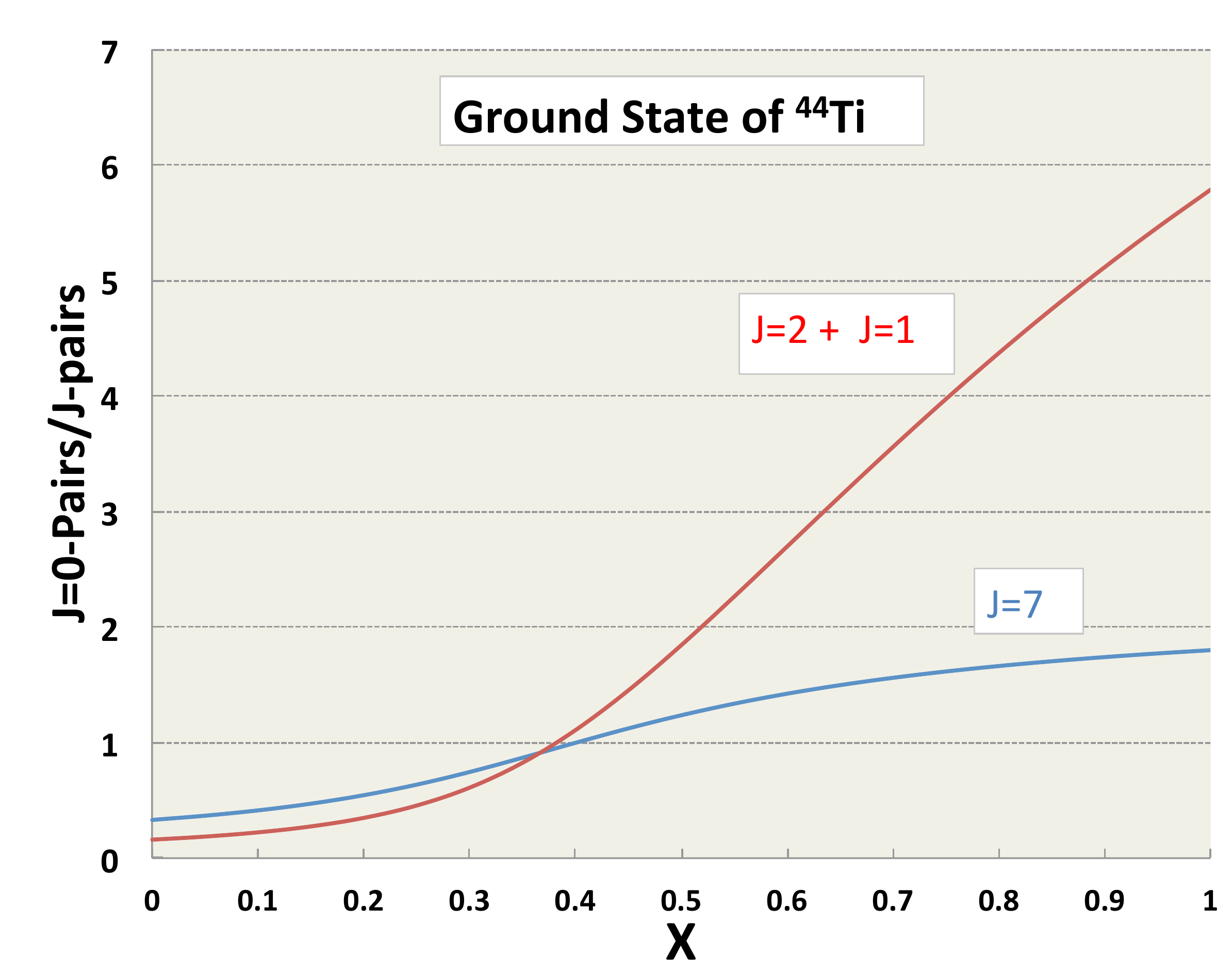,scale=0.35}
\caption{ Ratio of number of pairs of a given spin in the ground state of $^{44}$Ti as a function of   the relative strength of T=0 vs. T=1 pairing, measured by the parameter $x.$ Pure Isoscalar corresponds to x=0 and pure Isovector x=1.}
\label{pairs}
\end{minipage}
\end{center}
\end{figure}

 Earlier systematic analysis of two-neutron ($L=0$) transfer reactions using Ca and Ni  isotopes by Bayman and Hintz \cite{Bayman} found the data consistent with a picture involving configuration mixing induced by simple pairing degrees of freedom of the extra core neutrons.  The authors considered six shells:  2s$_{1/2}$ through 1f $_{5/2}$ for Ca and five shells:  1f$_{7/2}$ through 1g$_{9/2}$ for Ni, and diagonalized 
a pairing force between states of seniority zero. The resulting ground state wave-functions were used in DWBA calculations showing good agreement with single-particle data and $(p,t)$ and $(t,p)$ reactions. Further work by B\`es et al. provided conclusive evidence for isovector pairing in the form of pairing vibrations\cite{bes2,nathan}

As discussed earlier, the rapid quenching of $np$ pairs as one moves away from $N=Z$ suggests that the addition or removal of an $np$ pair from even-even to the lowest $0^+$ and $ 1^+$ states in 
odd-odd self conjugate nuclei stands out as the best tool to study $np$ pairing correlations  (cf. Fig. \ref{cartoon}).
This can be accomplished by stripping or pick-up reactions transferring a $np$ pair, like 
for example $(^3He,p)$. The enhancement of deuteron transfer reactions by $np$ pairing was first addressed by Fr\"obich \cite{frob} who predicted an increase of $\approx 2.5$ in the cross-section over the single-particle estimate.
More recently,  Van Isacker et al. \cite{piet}  studied the effect in the framework of the IBM model and independently concluded that the transfer intensities should reflect the nature of the ground state 
condensates as illustrated in Table \ref{piettable}, which shows the transfer amplitudes to scale like the number of bosons, $N_b$ for the dominant channel.

\begin{table}
\begin{center}
\epsfig{file=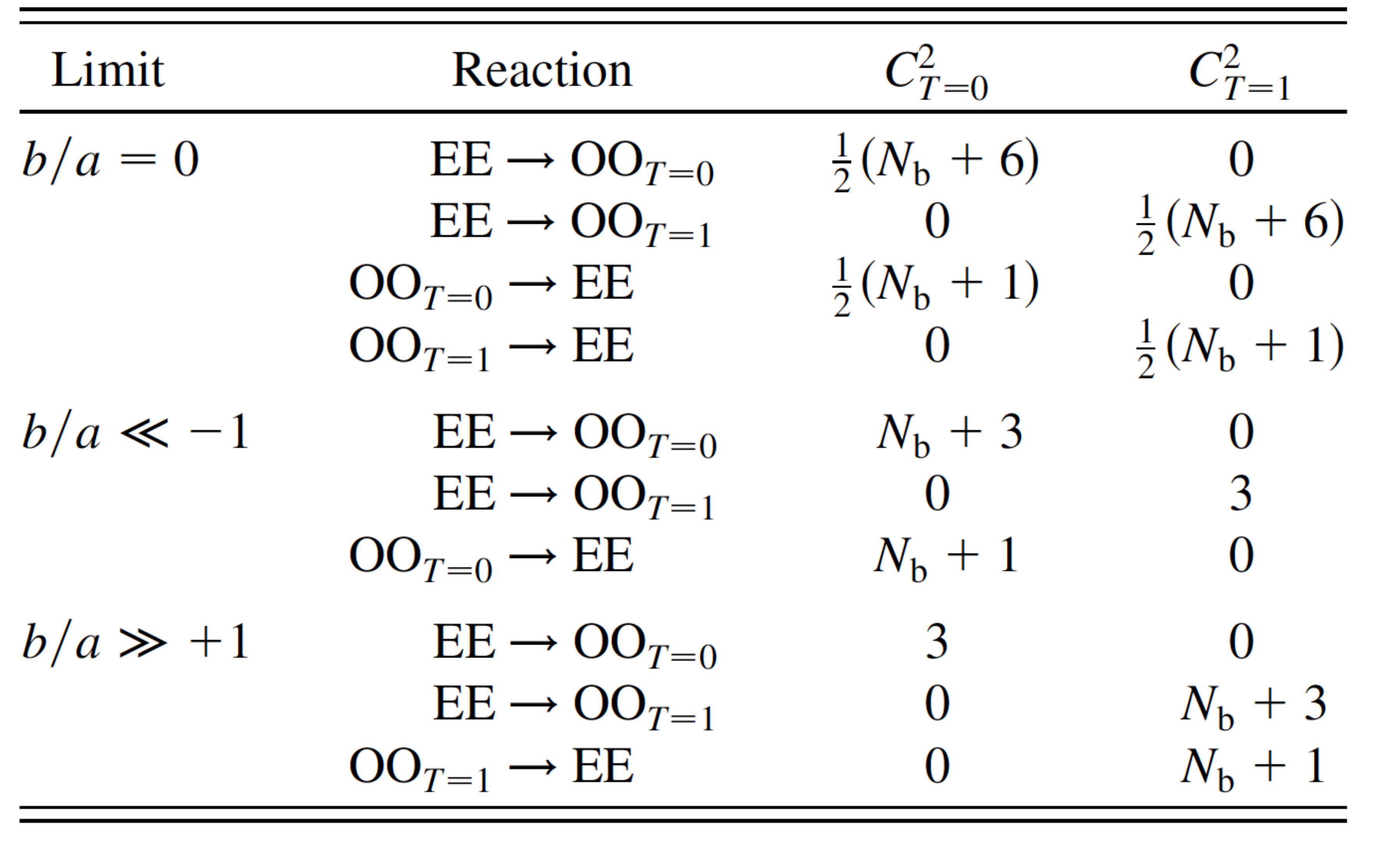,scale=0.3}
\end{center}
\caption{Expected dependence of the $np$ transfer amplitudes as calculated in Ref. \cite{piet}. The SU(4) limit corresponds to
$b/a=0$,   the isovector limit to $b/a >> 1$,  and the   isoscalar limit to $b/a << -1$.}
\label{piettable}
\end{table}

While absolute cross-section values are always desirable,  the ratio $\sigma(0^+)/\sigma(1^+)$ itself gives a measure of the pairing collectivity in the respective channels.  
Not requiring corrections due to kinematic effects,  the relative measurements can be compared directly with those expected for independent nucleon transfer.
This is illustrated schematically in Fig. \ref{pairs} showing the ratio in the number of pairs of angular momentum $J =0, 1$ and $ 2$ in the ground state of $^{44}$Ti as a function of a parameter $x$, that measures the relative strengths of the $T=0$ vs. $T=1$ pairing forces . As mentioned earlier, the calculations were performed following the formalism described in Ref. \cite{Zamick05}.   
Although not all completely due to the correlation aspects, there is a strong dependence of this ratio with $x$. Naturally, cross sections will scale with the number of pairs. 

%\footnote{ In relation to the discussion of the level structure of $^{92}$Pd,
%note the fact that the high-$j$ ($7^+$) pairs are rather insensitive to x. }

%\subsubsection{The $(^3He,p)$ reaction in reverse kinematics.}

With $^{40}$Ca being the last stable $N=Z$ nucleus,  the transfer  %$(^3He,p)$% 
reactions have to be studied in reverse kinematics as required for radioactive beams. 
Following from the argument that larger single-particle degeneracies $\Omega$  provide  a better opportunity to develop collective effects, the study  of two-nucleon transfer reactions in heavier $N=Z$ systems 
will be a unique program to elucidate the role of the collective properties of isoscalar  pairing.   Of possible reactions such as $(d,\alpha)$, $(\alpha,d)$, $(p, ^3He)$, $(^3He,p)$, etc. the latter offers clear advantages  due to energy and kinematic considerations but, more importantly, because $\Delta T=0,1$ are allowed.  In this way both low lying $0^+$ and $ 1^+$  states in odd-odd self conjugate nuclei will be populated.  
Here we briefly report on recent work by Macchiavelli et al. \cite{aom4}, carried out at the ATLAS facility in Argonne National Laboratory.   Similar experiments are planned at GANIL and HIE-ISOLDE \cite{didier,jenkins2}.

The proof of principle of the technique was done using stable beams of $^{28}$Si, $^{32}$S, $^{36}$Ar, and $^{40}$Ca, and a first successful application with a radioactive beam of $^{44}$Ti.  
\begin{figure}
\centerline{\psfig{file=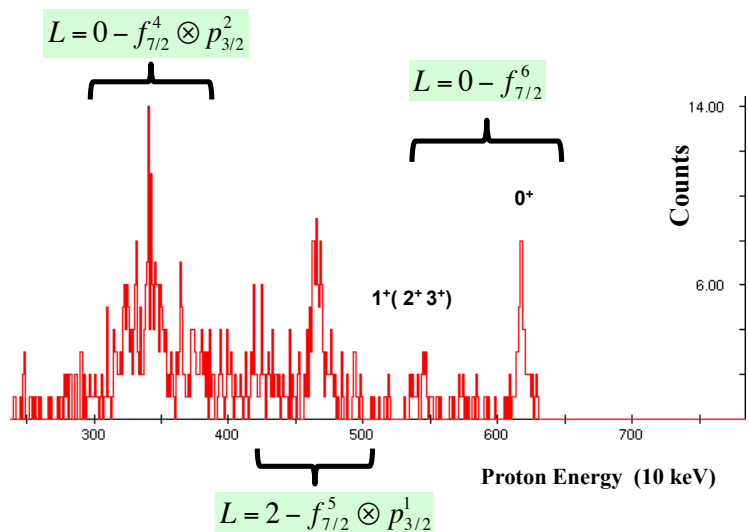,width=12cm}}
\caption{ Proton spectra for the $^{44}$Ti$(^3He,p)$$^{46}$V reaction, obtained by selecting  recoils with the FMA and after kinematic correction of the energies. }
\label{44Ti}
\end{figure}
Preliminary results are presented in Fig. \ref{44Ti} ,  showing the proton spectra to final states in $^{46}V$ observed in two groups of rings of the S1 detectors.   
We estimated a cross section for the transition to the ground state of  $\approx $ 1mb/str at forward angles in the center of mass system, which is in line with results in $^{40}$Ca. 
The labels indicate tentative assignments  for the different proton groups  within the fp shell,  as  suggested by their energies and relative intensities.

\begin{figure}
\centerline{\psfig{file=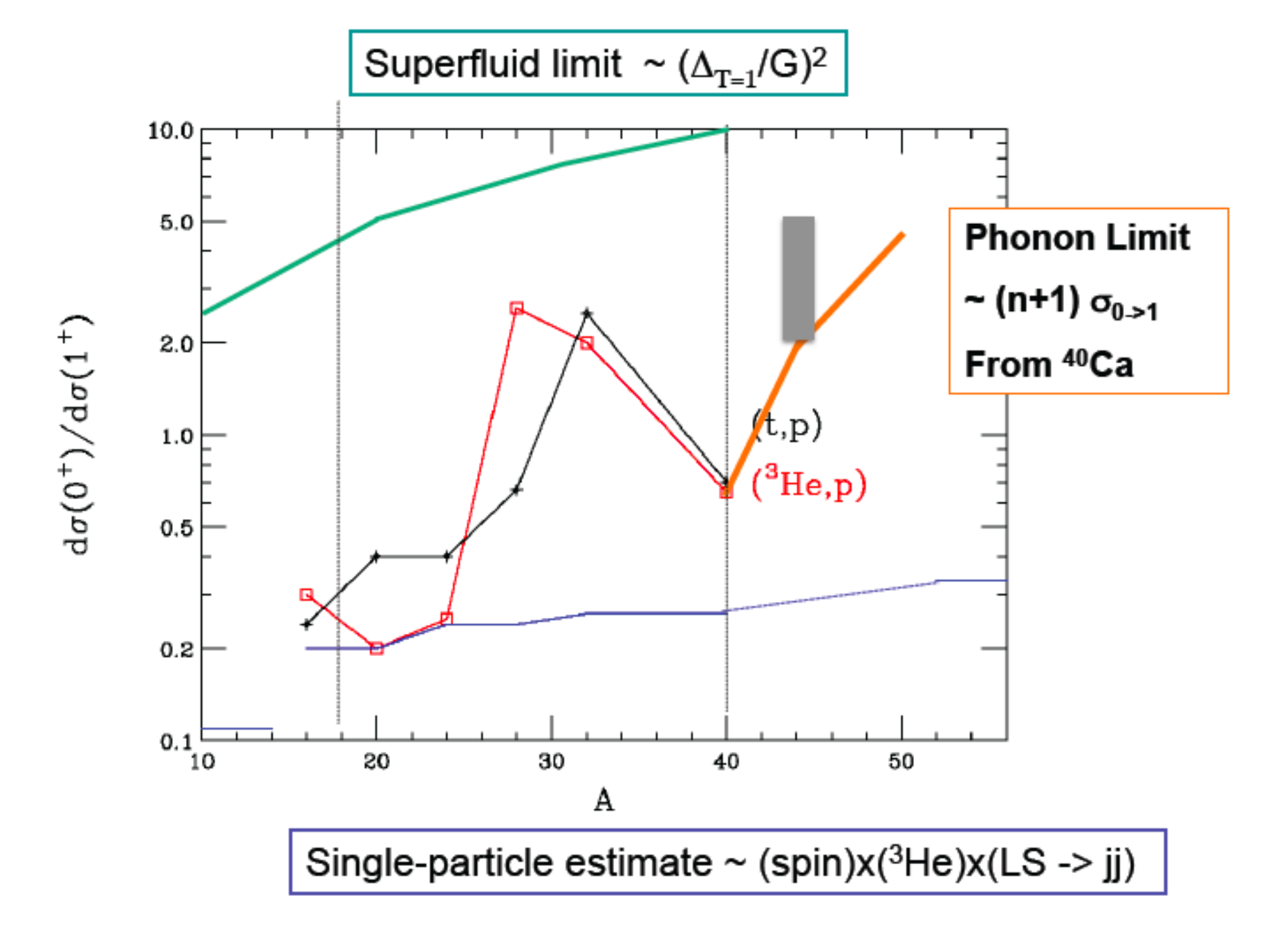,scale=0.7}}
\caption{ Systematics of the experimental  ratio $\sigma(0^+)/\sigma(1^+)$ between the cross sections for transfer
of a $T=1, ~ J=0$ and  a $T=0,~J=1$ $np$ pair for nuclei up to $^{40}$Ca.   {For comparison, the ratios derived from the
$(t,p)$ reactions are also shown, as they are related to each other due to isospin invariance.} 
The preliminary result for the  $^{44}$Ti  experiment is shown as a grey bar.  
Single-particle (blue) and superfluid ($T=1$) (green) limits are indicated as references.  
The orange line above $A=40$ shows the pair vibrational estimate,
which assumes collective isovector phonons and no collective enhancement for the isoscalar transfer amplitude.}
\label{syst}
\end{figure}

As discussed earlier,  we look at the ratio $\sigma(0^+)/\sigma(1^+)$. The systematics for nuclei up to $^{40}$Ca is shown in  Fig. \ref{syst}. The data for the $(^3He,p)$
and $(t,p)$ reactions was derived from ENSDF. 
The trend of the experimental data is shown by the red and black lines and included for reference, a Òsingle-particle estimateÓ  (blue line)
 and the isovector superfluid limit (green line). This is in analogy to a Weisskopf unit for electromagnetic transitions.  
  This estimate takes into account the spin factor of the final state (0 vs. 1),  the probability of finding the $np$ pair
in $^3He$ in a $T=0$ or $T=1$ state, and finally the probability that this pair is in an $L=0$ state in the $j^2$ configuration.
There is some enhancement of the ratio over the single-particle limit, which can be ascribed to the  $T=1$ pair correlations.
The grey bar  is the preliminary result for the $^{44}$Ti. The measured ratio lies close to the value expected within the vibrational scheme (orange line). 
With $^{44}$Ti interpreted as a two isovector- phonon state on the doubly magic $^{ 40}$Ca and the observed ratios in $^{42}$Sc the expected 
enhancement   is  $\sigma(0^+)/\sigma(1^+)_{^{46}V} \approx 3 \sigma(0^+)/\sigma(1^+)_{^{42}Sc} \approx 2$.  Also for reference,  the isovector superfluid limit is shown with the  green line.

%End AOM

An inspection of the available data on stable targets up to $^{40}$Ca recognizes the need for revisiting these earlier measurements. 
It is important that new, high quality data taken with the same equipment and kinematic conditions will become available.  A common 
analysis framework  could then be used  to establish a definite baseline of the effects on stable nuclei before embarking in a major 
program with radioactive beams.   A series of experiments  addressing these issues have recently started at the RCNP facility in Osaka-Japan \cite{jenny}.

\begin{figure}
\centerline{\psfig{file=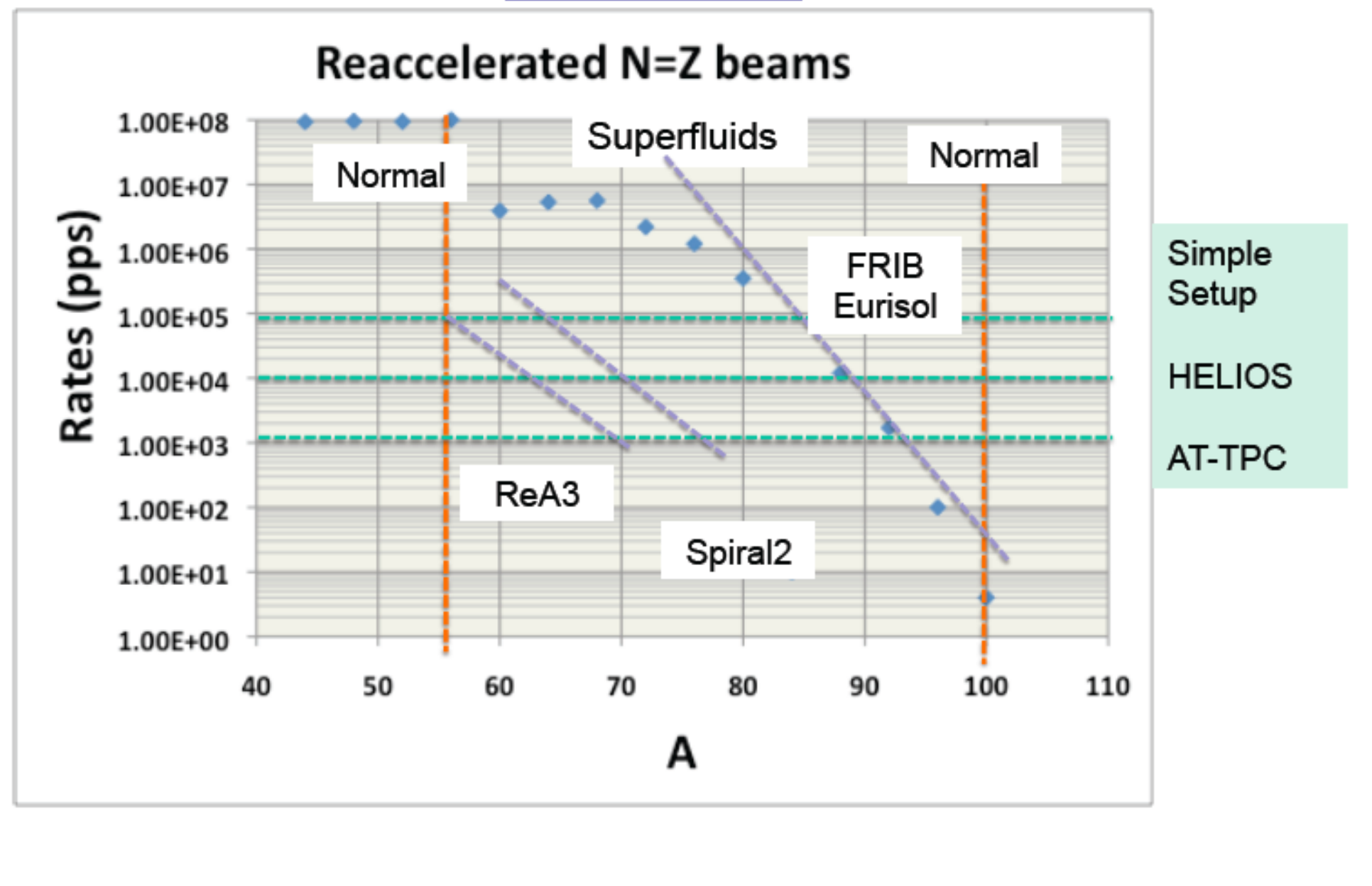,scale=0.7}}
\caption{ Expected rates of $N=Z$ beams at several facilities and the required intensities for these type of studies.  }
\label{rate}
\end{figure}

 Clearly, the study  of two-nucleon transfer reactions in heavier $N=Z$ systems will constitute a unique program to elucidate the role of the collective properties of isoscalar  pairing. 
With available beam intensities, these experiments are currently very challenging but 
studies of a few favorable cases are underway. Definitely, the study of these reactions from $^{40}$Ca to $^{100}$Sn 
will play a prominent part in the scientific portfolio
of the new generation of rare-isotope facilities around the world. 
 It is interesting to anticipate how far we can go along the $N=Z$ based on current expectations of beam intensities that will be available at
 new facilities.  This is illustrated in Fig.  \ref{rate} where the required intensity limits for typical setups are compared.  The region up to A=90 may be experimentally
 accessible. 

\begin{figure}
\centerline{\psfig{file=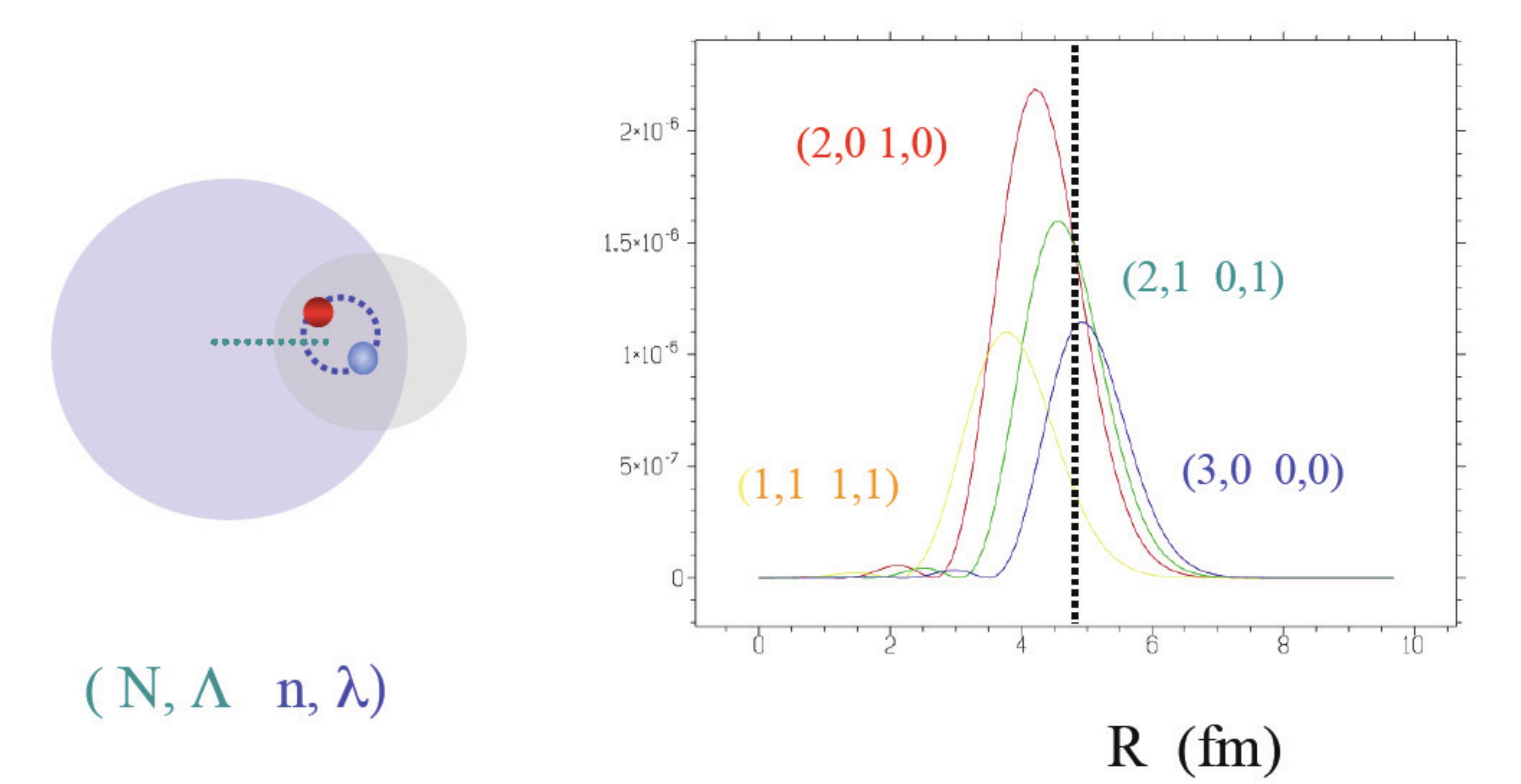,scale=0.4}}
\caption{ Schematic Form-factor for the knockout of an np pair.  The principal $(N,n)$ and angular momentum $(\Lambda, \lambda)$ numbers 
correspond to the center of mass and
relative motion of the pair. }
\label{formf}
\end{figure}

High-energy beams that will be available at fragmentation facilities in Japan (RIKEN), Europe (FAIR) and the US (FRIB) could further extend the reach
of $np$-pairing studies because of the higher luminosity for the secondary $N=Z$ beams.   It is interesting then to consider the use of  $np$ knock-out reactions 
as an additional tool. The contribution of Simpson and Tostevin in  Ref. \cite{ 50yearsBCS} addressed the role of pair correlations in this type of reactions.
Here, we briefly speculate on the complementary role to that of  two-nucleon transfer.   In such a reaction a spatially correlated  $np$ pair  in the projectile is knocked-out by the target (usually Be or C )
as illustrated on the left side of Fig. \ref{formf}.  To the right, the schematic form-factor for the knock-out of the $np$ pair shows a clear enhancement in the surface, due to  the strong absorption in the interior.    
However,  compared to the case of  $(^3He,p)$ or $(p, ^3He)$ where the pair is required to be in a relative $S$ state,   the knock-out process being less selective, allows for both $S-$ and $P-$wave contributions to the cross-section.
This is  of course relevant to the conclusions of Baroni et al.\cite{simone} discussed in Section \ref{sec:2jIVISD1} regarding the important (repulsive) role of the $P-$waves in modifying the pairing matrix elements in both $T=0$ and $T=1$ channels.

As pointed out earlier, it appears that the isoscalar pairing force  interaction is not strong enough to give rise to a $T=0$ isoscalar condensate. 
Thus,  the collective aspects may reveal themselves in the form of pair vibrational resonances, which will be fragmented among the surrounding quasi-particle states background. 
It would be desirable, for example,  to calculate the two-particle transfer strength within the QRPA approach, to guide the experimental exploration of this
pairing structure with the reactions discussed above.

 \section{Summary,  conclusions, and outlook}\label{sec:conc}
 Pairing in exotic nuclei is a subject of active research in nuclear physics, and one with  implications in
many areas of physics. Of particular interest for our community is to asses the role of  isovector and isoscalar  "Cooper pairs" composed of a neutron and a proton, which  are 
expected to occur in $N\approx Z$ nuclei. They represent a unique opportunity to study the competition of spin singlet and triplet pairing 
in fermion systems. Since  this is  an area of much activity it would have been a daunting task 
to cover all aspects in great detail in this manuscript.  
Rather, we have reviewed the current theoretical knowledge of the problem,  placing special emphasis on those
topics that have a direct connection with experimental data.  We discussed possible signatures of $np$ pairing
and confronted the available experimental data with state of the art theoretical expectations.  

There is   clear evidence for the existence of an
isovector $np$ pair  condensate in $N\approx Z$ with the strength required by the isospin invariance of the strong nuclear force. In $N=Z$ nuclei its strength 
is equal to the  strength of the $nn$ and $pp$ condensates, where the summed strength of the $pp$, $np$, and $nn$ channels
 is about same as in nuclei with $N>>Z$. This conclusion is based on the following qualitative observations that characterize a $T=1$, $J=0$
 $np$ pair condensate:
 \begin{itemize}
\item The binding energies show the characteristic $T(T+1)$ isorotational dependence on $T$, which signals the presence of an  isovector pair condensate 
that rotates  in isospace.  Such rotation implies the existence of a $np$ condensate on equal footing with the $nn$ and $pp$ condensates.
 \item  After subtraction of the isorotational energy, the intrinsic binding energies of lowest $T=1$ states in odd-odd $N=Z$ nuclei are approximately
 equal to the mean values of the ground states of their $N=Z$ even-even  neighbors, i. e. the $T=1$ $np$ pairs smoothly blend into the isovector condensate.   
 \item The matrix elements for transfer of an isovector $pn$ pair between $N=Z$ nuclei show the expected enhancement and the characteristic isorotational
 pattern.
 \item The excitation spectra of $N\approx Z$ nuclei can be interpreted as a combination of isorotational energy and intrinsic isovector quasiparticle energy.
 \item The binding energies of the even-$A$ nuclides around the double magic nuclei
  $^{40}_{20}$Ca$_{20}$ and $_{28}^{56}$Ni$_{28}$ organize into a pattern of soft $T=1$ $J=0$
 pair vibrational excitations, which represents the precursor of the isovector pair condensate appearing further in the open shell.  
 \item The moments of inertia of (spatial) rotational  bands in $N\approx Z$ nuclei show the characteristic reduction by the isovector pair condensate,
 which is quenched with increasing angular momentum. 
 \end{itemize}

These qualitative features of the isovector condensate are quantified by the  
  success of various approaches  (Phenomenological,  Cranked Mean Field,  Mean Field combined
with Random Phase Approximation, Projected Shell Model, Excited Vampir), which incorporate our knowledge    
about the like-particle isovector pair correlations in  $N>>Z$ nuclei. The analysis of the large-scale Shell Model calculations
lead to the same conclusions.  

The concept of spontaneous breaking of isospin symmetry by the isovector pair condensate is of central importance.
 It is reflected by the appearance of rotational motion in isospace and the possibility to describe the structure of $N\approx Z$ nuclei 
 in terms of isorotational excitations based on the intrinsic configurations of the quasiparticles that belong to the isovector pair field.
This considerably simplifies theoretical approaches. As a first approximation, one can stay on the level of the mean field. 
The isorotational energy is automatically included in the Cranked Hartree-Fock-Bogoliubov approximation, because the
standard particle number constrains realize cranking in isospace. The quantum corrections due to isospin conservation
can be taken into account in a qualitative way by  replacing the constraint $(N-Z)/2$ on the expectation value of the isospin
projection $T_z$  by $\sqrt{T(T+1)}$.  Within this frame, the rotational spectra of $N\approx Z$ nuclei have been described 
with the same accuracy as for $N>>Z$ nuclei.    Clearly there is need for improvement concerning the restoration of isospin.
The formalism for isospin projection has been worked out and tested against schematic models for which exact solutions are known.
So far only the approximate restoration of isospin in the framework of the Random Phase Approximation has been applied to realistic nuclei  
  
The $T(T+1)$  dependence of the symmetry energy on $T=M_T=\vert N-Z\vert/2$, which is a key element for the success 
of the various isovector only scenarios,  appears as a natural consequence of the existence of 
isorotational bands. This form of the symmetry energy approximates  very well the experimental binding energies, and accounts for the 
extra binding of the $N=Z$ nuclei (Wigner energy). 
Combining the symmetry energy with  isovector pairing explain the energy difference 
between the lowest $T=1$ and $T=0$ states in odd-odd $N=Z$ nuclei without the need of introducing additional correlations and 
adjustable parameters. Fluctuations of the binding energies around the average $T(T+1)$ form caused by the shell structure have been
fairly well reproduced in the framework of the Random Phase Approximation and a few-level configuration mixing approach, which
neglect isoscalar pair correlations.   

Though being rather successful, these  approaches do not derive  the   $T(T+1)$ form for the symmetry energy 
fully microscopically.  Part of it  comes from the isovector field,  but part has to be
phenomenologically  introduced. The origin of the latter requires further theoretical studies. 
The breaking of isospin invariance by the nuclear potential is a promising avenue to be explored.
Thereby the question  should be addressed to what extend the genuine  breaking of isospin symmetry 
 by the Coulomb interaction will modify the results obtained under the assumption of exact conservation of isospin symmetry. 

So far,  by contrasting experimental data with theory, we have not found
a "smoking gun"  for the presence of strong collectivity in the $T=0$ channel.  On the contrary,
the following evidence  speaks against the existence of an $T=0$ $J=1$ "deuteron-like" pair condensate:
\begin{itemize}
\item The energy of lowest $T=0$ state in odd-odd $N=Z$ nuclei is systematically larger than the average of the 
energies of their $N=Z$ even-even neighbors, i. e.  the added $T=0$ $pn$ pair cannot blend into a $T=0$ condensate,
because it does not exist.  The energy staggering is consistent with interpreting the lowest $T=0$ states as
two-quasiparticle excitations of the isovector pair vacuum.
\item The lowest $T=0$ state in most odd-odd $N=Z$ nuclei is "spin-aligned", i. e. its angular momentum is given by the sum
of the ground-state spins of the odd-$A$ neighbors, which is consistent with interpreting them as  two-quasiparticle configurations.
\item The "deuteron-like" $1^+$ states of odd-odd $N=Z$ nuclei are in most cases found at an excitation energy of 0.5 - 1 MeV.
Their magnetic moments and  transition probabilities for $M1$ $\gamma$-decay are well accounted for by interpreting them as 
two-quasiparticle excitations of the isovector pair vacuum.    
\item There is no evidence for a soft $T=0$ pair vibration in the nuclides  around  the double magic nuclei $^{40}_{20}$Ca$_{20}$ and $_{28}^{56}$Ni$_{28}$,
which were  expected as  precursors of an isoscalar condensate further out in the open shell.
\item The phenomenon of "gapless superfluidity", which characterizes the isoscalar condensate of spherical nuclei, is not observed.
All even-even $N=Z$ nuclei have a gap of the order 24 MeV$A^{-1/2}$ before the two-quasiparticle excitations start.
\item The strong enhancement of the Gamov-Teller $\beta$-decay of $N=Z$ nuclei,  predicted for a condensate of $T=0$ bosons, is not observed.

\end{itemize}

Although the nuclear force is stronger in the $T=0$ than in the $T=1$ channel it seems not give rise   to a "deuteron-like condensate"
or collective pair modes.   The spin-orbit splitting appears as the main culprit that conspires against the isoscalar force  and hinders
 its ability to generate collective pair motion. It becomes more effective with increasing mass number. 
Isoscalar pair correlations in the mass region $30<A<100$ may happen in the form of a pairing phonon at the best. 
For lighter nuclei the concepts of "pair condensate" and "pair vibrations" become questionable.

Mean field calculations using  a realistic ratio  of the coupling constants of  the $T=0$ and $T=1$ interactions demonstrate 
in a quantitate way that with  increasing mass the isovector solutions are favored     
as a consequence of the spin-orbit splitting. Surprisingly,  for $A>200$ the isoscalar solutions become favored, because 
the impact of the spin-orbit potential, which is located in the surface,  decreases.  An island of 
 isoscalar ground state solutions bordered by a shore of solutions with coexisting isoscalar and isovector pair 
fields is predicted in the region around $N=Z=65$.  As a possible experimental probe of this region, 
detailed spectroscopy following heavy-ion fusion  evaporation reactions with proton tagging could be considered. 
The nature of the low-lying excitation modes of these isospin mixed
condensates and whether they may decay  by the emission of a deuteron are interesting questions that theory 
should address.

At  present, the limitations of the available experimental data, in particular of good pair transfer cross sections, 
relegates the discussion of the strength of the isoscalar pair correlations to the realm of theory.    In contrast to  isovector pairing, where the 
strength of the $np$ correlations is tied to the strength of the like-particle correlations by isospin invariance, 
the strength of the isoscalar correlations depends on the $T=0$ effective interaction within the considered valence space,
which is an element of uncertainty. In our view, attempts to estimate it by adding a tunable $T=0$ interaction to models that take only
 isovector pairing into account and adjusting the coupling constant by fitting the experimental binding energies 
 have not  led to much new insight, because either only an upper limit for the coupling constant could be derived or 
 the models missed contributions from isovector pairing that are of the order of the isoscalar ones.

The fact that modern Shell Model codes are capable of calculating 
energies  and other properties of many of the $N\approx Z$ nuclei up to $A=100$ makes such calculations 
 an additional source of information,  complementary to experiment,  about nuclear pair correlations:
 a simulation of experiment on the computer. 
 Such "experiments" offer the opportunity to induce a phase transition from the unpaired to the paired regime
 by means of scaling the relevant matrix elements, i.e. to have a control parameter at disposal that is missing in experiments with real nuclei.
 The Invariant Correlation Entropy can be used as a signal for the cross-over between the phases.  
 Analyzing  the  
large-scale Shell Model calculations for real nuclei in this way and  by means of pair-counting operators indicated the following.
The isovector pair correlations dominate in the ground state. The relative strengths of the  isovector $np$, $nn$, and $pp$
correlations  are close to the isorotational ratios (CGC of the group SO(3) , even closer are 
the CGC of the group SO(5), which account for some modification of the condensate with $T$). They are rapidly
quenched by spatial rotation. Coexisiting, there are  "deuteron-like"  pair correlations, which weakly change with angular momentum.
In the ground state they are substantially weaker than the isovector correlations but 
eventually prevail when the latter are diminished at higher angular momentum.    

We have discussed the emergence of the isovector condensate and the ensuing consequences for energies and
pair transfer amplitudes in framework of simple models. The grand picture derived  has only partially confirmed by the 
existing analyses of the results of large-scale Shell Model calculations.
Thereby it seems important to come up with appropriate operators to be evaluated, which is yet an open field.
We consider the calculation of pair transfer amplitudes as essential. 
Another promising avenue could be 
carrying out HFB and QRPA  calculations using the same  the effective interaction as in the 
 Shell Model  calculations. In the case of the schematic models,  comparing the approximate and exact solutions 
allowed one to identify the pair vibrational regime, the region with the condensate present, and the cross-over  interval between them. 
Analyzing the large-scale  Shell Model results in the same way seems to be promising.

Mean field calculations using  a realistic microscopically  interactions
 found  competing solutions with exclusively $T=0$ and $T=1$ pair fields. Occasionally solutions
with coexisting pair fields of the two types were obtained as well.
The existence of distinct mean field  solutions must be interpreted
with care. There are substantial fluctuations of the pair fields, which become more and more important the 
lighter the nuclei are.  The existence of competing isovector and isoscalar mean field solutions should be
taken as  evidence that true wave function is a mixture of components with the character of these
solutions. These mean field results could be used for characterizing  the  pair correlations in large-scale  Shell Model calculations
along the mentioned lines. In the case of the lightest nuclei ($A<30$) beyond-mean field approaches are required
for a more quantitative understanding. The Generator Coordinate Method
starting from the competing mean field solutions  appears suitable.

Large-scale Shell Model calculations  based on an effective interaction that is well tuned to the experimental spectra are 
expected to provide the most  realistic estimate of the strength of the isoscalar pair correlations. 
However, there is no guarantee that this indeed the case. For example, the KB3 and the EPQQ interactions describe  
the low-lying states of $N\approx Z$ nuclei with comparable accuracy. The former induces the mentioned moderate isoscalar pair
correlations while the latter does not generate such correlations at all.
Additional criteria should be invoked that place the effective interaction on a more microscopic foundation.

The nature of correlations induced by the $T=0$ interaction needs further clarification.
For example, the correlation between the enhanced binding energy  of $N=Z$ nuclei and strength of
 the $T=0$ interaction found in  Shell Model calculations  does not necessarily signal the appearance of isoscalar
 pair correlations. Rather it turned out to be an indirect consequence of the reduction of the symmetry energy coefficient.
  Another example is the appearance of 
 a spin-aligned coupling scheme in $N= Z$ nuclei located at the bottom or top of a high-$j$ subshell. 
 It seems clear that the scheme is induced by the strong preference the aligned orientation of the
 spins of a $np$ pair by the $T=0$ interaction. However it is not clear whether and which 
 type of isoscalar correlations are involved. The  analysis of angular momentum dependence of
 the yrast energies and of  the Shell Model wave functions showed that spin-aligned coupling scheme coexists with
 isovector pair correlations, rather then replacing the latter as a kind of "new phase". Measuring  the yrast energies of $N=Z$ nuclei with
 $A>80$  to higher angular momentum and the lifetimes of electromagnetic  transitions between them would  
 be of great interest for better understanding the spin-aligned coupling scheme and its transition to the more conventional 
 rotation of $N=Z$ nuclei with $70<A<80$.

Microscopic studies of the  Gamov-Teller strength in $\beta$- and $\beta\beta$-decay and  charge exchange reactions seem to call for 
   moderate isoscalar pair correlations in order to account for the experimental data. Definitely, they  
do not require the presence of an isoscalar pair condensate. Rather, the comparison of QRPA calculations with data seem to point to the presence 
of dynamical isoscalar $np$ pair correlations, the strength of which is still strongly model dependent (in fact, some successful studies ignore them). 
The major element of uncertainty are the correlations in the particle - hole channel, which have a strong impact on the Gamov- Teller strength function.
 It is clear that more experimental data would be desirable to further constrain the parameters of the theoretical models.  

We have argued about the important role that two-nucleon transfer reactions and $np$ knockout reactions
 may play in the challenging endeavor of quantifying the strength of the isoscalar pair correlations.
  As new instrumentation and radioactive beams along  
the $N=Z$ line become available, it will be possible to study with these reactions the open-shell region between 
A=56 and A=100.  
In addition to the standard spectroscopic quantities as energies and electromagnetic and weak transition probabilities,
  beyond mean field studies should definitely  calculate two-particle transfer amplitudes, which has been seldom carried out in the 
  existing work.  It seems important to calculate the distribution of the transfer strength over excited states, 
  because from our present understanding the most likely scenario is that the isoscalar pair correlations will show up as 
  a pair vibrational phonon that is fragmented over surrounding quasiparticle excitations. Calculations of the transfer strength distribution
  in the framework of the Random Phase Approximation or large-scale Shell Model appear promising.

To conclude, pairing correlations in nuclei are a subject that still captures the imagination of the nuclear structure
community.  The last 20 years have seen considerable progress in our
understanding of the collective correlations of neutrons and protons,  and  we have not reached the end of the story 
as new and better data are coming along.  A  parallel development in both reaction and structure theory
will be needed  to firmly elucidate and further our knowledge of $np$ correlations in general and of the nature of the $T=0$ pair correlations
in particular.

{\bf Acknowledgement:}

This work has been supported by the US Department of Energy under grants DE-FG02-95ER40934 (Notre Dame) and DE-AC02-05CH11231(LBNL).
AOM should like to thank the members of the Nuclear Structure Group at LBNL, in particular Dr. Paul Falllon, for many illuminating discussions on 
several aspects of this review.


\begin{thebibliography}{99}
\itemsep -2pt 


\bibitem{bmp} A. Bohr, B.~R. Mottelson and D. Pines, 
\Journal{\PR} {110}{936}{1958}

\bibitem{BrogliaBrink}  For an in-depth study of
pairing in atomic nuclei,  the reader is referred to the excellent textbook by D.Brink and R. Broglia, Nuclear Superfluidity: Pairing in Finite Systems,  Cambridge University Press, Cambridge, 2005.

\bibitem{50yearsBCS}  {\em 50 years of BCS},  ed. R. A. Broglia and V. V. Zelevinsky, World Science Pub. 2012

\bibitem{DeanMorten} D.J.Dean and M. Hjorth-Jensen,
\Journal{\RMP} {75}{607} {2003}

\bibitem{Molinari75} {A. Molinari, M. B. Johnson, H. Behte,  W. M. Alberico,}
\Journal{\NPA} {239}{45}{1975}
\bibitem{Schiffer76} John P. Schiffer and William W. True,
\Journal{\RMP} {48}{191}{1976}

\bibitem{Dussel70} G. G. Dussel, E. Maqueda, and R. P. J. Perazzo,
\Journal{\NPA} {153}{469}{1970}
\bibitem{Flowers52} B. H. Flowers,
\Journal{\PRS} {212}{248} {1952} 
\bibitem{Evans81}J. A. Evans, G. G. Dussel, E. Maqueda, and R. P. J. Perazzo,
\Journal{\NPA} {367}{77}{1981}

\bibitem{Engel97} J. Engel, S. Pittel, M. Stoitsov, P. Vogel, and J. Dukelsky,
\Journal{\PRC} {55}{1781} {1997} 

\bibitem{AOM} A.O.Macchiavelli, P.Fallon, and Z.Harris, in  "Nuclear Physics on the 21st Century"
AIP Conference Proceedings 610 (2002)  Pag. 783.


\bibitem{Dussel86} G. G. Dussel, E. Maqueda, and R. P. J. Perazzo, J. A. Evans,
\Journal{\NPA} {450}{164}{1986}

\bibitem{Wigner37} E. Wigner,
\Journal{\PR}{51}{974}{1937}

\bibitem{Bes00}D. R. Bes, O. Civitarese, E. E. Maqueda, and N. N. Scoccola,
\Journal{\PRC} {61}{024315} {2000}


\bibitem{simone} Simone Baroni, Augusto O. Macchiavelli, and Achim Schwenk, \emph{Phys. Rev.}  C 81, 064308 (2010) 

\bibitem{Engel96} J. Engel, K. Langanke, P. Vogel,
\Journal{\PLB} {389}{211}{1996}
\bibitem{Engel98} J. Engel, K. Langanke, P. Vogel,
\Journal{\PLB} {429}{215}{1998}
\bibitem{Langanke95} K. Langanke, D. J. Dean, P. B. Radha, Y. Alhassid, and S. E. Koonin, 
\Journal{\PRC} {52}{718}{1995}
\bibitem{Langanke96} K. Langanke, D. J. Dean,  P. B. Radha, S. E. Koonin
\Journal{\NPA} {602}{244}{1996}
\bibitem{Langanke97a} K. Langanke, D. J. Dean, S. E. Koonin, P. B. Radha,
\Journal{\NPA} {613}{253}{1997}
\bibitem{Langanke97b} K. Langanke, P. Vogel, Dao-Chen Zheng,
\Journal{\NPA} {626}{735}{1997}
 \bibitem{Dean97} D.J. Dean, S.E. Koonin, K. Langanke, P.B. Radha,
\Journal{\PLB} {399}{1}{1997}
\bibitem{Poves98} A. Poves, G. Martinez-Pinedo,
\Journal{\PLB} {430}{203}{1998}
\bibitem{Martinez99} G. Martinez-Pinedo, K. Langanke, P. Vogel
\Journal{\NPA} {651}{379}{1999}

\bibitem{Dufor96} Marianne Dufour and Andr\'es P. Zuker,
\Journal{\PRC} {54}{1641}{1996}

\bibitem{Volya03} A. Volya, V. Zelevinsky,
\Journal{\PLB}{574}{27}{2003}
\bibitem{Brown88} B. A. Brown, B. Wildenthal,
\Journal{\ARN}{38}{29}{1988}

 \bibitem{Stoicheva06} G. Stoitcheva, W. Satu\l a, W. Nazarewicz, D. J. Dean, M. Zalewski, and H. Zdu\'nczuk,
\Journal{\PRC} {73}{061304(R)}{2006}
\bibitem{Qi11} C. Qi, J. Blomqvist, T. B\"ack, B. Cederwall, A. Johnson, R. J. Liotta, and R. Wyss, 
\Journal{\PRC} {84}{84, 021301(R)}{2011}
\bibitem{Lisetskiy04}A. F. Lisetskiy, B. A. Brown, M. Horoi \& H. Grawe,
\Journal{\PRC} {70}{044314}{2004}
\bibitem{Coraggio12}L. Coraggio, A. Covello,  A. Gargano, and N. Itaco,
\Journal{\PRC} {85}{034335}{2012}
\bibitem{Zerguine11}S. Zerguine and P. Van Isacker,
\Journal{\PRC} {83}{064314}{2011}
\bibitem{Kaneko99} K. Kaneko, M. Hasegawa, and Jing-ye Zhang,
\Journal{\PRC} {59}{740}{1999}
\bibitem{Hasegawa99} M. Hasegawa and K. Kaneko, 
\Journal{\PRC} {59}{1149}{1999}
\bibitem{Kaneko99b}K. Kaneko and M. Hasegawa,
\Journal{\PRC} {60}{024301}{1999}
\bibitem{Hasegawa00} M. Hasegawa, K. Kaneko , S. Tazaki,
\Journal{\NPA} {674}{411}{2000}A 674 (2000) 
\bibitem{Hasegawa01}M. Hasegawa, K. Kaneko , S. Tazaki,
\Journal{\NPA} {688}{765}{2001}
\bibitem{Kaneko01}K. Kaneko and M. Hasegawa,
\Journal{\PRO} {106}{1179}{2001}
\bibitem{Kaneko02}K. Kaneko, M. Hasegawa, and T. Mizusaki,
\Journal{\PRC} {66}{051306(R)}{2002}
\bibitem{Hasegawa04} M. Hasegawa, K. Kaneko, T. Mizusaki, S. Tazaki,
\Journal{\PRC} {69}{034324}{2004}
\bibitem{Hasegawa04b} M. Hasegawa, K. Kaneko, T. Mizusaki,
\Journal{\PRC} {70}{031301(R)}{2004}
\bibitem{Kaneko04}K. Kaneko, M. Hasegawa, and T. Mizusaki,
\Journal{\PRC} {70}{051301(R)}{2004}
\bibitem{Hasegawa05} M. Hasegawa, K. Kaneko, T. Mizusaki,
\Journal{\PRC} {71}{044301}{2005}
\bibitem{Hasegawa05a} M. Hasegawa,Y. Sun, K. Kaneko, T. Mizusaki,
\Journal{\PLB} {617}{150}{2005}
\bibitem{Hasegawa05b} M. Hasegawa, K. Kaneko, T. Mizusaki,
\Journal{\PRC} {72}{064320}{2005}
\bibitem{Hasegawa05c} M. Hasegawa and K. Kaneko,
\Journal{\NPA} {748}{393}{2005}
\bibitem{Hasegawa07}M. Hasegawa, K. Kaneko, T. Mizusaki, Y. Sun,
\Journal{\PLB} {656}{51}{2007}
\bibitem{Kaneko08}K. Kaneko, Y. Sun, M. Hasegawa, and T. Mizusaki,
\Journal{\PRC} {77}{064304}{2008}
\bibitem{Kaneko10}K. Kaneko, S. Tazaki, T. Mizusaki, Y. Sun, M. Hasegawa, and G. de Angelis,
\Journal{\PRC} {81}{061301(R)}{2010}
\bibitem{Aroua03} S. Aroua, L. Zamick, Y.Y. Sharon, E. Moya de Guerra, M.S. Fayache, P. Sarriguren, A.A. Raduta,
\Journal{\NPA} {728}{96}{2003}
\bibitem{Heyde10} K. Heyde, P. von Neumann-Cosel, and A. Richter, 
\Journal{\RMP} {82}{2365}{2010}

\bibitem{Frauendorf94} S. Frauendorf, J. Sheikh, and N. Rowley,
\Journal{\PRC} {50}{196} {1994}
\bibitem{Frauendorf99a} S. Frauendorf and J. Sheikh,
\Journal{\NPA} {645}{509}{1999}
\bibitem{Frauendorf99b} S. Frauendorf and J. Sheikh,
\Journal{\PRC} {80}{024301} {1999}
\bibitem{Frauendorf00} S. Frauendorf and J. Sheikh,
\Journal{\PST} {88}{162} {2000}
\bibitem{Sheikh00}Javid A. Sheikh and Ramon Wyss,
\Journal{\PRC} {62}{051302(R)}{2000}
\bibitem{Bentley13}I. Bentley and S. Frauendorf, 
\Journal{\PRC} {88}{014322 }{2013}

\bibitem{Camiz65} P. Camiz, A. Covello, and M. Jean,
\Journal{\NC}{36}{663}{1965}
\bibitem{Camiz66} P. Camiz, A. Covello, and M. Jean,
\Journal{\NC}{B XLII}{199}{1966}
\bibitem{Ginocchio68} J. N. Ginocchio and J. Weneser,
\Journal{\PREV} {170}{859} {1968}
\bibitem{Glowacz04} S. G\l owacz, W. Satu\l a and R.A. Wyss,
\Journal{\EPA} {19}{33} {2004}



\bibitem{Frauendorf01} S. Frauendorf,
\Journal{\RMP} {73}{463} {2001}
\bibitem{Satula01} W. Satu\l a and R. Wyss,
\Journal{\PRL} {86}{4488} {2001}


\bibitem{Dussel71} G. G. Dussel, R. P. J. Perazzo, D. B\`es, and R. Broglia ,
\Journal{\NPA} {175}{513}{1971}
\bibitem{Dussel72} G. G. Dussel, R. P. J. Perazzo, and D. B\`es,
\Journal{\NPA} {183}{298}{1972}




\bibitem{Satula01a} W. Satu\l a and R. Wyss,
\Journal{\PRL} {87}{52504} {2001}


\bibitem{Neergard02} K. Neergaard,
\Journal{\PLB} {537}{287}{2002}
\bibitem{Neergard03} K. Neergaard,
\Journal{\PLB} {572}{159}{2003}
\bibitem{Neergard09} K. Neergaard,
\Journal{\PRC} {80}{044313}{2009}


\bibitem{Goswami64} A. Goswami,
\Journal{\NP}{60}{228}{1964}
\bibitem{Goswami65} A. Goswami and L. S. Kisslinger,
\Journal{\PR}{140}{B26}{1965}
\bibitem{Chen67} M. T. Chen and  A. Goswami,
\Journal{\PLB}{24}{256}{1967}
\bibitem{Goodman68} A. L. Goodman, G. L. Struble and  A. Goswami,
\Journal{\PLB}{26}{20}{1968}
\bibitem{Wolter70} H. Wolter, A. Faessler, and  P. Sauer, 
\Journal{\PLB} {31}{516}{1970}
\bibitem{Wolter71} H. Wolter, A. Faessler, and  P. Sauer, 
\Journal{\NPA} {167}{108}{1971}







\bibitem{Goodman79} A. L. Goodman,
\Journal{\ANP}{11}{263}{1979}
\bibitem{Goodman70} A. L. Goodman, G. L. Struble, J. Bar-Touv, and  A. Goswami,
\Journal{\PRC}{2}{380}{1970}


\bibitem{Goodman99} A. L. Goodman, 
\Journal{\PRC} {60}{014311}{1999}

\bibitem{Bertsch10} G. F. Bertsch and Y. Luo, 
\Journal{\PRC} {81}{064320}{2010}
\bibitem{Gezerlis11} A. Gezerlis, G. F. Bertsch, and Y. L. Luo, 
\Journal{\PRL} {106}{252502}{2011}

\bibitem{Goodman98} A. L. Goodman, 
\Journal{\PRC} {58}{R3051}{1998}
\bibitem{Goodman01} A. L. Goodman, 
\Journal{\PRC} {63}{044325}{2001}




%\bibitem{Bertsch10b} G.F. Bertsch, J. Dukelsky, B. Errea, C. Esebbag,
%\Journal{\ANNP} {325}{1340}{2010}
\bibitem{Terasaki98}J. Terasaki, R. Wyss, P.-H. Heenen,
\Journal{\PRB} {437}{1}{1998}



\bibitem{Chen78}Hsi-Tseng Chen, H. M\"uhter and A. Faessler,
\Journal{\NPA} {297}{445}{1978}
\bibitem{Raduta00} A. A. Raduta and E. Moya de Guerra ,
\Journal{\ANNP} {284}{134}{2000}
\bibitem{Raduta01} A. A. Raduta, P. Sarriguren,  A. Faessler and E. Moya de Guerra ,
\Journal{\ANNP} {294}{182}{2001}
\bibitem{Raduta00a} A. A. Raduta, P. Sarriguren,  A. Faessler and E. Moya de Guerra ,
\Journal{\NPA} {675}{503}{2000}
\bibitem{Raduta03} A. A. Raduta, L. Zamick, E. Moya de Guerra,  A. Faessler and P. Sarriguren,
\Journal{\PRC} {68}{044317}{2003}
\bibitem{Raduta12} A. A. Raduta, M. I. Krivoruchenko,  and A. Faessler,
\Journal{\PRC} {68}{054314}{2012}


\bibitem{Petrovici96} A. Petrovici, K.W. Schmid, Amand Faessler ,
\Journal{\NPA} {605}{290}{1996}
\bibitem{Petrovici99} A. Petrovici, K.W. Schmid, Amand Faessler ,
\Journal{\NPA} {647}{197}{1999}
\bibitem{Petrovici00} A. Petrovici, K.W. Schmid, Amand Faessler ,
\Journal{\NPA} {665}{333}{2000}
\bibitem{Petrovici02a} A. Petrovici, K.W. Schmid, Amand Faessler ,
\Journal{\NPA} {708}{190}{2002}
\bibitem{Petrovici02b} A. Petrovici, K.W. Schmid, Amand Faessler ,
\Journal{\NPA} {710}{246}{2002}
\bibitem{Petrovici03} A. Petrovici, K.W. Schmid, Amand Faessler ,
\Journal{\NPA} {728}{396}{2003}
\bibitem{Schmidt87} K.W. Schmid, E Gr\"ummer and A. Faessler, 
\Journal{\ANNP} {180}{1}{1987}
\bibitem{Chasman02} R. R. Chasman,
\Journal{\PLB} {524}{81}{2002}
\bibitem{Chasman03a} R. R. Chasman,
\Journal{\PLB} {553}{204}{2003}
\bibitem{Chasman03b} R. R. Chasman,
\Journal{\PLB} {577}{47}{2003}
\bibitem{Chasman10} R. R. Chasman, P. Van Isaker,
\Journal{\PLB} {685}{55}{2010}
\bibitem{Satula97} W. Satu\l a and R. Wyss,
\Journal{\PLB} {393}{1}{1997}
\bibitem{Satula00} W. Satu\l a and R. Wyss,
\Journal{\NPA} {676}{120}{2000}

 \bibitem{Hara99} K. Hara, Y. Sun, T. Mizusaki,
\Journal{\PRL} {83}{1922}{1999}
\bibitem{Velaquez01} V. Vel\'azquez, J.G. Hirsch, Y. Sun,
\Journal{\NPA} {686}{129}{2001}
\bibitem{Sun01}Y. Sun, J.A. Sheikh,
\Journal{\PRC} {64}{ 031302(R)}{2001}
\bibitem{Palit01} R. Palit, J.A. Sheikh, Y. Sun, H.C. Jain, 
\Journal{\NPA} {686}{141}{2001}
\bibitem{Sun04} Y. Sun,
\Journal{\EPA} {20}{133}{2004}

\bibitem{JAN} J.Janecke, 
\Journal{\NPA} {73}{97}{1965}

\bibitem{Zeldes} N.Zeldes and S.Liran, 
\Journal{\PLB}{62}{12}{1976}

\bibitem{Vogel} P.Vogel, 
\Journal{\NPA}{662}{148}{2000}

\bibitem{aom1} A. O. Macchiavelli, P. Fallon, R. M. Clark, M. Cromaz, M. A. Deleplanque, R. M. Diamond, G. J. Lane, I. Y. Lee,
F. S. Stephens, C. E. Svensson, K. Vetter, and D. Ward, 
\Journal{\PRC} {61}{ 041303(R)}{2000}

\bibitem{BM} A.Bohr and B.R.Mottelson, Nuclear Structure, Benjamin-New York, Vol I , pag. 145.

\bibitem{Duflo} J. Duflo and A. P. Zuker, 
\Journal{\PRC}{52}{R23} {1995}


\bibitem{Myers82} W. D. Myers and W. J. Swiatecki,
{\em Annu. Rev. Nucl. Part. Sci.} 32 (1982) 309
\bibitem{Brenner90} D.S. Brenner, C. Wesselborg, R.F. Casten, D.D. Warner and J.-Y. Zhang,
\Journal{\PLB} {243}{1}{1997}
\bibitem{Satula97b} W. Satu\l a, D.J. Dean, J. Gary, S. Mizutori, W. Nazarewicz,
\Journal{\PLB} {407}{1}{1997}


\bibitem{Bentley14} I. Bentley, K. Neergaard, S. Frauendorf,
\Journal{\PRC}{89}{034302}{2014}, erratum
\Journal{\PRC}{89}{049901(E)}{2014}

\bibitem{ZR02}  L Zamick and J Q Robinson,
\Journal{\PAT}{65}{740}{2002} 
 
\bibitem{Satula00b} W. Satu\l a and R. Wyss,
\Journal{\PRL} {86}{4488}{2000}

%---------

\bibitem{BF79} R. Bengtsson, S. Frauendorf,
\Journal{\NPA} {327}{139}{1979}

\bibitem{deAngelis97} G. de Angelis, C. Fahlander, A. Gadea, E. Farnea, W. Gelletly,
A. Aprahamian, D. Bazzacco, F. Becker, P.G. Bizzeti, A. Bizzeti-Sona,
F. Brandolini, D. de Acufia, M. De Poli, J. Eberth, D. Foltescu, S.M. Lenzi,
S. Lunardi, T. Martinez, D.R. Napoli, P. Pavan, C.M. Petrache,
C. Rossi Alvarez, D. Rudolph, B. Rubio, W. Satu\l a, S. Skoda,
P. Spolaore, H.G. Thomas, C. A. Ur, R. Wyss,
\Journal{\PLB} {415}{217}{1997}

\bibitem{Freund93} S. Freund, j. Altmann, F. Becker, T. Burkardt,  j. Eberth, L. Funke, H. Grawe,
J. Heese, U. Hermkens, H. Kluge, K.H. Maier, T. Mylaeus, H. Prade, S. Skoda,
W. Teichert, H.-G. Thomas, A. v.d. Werth  and G. Winter, 
\Journal{PLB} {303}{167}{1993}

\bibitem{Rudolph96} D. Rudolph, C. J. Gross, J. A. Sheikh, D. D. Warner, I. G. Bearden, R. A. Cunningham, D. Foltescu,
W. Gelletly, F. Hannachi, A. Harder, T. D. Johnson, A. Jungclaus, M. K. Kabadiyski, D. Kast, K. P. Lieb,
H. A. Roth, T. Shizuma, J. Simpson, \"O. Skeppstedt, B. J. Varley, and M. Weiszflog,
\Journal{\PRL} {76}{376}{1996}

\bibitem{Satula94} W. Satu\l a and R. Wyss,
  \Journal{\PRC} {50}{2888}{1994}

\bibitem{Rudolph97} D. Rudolph,  C. Baktash,  C. J. Gross,  W. Satu\l a,  R. Wyss,  I. Birriel,  M. Devlin,  H.-Q. Jin,  D. R. LaFosse, 
F. Lerma,  J. X. Saladin,  D. G. Sarantites,  G. N. Sylvan,  S. L. Tabor,  D. F. Winchell, V. Q. Wood,
and C. H. Yu,
 \Journal{\PRC} {56}{R591}{1997}
 \bibitem{Gross97} C. J. Gross, D. Rudolph, W. Satu\l a, J. Alexander, C. Baktash, P. J. Coleman-Smith, D. M. Cullen,
R. A. Cunningham J. D. Garrett, W. Gelletly, A. Harder, M. K. Kabadiyski, I. Lazarus, K. P. Lieb, H. A. Roth,
D. G. Sarantites, J. A. Sheikh, J. Simpson, \"O  . Skeppstedt, B. J. Varley, and D. D. Warner,
 \Journal{\PRC} {56}{98}{1997}
 
 \bibitem{Fisher01} S. M. Fischer, C. J. Lister, D. P. Balamuth, R. Bauer, J. A. Becker, L. A. Bernstein, M. P. Carpenter, J. Durell,
N. Fotiades, S. J. Freeman, P. E. Garrett, P. A. Hausladen, R.V. F. Janssens, D. Jenkins, M. Leddy, J. Ressler,
J. Schwartz, D. Svelnys, D. G. Sarantites, D. Seweryniak, B. J. Varley, and R. Wyss, 
\Journal{\PRL} {87}{132501}{2001}
\bibitem{Marginean02} N. M\u{a}rginean, D. Bucurescu, C. Rossi Alvarez, C. A. Ur, Y. Sun, D. Bazzacco, S. Lunardi, G. de Angelis,
M. Axiotis, E. Farnea, A. Gadea, M. Ionescu-Bujor, A. Iord\u{a}chescu, W. Krolas, Th. Kr\"oll, S. M. Lenzi, T. Martinez,
R. Menegazzo, D. R. Napoli, P. Pavan, Zs. Podolyak, M. De Poli, B. Quintana, and P. Spolaore
\Journal{\PRC} {65}{051303(R)}{2002}    
 
\bibitem{Kelsall01} N. S. Kelsall, R. Wadsworth, A. N. Wilson, P. Fallon,  A. O. Macchiavelli,  R. M. Clark,  D. G. Sarantites,
D. Seweryniak, C. E. Svensson,2, S. M. Vincent, S. Frauendorf, J. A. Sheikh, and G. C. Ball,
\Journal{\PRC} {64}{024309}{2001}
\bibitem{Kelsall02} N. S. Kelsall, S. M. Fischer, D. P. Balamuth, G. C. Ball, M. P. Carpenter, R. M. Clark, J. Durell, P. Fallon,
S. J. Freeman, P. A. Hausladen, R. V. F. Janssens, D. G. Jenkins, M. J. Leddy, C. J. Lister, A. O. Macchiavelli,
D. G. Sarantites, D. C. Schmidt, D. Seweryniak, C. E. Svensson, B. J. Varley, S. Vincent , R. Wadsworth,
A. N. Wilson, A. V. Afanasjev, S. Frauendorf,. Ragnarsson,
\Journal{\PRC} {65}{044331}{2002}
\bibitem{OLeary03} C. D. O\`~Leary, C. E. Svensson, S. G. Frauendorf, A. V. Afanasjev, D. E. Appelbe, R. A. E. Austin, G. C. Ball,
J. A. Cameron,  R. M. Clark,  M. Cromaz,  P. Fallon,  D. F. Hodgson,  N. S. Kelsall,  A. O. Macchiavelli, 
I. Ragnarsson, D. Sarantites, J. C. Waddington, and R. Wadsworth,
\Journal{\PRC} {67}{021301(R)}{2003}
\bibitem{Kelsall04} N.S. Kelsall, C.E. Svensson, S. Fischer, D.E. Appelbe, R.A.E. Austin, D.P. Balamuth, G.C. Ball,
J.A. Cameron, M.P. Carpenter, R.M Clark , M. Cromaz, M.A. Deleplanque, R.M. Diamond, P. Fallon,
D.F. Hodgson, R.V.F. Janssens, D.G. Jenkins, G.J. Lane, C.J. Lister, A.O. Macchiavelli, C.D. O\`~Leary,
D.G. Sarantities, F.S. Stephens, D.C. Schmidt, D. Seweryniak, K. Vetter, J.C. Waddington, R. Wadsworth,
D. Ward, A.N. Wilson, A.V. Afanasjev, S. Frauendorf, and I. Ragnarsson,
\Journal{\EPA} {20}{131}{2004}
 \bibitem{Davies07}P. J. Davies, A. V. Afanasjev, R. Wadsworth, C. Andreoiu, R. A. E.,  M. P. Carpenter, D. Dashdorj,
S. J. Freeman,  P. E. Garrett, A. G\"orgen, J. Greene, D. G. Jenkins, F. L. Johnston-Theasby, P. Joshi,
A. O. Macchiavelli,  F. Moore, G. Mukherjee, W. Reviol, D. Sarantites, D. Seweryniak, M. B. Smith,
C. E. Svensson, J. J. Valiente-Dob\`on,  and D. Ward,
\Journal{\PRC} {75}{011302(R)}{2007}
\bibitem{Andreoiu07} C. Andreoiu, C. E. Svensson, A. V. Afanasjev, R. A. E. Austin, M. P. Carpenter, D. Dashdorj, P. Finlay,
S. J. Freeman, P. E. Garrett, J. Greene, G. F. Grinyer, A. G\"orgen, B. Hyland, D. Jenkins, F. Johnston-Theasby,
P. Joshi, A. O. Machiavelli, F. Moore, G. Mukherjee, A. A. Phillips, W. Reviol, D. G. Sarantites,
M. A. Schumaker, D. Seweryniak, M. B. Smith, J. J. Valiente-Dobon, and R. Wadsworth,
\Journal{\PRC} {75}{041301(R)}{2007}
\bibitem{Valiente08} J. J. Valiente-Dob\'on, C. E. Svensson, A. V. Afanasjev, I. Ragnarsson, C. Andreoiu, D. E. Appelbe, D. E.  Austin,
G. C. Ball, J. A. Cameron, M. P. Carpenter, R. M. Clark, M. Cromaz, D. Dashdorj, P. Fallon,1 S. J. Freeman,
P. E. Garrett, A. G\"orgen, G. F. Grinyer, D. F. Hodgson, B. Hyland, D. Jenkins,  F. Johnston-Theasby,  P. Joshi,
N. S. Kelsall, A. O. Macchiavelli, D. Mengoni, F. Moore, G. Mukherjee,  A. A. Phillips, W. Reviol, D. Sarantites,
M. A. Schumaker, D. Seweryniak, M. B. Smith, J. C. Waddington, R. Wadsworth, and D. Ward,
\Journal{\PRC} {77}{024312}{2008}
\bibitem{Johnston08} F. Johnston-Theasby, A. V. Afanasjev, C. Andreoiu, R. A. E. Austin, M. P. Carpenter, D. Dashdorj,
S. J. Freeman, P. E. Garrett, J. Greene, A. G\"orgen, D. G. Jenkins, P. Joshi, A. O. Macchiavelli, F. Moore,
G. Mukherjee, W. Reviol, D. Sarantites, D. Seweryniak, M. B. Smith, C. E. Svensson,
J. J. Valiente-Dob\'on, R. Wadsworth, and D. Ward,
\Journal{\PRC} {78}{034312}{2008}
\bibitem{Davies10} P. J. Davies, A. V. Afanasjev, R. Wadsworth, C. Andreoiu, R. A. E. Austin, M. P. Carpenter, D. Dashdorj, P. Finlay,
S. J. Freeman, P. E. Garrett, A. G\"orgen, J. Greene, G. F. Grinyer, B. Hyland, D. G. Jenkins, F. L. Johnston-Theasby,
P. Joshi, A. O. Macchiavelli, F. Moore, G. Mukherjee, A. A. Phillips, W. Reviol, D. Sarantites, M. A. Schumaker,
D. Seweryniak, M. B. Smith, C. E. Svensson, J. J. Valiente-D\'obon,4, and D. Ward,
\Journal{\PRC} {82}{061303(R)}{2010}
\bibitem{Cederwall11}
B. Cederwall, F. Ghazi Moradi, T. B\"ack, A. Johnson, J. Blomqvist, E. Cl\'ement, G. de France, R.Wadsworth, K. Andgren,
K. Lagergren, A. Dijon, G. Jaworski, R. Liotta, C. Qi, B. M. Nyak\'o, J. Nyberg, M. Palacz, H. Al-Azri, A. Algora, G. de
Angelis, A. Atac, S. Bhattacharyya, T. Brock, J. R. Brown, P. Davies, A. Di Nitto, Zs. Dombr\'adi, A. Gadea, J. G\'al,
B. Hadinia, F. Johnston-Theasby, P. Joshi, K. Juh\'asz, R. Julin, A. Jungclaus, G. Kalinka, S. O. Kara, A. Khaplanov,
J. Kownacki, G. La Rana, S. M. Lenzi, J. Moln\'ar, R. Moro, D. R. Napoli, B. S. Nara Singh, A. Persson, F. Recchia,
M. Sandzelius, J.-N. Scheurer, G. Sletten, D. Sohler, P.-A. S\"oderstr\"om, M. J. Taylor, J. Tim\'ar, J. J. Valiente-Dob\'on,
E. Vardaci \& S. Williams,
\Journal{\NAT} {469}{68}{2011}

\bibitem{Afanasjev04} A. V. Afanasjev and S. Frauendorf,
\Journal{\NPA} {746}{757c}{2004}
\bibitem{Afanasjev05} A. V. Afanasjev and S. Frauendorf
\Journal{\PRC} {71}{064318}{2005}
\bibitem{Afanasjev00} A. V. Afanasjev, J. K\"onig and P. Ring,
\Journal{\NPA} {676}{196}{2000}
\bibitem{Afanasjev96} A. V. Afanasjev, J. K\"onig and P. Ring,
\Journal{\NPA} {608}{107}{1996}
\bibitem{Afanasjev99} A. V. Afanasjev, D. B. Fossan, G. Lane and I. Ragnarsson,
\Journal{\PREP} {322}{1}{1999}
\bibitem{Afanasjev07} A. V. Afanasjev,
\Journal{\INT} {16}{275}{2007}

\bibitem{Langanke03} K. Langanke, D.J. Dean, W. Nazarewicz,
\Journal{\NPA} {728}{109}{2003}

\bibitem{Kaneko98} Kazunari Kaneko and Ying-ye Zhang,
\Journal{\PRC} {57}{1732}{1998}

\bibitem{Satula10} W. Satu\l a, J. Dobaczewski, W. Nazarewicz, and M. Rafalski,
\Journal{\PRC} {81}{054310}{2010}
%%%%
\bibitem{Qi10} C. Qi,
\Journal{\PRC} {81}{034318}{2010}
\bibitem{Qi11b}C. Qi, J. Blomqvist, T. B\"ack, B. Cederwall, A. Johnson, R. J. Liotta and R. Wyss,
\Journal{\PST} {150}{014031}{2011}
\bibitem{Danos67} M. Danos and V. Gillet,
\Journal{\PR} {161}{1034}{1967}
\bibitem{Frauendorf93} S. Frauendorf,
\Journal{\NPA}{557}{259c}{1993}
\bibitem{Clark00} R. M. Clark and A. O. Macchiavelli,
\Journal{\ARN} {50}{1}{2000}
\bibitem{privcom} { Private communications from C.Qi and K.Kaneko}.
\bibitem{Zamick05} L. Zamick, A. Escuderos, S. J. Lee, A. Z. Mekjian, E. Moya de Guerra, A. A. Raduta, and P. Sarriguren,
\Journal{\PRC} {71}{034317}{2005}
\bibitem{Neergard13} K. Neergaard,
\Journal{\PRC} {88}{034329}{2013} 
%AOM
\bibitem{Jenkins}   D. G. Jenkins, N. S. Kelsall, C. J. Lister, D. P. Balamuth, M. P. Carpenter, T. A. Sienko, S. M. Fischer, R. M. Clark, P. Fallon, A. Gšrgen, A. O. Macchiavelli, C. E. Svensson, R. Wadsworth, W. Reviol, D. G. Sarantites, G. C. Ball, J. Rikovska Stone, O. Juillet, P. Van Isacker, A. V. Afanasjev, and S. Frauendorf, 
\Journal{\PRC}{65}{064307}{2002}


\bibitem{Lisetzkiy99} A. F. Lisetskiy, R. V. Jolos, N. Pietralla, and P. von Brentano,
\Journal{\PRC} {60}{064310}{1999}


\bibitem{Paddy}A.B. Garnsworthy, P.H. Regan, L. C‡ceres, d, S. Pietri, Y. Sune, D. Rudolph, M. G—rska, Zs. Podoly‡k, S.J. Steer, R. Hoischen, A. Heinz, F. Becker, P. Bednarczyk,  P. Doornenbal, H. Geissel, J. Gerl, H. Grawe, J. Grebosz, A. Kelic, I. Kojouharov, N. Kurz, F. Montes, W. Prokopowicz, T. Saito, H. Schaffner, S. Tachenov, E. Werner-Malento, H.J. Wollersheim, G. Benzoni, B.B. Blank, C. Brandau,  A.M. Bruce, F. Camera, W.N. Catford, I.J. Cullen, Zs. Dombr‡di, E. Estevez, W. Gelletly, G. Ilie, J. Jolie, G.A. Jones, A. Jungclaus, M. Kmiecik, F.G. Kondev, T. Kurtukian-Nieto, S. Lalkovski, Z. Liu, A. Maj, S. Myalski, M. PfŸtzner,
S. Schwertel, T. Shizuma, A.J. Simons,  P.M. Walker, O. Wieland, and  F.R. Xu,   
\Journal{\PLB}{660}{326}{20008}


\bibitem{Ronen98} Y. Ronen and S. Shlomo,
\Journal{\PRC} {58}{884}{1998}

\bibitem{Vogel86} P. Vogel and M. R. Zirnbauer,
\Journal{\PRL} {57}{3148}{1986}

\bibitem{Engel88} J. Engel and P. Vogel, M. R. Zirnbauer,
\Journal{\PRC} {37}{731}{1988}

\bibitem{Raduta01a} A. A. Raduta, O. Haugh,  F. \v{S}imkovic, and A. Faessler,
\Journal{\JPG }{27}{2429}{2001}
\bibitem{Simkovic08}  \v{S}imkovic, A. Faessler, V. Rodin, P. Vogel, J. Engel, 
\Journal{\PRC }{77}{045503}{2008}
\bibitem{Homma96} H. Homma, E. Bender,  M. Hirsch,  K. Muto, H. V. Klapdor-Kleingrothaus,  and T. Oda,
\Journal{\PRC }{54}{2972}{1996}
\bibitem{Moller97} P. M\"oller, J. L. Nix and K. -L. Kratz,
\Journal{\ANUDT} {66}{131} {1997}
\bibitem{Sarriguren01} P. Sarriguren, E.Moya de Guerra, A. Escuderos,
\Journal{\NPA }{691}{631}{2001}
\bibitem{Sarriguren09} P. Sarriguren, 
\Journal{\PRC}{79}{044315}{2009}
\bibitem{Engel99} J. Engel, M. Bender, J. Dobaczewski, W. Nazarewicz, and R. Surman,
\Journal{\PRC} {60}{014302} {1999}
\bibitem{Kaneko01b} K. Kaneko and M. Hasegawa,
\Journal{\PRO} {105}{219} {2001}


\bibitem{FujitaCa}Y. Fujita, H. Fujita, T. Adachi, C. L. Bai, A. Algora, G. P. A. Berg, P. von Brentano, G. Col˜, M. Csatl—s,
J. M. Deaven, E. Estevez-Aguado, C. Fransen, D. De Frenne, K. Fujita, E. Ganio?lu,  C. J. Guess, J. Guly‡s,
K. Hatanaka, K. Hirota, M. Honma, D. Ishikawa, E. Jacobs, A. Krasznahorkay, H. Matsubara,
K. Matsuyanagi, R. Meharchand, F. Molina,K. Muto, K. Nakanishi, A. Negret, H. Okamura, H. J. Ong,
T. Otsuka, N. Pietralla, G. Perdikakis, L. Popescu, B. Rubio, H. Sagawa, P. Sarriguren, C. Scholl,
Y. Shimbara, Y. Shimizu, G. Susoy, T. Suzuki, Y. Tameshige, A. Tamii, J. H. Thies, M. Uchida, T.Wakasa,
M. Yosoi, R. G. T. Zegers, K. O. Zell,and J. Zenihiro,
\Journal{\PRL}{112}{112502}{2014}


\bibitem{gadea}  E. Grodner, A. Gadea, P. Sarriguren, S.M. Lenzi, J. Grebosz, J.J. Valiente-Dob\'on,
A. Algora, M. G\'orska, P.H. Regan, D. Rudolph, G. de Angelis, J. Agramunt, N. Alkhomashi,
L. Amon Susam, D. Bazzacco, J. Benlliure, G. Benzoni, P. Boutachkov, A. Bracco,
L. Caceres, R. B. \c{C}akirli, F. Crespi, C. Domingo-Pardo, M. Doncel, Zs. Dombradi,
P. Doornenbal, E. Farnea, E. Gani\v{o}glu, W. Gelletly, J. Gerl, A. Gottardo, T. H\"uy\"uk3, N. Kurz,
S. Leoni, D. Mengoni, F. Molina, A.I. Morales, R. Orlandi, Y. Oktem, R.D. Page,
D. Perez, S. Pietri, Zs. Podoly\'ak, A. Poves, B. Quintana, S. Rinta-Antila, B. Rubio,
B.S. Nara Singh, A.N. Steer, S. Verma, R. Wadsworth, O. Wieland, and H.J. Wollersheim, Submitted to  PRL
 \bibitem{bai}C.L. Bai, H. Sagawa, M. Sasano , T. Uesaka, K. Hagino , H.Q. Zhang, X.Z. Zhang, F.R. Xu,
 \Journal{\PLB}{719}{116}{2013} 
\bibitem{franco}F.Iachello, Proc. Int. Conf. on Shell Model and Nuclear Structure, A. Covello, ed., World Scientific, Singapore (1989), p. 407-419.
\bibitem{bruce} P. Halse, B.R. Barrett, Annals of Physics 192, 204-212.

\bibitem{sasano} M. Sasano, G. Perdikakis, R. G. T. Zegers, Sam M. Austin, D. Bazin, B. A. Brown, C. Caesar, A. L. Cole, J. M. Deaven, N. Ferrante, C. J. Guess, G. W. Hitt, R. Meharchand, F. Montes, J. Palardy, A. Prinke, L. A. Riley, H. Sakai, M. Scott, A. Stolz, L. Valdez, and K. Yako, 
\Journal{\PRL} {107}{202501} {2011}


\bibitem{Zhi13} Q. Zhi, E. Caurier, J. J. Cuenca-Garcia, K. Langanke, G. Martinez-Pinedo, and K. Sieja,
\Journal{\PRC} {87}{025803} {2013}
\bibitem{Caurier05} E. Caurier, G. Martinez-Pinedo, F. Nowacki,  A. Poves, A. P. Zuker,
\Journal{\RMP} {77}{427} {2005}
\bibitem{Petermann07} I. Petermann, G. Martinez-Pinedo, K. Langanke,  and E. Caurier,
\Journal{\EPA} {34}{319} {2007}
\bibitem{Petrovici11} A. Petrovici, K.W. Schmid, A. Faessler,
\Journal{\PPNP}{66}{2011}{287}
\bibitem{bes1} D.R.B\`es and R.A.Broglia,
\Journal{\NPA} {80}{289} {1966}
\bibitem{dussel} G.Dussel, E.Maqueda and D.R.B\`es, 
\Journal{\NPA} {153}{469} {1970}
\bibitem{bohr} A.Bohr, Proceedings of Dubna Symposium on Nuclear Structure, IAEA (1969), pag. 179.

\bibitem{nathan}O.Nathan,  Proceedings of Dubna Symposium on Nuclear Structure, IAEA (1969), pag. 191.

\bibitem{bes2}D.R.B\`es, R. A. Broglia, O. Hansen, O. Nathan, 
\Journal{\PREP} {34C}{1}{1977}, and references therein.

\bibitem{aom2} A.O. Macchiavelli, P. Fallon, R.M. Clark, M. Cromaz, M.A. Deleplanque,
R.M. Diamond, G.J. Lane, I.Y. Lee, F.S. Stephens, C.E. Svensson, K. Vetter, and
D. Ward, 
\Journal{\PLB}{480}{1}{2000}

\bibitem{bm1} A.Bohr and B.R.Mottelson, Nuclear Structure, 
Benjamin-New York, (1969 and 1975), Vol II, pag. 645.

\bibitem{BesGT} D.R.B\`es and O.Civitarese,
\Journal{\PRC} {78}{014317} {2008}

\bibitem{Yoshida14} K. Yoshida, arXiv:1308424

\bibitem{Broglia} R.Broglia, O.Hansen and C.Riedel,  
\Journal{\ANP}{6}{287}{1973}, and references therein.

\bibitem{Bayman} B. F. Bayman and N. M. Hintz, 
\Journal{\PR}{172}{1113}{1968}

\bibitem{frob} P. Fr\"obrich, 
\Journal{\PLB }{B37}{338}{1971}

\bibitem{piet} P.Van Isacker, D.D.Warner and A. Frank,  
\Journal{\PLB}{94}{162502}{2005}

\bibitem{aom4} A.O.Macchiavelli, E.Rehm, A.Gšrgen, P.Fallon, M.Cromaz, C.N.Davis, A.Heinz, 
C.L.Jiang, E.F.Moore, G.Mukherjee, R.Pardo, D.Seweryniak, J.P.Schiffer, J.Cizewski, J.Thomas, and  M.Paul, 
 ANL Physics Division Annual Report 2002 ANL-03/23, 21, and to be published.

\bibitem{didier}  D. Beaumel,  FUSTIPEN Topical meeting: "Neutron-proton pair correlations in N~Z nuclei", February 3, 2011, Caen, France

\bibitem{jenkins2} D. Jenkins,  private communication. 

\bibitem{ernst} K.E.Rehm, F. Borasi, C. L. Jiang, D. Ackermann, I. Ahmad, B. A. Brown, F. Brumwell, C. N. Davids, P. Decrock, S. M. Fischer, J. Gšrres, J. Greene, G. Hackmann, B. Harss, D. Henderson, W. Henning, R. V. F. Janssens, G. McMichael, V. Nanal, D. Nisius, J. Nolen, R. C. Pardo, M. Paul, P. Reiter, J. P. Schiffer, D. Seweryniak, R. E. Segel, M. Wiescher, and A. H. Wuosmaa , 
\Journal{\PRL}{80}{259}{1996}

\bibitem{chen}  K.E.Rehm, F Borasi, C.L Jiang, D Ackermann, I Ahmad, F Brumwell, C.N Davids, P Decrock, S.M Fischer, J Gšrres, J.P Greene, G Hackmann, B Harss, D Henderson, W Henning, R.V.F Janssens, G McMichael, V Nanal, D Nisius, J Nolen, R.C Pardo, M Paul, P Reiter, J.P Schiffer, D Seweryniak, R.E Segel, I Wiedenhšver, M Wiescher, A.H Wuosmaa, 
\Journal{\NIM}{449}{208}{2000}

\bibitem{jenny} Jenny Lee, N. Aoi, A. O. Macchiavelli, P. Fallon, D. Beaumel, I. J. Thompson, A. Tamii, K. Hatanaka, Y. Fujita, H. J. Ong, T. Suzuki, Y. Yasuda, J. Zenihiro, H. Fujita, H. Matsubara, T. Kawabata, N. Yokota E. Ganioglu, G. Susoy,  RCNP Experiment E365 (2011).

%
\end{thebibliography}
\end{document}